\documentclass[a4paper,11pt]{article}

\usepackage{graphicx}
\usepackage{tocloft}
\usepackage{amsmath, amssymb, amsfonts}
\usepackage{bbm}
\usepackage[top=20mm, bottom=20mm, left=20mm, right=20mm]{geometry}
\usepackage[bookmarks=false,pdfstartview=FitH,colorlinks=false]{hyperref}
\usepackage{fancyhdr}
\usepackage{tikz}
\usepackage{verbatim}
\usepackage{titlesec}
\usepackage{makecell}

\usetikzlibrary{snakes}
\usetikzlibrary{shapes.geometric}
\usetikzlibrary{shapes.misc}

\def\({\left(}
\def\){\right)}
\def\[{\left[}
\def\]{\right]}

\newcommand{\beq}{\begin{equation}}
\newcommand{\eeq}{\end{equation}}
\newcommand{\beqq}{\begin{equation*}}
\newcommand{\eeqq}{\end{equation*}}
\newcommand\beqa{\begin{eqnarray}}
\newcommand\eeqa{\end{eqnarray}}
\newcommand\bea{\begin{array}}
\newcommand\eea{\end{array}}

\DeclareMathOperator{\str}{STr}

\newcommand{\ket}[1]{|#1\rangle}

\newcommand{\tr}{\mathrm{Tr}}
\newcommand{\N}{\mathcal{N}}
\newcommand{\ord}[1]{\mathcal{O}\left(#1\right)}
\newcommand{\adsfive}{AdS_5 \times S^5}
\newcommand{\adscp}{AdS_{4}\times \mathbb{CP}^{3}}
\newcommand{\pmu}{{\bf P}\mu}

\newcommand{\mc}{\mathcal }
\newcommand{\nn}{\nonumber}

\def\bP{{\bf P}}
\def\bQ{{\bf Q}}

\newcommand{\sh}{{\rm sh}}
\newcommand{\ch}{{\rm ch}}

\newcommand{\M}{{\cal M}}

\newcommand{\pd}{\partial}

\newcommand{\eq}[1]{(\ref{#1})}
\newcommand{\alg}[2]{\mathfrak{#1}(#2)}

\newcommand{\Bigsbrk}[1]{\Bigl[#1\Bigr]}

\newcommand{\be}{\begin{equation}}
\newcommand{\ee}{\end{equation}}
\newcommand{\ba}{\begin{eqnarray}}
\newcommand{\ea}{\end{eqnarray}}

\newcommand{\ads}{$AdS_5\times S^5$\ }

\newcommand{\maldafive}{${\rm AdS}_{5}/{\rm CFT}_{4}$\ }

\def\d{\partial}
\newcommand{\ofrac}[1]{\frac{1}{#1}}

\makeatletter

\newenvironment{chapquote}[2][0em]
  {\setlength{\@tempdima}{#1}%
   \def\chapquote@author{#2}%
   \parshape 1 \@tempdima \dimexpr\textwidth-2\@tempdima\relax%
   \hfill\itshape}
  {\par\normalfont\hfill--\ \chapquote@author\hspace*{\@tempdima}\par\bigskip}
\makeatother

\hypersetup{pdfborder = {0 0 0}}

\numberwithin{equation}{section}

\titlespacing*{\section}{0pt}{5.5ex plus 1ex minus .2ex}{4.3ex plus .2ex}
\titlespacing*{\subsection}{0pt}{5.5ex plus 1ex minus .2ex}{4.3ex plus .2ex}

\begin{document}

\begin{titlepage}

\thispagestyle{empty}

\begin{center}

\begin{figure}[t]
	\centering
	\includegraphics[width=0.25\textwidth]{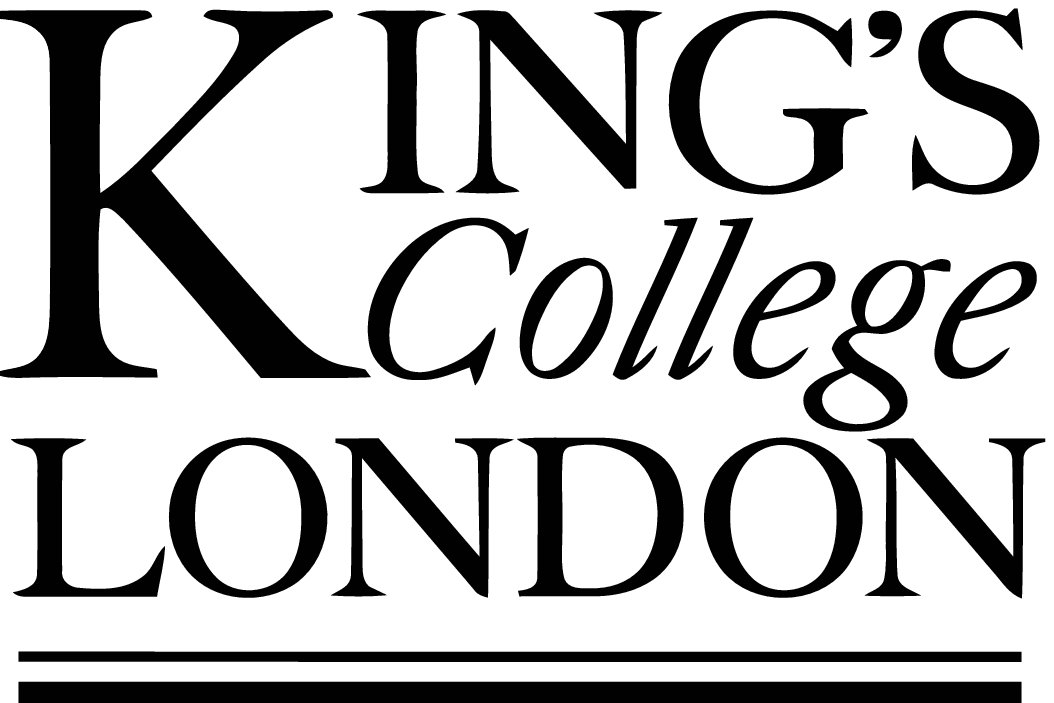}
\end{figure}

\vspace*{50pt}

\LARGE
\textbf{Exact Results in Supersymmetric Gauge Theories}

\vspace{40pt}

\Large
Saulius Valatka

\vspace{30pt}

\small
\it{Department of Mathematics, King's College London, \\
Strand, London WC2R 2LS, U.K.}

\vspace{230pt}

\Large\rm
Thesis supervisor Dr. Nikolay Gromov

\vspace{30pt}

\small
Thesis submitted in partial fulfilment of the requirements \\
of the Degree of Doctor of Philosophy

\vspace{40pt}

\large
September 2014

\end{center}

\end{titlepage}

\newgeometry{top=30mm, bottom=25mm, left=40mm, right=25mm}
\pagestyle{fancy}

\section*{Abstract}

\vspace{30pt}

In this thesis we discuss supersymmetric gauge theories, focusing on exact results achieved using methods of integrability. 
For the larger portion of the thesis we study the $\N=4$ super Yang-Mills theory in the planar limit, a recurring topic being the Konishi anomalous dimension, which is roughly the analogue for the mass of the proton in quantum chromodynamics.
The $\N=4$ supersymmetric Yang-Mills theory is known to be integrable in the planar limit, which opens up a wealth of techniques one can employ in order to find results in this limit valid at any value of the coupling. 

We begin with perturbation theory where the integrability of the theory first manifests itself.
Here we showcase the first exact result, the so-called slope function, which is the linear small spin expansion coefficient of the generalized Konishi anomalous dimension.
We then move on to exact results mainly achieved using the novel quantum spectral curve approach, the method allowing one to find scaling dimensions of operators at arbitrary values of the coupling.
As an example we find the second coefficient in the small spin expansion after the slope, which we call the curvature function.
This allows us to extract non-trivial information about the Konishi operator.

Methods of integrability are also applicable to other supersymmetric gauge theories such as ABJM, which in fact shares many similarities with $\N=4$ super Yang-Mills. We briefly review these parallel developments in the last chapter of the thesis.

\newpage 

\section*{Acknowledgements}

\vspace{30pt}

First of all I am immensely grateful to Nikolay Gromov for his brilliant supervision and collaboration throughout the years. 
I am also grateful to my colleagues Grigory Sizov and Fedor Levkovich-Maslyuk for all the invaluable discussions and collaborations. In addition I am very thankful to Fedor for thoroughly proofreading this thesis.
The staff and fellow students at King's have all contributed to the three wonderful years I spent there, for that I am very grateful.
I was also lucky enough to travel and visit various institutions around the world where I had the pleasure of meeting fellow scientists, I thank everyone for their hospitality and generosity.
Last but not least I want to thank my family and friends for their support and understanding. 
A special thanks goes to Ieva and Lara, your love and support have helped me stay sane throughout the years.

\newpage

\tableofcontents
\newpage
\listoffigures
\listoftables

\newpage


\section{Introduction}

\begin{chapquote}{Richard Feynman}
Everything is interesting if you go into it deeply enough.
\end{chapquote}

\noindent The title of this thesis is \emph{Exact Results in Supersymmetric Gauge Theories}. A reasonable question to ask is -- why would anyone care about that ? 
After all there is no evidence that supersymmetry is a true symmetry of nature and supersymmetric theories are mostly toy theories, we can not observe them in particle accelerators, as opposed to the Standard Model of particle physics. 
And indeed these are all valid points, however there are very good reasons for studying them. 

Consider $\N=4$ super Yang-Mills, from a pragmatic point of view it is the simplest non-trivial quantum field theory in four spacetime dimensions and since attempts at solving realistic QFTs such as the theory of strong interactions (QCD) have so far mostly resorted to numerical methods, it seems like a good starting point -- some go as far as calling it the harmonic oscillator of QFTs. 

Another (and probably the main) reason why $\N=4$ has been receiving so much attention in the last decades is the long list of mysterious and intriguing properties it seems to posses, making it almost an intellectual pursuit of understanding it. 
The theory has been surprising the theoretical physics community from the very beginning: it is a conformal theory in dimensions higher than two, it has a dual description in terms of a string theory and more recently it was discovered to be integrable in the planar limit. 
All of these properties give reasonable hope for actually solving the theory exactly, something that is highly non-trivial to achieve in any four dimensional interacting QFT.

In the remainder of the section we give a proper introduction to the subject from a historic point of view focusing on $\N=4$ SYM and its integrability aspect, for it is integrability that allows one to actually find exact results in the theory. We then give an overview of the thesis itself, emphasizing which parts of the text are reviews of known material and which parts constitute original work.

\subsection{Brief history of the subject}

Quantum field theory has been at the spotlight of theoretical physics since the middle of the last century when it was found that electromagnetism is described by the theory of quantum electrodynamics (QED). Since then people have been trying to fit other forces of nature into the QFT framework. 
Ultimately it worked: the theory of strong interactions, quantum chromodynamics or QCD for short, together with the electroweak theory, spontaneously broken down to QED, collectively make up the \emph{Standard Model} of particle physics, which has been extensively tested in particle accelerators since then. 

However nature did not give away her secrets without a fight. 
For some time it was thought that strong interactions were described by a theory of vibrating strings, as it seemed to incorporate the so-called Regge trajectories observed in experiments \cite{Veneziano:1968}. 
Even after discovering QCD, which is a Yang-Mills gauge theory, stringy aspects of it were still evident and largely mysterious. 
Most notably lattice gauge theory calculations at strong coupling suggested that surfaces of color-electric fluxes between quarks could be given the interpretation of stretched strings \cite{Wilson:1974}, thus an idea of a gauge-string duality was starting to emerge. 
It was strongly re-enforced by t'Hooft, who showed that the perturbative expansion of $U(N)$ gauge theories in the large $N$ limit could be rearranged into a genus expansion of surfaces triangulated by the double-line Feynman graphs, which strongly resembles string theory genus expansions \cite{THooft:1974}.

\vspace{20pt}
\newlength\yearposx
\begin{tikzpicture}[scale=1.82]

    \foreach \x in {1997,1998,2002,2003,2004}{
        \pgfmathsetlength\yearposx{(\x-1996)*1cm};
        \coordinate (y\x)   at (\yearposx,0);
        \coordinate (y\x t) at (\yearposx,+2pt);
        \coordinate (y\x b) at (\yearposx,-2pt);
    }
	
    \draw [->] (y1997) -- (y2004);
    \foreach \x in {1998,2002,2003} \draw (y\x t) -- (y\x b);

	\node at (y1998) [below=3pt] {1998}; 
		\node at (1.8cm, 0) [above=4pt] {AdS/CFT}; 
	\node at (y2002) [below=3pt] {2002}; \node at (y2002) [above=5pt] {BMN}; 
	\node at (y2003) [below=3pt] {2003}; \node at (y2003) [above=4pt] {Integrability}; 
\end{tikzpicture}
\vspace{20pt}

However it was the work of Maldacena in the end of 1997 that sparked a true revolution in the field \cite{Maldacena:1997re}. 
He formulated the first concrete conjecture, now universally referred to as \emph{AdS/CFT}, for a duality between a gauge theory, the maximally supersymmetric $\N=4$ super Yang-Mills, and type IIB string theory on $\adsfive$. 
Polyakov had already shown that non-critical string theory in four-dimensions describing gauge fields should be complemented with an extra Liouville-like direction thus enriching the space to a curved five dimensional manifold \cite{Polyakov:1997tj}. Furthermore the gauge theory had to be defined on the boundary of this manifold.
Maldacena's conjecture was consistent with this view, as the gauge theory was defined on the boundary of $AdS_5$, whereas the $S^5$ was associated with the internal symmetries of the gauge fields.
The idea of a higher dimensional theory being fully described by a theory living on the boundary was also considered before in the context of black hole physics \cite{'tHooft:1993gx, Susskind:1994vu} and goes by the name of holography, thus AdS/CFT is also referred to as a holographic duality.

The duality can be motivated by considering a stack of $N$ parallel D3 branes in type IIB string theory \cite{Polchinski:1995mt}. 
Open strings moving on the branes can at low energies be described by $\N=4$ SYM with the gauge group $U(N)$. 
Roughly, the idea is that there are six extra dimensions transverse to the stack of branes, thus a string stretching between two of them can be viewed as a set of six scalar fields ${(\Phi^i)^a}_b$ defined in four dimensional spacetime carrying two extra indices denoting the branes it is attached to. 
These are precisely the indices of the adjoint representation of $SU(N)$.
A similar argument can be put forward for other fields thus recovering the field content of $\N=4$ SYM.
Far away from the branes we have closed strings propagating in empty space. 
In the low energy limit these systems decouple and far away from the branes we are left with ten dimensional supergravity.

Another way of looking at this system is considering the branes as a defect in spacetime, which from the point of view of supergravity is a source of curvature. 
The supergravity solution carrying D3 brane charge can be written down explicitly \cite{Horowitz:1991cd}.
Far away from the branes it is obviously once again the usual flat space ten dimensional supergravity.
However near the horizon the geometry of the brane system becomes $\adsfive$.
Since both points of view end up with supergravity far away from the branes, one is tempted to identify the theories close to the branes, $\N=4$ SYM and type IIB string theory on $\adsfive$. 
This is exactly what Maldacena did in his seminal paper \cite{Maldacena:1997re}. 

By studying the supergravity solution one can identify the parameters of the theories, namely $\N=4$ SYM is parametrized by the coupling constant $g_{YM}$ and the number of colors $N$, whereas string theory has the string coupling constant $g_s$ and the string length squared $\alpha'$. 
These are identified in the following way
\beq
	 4\pi g_s = g_{YM}^2 \equiv \frac{\lambda}{N}, \;\;\;\;\;\;\; \frac{R^4}{\alpha'^2} = \lambda,
\eeq
where $\lambda$ is the t'Hooft coupling and $R$ is the radius of both $AdS_5$ and $S^5$, which is fixed as only the ratio $R^2/\alpha'$ is measurable.
A few things are to be noted here. First of all, the identification directly implements t'Hooft's idea of large $N$ expansion of gauge theory, since $g_s \sim 1/N$. 
In fact in the large $N$ limit only planar Feynman graphs survive and everything simplifies dramatically, a fact that we will take advantage of a lot in this thesis.
In this limit the effective coupling constant of the gauge theory is $\lambda$. 

The supergravity approximation is valid for low lying states when $\alpha' \ll R^2$, which corresponds to strongly coupled gauge theory, thus the conjecture is of the weak-strong type. 
This fact is a blessing in disguise, since initially it seems very restrictive as one can not easily compare results of the theories. 
However, it provides a possibility to access strongly coupled regimes of both theories, which was beyond reach before. 
Prescriptions for matching up observables on both sides of the correspondence were given in \cite{Gubser:1998bc, Witten:1998qj}. 
However, because of the weak-strong nature of the duality initial tests were performed only for BPS states, which are protected from quantum corrections.
The first direct match was observed in \cite{Witten:1998qj} where it was shown that the spectrum of half-BPS single trace operators matches the Kaluza-Klein modes of type IIB supergravity.
An important step in further understanding the conjecture was the formulation of type IIB string theory as a super-coset sigma model on the target space $PSU(2,2|4) / SO(2,4) \otimes SO(6)$, which of course has the same global symmetries as $\N=4$ SYM \cite{Metsaev:1998it}.

The situation changed dramatically in 2002 when Berenstein, Maldacena and Nastase devised a way to go beyond BPS checks \cite{Berenstein:2002jq}. 
The idea was to take an operator in gauge theory with large R-charge $J \sim \sqrt{N}$ and add some some impurities, effectively making it ``near-BPS''.
The canonical example of such an operator can be schematically written as $\tr(Z^J X^S)$, where $Z$ and $X$ are two complex scalar fields of $\N=4$ SYM, with $X$ being the impurities ($S \ll J$).
Since anomalous dimensions are suppressed like $\lambda/J^2$, perturbative gauge theory calculations are valid even at large $\lambda$, as long as $\lambda' \equiv \lambda / J^2 \ll 1$ and $N$ is large.
Keeping in mind potential problems with the order of limits, it is thus naively possible to compare gauge theory calculations with string theory results.
From the string theory point of view this limit corresponds to excitations of point-like strings with angular momentum $J$ moving at the speed of light around the great circle of $S^5$. 
The background seen by this string is the so-called pp-wave geometry \cite{Blau:2001ne} and string theory in this background is tractable \cite{Metsaev:2001bj}.

The discovery of the BMN limit was arguably the first time it was explicitly demonstrated how the world sheet theory of a string can be reconstructed by a physical picture of scalar fields dubbed as ``impurities'' propagating in a closed single trace operator of ``background'' scalar fields of the gauge theory. 
Shortly after this discovery Minahan and Zarembo revolutionized the subject once again by discovering Integrability at the end of 2002 \cite{Minahan:2002ve}.
They showed that in the large $N$ limit single-trace operators of scalar fields can be identified with spin chains and their anomalous dimensions at one-loop in weak coupling are given by the energies of the corresponding spin chain states.
These spin-chain systems are known to be integrable, which in practice allows one to solve the problem exactly using techniques such as the Bethe ansatz \cite{Bethe:1931}. 
This discovery sparked a very rapid development of integrability methods in AdS/CFT during the coming years \cite{Beisert:2010jr}.

\vspace{20pt}
\begin{tikzpicture}[scale=2.97]
    
	\coordinate (start) at (0,0);
	\coordinate (end) at (4.3cm,0);
	
    \foreach \x in {2003,2004,2005,2006,2007}{
        \pgfmathsetlength\yearposx{((\x-2003)*1cm + 0.1cm)};
        \coordinate (y\x)   at (\yearposx,0);
        \coordinate (y\x t) at (\yearposx,+0.5pt);
        \coordinate (y\x b) at (\yearposx,-0.5pt);
    }
	
    \draw [->] (start) -- (end);
    \foreach \x in {2003,2004,2005,2006,2007} \draw (y\x t) -- (y\x b);

	\node at (y2003) [below=3pt] {2003}; 
		\node at (0.5cm, 0) [above=4pt] {}; 
	\node at (y2004) [below=3pt] {2004}; 
		 \node at (y2004) [above=4pt] {KMMZ}; 
	\node at (y2005) [below=3pt] {2005}; 
		 \node at (1.6cm, 0) [above=4pt] {BDS}; 
	\node at (y2006) [below=3pt] {2006}; 
		 \node at (3.8cm, 0) [above=4pt] {BHL}; 
	\node at (y2007) [below=3pt] {2007}; 
\end{tikzpicture}
\vspace{20pt}

Solving a quantum field theory in principle means finding all $n$-point correlation functions of all physical observables. 
Since $\N=4$ SYM is conformal it is enough to find all 2-point and 3-point correlators, as all higher point correlation functions can be decomposed in terms of these basic constituents \cite{Belavin:1984ab}.
Due to conformal symmetry the two-point functions only depend on the scaling dimensions of operators, whereas for three point functions one also needs the so-called structure constants $C_{ijk}$ in addtion to the scaling dimensions.
Integrability methods from the very beginning mainly focused on solving the spectral problem in the large $N$ limit, that is finding the spectrum of operators with definite anomalous dimensions and their exact numeric values.
The initial discovery of \cite{Minahan:2002ve} was that the spectral problem was analogous to diagonalizing a spin chain Hamiltonian, which was identified with the dilatation operator of the superconformal symmetry of the theory.
The eigenstates correspond to operators in the gauge theory and the eigenvalues are their anomalous dimensions.

Soon after the initial discovery of integrability a spin chain formulation at one-loop was found for the full $PSU(2,2|4)$ theory, not only the scalar sector \cite{Beisert:2003jj}. 
The result was also extended to two and three loops \cite{Beisert:2003tq}. 
Integrability was also discovered at strong coupling as it was shown that the Metsaev-Tseytlin sigma model is classically integrable \cite{Bena:2003wd}. 
With integrability methods now being available at both weak and strong coupling it was possible to compare results in the BMN limit.
As expected, comparisons in the first two orders of the BMN coupling constant $\lambda'$ showed promising agreement \cite{Frolov:2003qc, Frolov:2003xy, Arutyunov:2003uj}, however an order of limits problem emerged at three loops \cite{Beisert:2003tq}, which puzzled the community for some time. 

All of these results seemed to suggest that integrability may be an all loop phenomenon, only the surface of it being scratched so far. 
This notion was strongly reinforced when classical string integrability was reformulated in the elegant language of algebraic curves by Kazakov, Marshakov, Minahan and Zarembo (KMMZ), which made the connection with weak coupling more manifest \cite{Kazakov:2004qf}. 
The algebraic curve was interpreted as the continuum limit of Bethe equations, which made it possible to speculate about all loop equations.
The first such attempt was made by Beisert, Dippel and Staudacher (BDS), who conjectured a set of Bethe equations and a dispersion relation which together successfully showcased some all-loop features \cite{Beisert:2004hm}.
This result was later extended to all sectors of the theory \cite{Beisert:2005fw}.
The BDS result was quickly shown to be incomplete as it was lacking a so-called dressing phase \cite{Arutyunov:2004vx}, a scalar function not constrained by symmetry of the problem.
It was found to leading order at strong coupling in \cite{Arutyunov:2004vx} and later to one-loop in \cite{Hernandez:2006tk}.
The discovery of the dressing phase also cured the infamous three loop discrepancy between weak and strong coupling results.
A crossing equation satisfied by the dressing phase was soon found \cite{Janik:2006dc} and eventually solved by Beisert, Hernandez and Lopez (BHL) \cite{Beisert:2006ib}.  
Collectively these results are often referred to as the asymptotic Bethe ansatz (ABA), since they are valid only for asymptotically long spin chains. 
When the states are short so-called wrapping effects become relevant. 
At weak coupling they manifest as long-range spin chain interactions wrapping around the chain, whereas at strong coupling they are due to virtual particles self-interacting across the circumference of the worldsheet \cite{Sieg:2005kd, Ambjorn:2005wa}.

\vspace{20pt}
\begin{tikzpicture}[scale=1.32]

	\coordinate (start) at (0,0);
	\coordinate (end) at (6.5cm,0);
	
    \foreach \x in {2009,2010,2012,2014}{
        \pgfmathsetlength\yearposx{(\x-2008)*1cm + 0.2cm};
        \coordinate (y\x)   at (\yearposx,0);
        \coordinate (y\x t) at (\yearposx,+2pt);
        \coordinate (y\x b) at (\yearposx,-2pt);
    }
    
    \draw [-] (0,0) -- (7cm,0);
	\draw [-, dashed] (7cm,0) -- (8.05cm,0);
	\draw [->] (8.05cm,0) -- (10cm,0);
	
	\foreach \x in {2009,2014} \draw (y\x t) -- (y\x b);
	
    \node at (y2009) [below=3pt] {2009}; \node at (y2009) [above=4pt] {TBA}; 
	\node at (y2014) [below=3pt] {2014}; 
		\node at (5.8cm, 0) [above=4pt] {QSC}; 
		
	\draw (8.9cm,-2pt) -- (8.9cm,+2pt);
	\node at (8.9cm,0) [below=4pt] {??}; \node at (8.9cm,0) [above=8pt] {$\N=4$ solved!}; 
	
\end{tikzpicture}
\vspace{20pt}

Once the asymptotic solution was found attention shifted to finite size corrections, which once resolved would in principle complete the solution to the spectral problem for single trace operators. 
Scattering corrections in finite volume for arbitrary QFTs were first addressed by L\"{u}scher \cite{Luscher:1986}, who derived a set of universal formulas, which were later generalized for non-relativistic theories \cite{Janik:2007wt}.
This approach, while very general and not directly related to integrability, was employed to calculate four \cite{Bajnok:2008bm} and five loop anomalous dimension coefficients \cite{Bajnok:2009vm} of the simplest non-BPS operator with length two, the Konishi operator.
The results agreed with available diagrammatic four-loop calculations \cite{Fiamberti:2007rj} and gave a new prediction for five loops.

An alternative approach more in line with integrability is the Thermodynamic Bete Ansatz (TBA).
Its origins can be traced back to Yang and Yang \cite{Yang}, however it was the work of Alexey Zamolodchikov \cite{Zamolodchikov1, Zamolodchikov2} that brought it to the mainstream. 
The idea is to consider the partition function of a two dimensional integrable QFT and it's ``mirror'' image found after exchanging length and time with a modular transformation.
At large imaginary times the partition function will be dominated by the ground state energy, whereas in the mirror theory large time means asymptotic length, which is under control using the asymptotic Bethe ansatz techniques.
Thus one can evaluate the partition function using the saddle point method and after rotating back to the original theory compute the exact ground state energy. 
Excited states can then be reached using analytic continuation. 
This approach was already proposed as an option for the AdS/CFT system in \cite{Ambjorn:2005wa} and was first discussed in depth in \cite{Arutyunov:2007tc}.
The TBA approach crystallized in 2009 with multiple groups publishing results almost simultaneously \cite{Gromov:2009tv, Bombardelli:2009ns, Gromov:2009bc, Arutyunov:2009ur}.
The Konishi anomalous dimension was initially checked at four \cite{Gromov:2009bc} and five \cite{Arutyunov:2010gb, Balog:2010xa} loops by linearising the TBA equations, showing precise agreement with results obtained using the L\"{u}scher method.
Ultimately the Konishi anomalous dimension was calculated numerically for a wide range of values of the t'Hooft coupling constant \cite{Gromov:2009zb}.

And so the spectral problem seemed to be solved, at least in the case of Konishi an exact and complete result was finally found, even if only numerically.
However it was increasingly becoming clear that the solution was not in its final and most elegant form.
Indeed the TBA equations are an infinite set of coupled integral equations, obviously one has to employ various numerical tricks to actually solve them and this mostly works in a case-by-case basis. 
Cases such as cusped Wilson lines \cite{Correa:2012hh, Gromov:2012eu} and the $\mathfrak{sl}(2)$ sector of the theory, containing the Konishi operator \cite{Gromov:2009bc}, have been worked out explicitly, however it still remains a hard problem in general.
From the very beginning alternative formulations of the solution were being proposed. 
An infinite set of non-linear functional equations, the so-called Y-system was proposed already in \cite{Gromov:2009bc}, later completed with analytical constraints coming from the TBA equations \cite{Cavaglia:2010nm}.
Connections of the \mbox{Y-system} with the Hirota bilinear relation were later explored in \cite{Gromov:2011cx} and the Y-system was reduced to a finite set of non-linear integral equations (FiNLIE).
The long sought beauty of the solution to the spectral problem was arguably uncovered with the formulation of the quantum spectral curve (QSC) approach \cite{Gromov:2013pga,Gromov:2014caa}, also referred to as the $\pmu$-system.
The whole TBA construction was ultimately reduced to a Riemann-Hilbert problem for eight $\cal Q$ functions, which generalize the quasi-momenta found in the algebraic curve construction, namely the classical spectral curve construction amounts to a WKB type approximation of the quantum spectral curve.
The QSC quickly showed its potential as previously known results were rederived almost without any effort and new results were being rapidly discovered \cite{Gromov:2013qga, Gromov:2014bva, Alfimov:2014bwa}.

Thus one can safely say that the spectral problem in planar $\N=4$ SYM is by now very well understood with numerical results readily available and deeper understanding of the structure being within reach.
Integrability methods have also been useful in other areas such as three point functions \cite{Escobedo:2010xs, Gromov:2012vu} and scattering amplitudes \cite{Drummond:2010km, Alday:2010kn}, however the situation there is still not as complete.
Having witnessed the successful resolution of the spectral problem it appears that $\N=4$ SYM is within reach of being solved completely.
If this programme were to be successfully carried out it would be the first example of a four dimensional interacting quantum field theory being solved exactly.
Undoubtedly this would provide a huge boost to our understanding of QFTs in general and hopefully bring us closer to solving QCD.

It turns out that $\N=4$ SYM is not the only example of an integrable supersymmetric gauge theory having a dual string description.
Probably the most famous other example involves the so-called ABJM theory, proposed by Aharony, Bergman, Jafferis and Maldacena \cite{Aharony:2008ug}, following \cite{Schwarz:2004yj, Bagger:2006sk,Gustavsson:2007vu,Gaiotto:2007qi}.
It is a three-dimensional superconformal Chern-Simons gauge theory with $\N=6$ supersymmetry.
This theory was conjectured to be the effective theory for a stack of M2 branes at a $Z_k$ orbifold
point. 
In the large $N$ limit its gravitational dual turns out to be M-theory on $AdS_4 \times S^7 / Z_k$. 
For large $k$ and $N$ with $\lambda = N/k$ fixed, the dual theory becomes type IIA superstring theory in $AdS_4 \times CP^3$.
This duality is also integrable \cite{Minahan:2008hf} and all of the developments outlined above have been reworked for it almost in parallel. 

\newpage
\subsection{Thesis overview}

The thesis consists of four core chapters, the first three of which cover $\N=4$ super Yang-Mills.
We start by introducing the theory in chapter \ref{sec:cft} where we give its field content, Lagrangian and talk briefly about symmetries and their representations.
We introduce the dual string theory in section \ref{sec:n4_strong} and talk briefly about its formulation as a super-coset sigma model.

Chapter \ref{sec:integrability} covers integrable structures found in $\N=4$ super Yang-Mills at strong and weak coupling, namely we discuss the spin chain picture at weak coupling in section \ref{sec:integrability_weak} and the classical spectral curve picture found at strong coupling in section \ref{sec:integrability_strong}.
A lot of focus in this chapter is put on the folded string solution described in section \ref{sec:folded_string}. 
It is the strong coupling dual to operators in the $\alg{sl}{2}$ sector of $\N=4$ super Yang-Mills, which are described in section \ref{sec:sl2_sector}. 
A key result of the chapter is the calculation of the Konishi anomalous dimension up to two loops at strong coupling achieved by boosting the one-loop result with the help of the exact slope function found in section \ref{sec:slope_function_aba}.

In chapter \ref{sec:exact_results} we move away from the perturbative regime and introduce exact solution methods for the spectral problem of $\N=4$ super Yang-Mills -- the thermodynamic Bethe ansatz and Y/T/$\mathcal{Q}$-systems in section \ref{sec:tba_y_system} and the novel quantum spectral curve construction in section \ref{sec:pmu_system}.
We then discuss exact solutions found using the quantum spectral curve starting with the slope function.
We rederive it in section \ref{sec:slope_pmu} and the calculation is then extended one order further to find the curvature function in \ref{sec:curvature}.
The Konishi anomalous dimension is revisited in section \ref{sec:konishi_three_loops} where using the curvature function we boost the previously obtained two-loop strong coupling result to three loops.
The chapter concludes with finding the anomalous dimension of a cusped Wilson line in the near-BPS limit in section \ref{sec:wilson_line} and addressing its classical limit.
 
Chapter \ref{sec:abjm} switches over from $\N=4$ super Yang-Mills to the ABJM theory and roughly follows the same path, however as most of the methods are very similar in spirit we move on much quicker.
We introduce the theory in section \ref{sec:abjm_intro} and discuss integrability in section \ref{sec:abjm_integrability}. 
Section \ref{sec:abjm_folded} describes the analogue of the folded string solution in ABJM, in particular the semi-classical quantization procedure of the solution.

\begin{figure}[h]
\centering
\begin{tikzpicture}[scale=1]
	
	
	\draw[thick] (0.5, 0) -- (11.5, 0) -- (11, 1) -- (1, 1) -- (0.5, 0); 
	\node at (6.5, 0) [above=5] {\Large 2};  
	
	\draw[thick] (11.7, 0) -- (11.7, 1) -- (12.7, 1) -- (12.7, 0) -- (11.7, 0);
	\node at (12.27, 0) [above=5] {\Large $5^{\,\star}$}; 
	
	\draw[thick] (2, 1) -- (2.5, 2) -- (4.5, 2) -- (5, 1); 
	\node at (3.5, 1) [above=5] {\Large 3.1}; 
	
	\draw[thick] (2.5, 2) -- (3, 3) -- (4, 3) -- (4.5, 2); 
	\node at (3.5, 2) [above=5] {\Large 3.2}; 
	
	\draw[thick] (5, 1) -- (5.5, 3); 
	\draw[thick] (6.5, 3) -- (7, 1); 
	\node at (6, 1.5) [above=5] {\Large 3.3}; 
	
	\draw[thick] (1, 3) -- (7, 3) -- (8, 5) -- (2, 5) -- (1, 3); 
	\node at (4.7, 3.5) [above=5] {\LARGE $3.4^{\,\dagger}$}; 
	
	\draw[thick] (8, 1) -- (8, 2) -- (9, 2) -- (9, 1); 
	\node at (8.5, 1) [above=5] {\Large 4.1}; 
	
	\draw[thick] (9.5, 1) -- (9, 2) -- (11.5, 2) -- (11, 1); 
	\node at (10.2, 1) [above=5] {\Large 4.2}; 
	
	\draw[thick] (8.5, 2) -- (8.5, 3.5); 
	\draw [decorate, decoration={snake, segment length=10, amplitude=1.5}, thick] (8.5, 3) -- (10.5, 3);
	\draw[thick] (10.5, 3.5) -- (10.5, 2); 
	\node at (9.6, 2) [above=5] {\Large $4.3^{\,\dagger}$}; 
	
	\draw[thick] (8.5, 3.5) -- (8.5, 5) -- (10.5, 5) -- (10.5, 3.5); 
	\node at (9.6, 3.3) [above=5] {\LARGE $4.4^{\,\dagger}$};

	\draw[thick] (10.5, 2) -- (11.5, 3.5) -- (12.5, 2) -- (11.5, 2); 
	\node at (11.5, 2) [above=5] {\Large $4.6^{\,\dagger}$}; 
	
	\draw[thick] (2.5, 5) -- (3, 6) -- (9.5, 6) -- (10, 5);
	\draw[thick] (7, 5) -- (8.5, 5); 
	\node at (6.5, 5) [above=5] {\LARGE $4.5^{\,\dagger}$}; 

\end{tikzpicture}
\caption[The interdependencies of the chapters and sections in the thesis]{The interdependencies of the chapters and sections in the thesis. Sections containing mostly original work are marked with daggers. Section 5 parallels the main text with original work on the folded string quantization.}
\label{fig:chapters}
\end{figure}
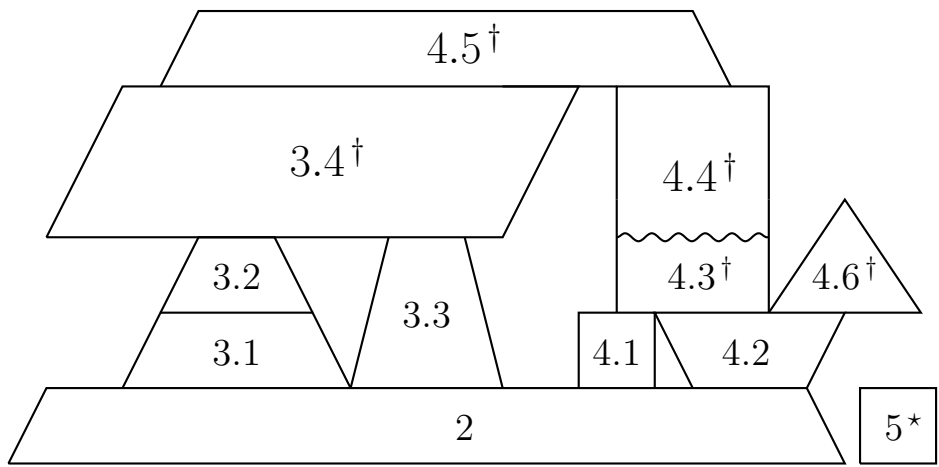

We end with conclusions and appendices containing some of the more technical details left out from the main text for brevity.
The interdependencies between the chapters and sections of the text are shown in figure \ref{fig:chapters}.

\newpage
\subsection{Original work}

The thesis contains original work by the author from five papers published in collaboration with fellow colleagues while working towards the PhD degree. 
Section \ref{sec:short_strings} is based on \cite{Gromov:2011bz}, where the two-loop strong coupling Konishi anomalous dimension was first calculated.
The calculation relied on semi-classical quantization of the folded string solution in $\N=4$ super Yang-Mills, the exact analogous calculation was then performed by the author in \cite{Beccaria:2012vb,Beccaria:2012qd} for the ABJM theory, which is the basis for section \ref{sec:abjm_folded}.
The subsections of \ref{sec:wilson_line} describing the classical limit of the cusped Wilson line are based on \cite{Sizov:2013joa}.
The remainder of chapter \ref{sec:exact_results} concerning the slope and curvature functions and their use to find the three-loop Konishi anomalous dimension at strong coupling are based on the work done in \cite{Gromov:2014bva}. 

Naturally in order to achieve a uniform flow throughout the text we introduced some filler sections outlining the basics of techniques we utilize later. 
These sections are kept short and are thoroughly filled with references to original work and/or reviews of the subject matter.
We hope the reader is not offended or annoyed by the inhomogeneous level of detail in various sections of the text and enjoys the thesis in its present form!

\vspace{60pt} 
 
\newpage 


\section{$\N=4$ super Yang-Mills}
\label{sec:cft}

\begin{chapquote}{German Proverb}
The devil is in the details.
\end{chapquote}

\noindent For the most part of this thesis we will be dealing with $\N=4$ super Yang-Mills theory. 
In this chapter we start off by defining it via its action and discussing its symmetries and observables. 
We also give an alternative formulation of the theory as a string theory, which is the core idea of the AdS/CFT correspondence. 
This formulation will later prove to be incredibly useful when discussing integrability and exact solutions.

\subsection{Action}

$\N=4$ super Yang-Mills theory is a quantum field theory much like the Standard Model of particle physics with a certain field content and interaction pattern.
It was first discovered by considering $\N=1$ super Yang-Mills theory in $9+1$ spacetime dimensions \cite{Brink:1977}, its action is given by
\begin{equation}
	S = \int d^{10} x \, \mathrm{Tr} \, \left( -\frac{1}{4}  F_{MN} F^{MN}  + \frac{1}{2} \bar{\Psi} \Gamma^M \mathcal{D}_M \Psi \right), \; \quad \; M = 1 \dots 10,
\end{equation}
where $\Psi$ is a Majorana-Weyl spinor in $9+1$ dimensions with $16$ real components and $\Gamma^M$ are the appropriate gamma matrices. 
The covariant derivative $\mathcal{D}_M$ is defined as
\begin{equation}
	\mathcal{D}_M = \partial_M - ig_{YM} \; [A_M, \, ],
\end{equation}
where $g_{YM}$ is the Yang-Mills coupling constant. 
The gauge group is in principle arbitrary, but we choose $SU(N)$ in anticipation of the AdS/CFT correspondence. 
By dimensionally reducing this theory on a flat torus $T^6$ one recovers the maximally supersymmetric $\N=4$ Yang-Mills gauge theory in $3+1$ spacetime dimensions.
The reduced action reads
\begin{equation}
\begin{split}
S = & \int d^4 x \, \tr \, \left( -\frac{1}{2} \mathcal{D}_\mu \Phi_I \mathcal{D}^\mu \Phi^I  + \frac{g_{YM}^2}{4} [\Phi_I, \Phi_J] [\Phi^I, \Phi^J] - \frac{1}{4}  F_{\mu\nu} F^{\mu\nu}  \right. \\
	 &\left. - \bar{\psi}^a \sigma^\mu \mathcal{D}_\mu \psi_a  + \frac{ig_{YM}}{2} \sigma^{ab}_I \psi_a [\Phi^I, \psi_b] + \frac{ig_{YM}}{2} \sigma_{ab}^I \bar{\psi}^a [\Phi_I, \bar{\psi}^b] \right).
\end{split}
\label{eq:n4_action}
\end{equation}
After dimensional reduction the gauge field $A_M$ decomposes to the four dimensional gauge field $A_\mu$ and to six real scalar fields $\Phi_I$ whereas the Majorana-Weyl spinor $\Psi_A$ breaks up into four copies of the left and right Weyl spinors in four dimensions 
\begin{equation}
	\Psi_A \, \, (A = 1, ..., 16) \, \, \, \rightarrow \, \, \,  \bar{\psi}^a_{\dot{\alpha}}, \; \psi_{a\alpha} \, \, (\alpha, \dot{\alpha} = 1, 2, \; \; a = 1, ..., 4).
\end{equation}
It also gives rise to the $SO(6) \simeq SU(4)$ symmetry called \emph{R-symmetry}, which originally was part of the ten dimensional Poincare group, but now acts as an internal symmetry of the supercharges.
The matrices $\sigma_I^{ab}$ in \eq{eq:n4_action} are the gamma matrices for this group.
It permutes the scalars, which live in the fundamental ${\bf 6}$ of $SO(6)$ and the spinors, which live in the fundamental of $SU(4)$, namely the lower index $a$ in $\psi_{a\alpha}$ transforms in ${\bf 4}$, while $\bar{\psi}_{\dot{\alpha}}^a$ transforms in ${\bf \bar{4}}$. 
From this it follows that we can combine the six real scalars $\Phi^I$ into three complex scalars $\Phi^{ab}$, often denoted as $X$, $Y$ and $Z$, which then transform under the second rank antisymmetric ${\bf 6}$ of $SU(4)$. 
The gauge field is a singlet under R-symmetry.

It is now a straightforward but rather tedious task to calculate the beta function for this theory. 
For any $SU(N)$ gauge theory at one loop level it is given by \cite{Gross:1973}
\begin{equation}
	\beta(g) = - \frac{g_{YM}^3}{16\pi^2} \left( \frac{11}{3} N - \frac{1}{6} \sum_s C_s - \frac{1}{3} \sum_f \tilde{C}_f \right)
\end{equation}
where the first sum is over the real scalars and the second one over the fermions. 
$C_s$ and $\tilde{C}_f$ are the quadratic Casimirs, which in our case are equal to $N$ since all fields are in the adjoint representation of the group. 
It is then easy to see that at least at one loop level the theory is conformally invariant. 
In fact the $\beta$ function was shown to be identically zero to all orders in perturbation theory \cite{Sohnius:1981ab,Mandelstam:1983, Brink:1983}, hence $\N=4$ super Yang-Mills is fully conformally invariant even after quantization. 
After discussing the full symmetry algebra of the theory and its representations we will give an elegant argument why this is true.

\subsection{Observables}

The theory has 16 on-shell degrees of freedom which make up the gauge multiplet of $\N=4$ supersymmetry, namely $(\Phi_I, \psi_a, A_\mu)$. 
Gauge invariant operators are then formed by taking traces over the gauge group. 
An important class of operators are the \emph{local operators}, which are traces of fields all evaluated at the same spacetime point. 
They have the general form
\begin{eqnarray}
	\mathcal{O}_{i_1 \mu i_2 \alpha \dots i_n \dots j_1 \nu \beta \dots j_n}(x) & = \tr \left[ \Phi_{i_1}(x) \mathcal{D}_\mu \Phi_{i_2}(x) \psi_\alpha(x) \dots \Phi_{i_n}(x) \right] \times \dots \nonumber \\
	& \dots \times \tr \left[ \Phi_{j_1}(x) \mathcal{D}_\nu \psi_\beta(x) \dots \Phi_{j_n}(x) \right]. 
\end{eqnarray} 
In this thesis we will be exclusively focusing on the planar limit, which is the limit when the number of colors $N$ is sent to infinity. 
Diagrams involving multi-trace operators are non-planar, hence suppressed in the large $N$ limit and therefore we will only be considering single trace operators.
An example of a non-local operator is the Wilson loop, given by \cite{Rey:1998bq, Maldacena:1998im} 
\beq
	W_L= \tr \, \( {\mathcal P}\exp\!\oint_C \! dt\(i  A\cdot\dot{x}+\vec\Phi\cdot\vec n\,|\dot x|\) \),
\eeq
which depends on the path $x^\mu(t)$ in spacetime, hence it is known as a \emph{line operator}. 
It also depends on the coupling to the scalar fields, which is encoded in the six-dimensional unit vector $\vec{n}(t)$. 
The scalar field term can also be understood by recalling that the scalar fields are a result of dimensional reduction from $9+1$ dimensions, thus the coupling vector $\vec{n}(t)$ together with the curve $x^\mu(t)$ make up a path $x^M(t)$ in $9+1$ dimensional spacetime. 
In later sections of the text we will be considering cusped Wilson lines with other operators inserted at the cusp. We will be mostly working in these two classes of operators, however in principle one could go on and define surface operators, etc.

\subsection{Symmetry}

Conformal symmetry, supersymmetry and R-symmetry are a part of a bigger group $PSU(2,2|4)$, which is also known as the \emph{$\N=4$ superconformal group} \cite{Nahm:1978ab}. 
It is the full symmetry group of $\N=4$ super Yang-Mills and is unbroken by quantum corrections. 
It is an example of a \emph{supergroup}, i.e. a graded group containing bosonic and fermionic generators. 
The theory of supergroups is highly developed \cite{Sohnius:1981ab,Kac:1977ab,Brink:1983,Beisert:2010kp} and much of the techniques from studying bosonic groups carry over to supergroups with some additional complications, i.e. Dynkin diagrams, root spaces, weights etc. 

$PSU(2,2|4)$ has the bosonic subgroup of $SU(2,2) \times SU(4)$, where $SU(2,2) \simeq SO(2,4)$ is the conformal group in four dimensions and $SU(4) \simeq SO(6)$ is the R-symmetry. 
The conformal group has the Poincar\'{e} group as a subgroup, which has a total of 10 generators including four translations $P_\mu$ and six Lorentz transformations $M_{\mu\nu}$, in addition there is the generator for dilatations $D$ and four special conformal generators $K_\mu$. 
Their commutation relations read
\begin{eqnarray}
 &[D, M_{\mu\nu}] = 0 \; \; \; [D, P_\mu] = -i P_\mu \; \; \; [D, K_\mu] = +i K_\mu,  \nonumber\\
 &[M_{\mu\nu}, P_\lambda] = -i(\eta_{\mu\nu} P_\nu - \eta_{\lambda\nu} P_\mu) \; \; \; [M_{\mu\nu}, K_\lambda] = -i(\eta_{\mu\lambda} K_\nu - \eta_{\lambda\nu} K_\mu),  \nonumber\\
 &[P_\mu, K_\nu] = 2i(M_{\mu\nu} - \eta_{\mu\nu} D).
 \label{eq:conformal_group}
\end{eqnarray}
$\N=4$ supersymmetry has 16 supercharges $Q_{a\alpha}$ and $\tilde{Q}^a_{\dot{\alpha}}$ where $\alpha, \dot{\alpha} = 1, 2$ are the Weyl spinor indices and $a = 1,...,4$ are the R-symmetry indices. 
These generators have the usual commutation and anti-commutation relations with the Poincar\'{e} generators given by 
\begin{eqnarray}
	& \{Q_{\alpha a}, \tilde{Q}^b_{\dot{\alpha}}\} = \gamma^\mu_{\alpha\dot{\alpha}} \delta_a^b P_\mu \; \; \; \{Q_{\alpha a}, Q_{\alpha b}\} = \{ \tilde{Q}^a_{\dot{\alpha}}, \tilde{Q}^b_{\dot{\alpha}} \} = 0, \nonumber\\
	& [M^{\mu\nu}, Q_{\alpha a}] = i \gamma^{\mu\nu}_{\alpha\beta} \epsilon^{\beta\gamma} Q_{\gamma a} \; \; \; [M^{\mu\nu}, \tilde{Q}^a_{\dot{\alpha}}] = i \gamma^{\mu\nu}_{\dot{\alpha}\dot{\beta}} \epsilon^{\dot{\beta}\dot{\gamma}} \tilde{Q}_{\dot{\gamma}}^a, \nonumber \\
	& [P_\mu, Q_{\alpha a}] = [P_\mu, \tilde{Q}^b_{\dot{\alpha}}] = 0,
\end{eqnarray} 
where $\gamma_{\alpha\beta}^{\mu\nu} = \gamma^{[\mu}_{\alpha\dot{\alpha}} \gamma^{\nu]}_{\beta\dot{\beta}} \epsilon^{\dot{\alpha}\dot{\beta}}$. 
Commutators between supercharges and the conformal generators are also non trivial and introduce new supercharges,
\begin{eqnarray}
	& [D, Q_{\alpha a}] = -\frac{i}{2} Q_{\alpha a} \; \; \; [D, \tilde{Q}_{\dot{\alpha}}^a] = -\frac{i}{2} \tilde{Q}_{\dot{\alpha}}^a, \nonumber \\
	& [K^\mu,  Q_{\alpha a}] = \gamma^\mu_{\alpha\dot{\alpha}} \epsilon^{\dot{\alpha} \dot{\beta}} \tilde{S}_{\dot{\beta} a} \; \; \; [K^\mu, \tilde{Q}_{\dot{\alpha}}^a] = \gamma^\mu_{\alpha\dot{\alpha}} \epsilon^{\alpha\beta} S_\beta^a,
	\label{eq:dq_commutators}
\end{eqnarray}
where $\tilde{S}_{\dot{\alpha} a}$ and $S_\alpha^a$ are the \emph{special conformal supercharges}. 
They have opposite \text{R-symmetry} representations compared to the usual supercharges. 
The special supercharges bring the total of supercharges to 32. 
The commutation and anti-commutation relations for the special conformal supercharges are very much like the ones for normal supercharges,
\begin{eqnarray}
	& \{S_{\alpha}^a, \tilde{S}_{\dot{\alpha} b}\} = \gamma^\mu_{\alpha\dot{\alpha}} \delta_b^a K_\mu \; \; \; \{S_{\alpha}^a, S_{\alpha}^b\} = \{ \tilde{S}_{\dot{\alpha} a}, \tilde{S}_{\dot{\alpha} b} \} = 0, \nonumber\\
	& [M^{\mu\nu}, S_{\alpha}^a] = i \gamma^{\mu\nu}_{\alpha\beta} \epsilon^{\beta\gamma} S_{\gamma}^a \; \; \; [M^{\mu\nu}, \tilde{S}_{\dot{\alpha} a}] = i \gamma^{\mu\nu}_{\dot{\alpha}\dot{\beta}} \epsilon^{\dot{\beta}\dot{\gamma}} \tilde{S}_{\dot{\gamma} a}, \nonumber \\
	& [K_\mu, S_{\alpha}^a] = [K_\mu, \tilde{S}_{\dot{\alpha} a}] = 0.
\end{eqnarray} 
Finally the anti-commutation relations between the special conformal and usual supercharges close the algebra,
\begin{eqnarray}
	\{ Q_{\alpha a}, S_\beta^b \} & = & - i \epsilon_{\alpha\beta} {{\sigma^{IJ}}_a}^b R_{IJ} + \gamma_{\alpha\beta}^{\mu\nu} {\delta_a}^b M_{\mu\nu} - \frac{1}{2} \epsilon_{\alpha\beta} {\delta_a}^b D \nonumber \\
	\{ \tilde{Q}_{\dot{\alpha}}^a, \tilde{S}_{\dot{\beta} b} \} & = & + i \epsilon_{\dot{\alpha}\dot{\beta}} {{\sigma^{IJ}}^a}_b R_{IJ} + \gamma_{\dot{\alpha}\dot{\beta}}^{\mu\nu} {\delta^a}_b M_{\mu\nu} - \frac{1}{2} \epsilon_{\dot{\alpha}\dot{\beta}} {\delta^a}_b D \nonumber \\
	\{ Q_{\alpha a}, \tilde{S}_{\dot{\beta} b} \} & = & \{ \tilde{Q}_{\dot{\alpha}}^a, S_\beta^b \} = 0
	\label{eq:qs_anticommutators}
\end{eqnarray}
where $R_{IJ}$ are the generators of R-symmetry with $I,J = 1, ..., 6$. 
All supercharges transform under the two spinor representations of the R-symmetry group and all other generators commute with it. 
All of the generators can be organized as follows
\beq
\(
	\begin{array}{c|c}
	K^\mu, P^\mu, M^{\mu\nu}, D & Q_{a\alpha}, \bar{S}_{a\dot{\alpha}} \\ \hline 
	S_{\alpha}^a, \bar{Q}_{\dot{\alpha}}^a  & R_{IJ}
	\end{array} 
\)
\eeq
where the generators in the diagonal blocks are bosonic and the ones in the anti-diagonal blocks are fermionic.
They have a definite dimensions, which are not modified by radiative corrections
\beq
	[D]=[R]=0\;, \quad [P]=1\;, \ [K]=-1\;, \quad [Q]=1/2\;,\  [S]=-1/2\;.
\eeq
In contrast, the classical dimensions of fields
\beq
	[\Phi^I] = [A_\mu] = 1\;, \quad [\psi_a] = \frac{3}{2},
\eeq
do receive radiative corrections and acquire \emph{anomalous dimensions}, which together with the bare dimension make up the conformal dimension
\beq
	\Delta = \Delta_0 + \gamma(g_{YM}).
\eeq
The name is justified by the fact that in conformal field theories all two point functions are determined by the scaling dimensions of the fields. 
More than that, together with the knowledge of all three point functions they are enough to determine any $n$-point function. 
This is why finding conformal dimensions of all operators, i.e. the spectrum of the theory is a very important step in solving it.

\subsubsection{Superconformal multiplets}

Fields of the theory can be organized in unitary representations of the superconformal symmetry group, which are labeled by quantum numbers of the bosonic subgroup
\beqa
	&SO(1,3) \times SO(1,1) \times SU(4) \nonumber \\
	&\quad \; (s_+, s_-) \quad \quad \Delta \quad \quad  [r_1, r_2, r_3]
\eeqa 
where $(s_+, s_-)$ are the usual positive half-integer spin labels of the Lorentz group, $\Delta$ is the positive conformal dimension that can depend on the coupling and $[r_1, r_2, r_3]$ are Dynkin labels of the $R$-symmetry.
All unitary representations of the superconformal group have been classified into four families \cite{Dobrev:1985ab,Dobrev:1985cd}, here we give a short description of the classification.
  
Looking at the commutation relations of the conformal subgroup (\ref{eq:conformal_group}), we see that the operators $P_\mu$ and $K_\mu$ act as raising and lowering operators for the dilatation operator $D$ -- this gives a hint as to how we could construct representations of the group. 
The dilatation operator $D$ is the generator of scalings, i.e. upon a rescaling $x \rightarrow \lambda x$ a local operator in a field theory scales as 
\begin{equation}
	\mathcal{O}(x) \rightarrow \lambda^{-\Delta} \mathcal{O}(\lambda x)
\end{equation}
where $\Delta$ is the conformal dimension of the operator $\mathcal{O}(x)$. 
Restricting to the point $x = 0$, which is a fixed point of scalings, we see that the conformal dimension is the eigenvalue of the dilatation operator,
\begin{equation}
	[D,\mathcal{O}(0)] = -i \Delta \mathcal{O}(0).
\end{equation}
It is now clear that acting on a field with $K_\mu$ should lower the dimension by one and acting with $P_\mu$ -- raise it by one. 
We can show this explicitly using the Jacobi identity as
\beq
	[D, [K_\mu, \mathcal{O}(0)]] = [[D, K_\mu], \mathcal{O}(0)] + [K_\mu, [D, \mathcal{O}(0)]] = -i (\Delta - 1) \; [K_\mu, \mathcal{O}(0)].
\eeq
Since operators in a unitary quantum field theory should have positive dimensions (aside from the identity operator), we should not be able to keep lowering the dimension indefinitely, i.e. there should always be an operator that satisfies
\begin{equation}
	[K_\mu, \tilde{\mathcal{O}}(0)] = 0.
\end{equation} 
We call such operators \emph{conformal primary operators}. 
Acting on these with $P_\mu$ keeps producing operators with a dimension one higher -- we call these the \emph{descendants} of $\tilde{\mathcal{O}}(0)$. 
We can also act with the supercharges and looking at the commutators in (\ref{eq:dq_commutators}) we see that they raise the dimension by $1/2$, while the special conformal supercharges lower it by $1/2$. 
Operators annihilated by special conformal supercharges are called \emph{superconformal primaries}, which is a stronger condition that being a conformal primary.

(Super-)conformal primaries and their descendants make up multiplets that constitute three families of discrete representations in the classification.
They are further distinguished by the number of supercharges the primary commutes with.
One example is a class of operators that satisfy the condition 
\beq
	\label{eq:halfBPS}
	\Delta = r_1 + r_2 + r_3,
\eeq 
a canonical representative would be a single-trace symmetrized scalar field operator such as 
\beq
	\mathcal{O}^{i j \dots k}(x) = \tr \( \Phi(x)^{(i} \Phi(x)^j \dots \Phi(x)^{k)} \).
\eeq
These operators commute with half of the supercharges, thus they are referred to as \text{1/2-BPS}. 
Similarly, operators commuting with a quarter and an eight of the supercharges are respectively denoted by 1/4-BPS and 1/8-BPS, they constitute the other two families of discrete representations.  
A key fact is that operators in the same representation must have the same anomalous dimension, because the generators of the group can only change it by half integer steps and there's only a finite number of generators. 
What is more, operators in the discrete BPS representations are protected from quantum corrections, because at any coupling the total dimension is always algebraically related to the Dynkin labels of the R-symmetry, e.g. as in \eq{eq:halfBPS}. 
Since charges of compact groups are quantized it must mean that the dimension can't continuously depend on the coupling and hence the anomalous dimension must vanish. 
This is however not true for the fourth family of representations, which is continuous and whose operators do not commute with any of the supercharges, hence they do acquire anomalous dimensions and are referred to as non-BPS.

Let us conclude the section with an elegant argument for why the beta function of $\N=4$ super Yang-Mills is zero.
One can use the algebra and shown that the operators $\tr \, F_+ F_+$ and $\tr \, F_- F_- $, where $F_\mp$ are the (anti-)self-dual field strengths, belong to the same multiplet as a superconformal primary \cite{Minahan:2010js}, meaning that the $\tr \, F_{\mu\nu} F^{\mu\nu}$ term in the Lagrangian is protected from quantum corrections, hence so is the coupling constant $g_{YM}$. 
This argument is valid to all orders in perturbation theory, which means that $\N=4$ super Yang-Mills is conformally invariant to all orders in perturbation theory and in fact even non-perturbatively.

\subsection{String description at strong coupling}
\label{sec:n4_strong}

As already briefly explained in the introduction, the AdS/CFT conjecture states that $\N=4$ super Yang-Mills is exactly dual to type IIB string theory on $\adsfive$, \cite{Maldacena:1997re,Gubser:1998bc,Witten:1998qj}. 
To be more precise, the gauge group of the Yang-Mills theory is taken to be $SU(N)$ and the coupling constant $g_{YM}$. 
The string theory is defined on $\adsfive$ where both $AdS_5$ and $S^5$ have radius $R$. 
The self-dual five-form field $F_5^+$ has integer flux through the sphere
\beq
	\int_{S^5} F_5^+ = N,
\eeq
and $N$ is identified with the number of colours in the gauge theory. 
The string theory is further parametrized by the string coupling $g_s$ and the string length squared $\alpha'$. The following relations are conjectured to hold
\beq
	 4\pi g_s = g_{YM}^2 \equiv \frac{\lambda}{N}, \;\;\;\;\;\;\; \frac{R^4}{\alpha'^2} = \lambda,
\eeq
where $\lambda$ is the t'Hooft coupling. We will be working in the planar limit $N \rightarrow \infty$ with $\lambda$ fixed. 
It is easy to see that in this limit $g_s \rightarrow 0$ and we are left with freely propagating strings.
Furthermore, the regime of strongly coupled gauge theory when $\lambda \rightarrow \infty$ corresponds to the regime of string theory where the supegravity approximation is valid, namely $\alpha' \ll R^2$.
The takeaway here is that one can formulate strongly coupled planar $\N=4$ super Yang-Mills as a classical theory of free strings on $\adsfive$.

\subsubsection{Sigma model formulation}

A very useful formulation of string theory on $AdS_5 \times S^5$ is the coset space sigma model \cite{Henneaux:1984mh, Metsaev:1998it} with the target superspace of
\begin{equation}
	\frac{PSU(2,2|4)}{SO(1,4) \times SO(5)}.
\end{equation}
\vspace{2pt}
The bosonic part of the supercoset where the string moves is given by
\beq
	\frac{SO(2,4) \times SO(6)}{SO(1,4) \times SO(5)} = \adsfive,
\eeq
which is constructed as the coset between the isometry and isotropy groups of $\adsfive$. 
The action is then written in terms of the algebra of $PSU(2,2|4)$.

The superalgebra $\alg{psu}{2,2|4}$ has no realization in terms of matrices, instead it is the quotient of $\alg{su}{2,2|4}$ by matrices proportional to the identity. 
On the other hand $\alg{su}{2,2|4}$ is a matrix superalgebra spanned by $8\times8$ supertraceless matrices
\beq
M = \( {\begin{array}{c|c}
 A & B  \\
 \hline
 C & D  \\
 \end{array} } \),
\eeq
where the supertrace is defined as 
\beq
\str M = \tr A - \tr D.
\eeq
$A$ and $D$ are elements of $\alg{su}{2,2}$ and $\alg{su}{4}$ respectively, whereas the fermionic components are related by
\beq
	C = \( {\begin{array}{c|c}
 +\mathbbm{1}_{2\times2} & 0  \\
 \hline
 0 & -\mathbbm{1}_{2\times2}  \\
 \end{array} } \) B^\dagger.
\eeq
An important feature of this algebra is the following automorphism
\beq
\Omega \circ M = \( {\begin{array}{c|c}
 E A^T E & -E C^T E  \\
 \hline
 E B^T E & E D^T E \\
 \end{array} } \),
 \; \quad  \;
E = \( {\begin{array}{cccc}
 0 & -1 & 0 & 0  \\
 1 & 0 & 0 & 0 \\
 0 & 0 & 0 & -1  \\
 0 & 0 & 1 & 0 \\
 \end{array} } \),
\eeq
which endows the algebra with a $\mathbb{Z}_4$ grading \cite{Berkovits:1999zq}, since one can easily check that $\Omega^4 = 1$.
This in turn means that any element of the algebra can be decomposed under this grading as
\beq
	M = \sum_{i=0}^3 M^{(i)},
\eeq
where
\beqa
	\label{eq:M_components}
	M^{(0,2)} &= \frac{1}{2} \( {\begin{array}{c|c}
 A \;\, \pm E A^T E & 0  \\
 \hline
 0 & D \pm E D^T E  \nonumber \\
 \end{array} } \) \\
 M^{(1,3)} &= \frac{1}{2} \( {\begin{array}{c|c}
 0 & B \pm i E C^T E  \\
 \hline
 C \mp i E B^T E & 0  \\
 \end{array} } \)
\eeqa 
and the morphism then acts on the elements of the decomposition as
\beq
	\Omega \circ M^{(n)} = i^n M^{(n)}.
\eeq
The Metsaev-Tseytlin action for the Green-Schwarz superstring is then given by
\begin{equation}
	\label{eq:mt_action}
	S = \frac{\sqrt{\lambda}}{4 \pi} \int \str \left( J^{(2)} \wedge * J^{(2)} - J^{(1)} \wedge J^{(3)} + \Lambda \wedge J^{(2)} \right),
\end{equation}
which is written down in terms of the graded elements of the algebra current
\begin{equation}
	J = -g^{-1} \mathrm{d} g 
	\label{eq:j_current}
\end{equation}
where $g(\sigma, \tau) \in PSU(2,2|4)$ is a map from the string worldsheet to the supergroup $PSU(2,2|4)$. 
The last term contains a Lagrange multiplier $\Lambda$, which ensures that $J^{(2)}$ is supertraceless, whereas all other components are manifestly traceless as seen from \eq{eq:M_components}. 
Since the target space is the coset of $PSU(2,2|4)$ by $SO(1,4) \times SO(5)$, the map $g$ has an extra gauge symmetry
\begin{equation}
	g \rightarrow gH, \,\,\,\,\, H \in SO(1,4) \times SO(5)
\end{equation}
under which the components of the supercurrent transform as
\beqa
	& J^{(0)} \rightarrow H^{-1} J^{(0)} H - H^{-1} \mathrm{d} H \\
	& J^{(i)} \rightarrow H^{-1} J^{(i)} H, \quad i=1,2,3
\eeqa
The equations of motion read
\beq
	d * k = 0,
\eeq
where $k = gKg^{-1}$ and
\beq
	K = J^{(2)} + \frac{1}{2} * J^{(1)} - \frac{1}{2} * J^{(3)} - \frac{1}{2} * \Lambda.
\eeq
They are equivalent to the conservation of the Noether current associated to the global left $PSU(2,2|4)$ multiplication symmetry.

Finally let us briefly remark on how the action reduces to the usual sigma model action if one restricts to bosonic fields. 
A purely bosonic representative of $PSU(2,2|4)$ has the form
\beq
	g = \( {\begin{array}{c|c}
 A & 0  \\
 \hline
 0 & D  \\
 \end{array} } \),
\eeq
where $A \in SO(6) \simeq SU(4)$ and $D \in SO(2,4) \simeq SU(2,2)$. Then we see that $A E A^T$ is a good parametrization of 
\beq
	\frac{SO(6)}{SO(5)} \simeq \frac{SU(4)}{SP(4)} = S^5,
\eeq
since it is invariant under $A \rightarrow A H$ with $H \in SP(4)$. 
Similarly $D E D^T$ parametrizes $AdS_5$. 
If we now define the coordinates $u^i$ and $v^i$ in the following way
\beq
	u^i \Gamma^S_i = A E A^T, \quad \quad v^i \Gamma^A_i = D E D^T,
\eeq
with $\Gamma^S$ and $\Gamma^A$ being the gamma matrices of $SO(6)$ and $SO(2,4)$ respectively, then by construction they will satisfy the following constraints
\beqa
	1 &= u \cdot u \equiv +u_1^2 + u_2^2 + u_3^2 + u_4^2 + u_5^2 + u_6^2 \nonumber \\
	1 &= v \cdot v \equiv -v_1^2 - v_2^2 - v_3^2 - v_4^2 + v_5^2 + v_6^2,
\eeqa
and the action \eq{eq:mt_action} will read
\beq
	S_b = \frac{\sqrt{\lambda}}{4\pi} \int_0^{2\pi} d\sigma \int d\tau \; \sqrt{h} \( h^{\mu\nu} \pd_\mu u \cdot \pd_\nu u + \lambda_u \( u \cdot u - 1 \) - \( u \rightarrow v \) \),
\eeq
which is just the usual non-linear sigma model for a string moving in $\adsfive$.

\newpage 


\section{Perturbative results}
\label{sec:integrability}

\begin{chapquote}{Old Chinese Proverb}
It is better to take many small steps in the right direction than to make a great leap forward only to stumble backward.
\end{chapquote}

\noindent In this section we start attacking the problem of finding the spectrum and as expected we begin with perturbation theory.
Starting at weak coupling we quickly stumble upon an amazing feature of the theory, so-called integrability, which allows one to apply numerous techniques that greatly simplify the problem.
We demonstrate integrability from the string theoretic perspective at strong coupling as well, which suggests a unified picture of the integrable structure embedded in the theory persisting to all loops.
After discussing results achievable via integrability in the perturbative regime we finish off with our first exact result, the slope function, which in turn allows one to extract novel information about the spectrum.

\subsection{One loop at weak coupling}
\label{sec:integrability_weak}

We begin with two point correlation functions of local operators.
In any conformal field theory they are constrained by symmetry, namely for operators that are eigenvalues of dilatations they have the following form at all loop levels up to unphysical normalization factors \cite{Belavin:1984ab, Osborn:1993cr}
\begin{equation}
	\left< \mathcal{O}(x) \, \tilde{\mathcal{O}}(y) \right> = \frac{1}{|x-y|^{2\Delta}},
\end{equation}
where $\Delta$ is the scaling dimension of the operator. 
Classically $\Delta = \Delta_0$ is simply the mass dimension, but at the quantum level it receives radiative corrections and acquires an anomalous dimension $\gamma$, such that $\Delta(g_{YM}) = \Delta_0 + \gamma(g_{YM})$, where the anomalous dimension depends on the coupling. 
Usually the corrections are small and the correlator can be expanded perturbatively.
Of course one has to be careful here, as expanding in $\gamma$ would result in expressions like $\log|x-y|$, which do not make sense. 
To that end we introduce a scale $\mu$ and expand the following quantity instead 
\begin{equation}
	\mu^{-2\gamma} \left< \mathcal{O}(x) \, \tilde{\mathcal{O}}(y) \right> = \frac{1}{|x-y|^{2\Delta_0}} \left(1 - \gamma \log \mu^2 |x-y|^2 \right),
	\label{eq:anomdim_expansion}
\end{equation}
however we will formally assume that the factor $\mu^{-\gamma}$ is absorbed into the field definition and thus we will ignore it from now on. We can now take some explicit local operator $\mathcal{O}(x)$, calculate the correlator using perturbation theory and read off the anomalous dimension $\gamma$.
Let us start with a very simple chiral primary operator 
\begin{equation}
	\Psi = \tr \, Z^L  = {Z^a}_b {Z^b}_c \dots {Z^l}_a,
	\label{eq:primary_op}
\end{equation}
where the complex scalar field $Z$ and its conjugate $\tilde{Z}$ have the standard tree level correlators
\begin{equation}
	\left< {Z^a}_b(x) {{\tilde{Z^{b'}}}}_{a'}(y) \right>_{\mathrm{tree}} = \frac{{\delta^a}_{a'} {\delta_b}^{b'}}{|x-y|^{2}}.
	\label{eq:z_correlator}
\end{equation}  
In order to find the anomalous dimension of the operator $\Psi$ we must calculate the correlator $\langle\Psi(x) \tilde{\Psi}(x)\rangle$. We do this by using Wick's theorem and plugging in the two-point correlator (\ref{eq:z_correlator}), which produces a lot of terms with delta function contractions between the adjoint indices. Some examples are
\begin{subequations}
	\begin{equation} 
	\dots {\delta^{a'}}_{a} \; {\delta^{a}}_{a'} \; {\delta^{b'}}_{b} \; {\delta^{b}}_{b'} \; {\delta^{c'}}_{c} \;  {\delta^{c}}_{c'} \; \dots 
	\end{equation}
	\vspace{-22pt}
	\begin{equation}
	\dots {\delta^{a'}}_{c} \; {\delta^{c}}_{a'} \; {\delta^{b'}}_{a} \; {\delta^{a}}_{b'} \; {\delta^{c'}}_{b} \;  {\delta^{b}}_{c'} \; \dots 
	\end{equation}
	\begin{equation}
	\dots {\delta^{a'}}_{a} \; {\delta^{a}}_{b'} \; {\delta^{c'}}_{b} \; {\delta^{b}}_{a'} \; {\delta^{b'}}_{c} \;  {\delta^{c}}_{c'} \; \dots 
	\end{equation}
	\label{eq:delta_contractions}
\end{subequations}
\begin{figure}[t]
\centering
\begin{tikzpicture}[scale=0.5,every node/.style={draw,shape=circle,fill=black,scale=0.4}]
	

	\def\height{4}
	\def\width{7}
	\def\gap{2}
	
	\begin{scope}[shift={(\gap,0)}]
	
		\draw[line width=0.5mm] (0,0) -- (\width,0);
		\draw[line width=0.5mm] (0,\height) -- (\width,\height);
		
		\foreach \x in {1,2,3,4,5,6} {
			\node at (\x, 0) {}; \node at (\x, \height) {};
			\draw (\x,0) -- (\x,\height);
		}

	\end{scope}
	
	\begin{scope}[shift={(\width+2*\gap,0)}]
	
		\draw[line width=0.5mm] (0,0) -- (\width,0);
		\draw[line width=0.5mm] (0,\height) -- (\width,\height);
		
		\foreach \x in {1,2,3,4,5} {
			\node at (\x, 0) {}; \node at (\x, \height) {};
			\draw (\x,0) -- (\x+1,\height);
		}
		
		\node at (6, 0) {}; ;
		
		\draw (0.5, \height / 2) -- (1,\height);
		\draw (6,0) -- (6.5,\height / 2);

	\end{scope}
	
	\begin{scope}[shift={(2*\width+3*\gap,0)}]
	
		\draw[line width=0.5mm] (0,0) -- (\width,0);
		\draw[line width=0.5mm] (0,\height) -- (\width,\height);
		
		\foreach \x in {1,2,5,6} {
			\node at (\x, 0) {}; \node at (\x, \height) {};
			\draw (\x,0) -- (\x,\height);
		}
		
		\draw (3,0) -- (3.4, 0.4 * \height); \draw (3.6, 0.6 * \height) -- (4, \height);
		\draw (4,0) -- (3,\height);
		
		\node at (3, 0) {}; \node at (3, \height) {};
		\node at (4, 0) {}; \node at (4, \height) {};

	\end{scope}
	
\end{tikzpicture}
\caption[Possible types of Wick contractions between single trace operators]{Possible types of Wick contractions (vertical lines) between single trace operators. The constituent scalar fields are represented by dots in the horizontal lines, which represent the successive index contractions due to the trace. First two figures are examples of planar contractions while the last one is an example of a non-planar contraction.}
\label{fig:planar_nonplanar}
\end{figure}
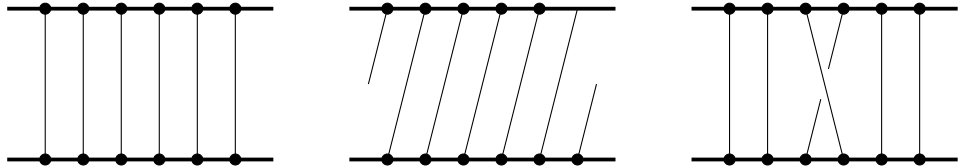
These contractions have a graphical interpretation. Consider the scalar field ${Z^a}_b$ as a dot and each contraction of the adjoint indices as a line connecting these dots, then the chiral primary operator $\Psi$ is simply a circle due to the trace. Wick's theorem says that in order to find the correlator $<\Psi(x) \tilde{\Psi}(x)>$ we must sum all possible ways we can connect the dots in the circle of $\Psi$ to the dots in the circle of $\tilde{\Psi}$. All the delta function contractions that we get after expanding the correlator represent precisely all the possible ways we can contract the dots in the circles. The three excerpts of contractions shown in (\ref{eq:delta_contractions}) can be represented graphically as shown in figure \ref{fig:planar_nonplanar}. One can immediately notice that the first two are planar, while the third one is intersecting itself. Evaluating the three contractions we immediately see that planar ones produce a factor of $N^3$ while the non-planar one produces a factor of $N$, i.e. non-planar diagrams are suppressed and we can discard them once we take the planar limit $N \rightarrow \infty$. 
All that's left then are cyclic permutations of lines by shifting all of them as seen in \mbox{figure \ref{fig:planar_nonplanar}} while going from the left to the middle diagram. 
There are $L-1$ shifts that can be done in this way, since after making a full circle we return to the initial configuration. 
Thus finally for the chiral primary correlator at tree level we find
\begin{equation}
	\left< \Psi(x) \, \tilde{\Psi}(y) \right>_{\mathrm{tree}} = \frac{L N^L}{|x-y|^{2L}},
\end{equation}
where $N^L$ comes from the contractions and $L$ from all the possible planar ways we can contract. 
This can easily be generalized for correlators of operators with arbitrary scalar fields $\Phi_{I_1 I_2 \dots I_L}(x) = \tr \left[ \Phi_{I_1}(x) \Phi_{I_2}(x) \dots \Phi_{I_L}(x) \right]$ to
\begin{equation}
	\left< \Phi_{I_1 I_2 \dots I_L}(x) \, \tilde{\Phi}^{J_1 J_2 \dots J_L}(y)  \right>_{\mathrm{tree}} = \frac{1}{|x-y|^{2L}} \left( \delta_{I_1}^{J_1} \delta_{I_2}^{J_2} \dots \delta_{I_L}^{J_L} + \mathrm{cycles} \right),
	\label{eq:tree_correlator}
\end{equation}
where ``cycles'' refers to terms with the $J$ indices pushed. 
$I$ and $J$ are flavor indices, the color indices are suppressed.
 
So far so good, but in order to calculate anomalous dimensions we have to go beyond tree level. 
This may seen like a highly non-trivial thing to do, since we expect not only scalar interactions, but also gluon exchanges and fermion loops appearing. 
Luckily the symmetry of the theory allows one to calculate all gluon and fermion effects in one go. 
First let's concentrate on the bosonic sector of the theory ignoring gluons. 
The action (\ref{eq:n4_action}) contains a single scalar-only interaction term
\begin{eqnarray}
S_{\Phi} & = &  \frac{g_{YM}^2}{4} \sum_{I,J} \int d^4 x \, \tr \, [\Phi_I, \Phi_J] [\Phi_I, \Phi_J] \nonumber \\
& = & - \frac{g_{YM}^2}{2} \sum_{I,J} \int d^4 x \, \left( \tr \, \Phi_I\Phi_I\Phi_J\Phi_J - \tr \, \Phi_I\Phi_J\Phi_I\Phi_J \right).
\end{eqnarray} 
In order to calculate the correlator (\ref{eq:tree_correlator}) at one-loop level, one should insert this term and Wick contract. 
Just like in tree level, we only have to keep planar diagrams. 
For the interaction terms this means that only neighbouring fields can interact. 
This drastically reduces the number of terms we get after Wick contracting. 
Because of that it is enough to consider a length two operator $\Phi_{I_k I_{k+1}}$ and with a bit of work one can show that at one-loop level we get
\begin{eqnarray}
	 & \left< \Phi_{I_k I_{k+1}}(x) \, \tilde{\Phi}^{J_k J_{k+1}} \right>_{\mathrm{one-loop}}  = \frac{\lambda}{16\pi^2} \frac{\log (\mu^2|x-y|^2)}{|x-y|^{2L}} \times \nonumber \\ 
	& \times  \left( 2 {\delta_{I_k}}^{J_{k+1}} {\delta_{I_{k+1}}}^{J_{k}} - \delta_{I_k I_{k+1}} \delta^{J_k J_{k+1}} - {\delta_{I_k}}^{J_{k}} {\delta_{I_{k+1}}}^{J_{k+1}} \right),
	\label{eq:loop_correlator}
\end{eqnarray}
where $\lambda = g_{YM}^2 N$ is the t'Hooft coupling. 
Comparing this to (\ref{eq:tree_correlator}) we see that effectively the interactions permute and contract the delta function indices. 
We can introduce exchange and trace operators to make this explicit. 
The permutation operator, also called the exchange operator, $\mathcal{P}_{l,l+1}$ is defined by it's action on a set of delta functions,
\begin{equation}
	\label{eq:spin_perm}
	\mathcal{P}_{l,l+1} \; {\delta_{I_{1}}}^{J_{1}} \dots {\delta_{I_{l}}}^{J_{l}} {\delta_{I_{l+1}}}^{J_{l+1}} \dots {\delta_{I_{L}}}^{J_{L}} = {\delta_{I_{1}}}^{J_{1}} \dots {\delta_{I_{l}}}^{J_{l+1}} {\delta_{I_{l+1}}}^{J_{l}} \dots {\delta_{I_{L}}}^{J_{L}}
\end{equation}
and the trace operator $\mathcal{K}_{l,l+1}$ is defined as
\begin{equation}
	\label{eq:spin_contract}
	\mathcal{K}_{l,l+1} \; {\delta_{I_{1}}}^{J_{1}} \dots {\delta_{I_{l}}}^{J_{l}} {\delta_{I_{l+1}}}^{J_{l+1}} \dots {\delta_{I_{L}}}^{J_{L}} = {\delta_{I_{1}}}^{J_{1}} \dots {\delta_{I_l I_{l+1}}} {\delta}^{J_{l} J_{l+1}} \dots {\delta_{I_{L}}}^{J_{L}}.
\end{equation}
Using these operators we can rewrite the correlator in (\ref{eq:loop_correlator}) in a more compact notation
\begin{eqnarray}
	  \left< \Phi_{I_k I_{k+1}}(x) \, \tilde{\Phi}^{J_k J_{k+1}} \right>_{\mathrm{one-loop}} = &  \nonumber \\
	  = \frac{\lambda}{16\pi^2} \frac{\log (\mu^2|x-y|^2)}{|x-y|^{2L}} 
	 & \left( 2 \; \mathcal{P}_{k,k+1} - \mathcal{K}_{k,k+1} - 1 \right) \delta_{I_k}^{J_{k}} \delta_{I_{k+1}}^{J_{k+1}}.
\end{eqnarray}
This result includes only interactions with four scalars, however as mentioned before at one-loop level we can also have gluon interactions and fermion loops in scalar propagators. 
The nice thing about these is that such interactions don't alter the flavor index structure, i.e. there are no permutations or traces. 
Basically this happens because the gluon transforms trivially under R-symmetry and hence can't change the flavor index (which transforms under R-symmetry). 
Similarly, fermions can only appear in loops altering scalar self-energies, hence they also leave the flavor structure intact.
Thus all of these interactions contribute a constant term $C$, which we can determine later. We can generalize our one-loop result with all interactions included for operators of arbitrary length,
\begin{equation}
\begin{split}
	 & \left< \Phi_{I_1 I_2 \dots I_L}(x) \, \tilde{\Phi}^{J_1 J_2 \dots J_L}(y)  \right>_{\mathrm{one-loop}}
	  = \frac{\lambda}{16\pi^2} \frac{\log (\mu^2|x-y|^2)}{|x-y|^{2L}} \times \nonumber \\
	 & \times \sum_{l=1}^L \left( 2 \; \mathcal{P}_{l,l+1} - \mathcal{K}_{l,l+1} - 1 + C\right)
	  \left( \delta_{I_1}^{J_1} \delta_{I_2}^{J_2} \dots \delta_{I_L}^{J_L} + \mathrm{cycles} \right).
\end{split}
\end{equation}
Combining this with the tree level result (\ref{eq:tree_correlator}) and comparing to the general expression of a two-point function at one-loop level (\ref{eq:anomdim_expansion}) we can deduce the anomalous dimension $\gamma$, which now becomes an operator $\Gamma$ because of the flavor mixing. 
It is given by
\begin{equation}
	\Gamma = \frac{\lambda}{16\pi^2}\sum_{l=1}^L \left( - 2 \; \mathcal{P}_{l,l+1} + \mathcal{K}_{l,l+1} + 1 - C\right).
\end{equation}
At first sight it may seem strange that what was supposed to be a number, i.e. a correction to the mass dimension of an operator has turned out to be an operator acting on the flavor space, i.e. a matrix. 
But this is very natural and in fact expected, since interactions can change the flavor of fields and we can't be sure that an operator at the quantum level has the same flavor indices as it does at the classical level. 
This line of thinking may lead to a natural question, why do we have mixing between the scalars only and not between all the fields in the theory including fermions, which miraculously do not appear. 
It turns out that this is a one-loop feature only and mixing becomes a problem at higher loop levels \cite{Minahan:2002ve}. 
In fact it is already a problem even at one loop if one considers the eigenstates of the dilatation operator.  

Now that we have acknowledged that the anomalous dimension is a matrix and found an expression for it, the next logical step would be diagonalizing it and finding the flavor eigenstates. 
One example of such an eigenstate is the chiral primary operator $\Psi$. 
Since it contains scalar fields of only one type, the permutation and trace operators act trivially on it. Thus we see that
\begin{equation}
	\Gamma \, \Psi = \frac{\lambda}{16\pi^2}\sum_{l=1}^L \left( -2 + 1 - C \right) \Psi,
\end{equation}
but we already saw that a chiral primary has an anomalous dimension of zero, which then fixes the constant $C$ to $-1$. 
And finally we get
\begin{equation}
	\label{eq:so6_dimension}
	\Gamma = \frac{\lambda}{16\pi^2}\sum_{l=1}^L \left(2 - 2 \; \mathcal{P}_{l,l+1} + \mathcal{K}_{l,l+1} \right).
\end{equation}
A keen eye might already notice that this expression resembles a Hamiltonian of a spin chain. 
In fact, this is hardly surprising, since from the very beginning we were talking about fields as objects in some closed line, which indeed resembles a spin chain. 
Furthermore the correlators that we were calculating are nothing more that propagators from one state of the chain to another, hence no wonder that the operator describing this evolution looks like a Hamiltonian for a spin chain. 
This identification is very useful, because the spin chains that appear in AdS/CFT are integrable and can be solved exactly, which gives us hope that we can apply the same techniques here and solve the spectral problem in $\N=4$ super Yang-Mills exactly. 
The first steps towards this goal were outlined in the seminal paper \cite{Minahan:2002ve}, which launched the integrability program in AdS/CFT. 
However saying that the spectral problem can be solved exactly in this particular case is too strong, since we are only at one-loop level. 
Nevertheless one can apply the same techniques going beyond one-loop level, as we shall soon see in the coming sections.

\subsubsection{The $\mathfrak{su}(2)$ sector}

In the previous section we considered single trace operators potentially containing all six scalar fields, we also mentioned that at higher loops the remaining fields of the theory start mixing in, i.e. the scalar sector is only closed at one-loop level.
However it is easy to see that there exist sectors that are closed at all loops.
The anomalous dimension matrix is simply the dilatation operator minus the bare dimension and from the algebra of the theory we know that dilatations commute with Lorentz and R-symmetries at any value of the coupling.
We can thus conclude that only operators with the same bare dimensions, Lorentz charges and R-charges can mix when acting with the anomalous dimension matrix.
Furthermore, since this is true at any value of the coupling it must follow that all the coefficients in
\beq
	D = \sum_n \lambda^n D^{(2n)},
\eeq
commute with  Lorentz and R-symmetry generators, here $D^{(2)} \equiv \Gamma$ is the one-loop dilatation operator found in the last section. 

Arguably the simplest possible closed sector is the so-called $\mathfrak{su}(2)$ sector, containing only two scalar fields $X$ and $Z$. 
An operator with $M$ and $L-M$ scalars $X$ and $Z$ has the charges $(0, 0, L; M, L-M, 0)$, the only other operators with these charges are permutations of this operator, hence the sector is closed. 
The anomalous dimension operator in this sector is given by
\begin{equation}
	\Gamma = \frac{\lambda}{8\pi^2}\sum_{l=1}^L \left(1 - \; \mathcal{P}_{l,l+1} \right),
\end{equation}
which lacks the contraction operator term compared to \eq{eq:so6_dimension}. 
We neglect it since operators in this sector do not contain both scalars and their conjugates, thus no contractions are possible.
Up to a constant factor this is the same as the Hamiltonian for the Heisenberg spin chain (also called the XXX spin chain), which is a quantum description of a one dimensional magnet. 
The Hamiltonian is given by
\begin{equation}
	\label{eq:su2_hamiltonian}
	\mathbf{H} = \sum_{l=1}^L \left(1 - \mathcal{P}_{l,l+1} \right),
\end{equation}
which can also be rewritten in terms of Pauli matrices as
\begin{equation}
	\mathbf{H} = 2 \sum_{l=1}^L \left( \frac{1}{4} - \vec{S}_l \cdot \vec{S}_{l+1} \right), \;\;\;\; \vec{S}_l = \frac{1}{2} \vec{\sigma}_l.
\end{equation}
Hence solving the spectral problem in this sector translates into solving the Schr\"{o}dinger equation
\begin{equation}
	\mathbf{H} \, \ket{\psi} = E \, \ket{\psi},
	\label{eq:Schrodinger}
\end{equation}
where we now seek to find the energy eigenvalues for the Hamiltonian of the spin chain. 
If the chain is short, this is a trivial diagonalization problem that can be easily solved by a present day computer. 
However this problem was first solved analytically by Hans Bethe \cite{Bethe:1931ab} in a time when computers were still in their infancy. 
The original solution now goes by the name of \emph{coordinate Bethe ansatz} and it is by far one of the most important and beautiful solutions in physics in the past century, which is still very widely used even to this day. 
The idea is to make an educated guess for the wave function $\ket{\psi}$, plug it in to the Schr\"{o}dinger equation and determine when does it actually hold. 
This produces a set of algebraic Bethe ansatz equations for a set of variables unimaginatively called the Bethe roots. 
All observables can then be expressed in terms of these numbers as simple algebraic functions, thus transforming a diagonalization problem to an algebraic problem. 
This has an enormous advantage, since in the asymptotic limit, when the spin chain is very large, instead of diagonalizing an infinite matrix, the set of algebraic equations actually simplify and produce integral equations, which can be solved.

In the spin chain language the scalar fields can be treated as up and down spin states, i.e.
\begin{equation}
	\ket{\uparrow\,} = Z = \left( \begin{matrix} 1 \\ 0 \end{matrix} \right), \,\,\,\, \ket{\downarrow\,} = X = \left( \begin{matrix} 0 \\ 1 \end{matrix} \right),
\end{equation}
thus local single trace operators can be treated as states of a spin chain, e.g.
\begin{equation}
	\tr \left( XXZXXZX \right) \equiv \ket{\downarrow \, \downarrow \, \uparrow \, \downarrow \, \downarrow \, \uparrow \, \downarrow \,}.
\end{equation}
Due to the cyclicity of the trace all rotations of the chain are equivalent. 
We should also specify the periodicity boundary condition
\begin{equation}
	\vec{S}_{L+1} = \vec{S}_1.
\end{equation}
The operators $\vec{S}_l$ act as Pauli matrices on the $l$'th spin site and trivially on all the others. 
Since a spin ``chain'' with a single site would have a state space $\mathbb{C}^2$, a spin chain of length $L$ has a state space $(\mathbb{C}^2)^{\otimes_L}$, which has $2^L$ basis vectors and the Hamiltonian is then a $2^L \times 2^L$ matrix, which we need to diagonalize. 
Of course, technically the state space is smaller due to the cyclicity of the chain, however as is common in physics we stick with the redundant description for simplicity.
Working directly with Pauli matrices one can find some simple results directly, e.g. it is trivial to show that the chiral primary operator
\begin{equation}
	\ket{\Psi} = \tr \; Z^L = \ket{\uparrow \, \uparrow \, \dots \, \uparrow \,}
\end{equation}
is an eigenstate of the Hamiltonian with zero energy, i.e. it is the ferromagnetic ground state of the spin chain, which we will denote as $\ket{0}$ from now on. 
This is expected, since we know that chiral primaries have zero anomalous dimensions. 
Another eigenstate of the Hamiltonian is the \emph{single magnon} state, defined as
\begin{equation}
	\ket{\psi^{(p)}} = \sum_{n=1}^L e^{ipn} \ket{n},
\end{equation} 
where $\ket{n}$ is the ground state with the $n$'th spin flipped,
\begin{equation}
	\ket{n} = S^-_n \, \ket{0} = \ket{\uparrow \, \uparrow \, \uparrow \, \dots \, \downarrow \, \dots \, \uparrow \, \uparrow \, \uparrow \,},
\end{equation}
here $p$ is formally just a parameter, but it can be interpreted as the momentum of the excitation travelling in the spin chain. 
Due to the cyclicity of the chain the momentum is quantized,
\begin{equation}
	p = \frac{2\pi}{L} n, \,\,\,\,\, n \in \mathbb{Z},
\end{equation}
where $n$ is the mode number. 
The energy of the excitation is given by the dispersion relation, which we find using the Hamiltonian \eq{eq:su2_hamiltonian},
\begin{equation}
	E(p) = 4 \, \mathrm{sin}^2 \, \frac{p}{2}.
	\label{eq:magnon_energy}
\end{equation}
Now consider a two magnon state
\begin{equation}
	\ket{\psi^{(k,p)}} = \sum_{n<m} \psi^{(k,p)}(n, m) \, \ket{n,m},  \,\,\,\,\, \ket{n,m} = S^-_n \, S^-_m \, \ket{0},
\end{equation}
where $k$ and $p$ are the momenta of the excitations. The situation is not so trivial this time, since the two magnons might scatter among themselves. We now plug this into (\ref{eq:Schrodinger}) and find the conditions for $\psi^{(k,p)}(n,m)$, which are
\begin{equation}
\begin{split}
	E \, \psi^{(k,p)}(n,m) = 4 \, \psi^{(k,p)}(n,m) & -  \psi^{(k,p)}(n+1,m)  -  \psi^{(k,p)}(n-1,m) \\ 
	                              & -  \psi^{(k,p)}(n, m+1)  -  \psi^{(k,p)}(n, m-1),
\end{split}
\end{equation}
when $m > n+1$ and
\begin{equation}
	E \, \psi^{(k,p)}(n, n+1) = 2 \, \psi^{(k,p)}(n, n+1) - \psi^{(k,p)}(n-1, n+1) - \psi^{(k,p)}(n, n+2),
\end{equation}
when $m = n+1$, i.e. when the two magnons scatter. 
The solution is now a superposition of single magnon states
\begin{equation}
	\psi^{(k,p)}(n,m) = e^{ikn + ipm} + S(k,p) e^{ipn + ikm},
\end{equation}
where
\begin{equation}
	S(p,k) = \frac{\frac{1}{2} \mathrm{cot} \frac{k}{2} - \frac{1}{2} \mathrm{cot} \frac{p}{2} - i}{\frac{1}{2} \mathrm{cot} \frac{k}{2} - \frac{1}{2} \mathrm{cot} \frac{p}{2} + i}
\end{equation}
is the scattering matrix. 
As required, such a state is an eigenstate and the energy is given by
\begin{equation}
	E = E(p) + E(k),
\end{equation}
i.e. it is simply the sum of the single magnon energies. 
Finally the spin chain periodicity condition imposes the following equations
\begin{equation}
	e^{ikL} \, S(p,k) = e^{ipL} \, S(k, p) = 1.
\end{equation}
It is now straightforward to generalize this procedure, which is exactly what Bethe did. 
The wave function for $M$ spins down can be written as
\begin{equation}
	\ket{\psi^{(p_1,p_2,\dots,p_M)}} = \sum_{1 \leq l_1 < l_2 < \dots < l_M \leq L} \psi^{(p_1,p_2,\dots,p_M)}(l_1, l_2, \dots, l_M) \, S^-_{l_1} \, S^-_{l_2} \, \dots \, S^-_{l_M} \, \ket{0}.
\end{equation}
The sum is chosen in a way so as not to over count states. The Bethe ansatz is the educated guess of the wave function
\begin{equation}
	\psi^{(p_1,p_2,\dots,p_M)}(l_1, l_2, \dots, l_M) = \sum_{\sigma \, \in \; perm(1,2,\dots\,M)} A(p) \, e^{ip_{\sigma_1} l_1 + ip_{\sigma_2} l_2 + \dots + ip_{\sigma_M} l_M}, 
\end{equation}
where the sum runs over all permutations of the down spin labels $1, 2, \dots, M$. 
$p_i$ are the momenta of the down spins, which can be treated as excitations moving in the vacuum state of the spin chain. 
The ansatz then looks like a superposition of plane waves. 
As in the two magnon case, one should now plug in the ansatz and find the conditions that make it work. 
The result is a set of algebraic equations, called the \emph{Bethe equations}
\begin{equation}
	e^{ip_k L} = - \prod_{\substack{j=1 \\ j \neq k}}^M \frac{e^{ip_j} - e^{ip_k} + 1}{e^{ip_k} - e^{ip_j} + 1} \,\,\,\, \mathrm{for} \,\, k = 1,2, \dots, M
	\label{eq:bethe_coordinate}
\end{equation}
and the amplitude is given by
\begin{equation}
	A(r) = \mathrm{sign}(\sigma) \prod_{j<k} \left( e^{ip_j} - e^{ip_k} + 1 \right).
\end{equation}
These equations can be interpreted physically once rewritten as
\begin{equation}
	e^{ip_k L} \prod_{\substack{j=1 \\ j \neq k}}^M S(p_j, p_k) = 1, \,\,\,\, \mathrm{where} \,\, S(p_j, p_k) = -\frac{e^{ip_k} - e^{ip_j} + 1}{e^{ip_j} - e^{ip_k} + 1}.
	\label{eq:CBA}
\end{equation}
This is simply saying that if we take a magnon, carry it around the spin chain, the total phase change which is a result of free propagation (represented by $e^{ip_k L}$) and scattering with other magnons (due to $S(p_j,p_k)$) must be trivial.  
Changing variables to
\begin{equation}
	\label{eq:up_param}
	e^{ip_k} = \frac{u_k + i/2}{u_k - i/2}, \;\;\;\; u_k = \frac{1}{2} \, \mathrm{cot} \, \frac{p_k}{2},
\end{equation}
brings the Bethe equations (\ref{eq:bethe_coordinate}) to a more familiar form
\begin{equation}
	\label{eq:su2_bae}
	\left( \frac{u_k + i/2}{u_k - i/2} \right)^L = \prod_{\substack{j=1 \\ j \neq k}}^M \frac{u_k - u_j + i}{u_k - u_j - i},
\end{equation}
where now one solves for the Bethe roots $u_k$, also known as magnon rapidities. 
It is now straightforward to see that this general solution reproduces the two magnon scenario we discussed earlier. 
The energy of the $M$ magnon state is given by
\begin{equation}
	E = \sum_{k=1}^M \frac{1}{u_k^2 + 1/4},
\end{equation}
 which also agrees with the single and two magnon examples. 
 
They key thing worth noting in (\ref{eq:CBA}) is that the spin chain can be fully described in terms of the scattering matrix for just two particles, i.e. the full $M$ particle scattering matrix factorizes.
This is the defining property of integrability, since factorized scattering means that individual momenta are conserved in each two particle scattering producing a tower of conserved quantities -- just the thing one would want in an integrable system. 
 
\subsubsection{The $\mathfrak{sl}(2)$ sector}
\label{sec:sl2_sector}

The $\mathfrak{su}(2)$ sector has a finite dimensional state space for a given length $L$ of the spin chain since we are dealing with finite dimensional representations of a compact group.
The simplest non-compact closed sector is the $\mathfrak{sl}(2)$ sector, which consists of operators of the form \cite{Beisert:2003yb}
\beq
	\label{eq:sl2_operators}
	\mathcal{O} = \tr \( Z^{J-1} \, \mathcal{D}_+^S \, Z \) + \mathrm{permutations},
\eeq
where $\mathcal{D_+} = \mathcal{D}_1 + i\, \mathcal{D}_2$ is the lightcone covariant derivative with global charges given by $(\frac{1}{2},\frac{1}{2}, 1; 0,0,0)$. Mixing simply redistributes the $S$ covariant derivative applications among the $J$ scalars. In this case we are dealing with infinite dimensional representations of $\mathfrak{sl}(2)$, namely the number of covariant derivatives is in principle unlimited.

It is convenient to introduce a creation-annihilation operator algebra by defining 
\beq
	{(\mathbf{a}^\dagger)}^n \, \ket{0} \equiv \frac{1}{n!}(\mathcal{D}_+)^n \, Z,  
\eeq
where $\ket{0}$ is the state annihilated by $\mathbf{a}$. The canonical commutator is defined as usual with $[\mathbf{a}, \mathbf{a}^\dagger] = 1$. The sector is then invariant under the $\mathfrak{sl}(2)$ subalgebra of the full supercoformal algebra given by 
\beq
	J'_{-} = \mathbf{a}^\dagger, \;\;\; J'_3 = \frac{1}{2} + \mathbf{a}^\dagger \mathbf{a}, \;\;\; J'_{+} = \mathbf{a} + \mathbf{a}^\dagger \mathbf{a}^\dagger \mathbf{a},
\eeq
with the defining commutation relations among them
\beq
	[J'_+, J'_-] = -2 \, J'_3, \;\;\; [J'_3, J'_\pm] = \pm J'_\pm.
\eeq
Taking a trace of $J$ operators with $S$ covariant derivatives is then equivalent to an $\mathfrak{sl}(2)$ spin chain with $s=-1/2$ representations at each site. 
The Hamiltonian density is given by \cite{Beisert:2003yb}
\beq
	\mathbf{H}_{ij} \; {(\mathbf{a}^\dagger_i)}^{k} {(\mathbf{a}^\dagger_j)}^{m} \ket{00}  = \sum_{k'=0}^{m+k}  \( \delta_{k=k'} \( h(k) + h(m) \) - \frac{\delta_{k \neq k'}}{|k-k'|} \) {(\mathbf{a}^\dagger_i)}^{k'} {(\mathbf{a}_j^\dagger)}^{m+k-k'} \ket{00},
\eeq
where $h(k)$ is the $k$'th harmonic number given by $\sum_{i=1}^k 1/i$. 
The Hamiltonian is a sum of nearest neighbour interactions
\beq
	\label{eq:sl2_hamiltonian}
	\mathbf{H} = \sum_{i=1}^J \mathbf{H}_{i,i+1}.
\eeq
The spin chain also admits a set of Bethe ansatz equations for the spectrum given by
\begin{equation}
	\label{eq:sl2_bae}
	\left( \frac{u_k + i/2}{u_k - i/2} \right)^J = \prod_{\substack{j=1 \\ j \neq k}}^S \frac{u_k - u_j - i}{u_k - u_j + i},
\end{equation}
which are remarkably similar to the $\mathfrak{su}(2)$ equations \eq{eq:su2_bae}. Once the Bethe roots are found the energy of the state can be found as
\begin{equation}
	\label{eq:sl2_E}
	E = \sum_{k=1}^S \frac{1}{u_k^2 + 1/4}.
\end{equation}

The most famous $\mathfrak{sl}(2)$ operator is the so called Konishi operator
\beq
	\mathcal{O}_{K} \equiv \tr \( \mathcal{D}_+ Z \, \mathcal{D}_+ Z \) - \tr \( Z \, \mathcal{D}_+^2 Z  \).
\eeq
It has the classical dimension $\Delta_0 = 4$, which is obvious from dimensionality. 
A simple calculation shows that it is an eigenstate of the Hamiltonian \eq{eq:sl2_hamiltonian} with eigenvalue $12$. 
The same result can also be found from the Bethe ansatz equations \eq{eq:sl2_bae}, \eq{eq:sl2_E}.
It turns out that the Konishi operator is an eigenstate of the dilatation operator at all loops \cite{Ryzhov:2003kk}, thus it is a very convenient object to study.
So far we can summarize our knowledge of its anomalous dimension as a weak coupling expansion
\beq
	\Delta = 4 + 12 g^2 + \ord{g^4}.
\eeq
Later sections of this thesis will be mostly concerned with the strong coupling expansion of this anomalous dimension.

\subsubsection{Arbitrary sectors}

The Bethe ansatz equations for the $\mathfrak{su}(2)$ sector \eq{eq:su2_bae} and for the $\mathfrak{sl}(2)$ sector \eq{eq:sl2_bae} look remarkably similar, suggesting that there might be a generalization for arbitrary algebras and representations.
And indeed such equations exist, they are given by \cite{Saleur:2000kk} 
\begin{equation}
	\left( \frac{u_{i,k} + \frac{i}{2} V_{k}}{u_{i,k} - \frac{i}{2} V_{k}} \right)^L = \prod_{l=1}^r \prod_{\substack{j=1 \\ j \neq i}}^{J_l} \frac{u_{i,k} - u_{j,l} + \frac{i}{2} M_{kl}}{u_{i,k} - u_{j,l} - \frac{i}{2} M_{kl}},
	\label{eq:general_bae}
\end{equation}
where $M_{kl}$ is the Cartan matrix of the symmetry algebra and $V_{k}$ is the vector of highest weights for the representation that the spin sites live in. This is a set of equations for the Bethe roots $u_{k,i}$, where $k=1,\dots,\mathrm{rank}(G)$ and $i = 1,\dots,J_{k}$ with $J_k$ being the number of excitations of type $k$ (each type corresponds to a different node of the Dynkin diagram, hence $k$ has $\mathrm{rank}(G)$ possible values). The total number of excitations is then $J = \sum J_k$. All of the conserved charges of the system can now be given in terms of the Bethe roots as
\begin{equation}
	Q_r = \frac{i}{r-1} \sum_{l=1}^{r} \sum_{j=1}^{J_r} \left( \frac{1}{\left(u_{j,l} + \frac{i}{2} V_{l}\right)^{r-1}} - \frac{1}{\left(u_{j,l} - \frac{i}{2} V_{l}\right)^{r-1}} \right).
\end{equation} 
In particular energy is simply the second conserved charge, $E=Q_2$.
It is now trivial to check that these equations reproduce all of the Bethe equations discussed so far. 
It is also a matter of simple algebra to derive them for other closed sectors, such as  $\mathfrak{su}(2|3)$ or even the full superconformal algebra $\mathfrak{psu}(2,2|4)$.


\subsection{Higher loops and asymptotic length}

The next step in solving the spectral problem is increasing the loop level. 
For the $\mathfrak{su}(2)$ sector this has first been done for two-loops by using symmetry constraints to fix the structure of the operator. 
The resulting dilatation operator is given by \cite{Beisert:2003tq}
\begin{equation}
	\Gamma_{2-loop} = \frac{\lambda}{8\pi^2}\sum_{l=1}^L \left(-4 + 6 \, \mathcal{P}_{l,l+1} - \left( \mathcal{P}_{l,l+1} \, \mathcal{P}_{l+1,l+2} + \mathcal{P}_{l+1,l+2} \, \mathcal{P}_{l,l+1} \right) \right).
\end{equation}
In the spin chain picture this corresponds to a Hamiltonian for a long range spin chain with two nearest neighbour interactions. 
This is hardly surprising, in fact one can expect the range of the spin chain to increase together with the loop level, as can be easily seen from diagrammatic arguments.
This long range spin chain is known to be integrable.
In fact one can reverse the problem and ask what is the most general form of a long range spin chain Hamiltonian that is still integrable, given its nearest neighbour truncation. 
The result, up to unknown constant factors, has been worked out \cite{Bargheer:2009xy} and the structure of the Hamiltonian matches loop calculations that are currently available up to five loops. 
It is now widely believed that the dilatation operator is integrable to all loops.

The method of long range spin chain deformations also predicts how the Bethe ansatz equations get modified at higher loops. 
Surprisingly the only changes that have to be introduced are the rapidity map
\beq
	u_k + \frac{i}{2} V_k \rightarrow x\(u_k + \frac{i}{2} V_k\), \;\;\; u(x) = x + \sum_{k=3}^\infty \frac{\alpha_k}{x^{k-2}},
\eeq
and the dressing phase $\sigma^2(u_k, u_j)$ for the scattering matrix
\beq
	\label{eq:deformed_dressing}
	S(u_k, u_j) \rightarrow S(u_k, u_j) \, \sigma^2(u_k, u_j) \equiv S(u_k, u_j) \exp\( 2i\theta(u_k, u_j) \) ,
\eeq
with
\beq
	\label{eq:bes_phase}
	\theta(u_k, u_j) = \sum_{s>r=2}^\infty \beta_{r,s} \( q_r(u_k)q_s(u_j) - q_s(u_k)q_r(u_j) \),
\eeq
where $q_r(u_k)$ is the eigenvalue of the conserved charge $\mathbf{Q}_r$ on the single magnon state $\ket{u_k}$.
The constants $\alpha_k$ and $\beta_{r,s}$ contain dynamical information about the theory and should be determined by other means, such as loop calculations \cite{Bargheer:2009xy}.

Going higher up in loops poses an additional complication, namely the fact that interactions get long ranged and can start wrapping around short operators, this is the so-called \emph{wrapping problem}.
For starters it is easiest to avoid it by considering asymptotically long operators. 
The rapidity map for $\N=4$ SYM has been conjectured to be \cite{Beisert:2004hm}
\begin{equation}
	\label{eq:xox_rapidity}
	x + \frac{1}{x} = \frac{4 \pi}{\sqrt{\lambda}} u, \,\,\,\,\,\,\,\,\, x^{\pm} + \frac{1}{x^{\pm}} = \frac{4 \pi}{\sqrt{\lambda}} \left( u \pm \frac{i}{2} \right),
\end{equation}
whereas the dressing phase only appears at four loops \cite{Beisert:2006ez}. 
The most general form of the Bethe ansatz equations \eq{eq:general_bae} modified by the rapidity map and the dressing phase are referred to as the \emph{asymptotic Bethe ansatz} equations.
They have been extensively verified \cite{Arutyunov:2003rg,Giombi:2010zi} since their original proposal.


\subsubsection{A glimpse ahead: the slope function}
\label{sec:slope_function_aba}

It is now a simple exercise to write down the asymptotic Bethe ansatz equations for the $\mathfrak{sl}(2)$ sector, which are \cite{Arutyunov:2004vx}
\beq
	\label{eq:sl2_aba}
	\( \frac{x_k^+}{x_k^-} \)^J = \prod_{j \neq k}^S \frac{x_k^- - x_j^+}{x_k^+ - x_j^-} \, \frac{1 - 1/(x_k^+ x_j^-)}{1 - 1/(x_k^- x_j^+)} \; \sigma^2(u_k, u_j), \;\;\; k = 1, \dots S
\eeq
where 
\beq
	\Delta = J + S + \gamma(g), \;\;\; \gamma(g) = \frac{i \sqrt{\lambda}}{2\pi} \sum_{j=1}^S \( \frac{1}{x_j^+} - \frac{1}{x_j^-} \).
\eeq
This asymptotic Bethe ansatz \eq{eq:sl2_aba} is the first non-trivial exact result we encountered so far, even if only valid in the asymptotic limit. In this short paragraph we will demonstrate how it can be used to find the exact \emph{slope function} $\gamma^{(1)}(g)$, which is defined as the linear term in the small $S$ expansion of the anomalous dimension, namely
\beq
	\label{eq:slope_definition}
	\gamma(g) = \gamma^{(1)}(g) \; S + \gamma^{(2)}(g) \; S^2 + \ord{S^3}.
\eeq
The subleading coefficient $\gamma^{(2)}(g)$ is called the \emph{curvature function} and it will be the main study object of section \ref{sec:curvature}.
We address the question of what it actually means to send an integer quantity $S$ to zero in section \ref{sec:ancont}. 

The slope function was initially conjectured in \cite{Basso:2011rs} and later independently derived in \cite{Gromov:2012eg} and \cite{Basso:2012ex}, our derivation will follow the former reference. The starting point is the logarithm of the asymptotic Bethe ansatz \eq{eq:sl2_aba}, given by
\beq
	\label{eq:log_sl2_aba}
	\frac{J}{i} \log \( \frac{x_k^+}{x_k^-} \) - \sum_{j \neq k}^S \frac{1}{i} \log \( \frac{x_k^- - x_j^+}{x_k^+ - x_j^-} \, \frac{1 - 1/(x_k^+ x_j^-)}{1 - 1/(x_k^- x_j^+)} \; \sigma^2(u_k, u_j) \) = 2 \pi n_k,
\eeq
where $n_k$ is the mode number of the $k$'th Bethe root. In the small $S$ limit the number of Bethe roots also tends to zero and in this regime they stop interacting \cite{Basso:2011rs}, thus we will consider the case when $n_k = n$ and the general result will simply be a linear combination of terms with different values of $n_k$. The key idea of the derivation is assuming that the result only depends on the combination $\Lambda \equiv n \sqrt{\lambda}$ and taking the small $n$ limit. Obviously this limit is also the strong coupling limit, as $\lambda \sim 1/n^2 \rightarrow \infty$. This considerably simplifies the derivation, for starters we only need the strong coupling expansion of the dressing phase, which is given by \cite{Arutyunov:2004vx}
\beq
	\log \, \sigma(u_k, u_j) \simeq -\log \( \frac{1-1/(x_k^+ x_j^-)}{1-1/(x_k^- x_j^+)} \) + i\,(u_j - u_k) \log \( \frac{x_j^- x_k^- - 1}{x_j^- x_k^+ - 1} \frac{x_j^+ x_k^+ - 1}{x_j^+ x_k^- - 1} \).
\eeq
Also, since $u_k \sim 1/n$ we can simplify the shifts in the rapidities $u_k$, namely
\beq
	x_k^\pm = x\(u_k \pm \frac{i}{2}\) = x\( \frac{1}{g} \(x_k + \frac{1}{x_k}\) \pm \frac{i}{2} \) = x_k \pm \frac{i}{2g}\frac{x_k^2}{x_k^2 - 1} + \ord{\frac{1}{g^2}}.
\eeq
Plugging in the leading order dressing phase expansion and getting rid of the shifts in the rapidities reduces the asymptotic Bethe ansatz equations \eq{eq:log_sl2_aba} to
\beq
	\label{eq:bae_largen}
	\sum_{j \neq k} \frac{2}{x_k - x_j} + \frac{1}{x_k} \( J + \gamma + \frac{2}{1-x_k^2} \) = \frac{\Lambda (x_k^2 - 1)}{2x_k^2},
\eeq
which are now starting to resemble equations found in matrix models.
In anticipation of this we introduce the resolvent
\beq
	\label{eq:slope_resolvent}
	G(x) = \sum_{j=1}^S \frac{1}{x-x_j},
\eeq
the anomalous dimension is then given by
\beq
	\label{eq:gamma_resolvent}
	\gamma = G(1) - G(-1).
\eeq
We now multiply \eq{eq:bae_largen} by $(x-x_k)^{-1}$ and sum over $k$, which yields
\beq
	\label{eq:Geq}
	G^2(x) + G'(x) + \( \frac{J + \gamma + 2}{x} - \frac{2x}{x^2-1} + \frac{\Lambda}{2} \frac{1-x^2}{x^2} \) G(x) = F(x),
\eeq
where
\beq
	F(x) = \frac{\Lambda}{2}\frac{G(0)+G'(0)x}{x^2} + \( J + \gamma + 2 \)\frac{G(0)}{x} - \frac{G(1)}{x-1} - \frac{G(-1)}{x+1}
\eeq
and we used the following well known identity from matrix model literature
\beq
	\sum_{j\neq k} \frac{2}{(x-x_k)(x_k-x_j)} = G^2(x) + G'(x).
\eeq
Next we expand \eq{eq:Geq} at large $x$, obviously it still has to be satisfied order by order.
In this limit $G(x) \sim S/x$ and at second to leading order we find an equation for $G'(0)$,
\beq
	\Lambda G'(0) = 2G(1) + 2G(-1) - 2G(0)(J + \gamma + 2) - \Lambda S,
\eeq
which we then stick back into $F(x)$ to produce
\beq
	F(x) = \frac{\Lambda}{2}\( \frac{G(0)}{x^2} - \frac{S}{x} \) + \frac{G(-1)}{x(x+1)} - \frac{G(1)}{x(x-1)},
\eeq
thus introducing the parameter $S$, which in principle could now be non-integer.
Finally we take the small $S$ limit by noting that $G(x) \sim S$, keeping only the leading $S$ term $\gamma^{(1)}$ in the anomalous dimension and dropping all sub-leading $S$ terms. 
What remains is a first order linear differential equation, which after integrating gives
\beq
	G(x) = \frac{x^2-1}{x^{J+2}} e^{\Lambda \frac{x^2+1}{2x}} \int_{x_0}^x dy \; F(y) \frac{y^{J+2}}{y^2-1} e^{-\Lambda \frac{y^2+1}{2y}},
\eeq
where $x_0$ is the integration constant.
We can immediately set it to zero by requiring the resolvent to be finite at the origin.
Furthermore we can fix the remaining unknown constants $G(0)$ and $G(\pm 1)$ by requiring analyticity of the resolvent, which is manifest in the definition \eq{eq:slope_resolvent}.
The integrand has poles at $\pm 1$, which may lead to logarithmic singularities after integrating, unless the residues are zero.
This provides the following constraints
\begin{subequations}
  \begin{align}
	\Lambda(G(0)-S) - G(+1)(2J+1)+G(-1) &= 0, \\
	\Lambda(G(0)+S) + G(-1)(2J+1)-G(+1) &= 0.
  \end{align}
\end{subequations}
Together with \eq{eq:gamma_resolvent} we can solve this system of equations for the unknowns $G(0)$ and $G(\pm 1)$ in terms of $S$, $J$, $\Lambda$ and $\gamma^{(1)}$.
Plugging everything in and integrating by parts we get
\beq
\label{eq:Gfinal}
G(x) = -\frac{\Lambda}{2}\frac{S}{J}-\frac{\gamma^{(1)}}{2x}-\frac{\Lambda}{4J} \frac{x^2-1}{x^{J+2}} e^{\Lambda \frac{x^2+1}{2x}} \int_0^x dy \( \gamma^{(1)} J y^{J-1} + \Lambda S y^J \) e^{-\Lambda \frac{y^2+1}{2y}}.
\eeq
We fix the final unknown $\gamma^{(1)}$ by requiring analyticity at the origin, since in general it can be a branch point. To that end we recall the integral representation of the modified Bessel function
\beq
	I_\nu(\Lambda) = \frac{(-1)^{-\nu}}{2\pi i} \oint dy \; y^{\nu-1} e^{\Lambda \frac{y^2+1}{2y}},
\eeq
where the integration contour goes around the origin counter-clockwise. 
Since this integral has no branchpoints at the origin, we see that if we tune the integrand of \eq{eq:Gfinal} in the following way
\beq
	\gamma^{(1)} J (-1)^J I_J(\Lambda) + \Lambda S (-1)^{J+1} I_{J+1}(\Lambda) = 0,
\eeq
then $G(x)$ is also regular at the origin. 
This condition fixes the leading small $S$ coefficient of the anomalous dimension, i.e. the slope function $\gamma^{(1)}$ to be
\beq
	\label{eq:slope_function}
	\gamma^{(1)}(\Lambda) = \frac{\Lambda}{J} \frac{I_{J+1}(\Lambda)}{I_J(\Lambda)}.
\eeq
This result is exact in the sense that it is valid at any value of the coupling.
We will be making use of it later in the text to extract non-trivial information about the Konishi anomalous dimension.

\subsection{Strong coupling}
\label{sec:integrability_strong}

Since integrability seems to be an all loop phenomenon as exemplified by the asymptotic Bethe ansatz, one would expect to find it at strong coupling as well where $N=4$ super Yang-Mills admits a dual string theory formulation as discussed in section \ref{sec:n4_strong}. 
In this section we showcase integrability of the classical $\adsfive$ string using the elegant language of classical spectral curves \cite{Kazakov:2004qf,Beisert:2005bm}. 

\subsubsection{The classical spectral curve}
\label{sec:spectral_curve}

Recall that classical string theory on $\adsfive$ can be formulated as a super-coset sigma model, which is defined in terms of the algebra current $J = -g^{-1} \mathrm{d} g$. 
This current has the property of being flat,
\begin{equation}
	dJ - J \wedge J = 0,
\end{equation}
what is more, one can find a one parameter family of connections \cite{Bena:2003wd}
\begin{equation}
\begin{split}
	\label{eq:lax_monodromy}
	L(x) = J^{(0)} & + \,\,\, \frac{x^2 + 1}{x^2 - 1} \; J^{(2)}  - \,\,\,\, \frac{2x}{x^2 - 1} \; \left( * J^{(0)} - \Lambda \right) \\ 
	 & + \sqrt{\frac{x+1}{x-1}} \; J^{(1)} + \sqrt{\frac{x-1}{x+1}} \; J^{(3)},
\end{split}
\end{equation} 
which are flat for any $x$,
\begin{equation}
	dL(x) - L(x) \wedge L(x) = 0.
\end{equation}
Here $L(x)$ is the \emph{Lax connection} and $x$ is the spectral parameter. 
The existence of such a set of connections signals that the theory is at least classically integrable. 
This can be shown by constructing the monodromy matrix
\begin{equation}
	\label{eq:monodromy}
	\Omega(x) = \mathcal{P} \; \mathrm{exp} \oint_\gamma L(x),
\end{equation}
where $\gamma$ is any path wrapping the worldsheet cylinder. 
Since the connection is flat, by definition it is path independent and we can evaluate the integral along any constant $\tau$ loop. 
Furthermore, shifting the $\tau$ value corresponds to doing a similarity transformation on the monodromy matrix, meaning that the eigenvalues must be time independent. 
Thus we have an infinite tower of conserved charges, hinting that the theory is integrable. 
Technically to prove classical integrability one has to also show that the conserved charges are local and that they are in involution with each other \cite{Das:2004hy,Berkovits:2004jw}.
In order to find the eigenvalues of the monodromy matrix one has to solve the characteristic equation, which in this case is a polynomial of order eight.
This equation in turn defines an eight-sheeted Riemann surface where the sheets can be connected with branch cuts of the square root type.
We refer to this surface as the \emph{classical spectral curve} or alternatively the algebraic curve. 
Denote the eigenvalues of the monodromy matrix as
\begin{equation}
	\{ e^{i\hat{p}_1(x)}, \; e^{i\hat{p}_2(x)}, \; e^{i\hat{p}_3(x)}, \; e^{i\hat{p}_4(x)} \; | \; e^{i\tilde{p}_1(x)}, \; e^{i\tilde{p}_2(x)}, \; e^{i\tilde{p}_3(x)}, \; e^{i\tilde{p}_4(x)} \},
\end{equation} 
where the quantities $p(x)$ are called \emph{quasi-momenta} and we use the convention where hatted quantities correspond to $AdS_5$ variables and quantities with tildes correspond to $S^5$. 
The quasi-momenta $p(x)$ being logarithms of the eigenvalues live on the sheets of this Riemann surface.
The key idea here is that the quasi-momenta provide an alternative representation of classical string solutions which lends itself better to generalization. 
Namely, this description of classical solutions is more natural in light of integrability, furthermore it is better suited for quantization.

\begin{figure}[t]
	\centering
		\includegraphics[width=0.75\textwidth]{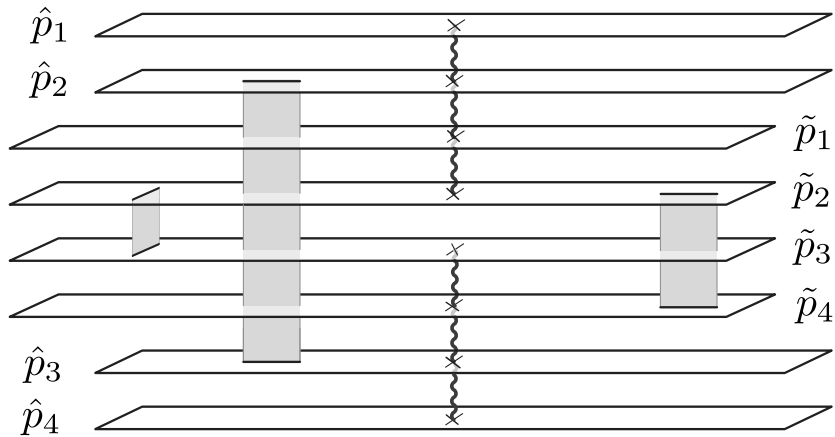}
	\caption[Example of a classical spectral curve]{Examples of cuts connecting the eight sheets of the Riemann surface corresponding to the classical spectral curve for strings in $AdS_5 \times S^5$. The wavy line corresponds to the pole at $x = 1$.}
	\label{fig:cuts}
\end{figure}

The physical picture is that each cut between two sheets represents an excitation whose polarization is determined by the sheets it connects, an example is shown in figure \ref{fig:cuts}. 
Four of the eight sheets correspond to the $AdS_5$ part of the string target space and the other four to the $S^5$ part. 
$\adsfive$ has $16 = 8_B + 8_F$ types of excitations, in the algebraic curve language this is implemented as a requirement that only sheets from the following sets be connected
\begin{equation}
	\label{eq:polarizations}
	i \in \{ \, \tilde{1}, \, \tilde{2}, \, \hat{1}, \, \hat{2} \, \}, \;\;\;\; j \in \{ \, \tilde{3}, \, \tilde{4}, \, \hat{3}, \, \hat{4} \, \},
\end{equation}
furthermore bosonic excitations correspond to cuts between sheets of the same type (hat or tilde), whereas fermionic excitations connect sheets of different types. 
Obviously fermions do not exist at the classical level, thus cuts can only represent the 8 types of bosonic excitations.
Fermionic excitations start appearing as microscopic cuts, i.e. poles during quantization. 
Solutions in closed sectors, e.g. strings moving in the $\mathbb{R} \times S^3$ submanifold of the target space will be limited to cuts between a subset of the eight sheets.

Denote the branch cut between sheets $i$ and $j$ as $\mathcal{C}^{ij}$, the quasi-momenta on these sheets have discontinuities when going through the cut given by
\begin{equation}
	p_i(x + i\epsilon) - p_j(x - i\epsilon) = 2 \pi n_{ij},
	\label{eq:cut_condition}
\end{equation}  
where $n_{ij}$ is an integer mode number arising due to the logarithm.
For each cut we associate the so called \emph{filling fraction} defined by
\begin{equation}
	\label{eq:filling}
	S_{ij} = \pm \frac{\lambda}{8 \pi^2 i} \oint_{\mathcal{C}^{ij}} \left( 1 - \frac{1}{x^2} \right) p_i(x) dx,
\end{equation}
where the sign is $+1$ for $i=\hat{1},\hat{2}$ and $-1$ for $i=\tilde{1},\tilde{2}$. These are the action variables of the theory \cite{Dorey:2006zj}, roughly they measure the length of the cut and in the physical picture they correspond to the amplitude of the excitation. 
They can be shown to take on integer values, which is natural since we anticipate the classical cuts to be collections of large numbers of poles which condense in the classical limit.
Thus we see that the algebraic curve construction acts like a Fourier decomposition -- string solutions are described as collections of excitations each having definite polarizations, mode numbers and amplitudes.

Let us now review some of the analyticity properties of the quasi-momenta. 
Since the Lax connection has poles at $x = \pm 1$, so do the quasi-momenta (as shown in figure \ref{fig:cuts}). 
Due to the Virasoro constraint, which comes about from the diffeomorphism invariance of the worldsheet, the residues of the quasi-momenta are constrained to
\begin{equation}
	\label{eq:residue_sync}
	\{ \hat{p}_1, \; \hat{p}_2, \; \hat{p}_3, \; \hat{p}_4 \; | \; \tilde{p}_1, \; \tilde{p}_2, \; \tilde{p}_3, \; \tilde{p}_4 \} = \frac{\{ \, \alpha_{\pm}, \, \alpha_{\pm}, \, \beta_{\pm}, \, \beta_{\pm}, \, | \, \alpha_{\pm}, \, \alpha_{\pm}, \, \beta_{\pm}, \, \beta_{\pm} \}}{x \pm 1}.
\end{equation}
An additional constraint on the quasi-momenta comes from the fact that the algebra $\alg{psu}{2,2|4}$ has an automorphism, which is the cause for the $\mathbb{Z}_4$ grading. 
The constraints are given by \cite{Gromov:2008ec}
\begin{eqnarray}
	\label{eq:quasi_inversion}
	\tilde{p}_{1,2}(x) & = & -\tilde{p}_{2,1}(1/x) - 2 \pi m \nonumber \\
	\tilde{p}_{3,4}(x) & = & -\tilde{p}_{4,3}(1/x) + 2 \pi m \nonumber \\
	\hat{p}_{1,2,3,4}(x) & = & -\hat{p}_{2,1,4,3}(1/x).
\end{eqnarray}
These relations define an inversion symmetry $x \rightarrow 1/x$. 
Finally one can look at the asymptotics of the quasi-momenta as the spectral parameter becomes infinite. In this limit the Lax connection becomes related to the Noether currents of the theory and hence one can relate the quasi-momenta to the charges of the global symmetry algebra by \cite{Arutyunov:2004yx,Beisert:2005bm}
\begin{equation}
\left(
\begin{array}{c}
  \hat{p}_1 \\
  \hat{p}_2 \\
  \hat{p}_3 \\
  \hat{p}_4 \\
  \hline
  \tilde{p}_1 \\
  \tilde{p}_2 \\
  \tilde{p}_3 \\
  \tilde{p}_4 \\
\end{array}
\right) = \frac{2 \pi}{x}
\left(
\begin{array}{c}
  + \mathcal{E} - \mathcal{S}_1 + \mathcal{S}_2 \\
  + \mathcal{E} + \mathcal{S}_1 - \mathcal{S}_2 \\
  - \mathcal{E} - \mathcal{S}_1 - \mathcal{S}_2 \\
  - \mathcal{E} + \mathcal{S}_1 + \mathcal{S}_2 \\
  \hline
  + \mathcal{J}_1 + \mathcal{J}_2 - \mathcal{J}_3 \\
  + \mathcal{J}_1 - \mathcal{J}_2 + \mathcal{J}_3 \\
  - \mathcal{J}_1 + \mathcal{J}_2 + \mathcal{J}_3 \\
  - \mathcal{J}_1 - \mathcal{J}_2 - \mathcal{J}_3 \\
\end{array}
\right),
\label{eq:quasi_asymptotics}
\end{equation}
where the charges are rescaled by $\mathcal{Q} = Q / \sqrt{\lambda}$. 
Thus we see that we can characterize the quasi-momenta by describing their behaviour at poles, under symmetries, by their asymptotics and their filling fractions. 

Finally one may ask how this picture of Riemann surfaces with cuts emerges from the gauge theory perspective where the spectrum is described by the Bethe ansatz. 
In the scaling limit when lengths of operators become large the Bethe roots $u_i$ start condensing in the complex plane and start looking like cuts. 
Naturally it is tempting to interpret the cuts of the algebraic curve as collections of very large numbers of poles.
Ultimately a string Bethe ansatz was proposed by Arutyunov, Frolov and Staudacher describing the distribution of these poles \cite{Arutyunov:2004vx},
\beq
	\(\frac{x_j^+}{x_j^-}\)^L = \prod_{j\neq i}^M \frac{u_i - u_j + i}{u_i - u_j - i} \, \sigma^2_{AFS}(u_i, u_j),
\eeq
where $\sigma_{AFS}$ is the dressing phase found at strong coupling.
Compared to the asymptotic Bethe ansatz for the $\alg{su}{2}$ sector
\beq
	\(\frac{x_j^+}{x_j^-}\)^L = \prod_{j\neq i}^M \frac{u_i - u_j + i}{u_i - u_j - i},
\eeq
it is natural to assume that there should be an interpolating Bethe ansatz valid to all orders of the coupling constant.
Indeed the all-loop asymptotic Bethe ansatz for the full superconformal algebra was formulated \cite{Beisert:2005fw} as we described briefly in the previous sections, which interpolated nicely between gauge theory and the algebraic curve. 
In particular the dressing phase $\sigma_{AFS}$ is a strong coupling limit of the full dressing phase \eq{eq:deformed_dressing} one finds when deforming to long range spin chains.

\subsubsection{Quantization and semi-classics}

Consider a classical string solution characterized by some conserved charges, expanding the superstring action around this solution produces a quadratic lagrangian whose quantization yields the semiclassical spectrum
\beq
	\label{eq:quant_energy_full}
	E(\{N_{ij,n}\}) = E_{cl} + E_0 + \sum_{ij,n} N_{ij,n} \mathcal{E}_{ij,n},
\eeq
where $N_{ij,n}$ is the number of excited quanta with energy $\mathcal{E}_{ij,n}$. Here $ij$ label the different polarizations and $n$ the mode numbers of the excitations. The classical energy is $E_{cl}$ whereas $E_0$ is the ground state energy coming from quantization, the last two terms in \eq{eq:quant_energy_full} are analogues of $\frac{1}{2} \omega$ and $N \omega$ for the harmonic oscillator. Just like in the case of the harmonic oscillator we can infer the ground state energy given the level spacings, it is simply
\beq
	E_0 = \frac{1}{2} \sum_{ij,n}(-1)^{F_{ij}} \mathcal{E}_{ij,n},
\eeq
where $(-1)^{F_{ij}} = \pm 1$ for bosonic/fermionic excitations. In this section we will review the quantization procedure in the algebraic curve formalism \cite{Gromov:2008ec}, which is equivalent to the semi-classical computation of quadratic fluctuations in the sigma-model \cite{Frolov:2002av,Frolov:2003tu,Park:2005ji}, yet is significantly more efficient.

Roughly the idea is that given a classical string solution represented by some cuts between sheets, as seen for example in figure \ref{fig:cuts}, we perturb it by adding microscopic cuts, which can be treated as a finite number of poles.
Just like before the indices $ij$ denoting the connected sheets represent the polarization of the excitation, they can take on values given in \eq{eq:polarizations}, however unlike in the classical setting the excitations can be fermionic as well.
The introduction of these fluctuations backreacts on the classical quasi-momenta $p_k(x)$ shifting them slightly to
\beq
	p_k(x) \rightarrow p_k(x) + \delta^{ij}_n p_k(x).
\eeq
The shifted quasi-momenta still have to satisfy \eq{eq:cut_condition}, which determines the positions $x_n^{ij}$ of the poles
\beq
	\label{eq:pole_pos}
	p_i(x_n^{ij}) - p_j(x_n^{ij}) = 2 \pi \, n_{ij}.
\eeq
The fluctuation will add a pole to the quasi-momentum at this position
\beq
	\delta^{ij}_n p_k = \epsilon_k \frac{\alpha(x^{ij}_n)}{x-x_n^{ij}},
\eeq
where the signs are
\beq
	\epsilon_{\hat{1}} = \epsilon_{\hat{2}} = -\epsilon_{\hat{3}} = -\epsilon_{\hat{4}} = -\epsilon_{\tilde{1}} = -\epsilon_{\tilde{2}} = \epsilon_{\tilde{3}} = \epsilon_{\tilde{4}} = 1
\eeq
and the residue is chosen such that the filling fraction \eq{eq:filling} increases by one, namely
\beq
	\alpha(x) = \frac{4\pi}{\sqrt{\lambda}} \frac{x^2}{x^2 - 1}.
\eeq
The total shifted quasi-momentum is obtained by summing over all fluctuations
\beq
	\delta p_k \sim \sum_{ij} \epsilon_k N_n^{ij} \frac{\alpha(x_n^{ij})}{x-x_n^{ij}}.
\eeq
It still has to satisfy all the analyticity properties outlined in the previous section, this in turn imposes a lot of constraints of the shifts themselves.
The Virasoro constraint implies the synchronization of residues \eq{eq:residue_sync}, which for the shifts translates to
\begin{equation}
	\{ \delta\hat{p}_1, \, \delta\hat{p}_2, \, \delta\hat{p}_3, \, \delta\hat{p}_4 \, | \, \delta\tilde{p}_1, \, \delta\tilde{p}_2, \, \delta\tilde{p}_3, \, \delta\tilde{p}_4 \} = \frac{\{ \, \delta\alpha_{\pm}, \, \delta\alpha_{\pm}, \, \delta\beta_{\pm}, \, \delta\beta_{\pm}, \, | \, \delta\alpha_{\pm}, \, \delta\alpha_{\pm}, \, \delta\beta_{\pm}, \, \delta\beta_{\pm} \}}{x \pm 1}.
\end{equation}
Similarly the asymptotics of the quasi-momenta encode the global charges as seen in \eq{eq:quasi_asymptotics}, for the shifts this translates to
\begin{equation}
\left(
\begin{array}{c}
  \delta \hat{p}_1 \\
  \delta \hat{p}_2 \\
  \delta \hat{p}_3 \\
  \delta \hat{p}_4 \\
  \hline
  \delta \tilde{p}_1 \\
  \delta \tilde{p}_2 \\
  \delta \tilde{p}_3 \\
  \delta \tilde{p}_4 \\
\end{array}
\right) = \frac{4 \pi}{x \sqrt{\lambda}}
\left(
\begin{array}{rl}
  + \delta \Delta/2 &+ N_{\hat{1}\hat{4}} + N_{\hat{1}\hat{3}} + N_{\hat{1}\tilde{3}} + N_{\hat{1}\tilde{4}} \\
  + \delta \Delta/2 &+ N_{\hat{2}\hat{3}} + N_{\hat{2}\hat{4}} + N_{\hat{2}\tilde{4}} + N_{\hat{2}\tilde{3}} \\
  - \delta \Delta/2 &- N_{\hat{2}\hat{3}} - N_{\hat{1}\hat{3}} - N_{\tilde{1}\hat{3}} - N_{\tilde{2}\hat{3}} \\
  - \delta \Delta/2 &- N_{\hat{1}\hat{4}} - N_{\hat{2}\hat{4}} - N_{\tilde{2}\hat{4}} - N_{\tilde{1}\hat{4}} \\
  \hline
   &- N_{\tilde{1}\tilde{4}} - N_{\tilde{1}\tilde{3}} - N_{\tilde{1}\hat{3}} - N_{\tilde{1}\hat{4}} \\
   &- N_{\tilde{2}\tilde{3}} - N_{\tilde{2}\tilde{4}} - N_{\tilde{2}\hat{4}} - N_{\tilde{2}\hat{3}} \\
   &+ N_{\tilde{2}\tilde{3}} + N_{\tilde{1}\tilde{3}} + N_{\hat{1}\tilde{3}} + N_{\hat{2}\tilde{3}} \\
   &+ N_{\tilde{1}\tilde{4}} + N_{\tilde{2}\tilde{4}} + N_{\hat{2}\tilde{4}} + N_{\hat{1}\tilde{4}} \\
\end{array}
\right) + \ord{\frac{1}{x^2}},
\label{eq:quasi_shift_asymptotics}
\end{equation}
where $\delta\Delta$ is the energy shift due to the additional excitations. From here one can read off the individual fluctuation frequencies
\beq
	\label{eq:quasi_energy}
	\Omega_n^{ij} = -2 \delta_{i,\hat{1}} + \frac{\lambda}{2\pi} \lim_{x\rightarrow \infty} x \, \delta_n^{ij} p_{\hat{1}}(x)
\eeq
and the energy shift is then a sum over all frequencies
\beq
	\delta \Delta = \sum_{ij,n} N_{ij}^n \Omega_n^{ij}.
\eeq
The description outlined above is fully sufficient to calculate the semi-classical spectrum around a classical solution -- one has to find the locations of poles, find shifts to the quasi-momenta by utilizing their analyticity properties and finally calculate the 16 fluctuation frequencies and sum them up. 
This produces the energy shift
\beq
\delta \Delta = E(\{N_{ij,n}\}) - E(\{\}) = \sum_{ij,n} N_{ij}^n \Omega_n^{ij},
\eeq
Another quantity of interest is the one-loop shift
\beq
	\label{eq:one_loop_shift}
	E_0 = \frac{1}{2} \sum_{ij,n}(-1)^{F_{ij}} \Omega_n^{ij}
\eeq
appearing in the loop expansion of the energy of a string state as
\beq
	E(\{\}) = E_{cl} + E_0 + \ord{1/\sqrt{\lambda}},
\eeq
where the classical energy $E_{cl}$ is of order $\sqrt{\lambda}$ and $E_0$ is of order $1$.
One can of course proceed with semi-classical quantization, find all the fluctuation frequencies and sum them up by hand to find the one-loop shift, however it would be nicer to find the result in one go.
To that end we introduce the off-shell fluctuations $\delta^{ij} p_k(x;y)$ which are defined by the same asymptotics as the on-shell fluctuations $\delta^{ij}_n p_k(x)$ but the position of the pole is left unspecified, namely
\beq
	\delta^{ij}_n p_k(x) = \delta^{ij} p_k(x;y) \vert_{y=x_n^{ij}}.
\eeq
Note that the off-shell quantity depends on the mode number $n$, which is a function of the pole position via \eq{eq:pole_pos}, which we simply left unspecified as $y$ in the off-shell quantity.
Similarly we introduce off-shell fluctuation energies
\beq
	\Omega_n^{ij} = \Omega^{ij}(y)\vert_{y=x_n^{ij}},
\eeq
which can easily be found if the on-shell frequencies are known by
\beq
	\label{eq:off_shell_freq}
	\Omega^{ij}(y) = \Omega_n^{ij}\vert_{n\rightarrow \frac{p_i(y) - p_j(y)}{2\pi}}.
\eeq
The main advantage of introducing the off-shell frequencies is that due to the $\mathbb{Z}_4$ grading of the $\alg{psu}{2,2|4}$ algebra the quasi-momenta enjoy an inversion symmetry under $x \rightarrow 1/x$ as seen in \eq{eq:quasi_inversion}.
This constrains the off-shell frequencies as well.
Consider symmetric classical configurations that have pairwise symmetric quasi-momenta
\beq
	p_{\hat{1},\hat{2},\tilde{1},\tilde{2}} = -p_{\hat{4},\hat{3},\tilde{4},\tilde{3}},
\eeq
it is known to be the case for all rank one solutions \cite{Gromov:2008ec}, which in particular are dual to states in the $\alg{su}{2}$ and $\alg{sl}{2}$ sectors of $\N=4$ super Yang-Mills. 
All of the off-shell frequencies can then be expressed in terms of two, namely $\Omega^{\tilde2 \tilde3}$ and $\Omega^{\hat2 \hat3}$, which we denote as the basis of frequencies.
The rest of the fluctuations are given by
\beq
\begin{aligned}
\label{eq:freq_relations}
\Omega^{\tilde1\tilde4} (y) &= -  \Omega^{\tilde2 \tilde3}(1/y) +\Omega^{\tilde2\tilde3}(0)\cr 
\Omega^{\tilde2 \tilde4} (y) =
\Omega^{\tilde1\tilde3}(y)  &= {1\over 2} \left( \Omega^{\tilde2 \tilde3}(y) + \Omega^{\tilde1\tilde4} (y) \right)
                               = {1\over 2} \left( \Omega^{\tilde2 \tilde3}(y) - \Omega^{\tilde2\tilde3} (1/y)+\Omega^{\tilde2\tilde3}(0) \right)
                                 \cr
 \Omega^{\hat1\hat4} (y) & = - \Omega^{\hat2 \hat3}(1/y) -2 \cr
\Omega^{\hat2 \hat4} (y) =
\Omega^{\hat1 \hat3} (y) &= {1\over 2}  \left( \Omega^{\hat2 \hat3}(y) + \Omega^{\hat1\hat4} (y) \right)
				    = {1\over 2}  \left( \Omega^{\hat2 \hat3}(y) - \Omega^{\hat2\hat3} (1/y) \right) -1 \cr
\Omega^{\hat2\tilde4}(y) =
\Omega^{\tilde1 \hat3} (y) &= {1\over 2} \left(\Omega^{\hat2 \hat3}(y)+ \Omega^{\tilde1\tilde4} (y)  \right)
				   = {1\over 2} \left( \Omega^{\hat2 \hat3}(y)- \Omega^{\tilde2 \tilde3}(1/y)+\Omega^{\tilde2\tilde3}(0) \right)  \cr				
\Omega^{\tilde2\hat4}(y)=
\Omega^{\hat1 \tilde3} (y) &= {1\over 2} \left( \Omega^{\tilde2\tilde3} (y) + \Omega^{\hat1 \hat4}(y)\right)
				 = {1\over 2}  \left( \Omega^{\tilde2\tilde3} (y) - \Omega^{\hat2 \hat3}(1/y)\right)   -1 \cr
\Omega^{\tilde1 \hat4} (y) =
\Omega^{\hat1 \tilde4} (y) &= {1\over 2} \left( \Omega^{\tilde1 \tilde4}(y) + \Omega^{\hat1\hat4} (y) \right)
				   = {1\over 2} \left(- \Omega^{\tilde2\tilde3}(1/y) - \Omega^{\hat2\hat3}(1/y)+\Omega^{\tilde2\tilde3}(0) \right)
				   -1  \cr	
\Omega^{\hat2\hat3}(y)=
\Omega^{\tilde2 \hat3} (y) &= {1\over 2} \left( \Omega^{\tilde2 \tilde3}(y) + \Omega^{\hat2\hat3} (y) \right)	 \,.
\end{aligned}
\eeq
Knowing the off-shell frequencies and the quasi-momenta one can express the one-loop shift as a contour integral
\beq
	\label{eq:one_loop_integral}
	E_0 = \frac{1}{2} \sum_{ij} (-1)^{F_{ij}} \oint \frac{dx}{2\pi i} \( \Omega^{ij}(x) \d_x \log \sin \frac{p_i - p_j}{2} \),
\eeq
where the integrand is chosen carefully such that it contains poles at each fluctuation insertion point $x_n^{ij}$ with residues $\Omega^{ij}(x_n^{ij})$, so that the result is equivalent to \eq{eq:one_loop_shift}.
There are three contributions to one loop energy shift that are different by their nature.
They can be separated into an ``anomaly" contribution, a contribution from the dressing phase
and a wrapping contribution, which is missing in the asymptotic Bethe ansatz.
\beq
E_0=\delta \Delta_{\rm anomaly} +\delta \Delta_{\rm dressing} + \delta \Delta_{\rm wrapping}\;,
\eeq
where each of these contributions is simply an integral of some closed form expression,
\beqa
\label{eq:delta_E_3}
  \delta \Delta_{\rm anomaly} &=& -\frac{4}{ab-1}\int_a^b \frac{dx} {2\pi i}
  \frac{y(x)}{x^2-1} \partial_x \log \sin p_{\hat 2}\;,\\
  \label{eq:delta_E_1}
  \delta \Delta_{\rm dressing} &=& \sum_{ij} (-1)^{F_{ij}} \int\limits_{-1}^{1} \frac{dz} {2\pi
    i} \left( \Omega^{ij}(z) \, \partial_z \frac{i (p_i -p_j)}{2} \right)\;,\\
  \label{eq:delta_E_2}
  \delta \Delta_{\rm wrapping} &=& \sum_{ij} (-1)^{F_{ij}} \int\limits_{-1}^{1} \frac{dz} {2\pi
    i} \left( \Omega^{ij}(z) \, \partial_z \log (1- e^{-i(p_i -p_j)}) \right)\;.
\eeqa
As always $i$ takes values $\hat 1,\hat 2,\tilde 1,\tilde 2$ whereas $j$ runs over $\hat 3,\hat 4,\tilde 3,\tilde 4$.

\subsubsection{Folded string}
\label{sec:folded_string}

Operators \eq{eq:sl2_operators} from the $\alg{sl}{2}$ sector are known to be dual to folded rotating string solutions in $\adsfive$, these are closed strings rotating around their center of mass in an $AdS_3$ subspace of $AdS_5$ with spin $S$ \cite{Frolov:2002av}.
Additionally they orbit the big circle of $S^5$ with angular momentum $J$, also referred to as twist.
These parameters as expected correspond to the number of scalars $J$ and number of derivatives $S$ in the gauge theory operators. 
In the classical regime these are assumed to scale as $\sqrt{\lambda}$, thus we use $\mathcal{S} = S/ n \sqrt{\lambda}$ and $\mathcal{J} = J / n \sqrt{\lambda}$ when describing the classical solution, which corresponds to long operators in gauge theory. 
The number of spikes $n$ corresponds to the mode number $n$ in the language of Bethe states.

Given the explicit string solution (see \cite{Tseytlin:2010jv} for details) one could follow the steps outlined in the classical spectral curve construction, namely calculate the monodromy matrix \eq{eq:monodromy}, diagonalize it and extract the quasi-momenta. 
While it is indeed possible to do, we will present an alternative method based on analyticity properties of the quasi-momenta when we discuss the classical limit of cusped Wilson lines in section \ref{sec:wilson_classical}.
Here we present the result \cite{Gromov:2011de}, which consists of two ``basis'' quasi-momenta
\beqa
  \label{eq:p_a}\nn
  p_{\hat 2} &=& \pi n - 2\pi n{\cal J} \left( \frac{a}{a^2-1} -
    \frac{x}{x^2-1} \right) \sqrt{\frac{(a^2-1)
      (b^2-x^2)}{(b^2-1)(a^2-x^2)}} \\ \nn &+& \frac{8\pi\, n\, a b\, {\cal S} F_1(x)}{(b-a)(ab+1)} +
  \frac{2\pi n {\cal J} (a-b) F_2(x)}{\sqrt{(a^2-1)(b^2-1)}},\\
  \label{eq:p_s}
  p_{\tilde 2} &=& \frac{2\pi {\cal J}x}{x^2-1},
\eeqa
while the remaining functions can be determined by utilizing the $x \rightarrow 1/x$ inversion symmetry as shown in \eq{eq:quasi_inversion}, resulting in
\beqa
  \label{eq:quasi-momenta_symmetry_A}
  p_{\hat{2}} (x) &=& -p_{\hat{3}}(x) = -p_{\hat{1}}(1/x) = p_{\hat{4}}
  (1/x)\;,\\
  \label{eq:quasi-momenta_symmetry_S}
  p_{\tilde{2}} (x) &=& -p_{\tilde{3}} (x) = p_{\tilde{1}} (x) =
  -p_{\tilde{4}}(x)\;.
\eeqa
The functions $F_1(x)$ and $F_2(x)$ can be expressed
in terms of the elliptic integrals:
\begin{eqnarray}
 \label{eq:not}
 F_1(x) &=& i F \left( i \sinh^{-1}
 \left. \sqrt{\frac{(b-a)(a-x)}{(b+a)(a+x)}} \right\vert\frac{(a+b)^2}{(a-b)^2} \nonumber
 \right)\;, \\
\nn
 F_2(x) &=& i E \left( i \sinh^{-1}
 \left.\sqrt{\frac{(b-a)(a-x)}{(b+a)(a+x)}} \right\vert \frac{(a+b)^2}{(a-b)^2}
 \right)\;.
 \end{eqnarray}
This is a two-cut solution with symmetric cuts on the real axis given by the branch points $a < b$ (and $-b < -a$).
 The classical energy $\Delta$ of the folded string is a function of the Lorentz spin $\mathcal{S}$,
twist $\mathcal{J}$ and the mode number $n$. This function can be written in a parametric form
in terms of the branch points $a$ and $b$ \cite{Gromov:2011de,Beisert:2003ea,Kazakov:2004qf,Kazakov:2004nh}:
\beqa
\nn{2\pi {\cal S}}&=&\frac{ab+1}{ab}\[b E\(1-\tfrac{a^2}{b^2}\)-aK\(1-\frac{a^2}{b^2}\)\]\;,\\
{2\pi {\cal J}}&=&\frac{2\sqrt{(a^2-1)(b^2-1)}}{b}K\(1-\frac{a^2}{b^2}\)\;,\\
\nn{2\pi {\cal D}}&=&\frac{ab-1}{ab}\[b E\(1-\tfrac{a^2}{b^2}\)+aK\(1-\frac{a^2}{b^2}\)\]\;.
\eeqa
where ${\cal D}=\Delta / n\sqrt\lambda$ and $E$, $K$ are elliptic integrals of the first kind.

Next we proceed to the semi-classical quantization of the folded string solution, more precisely we are after the one-loop shift.
Instead of calculating the 16 fluctuation frequencies $\Omega_n^{ij}$ we adopt the off-shell frequency formalism outlined above.
Once again, the exact formulae for the off-shell frequencies can be found solely from analyticity constraints by first determining the off-shell shifts in quasi-momenta and then using the definitions \eq{eq:quasi_energy} and \eq{eq:off_shell_freq}.
The answer turns out to be surprisingly simple and is given by the two basis frequencies
\beqa
\label{oma}
\Omega^{\tilde 2\tilde 3}(x)&=&\frac{2}{a
  b-1}\frac{\sqrt{a^2-1}\sqrt{b^2-1}}{x^2-1}\;,\\
\label{oms}
\Omega^{\hat 2\hat 3}(x)&=&\frac{2}{a b-1}\(1-\frac{y(x)}{x^2-1}\)\;.
\eeqa
where $y(x)=\sqrt{x-a} \sqrt{a+x} \sqrt{x-b} \sqrt{b+x}$.
The remaining frequencies can be read off from the relations \eq{eq:freq_relations}.
What remains in order to find the one-loop shift is performing the sum \eq{eq:one_loop_shift} or alternatively evaluating the integral \eq{eq:one_loop_integral} numerically.
The answer involves an infinite sum over all mode numbers hence it is not very illuminating at this point.

Let us now consider the $\mathcal{S} \rightarrow 0$ limit.
The square of the classical energy has a very nice expansion in this limit
\beqa
	\label{eq:classical_energy}
 {\cal D}^2&=&{\cal J}^2+2 \, {\cal S} \, \sqrt{{\cal J}^2+1}+{\cal S}^2 \, \frac{2 {\cal J}^2+3}{2
   {\cal J}^2+2}-{\cal S}^3 \, \frac{{\cal J}^2+3}{8
   \left({\cal J}^2+1\right)^{5/2}}
   +{\cal O}\left({\cal S}^4\right)\;.\qquad
\eeqa
The one-loop shift expanded up to two orders in $\mathcal{S}$ reads
\small
\beqa
\label{eq:delta_oneloop_slope}
\Delta&\simeq&
\frac{-{\cal S}}{2 \left({\cal J}^3+{\cal J}\right)}+{\cal S}^2\[\frac{3 {\cal J}^4+11 {\cal J}^2+17
   }{16 {\cal J}^3 \left({\cal J}^2+1\right)^{5/2}}
\!-\!\sum_{\substack{m>0 \\ m \neq n}}\frac{n^3m^2  \left(2 m^2+n^2 {\cal J}^2-n^2\right)}{{\cal J}^3 \left(m^2-n^2\right)^2
   \left(m^2+n^2 {\cal J}^2\right)^{3/2}}\]. \;\;\;\;\;\;\;\;
   \;.
\eeqa
\normalsize
The sum is nothing but a sum over the fluctuation energies,
whereas the remaining terms originate from the ``zero"-modes
$m=n$, which have to be treated separately.
The sum can be very easily expanded for small ${\cal J}$.
It is easy to see that the expansion coefficients will be certain combinations
of zeta-functions. It is also easy to see that
the dependence on the mode number $n$ is rather nontrivial.
The expansion of the one loop energy first in small ${\cal S}$
up to a second order and then in small ${\cal J}$ reads
\beq
\label{delta_oneloop_sj}
\Delta_{\rm1-loop}\simeq
\left\{
\begin{array}{ll}
 -\frac{\mathcal{S}}{2 \mathcal{J}}+
 \mathcal{S}^2 \left(+\frac{1}{2\mathcal{J}^3}-\frac{3 \zeta_3}{2 \mathcal{J}}-\frac{1}{16 \mathcal{J}}\right) & \;\;\;\;\mathrm{for}\;n=1, \\
 -\frac{\mathcal{S}}{2 \mathcal{J}}+
 \mathcal{S}^2 \left(+\frac{1}{2
   \mathcal{J}^3}-\frac{12 \zeta_3}{\mathcal{J}}-\frac{17}{16 \mathcal{J}}\right) & \;\;\;\;\mathrm{for}\; n=2, \\
 -\frac{\mathcal{S}}{2 \mathcal{J}}+
 \mathcal{S}^2 \left(-\frac{5}{8
   \mathcal{J}^3}-\frac{81 \zeta_3}{2 \mathcal{J}}-\frac{7}{4 \mathcal{J}}\right) & \;\;\;\;\mathrm{for}\;n=3. 
\end{array}
\right.
\eeq
We note that the contributions ${\cal S}^2/{\cal J}^3$ are universal for $n=1$ and $n=2$, however starting from $n=3$ we get different coefficients.

\subsection{Short strings}
\label{sec:short_strings}

Even though semi-classical results are technically only valid for the charges scaling as $\sqrt{\lambda}$, there is evidence \cite{Roiban:2009aa} that re-expanding the classical energy plus the one-loop shift yields a reasonable result in terms of the unscaled charges $J$ and $S$, thus providing one the possibility to probe so-called short string states, the Konishi operator being a prime example.
Namely summing up the square root of \eq{eq:classical_energy} and \eq{delta_oneloop_sj}, introducing the unscaled charges $S = \sqrt{\mu}\, \mathcal{S}$, $J = \sqrt{\mu}\, \mathcal{J}$ and re-expanding in terms of $\mu \equiv n^2 \lambda$ yields the following one-loop energy for the folded string solution \cite{Gromov:2011de}
\beq
\Delta_{S,J,n}\simeq\sqrt{2S}\mu^{1/4}
+\frac{2J^2+3S^2-2S}{4\,(2 S)^{1/2}\,\mu^{1/4}}.
\eeq
For the Konishi operator we get
\beq
\Delta_{2,2,1}=2 \, \lambda^{1/4} + \frac{2}{\lambda^{1/4}}+ \ord{\lambda^{-3/4}},
\eeq
where the first coefficient of the expansion was first derived in \cite{Gubser:2002tv}.
The sub-leading coefficient was first suggested to be $1$ in \cite{Roiban:2009aa}, however a possible issue with the argument was soon pointed out in \cite{Arutyunov:2009ax}, where numerical calculations indeed confirmed the correct answer to be $2$.

In this section we will show how combining the knowledge of the semi-classical one-loop shift \eq{eq:delta_oneloop_slope} and the slope function \eq{eq:slope_function} can yield further information about the Konishi anomalous dimension.
Later in the text when we find the next small spin coefficient after the slope, which we call the curvature function, we will revisit the techniques presented here in order to boost the obtain results one order further.

\subsubsection{Structure of small spin expansions}
\label{sec:small_spin_structure}

Consider the classical string energy \eq{eq:classical_energy}, the re-expansion of $\Delta^2$ in the large $\mu \equiv \lambda n^2$ limit with $S$ and $J$ fixed has a particularly nice structure
\small
\beq
\Delta^2=J^2+S
\(
2\, \sqrt{\mu}+\frac{J^2}{\sqrt{\mu}}+\dots
\)
+S^2
\(
\frac{3}{2}-\frac{J^2}{2\sqrt{\mu}}
+\dots
\)
-S^3
\(
\frac{3}{8\sqrt{\mu}}
-\frac{13 J^2}{16\sqrt[3]{\mu}}
+\dots
\)
+{\cal O}({ S}^4)
\label{dsquare_tree}
\eeq
\normalsize
where each next term in $S$ gets more and more suppressed for large $\mu$. 
It was conjectured \cite{Basso:2011rs} that a generalization of this result holds for any operator at all values of the coupling constant, more precisely the statement was that making expansions of the scaling dimension squared first in $S\to 0$ and then in $\mu\to \infty$ should reveal the following structure
\small
\beq\label{eq:delta_squared_basso}
\Delta^2=J^2+S
\(
A_1\sqrt{\mu}+A_2+\dots
\)
+S^2
\(
B_1+\frac{B_2}{\sqrt\mu}
+\dots
\)
+S^3
\(
\frac{C_1}{\mu^{1/2}}
+\frac{C_2}{\mu^{3/2}}
+\dots
\)
+{\cal O}({ S}^4)\;,
\eeq
\normalsize
where the coefficients $A_i,\;B_i,\;C_i$, etc. are some functions of $J$.

\begin{figure}[t]
    \begin{tabular}{cc}
    \includegraphics[scale=0.7]{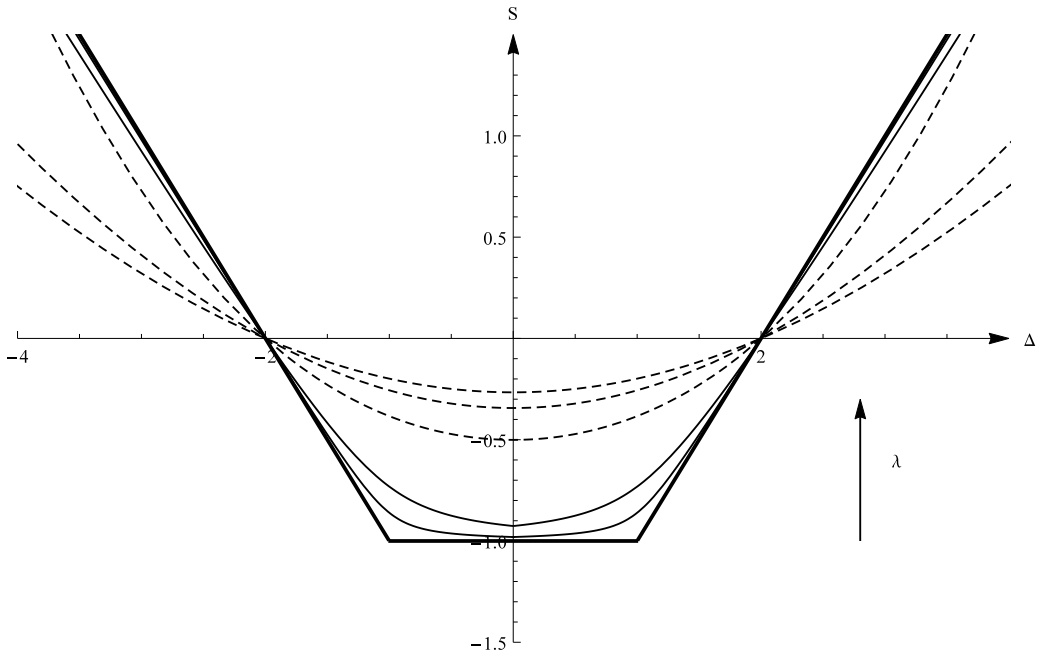}
    \end{tabular}
\caption[The BFKL trajectories $S(\Delta)$ for the twist-2 operator]{The BFKL trajectories $S(\Delta)$ for the twist-2 operator at various values of the coupling. Solid black lines are obtained using the known two loop weak coupling expansion \cite{Brower:2006ea,Kotikov:2002ab} and dashed lines are obtained using the strong coupling expansion \cite{Costa:2012cb,Kotikov:2013xu,Brower:2013jga}.}
\label{fig:bfkl}
\end{figure}

Indeed it is not hard to understand this constraint.
A good entry point is considering the inverse relation $S(\Delta)$, frequently encountered in the context of BFKL \cite{bfkl,Alfimov:2014bwa}, where one also usually sets $n=1$. 
It satisfies a few basic properties, namely the curve $S(\Delta)$ goes through the points $(\pm J, 0)$ at any coupling, because at $S=0$ the operator is BPS. 
At the same time for non-BPS states one should have $\Delta(\lambda)\propto \lambda^{1/4}\to\infty$ \cite{Gubser:1998bc} which indicates that if $\Delta$ is fixed, $S$ should go to zero, thus combining this with the knowledge of fixed points $(\pm J, 0)$ we conclude that at infinite coupling $S(\Delta)$ is simply the line $S = 0$. 
As the coupling becomes finite $S(\Delta)$ starts bending from the $S=0$ line and starts looking like a parabola going through the points $\pm J$, as shown in figure~\ref{fig:bfkl}. 
Based on this qualitative picture and the scaling $\Delta(\lambda)\propto \lambda^{1/4}$ at $\lambda\rightarrow\infty$ and fixed $J$ and $S$, one can write down the following ansatz,
\beqa
\label{eq:sofdelta}
	S(\Delta) &=& \left( \Delta^2 - J^2\right)\Bigl( \alpha_1 \frac{1}{\lambda^{1/2}} + \alpha_2 \frac{1}{\lambda} + (\alpha_3 + \beta_3 \Delta^2) \frac{1}{\lambda^{3/2}} + (\alpha_4 + \beta_4 \Delta^2) \frac{1}{\lambda^{2}}    \Bigr. \nn\\
	&+&
	\Bigl.
	(\alpha_5 + \beta_5 \Delta^2+\gamma_5\Delta^4) \frac{1}{\lambda^{5/2}}
	+(\alpha_6 + \beta_6 \Delta^2+\gamma_6\Delta^4) \frac{1}{\lambda^{3}}
	+\dots
	\Bigr).
\eeqa
The reason for omitting odd powers of the scaling dimension from the ansatz will become clear later when we discuss the $\pmu$-system, where we will see that only the square of $\Delta$ enters the equations. 
We can now invert the relation and express $\Delta$ in terms of $S$ at strong coupling, which exactly reproduces \eq{eq:delta_squared_basso}. There exists a one-to-one mapping between the coefficients $\alpha_i$, $\beta_i$, etc. and $A_i$, $B_i$ etc, which is rather complicated but easy to find.
The pattern in \eq{eq:delta_squared_basso} continues to higher orders in $S$ with further coefficients $D_i$, $E_i$, etc. and powers of $\mu$ suppressed incrementally. This structure is a non-trivial constraint on $\Delta$ itself as one easily finds from \eq{eq:delta_squared_basso} that
\small
\beqa\label{eq:delta_small_s}
\Delta&\simeq&J+\frac{S}{2J}
\(
A_1\sqrt{\mu}+A_2+\frac{A_3}{\sqrt{\mu}}+\dots
\)\\
\nn&+&S^2
\(
- \frac{A_1^2}{8J^3} \, \mu
-  \frac{A_1A_2}{4J^3} \, \sqrt{\mu}
+\[\frac{B_1}{2J}-\frac{A_2^2+2A_1 A_3}{8J^3}\]
+
\[
\frac{B_2}{2J}
-\frac{A_2A_3+A_1A_4}{4J^3}
\]  \frac{1}{\sqrt\mu}
+\dots
\),
\eeqa
\normalsize
where we introduced the mode number by the naive replacement $\lambda \to \mu \equiv n^2 \lambda$.
By definition the coefficients of $S$ and $S^2$ are the slope and curvature functions respectively as defined in \eq{eq:slope_definition}, so now we have their expansions at strong coupling in terms of $A_i,\;B_i,\;C_i$, etc. 
Since the $S$ coefficient only contains the constants $A_i$, we can find all of their values by simply expanding the slope function \eq{eq:slope_function} that we found earlier at strong coupling. 
We get
\beq
\label{eq:bassos_as}
A_1=2\;\;,\;\;
A_2=-1\;\;,\;\;
A_3=J^2-\frac{1}{4}\;\;,\;\;
A_4=J^2-\frac{1}{4}\dots\;.
\eeq
Note that in this series the power of $J$ increases by two at every other member, which is a direct consequence of omitting odd powers of $\Delta$ from \eq{eq:sofdelta}. 
We also expect the same pattern to hold for the coefficients $B_i$, $C_i$, etc.
It is not hard to see that the sub-leading coefficient in the spin $S$ only contains $A_i$ and $B_i$ coefficients, thus given the $A$'s we could in principle fix all $B$'s.
Hence we conclude that the letters in $A_i$, $B_i$, $C_i$ etc. are directly linked to the generalizations of the slope function $\gamma^{(n)}$.

\subsubsection{Two-loop prediction}

Let us now consider the case $n=1$.
We are interested in the coefficients of the strong coupling expansion of $\Delta$, namely
\beq
	\Delta = \Delta^{(0)} \lambda^\frac{1}{4} + \Delta^{(1)} \lambda^{-\frac{1}{4}}  + \Delta^{(2)} \lambda^{-\frac{3}{4}} + \Delta^{(3)} \lambda^{-\frac{5}{4}} + \dots
\eeq
First, we utilize the structure \eq{eq:delta_squared_basso} and by fixing $S$ and $J$ we re-expand the square root of $\Delta^2$ at strong coupling to find
\beq
	\label{eq:delta_abc}
	\Delta = \sqrt{A_1 S} \, \sqrt[4]{\lambda}  + \frac{\sqrt{A_1} \left( J^2 + A_2 S + B_1 S^2 \right)}{2 A_1 \sqrt{S}} \, \frac{1}{\sqrt[4]{\lambda}} + {\cal O}\(\frac{1}{\lambda^\frac{3}{4}}\).
\eeq
Thus we reformulate the problem entirely in terms of the coefficients $A_i$, $B_i$, $C_i$, etc. For example, the next coefficient in the series, namely the two-loop term is given by
\beq
	\label{eq:delta_2loops_abc}
	\Delta^{(2)} = -\frac{\left(2 A_2 + 4 B_1+J^2\right)^2-16 A_1 (A_3+2 B_2+4 C_1)}{16 \sqrt{2} A_2^{3/2}}.
\eeq
Further coefficients become more and more complicated, however a very clear pattern can be noticed after looking at these expressions: we see that the term $\Delta^{(n)}$ only contains coefficients with indices up to $n+1$, e.g. the tree level term $\Delta^{(0)}$  only depends on $A_1$, the one-loop term depends on $A_1$, $A_2$, $B_1$, etc. Thus we can associate the index of these coefficients with the loop level. Conversely, from the last section we learned that the letter of $A_i$, $B_i$, etc. can be associated with the order in $S$, i.e. the slope function fixed all $A_i$ coefficients and the curvature function in principle fixes all $B_i$ coefficients.

Looking at \eq{eq:delta_abc} we see that knowing $A_i$ and $B_i$ only takes us to one loop, in order to proceed we need to know some coefficients in the $C_i$ and $D_i$ series. This is where the knowledge of the classical energy \eq{eq:classical_energy} and its semi-classical correction \eq{eq:delta_oneloop_slope} come in handy. We add up the classical and the 1-loop contributions, take $S$ and $J$ fixed at strong coupling and compare the result to \eq{eq:delta_squared_basso}. By requiring consistency we are able to extract the following coefficients for $n=1$,
\beqa
 \label{eq:abcd2}
 \begin{array}{rcrlrlrcl}
  A_1 &=&  &2, &A_2&  &=& -&1  \\
  B_1 &=&  &3/2, &B_2&  &=& -&3\,\zeta_3+\frac{3}{8}  \\
  C_1 &=& -&3/8, &C_2& &=& &\frac{1}{16} \, (60 \, \zeta_3 + 60 \, \zeta_5 - 17) \\
  D_1 &=&  &31/64, &D_2& &=& & \frac{1}{512} (-5520 \, \zeta_3 - 5120 \, \zeta_5 -3640 \, \zeta_7 +901). 
 \end{array}
\eeqa
As discussed in the previous section, we can in principle extract all coefficients with indices $1$ and $2$. In order to find e.g. $B_3$ we would need to extend the quantization of the classical solution to the next order. 
For general mode numbers $n$ one can extract the following values for $B_i$,
\beq\label{BB}
B_1=\frac{3}{2}\;\;,\;\;
B_2=
\left\{
\bea{ll}
-3\,\zeta_3+\frac{3}{8}&\;\;,\;\;n=1\\
-24\,\zeta_3-\frac{13}{8}&\;\;,\;\;n=2\\
-81\,\zeta_3-\frac{24}{8}&\;\;,\;\;n=3
\eea
\right..\vspace{10pt}
\eeq
Combining all of this information we find the following result for spin 2 operators
\beq
\Delta_{2,J,1}=2 \, \lambda^{1/4}+
\frac{\frac{J^2}{4}+1}{\lambda^{1/4}}+\frac{-\frac{J^4}{64}+\frac{3 J^2}{8}-3\, \zeta
   (3)-\frac{3}{4}}{\lambda^{3/4}} + \ord{\frac{1}{\lambda^{5/4}}}\;,
\eeq
which for Konishi reads
\beq
\Delta_{2,2,1}=2 \, \lambda^{1/4}+
\frac{2}{\lambda^{1/4}}+\frac{\frac{1}{2} -3\, \zeta_3}{\lambda^{3/4}} + \ord{\frac{1}{\lambda^{5/4}}}\;.
\eeq
Generalizing the discussion above we conclude that this procedure yields the $n$-loop scaling dimension given the values of 
\beq
	A_{1,2,\dots,n+1}, \;\;\; B_{1,2,\dots,n}, \;\;\; C_{1,2,\dots,n-1}, \;\;\; \dots, \;\;\; A^{(n+1)}_1,
\eeq
thus e.g. to go to three loops we would need to know $A_{1,2,3,4}$, $B_{1,2,3}$, $C_{1,2}$, $D_1$, where the only unknown at this point is $B_3$.
In order to find it we would either need to find the curvature function $\gamma^{(2)}$ or quantize the classical string to one more order.

\subsubsection{Inconsistencies for higher mode numbers}
\label{sec:inconsistencies}

The analysis in the previous sections was done only up to second order in the small spin expansion, namely using the one-loop semi-classical energy given by \eq{delta_oneloop_sj}. 
Extending our results to higher orders in the spin we found perfect agreement with the conjectured structure \eq{eq:delta_squared_basso} for mode number $n=1$, yet for cases with $n>1$ there are inconsistencies.
For $n=2$ the first inconsistency appears in the $\frac{S^3\mu}{J^4}$ term and for $n=3$ there are already inconsistencies at order $S^2$. We found that for $n>1$ one has to modify the structure in \eq{eq:delta_squared_basso} by
including negative coefficients in order for it to be consistent with our one-loop results. E.g. for $n=2$ the structure has to be modified starting with the $S^3$ term, which now becomes
\beq
\(
C_{-2}\;\mu+\frac{C_1}{\sqrt\mu}+\frac{C_2}{\mu}+\dots
\) S^3
\eeq
with $C_{-2}=12/J^4$. To the next order in $S$ we find
\beq
\(
D_{-4}\;\mu^{3/2} +D_{-2}\;\sqrt{\mu}+ \frac{D_{0}}{\sqrt\mu}+\frac{D_1}{\mu}+\dots
\) S^4
\eeq
where $D_{-4}=-\frac{78}{J^6},\;D_{-2}=-\frac{36}{J^4},\;D_0=\frac{21}{2J^2}$.
For $n=3$ the first modification already occurs at order $S^2$ and it can be resolved if the term $-\frac{9 S^2\sqrt\mu}{4 J^2}$ is added to \eq{eq:delta_squared_basso}.
Thus effectively the conjectured structure \eq{eq:delta_squared_basso} has to be modified as in the $n=2$ case by including negative coefficients, which now depend on $n$ in a non-trivial way.
It is also worth noticing that since inconsistencies start appearing at orders of $\frac{S^2}{J^2}$ and $\frac{S^3}{J^4}$ for $n=3$ and $n=2$ respectively, one might guess that
there should be an inconsistency at order $\frac{S^4}{J^6}$ for $n=1$, however we found no such thing.

We will revisit this topic in section \ref{sec:curvature_inconsistencies} when we try to perform this procedure again after having found the curvature function.
As one might expect, similar issues appear in that case as well.



\newpage 


\section{Exact results}
\label{sec:exact_results}

\begin{chapquote}{Bruce Lee}
The less effort, the faster and more powerful you will be.
\end{chapquote}

\noindent In this chapter we leave the perturbative regime (weak and strong coupling) behind and move on towards exact results in planar $\N=4$ super Yang-Mills, where by ``exact'' we mean available at any value of the coupling constant.
In principle such results are the ultimate goal of the whole AdS/CFT programme and being able to extract them is a remarkable achievement.

One can argue that the spectral problem has been worked out exactly, at least conceptually, as we will review shortly.
However in practice one immediately runs into technical difficulties with finding solutions, thus at the moment only certain calculations have been carried out explicitly.
In this section we will present a few examples of exact results, most notably the slope and curvature functions, which are the two leading coefficients in the small spin expansion of the anomalous dimension of the folded string.
As mentioned before they are indeed exact in the coupling constant, yet they are somewhat abstract and one might say of little use.
We will argue the opposite by demonstrating how one can use them in order to find new information about physically relevant quantities such as the Konishi anomalous dimension.

We will start the chapter by reviewing the exact solution to the spectral problem of AdS/CFT, first discussing historic approaches and quickly moving on to the novel quantum spectral curve approach.
We then devote the rest of the chapter to various exact results, mostly achieved using the quantum spectral curve construction.

\subsection{Solution to the spectral problem}
\label{sec:tba_y_system}

Historically the first solution to the spectral problem that fully incorporated finite length corrections was the thermodynamic Bethe ansatz (TBA) \cite{Gromov:2009tv,Bombardelli:2009ns,Gromov:2009bc,Arutyunov:2009ur}, which can alternatively be reformulated as an infinite set of functional relations, the so called Y-system.
Being infinite sets of integral or functional equations they are obviously very hard to use in practice.
Both of these formulations can now be seen as intermediary steps towards the more elegant quantum spectral curve construction, which we cover in more depth in the next section.

\subsubsection{Thermodynamic Bethe ansatz}

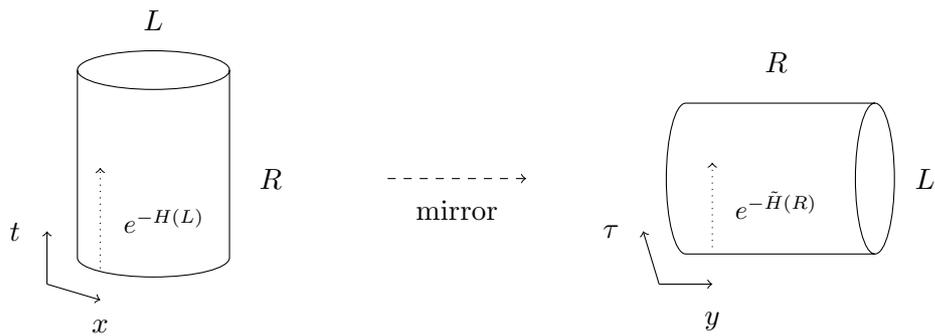
\begin{figure}[t]
\centering
\begin{tikzpicture}[scale=0.7]
	
	
	\begin{scope}[shift={(-1,0)}]

		\node[draw,cylinder,black,rotate=90,minimum height=3cm,minimum width=2cm,aspect=2] (c) at (0, 0) {};
		
		\draw[->] (-2, -2) -- (-2, -1); \node at (-2.6, -1) {$t$};
		\draw[->] (-2, -2) -- (-1, -2.3); \node at (-1, -2.8) {$x$};
		
		\node at (2.2, 0) {$R$};
		\node at (0, 3) {$L$};
		
		\draw[->,dotted] (-1, -1.7) -- (-1, 0.2);
		\node at (0.2, -0.8) {\small $e^{-H(L)}$};

	\end{scope}
	
	\draw[->,dashed] (3.4, 0) -- (6, 0);
	\node at (4.7, -0.6) {mirror};
	
	\begin{scope}[shift={(10.5,0)}]
		
		\node[draw,cylinder,black,rotate=0, minimum height=3cm,minimum width=2cm,aspect=2] (c) at (0, 0) {};
		
		\draw[->] (-2, -2) -- (-2.3, -1); \node at (-1, -2.7) {$y$};
		\draw[->] (-2, -2) -- (-1, -2); \node at (-2.9, -1) {$\tau$}; 
		
		\node at (3, 0) {$L$};
		\node at (0.2, 2.2) {$R$};
		
		\draw[->,dotted] (-1, -1.3) -- (-1, 0.3);
		\node at (0.2, -0.5) {\small $e^{-\tilde{H}(R)}$};
		
	\end{scope}

\end{tikzpicture}
\caption[A high level illustration of the idea behind the TBA method]{A high level illustration of the idea behind the TBA method: by double Wick rotating the theory one can exchange time with length and thus finite volume scattering with the ground state scattering at finite length.}
\label{fig:tba_cylinders}
\end{figure}

The technical details of the thermodynamic Bethe ansatz (TBA) approach are overwhelmingly difficult, in fact the lack of elegance and simplicity in this approach led many to believe that it might be only an intermediary step towards a more satisfactory solution. 
That is why we will only illustrate the basic ideas behind the TBA at a conceptual level, the full gory technical details can be found in numerous literature reviews (see \cite{Bajnok:2010ke} for an overview). 

In the spin chain picture we identify operators with states of the chain and the dilatation operator with the Hamiltonian. 
This mapping enables one to use the physical language of magnons with momenta propagating on a closed chain with the time evolution being determined by the Hamiltonian.
On the string theory side this naturally corresponds to excitations of the string evolving in time with the string Hamiltonian. 
The key idea of TBA is shown in figure \ref{fig:tba_cylinders}: consider this system with time and space interchanged, more precisely we analytically continue in time by introducing $y \equiv i t$ and consider a theory where $y$ plays the role of space and $\tau$ defined via $x \equiv i \tau$ is time. We call this double Wick rotated theory the \emph{mirror} of the original.
Obviously it is a completely different theory, for example the dispersion gets inverted by switching $(E,p) \to i (\tilde{p}, \tilde{E})$, thus if we start from the asymptotic dispersion $E(p)$ for the $\alg{sl}{2}$ sector we end up with 
\begin{equation}
	E(p) = \sqrt{1+4g^2 \sin^2 \frac{p}{2}}  \;\;\;\; \Longleftrightarrow \;\;\;\; 
	\tilde{E}(\tilde{p}) = 2 \, \mathrm{arcsinh} \( \frac{1}{2g} \sqrt{\tilde{p}^2 + 1} \).
\end{equation}
The scattering matrix also has a different pole structure, meaning that the theory has different bound states from the original ones. 
But the remarkable thing is that mirroring a theory preserves integrability, meaning that one can solve it with an asymptotic Bethe ansatz. 
One can use this fact, since the partition functions for these theories satisfy the identity
\begin{equation}
	Z(L,R) = \tilde{Z}(R,L),
\end{equation}
where $L$ is the length scale of the original theory and $R$ is the time scale. 
At asymptotic time scales the original partition function is dominated by contributions from the ground state, this applies to any length scale of the system, thus in the asymptotic time limit 
\begin{equation}
	Z(L,R) = \tr \, e^{-R \, H(L)} \xrightarrow[\text{$R \rightarrow \infty$}]{} e^{-R \, E_0(L)}.
\end{equation}
This limit corresponds to the infinite length limit for the mirror model where the spectrum is controlled by the asymptotic Bethe ansatz.
Here we have
\begin{equation}
	\tilde{Z}(R,L) = \tr \, e^{-L \, \tilde{H}(R)} \xrightarrow[\text{$L \rightarrow \infty$}]{}  \sum_n e^{-L \, \tilde{E}_n(R)},
\end{equation}
where $\tilde{H}$ is the Hamiltonian of the mirror theory. 
In the large $R$ limit we introduce the density of particles in momentum space $\rho(\tilde{p}) = \Delta n / (R \Delta \tilde p)$ and the logarithm of the mirror asymptotic Bethe equations reads
\beq
	\tilde{p}(u) + \int du' \( -i \log S(u, u') \) \rho(u') = \frac{2\, \pi n}{R},
\eeq
where we introduce a parametrization for the momentum $\tilde{p}(u)$, just as in \eq{eq:up_param}. 
Particles with momentum $\tilde{p}$ that are not excited but nevertheless satisfy the Bethe equations are called holes and their density is denoted by $\tilde{\rho}$.
The relation between the two densities is given by
\beq
	\partial_u \tilde{p} - 2\pi \( \rho + \tilde \rho \) = - \int du' K(u, u') \rho(u') \equiv -K \star \rho,
\eeq
where the convolution kernel is
\beq
	K(u, u') = -i \partial_u \log S(u, u').
\eeq
Finally the partition function can be written as
\beq
	Z(L,R) = \int \mathcal D \rho \, e^{-L \tilde{E}[\rho] + S[\rho, \tilde{\rho}] },
\eeq
where $S[\rho, \tilde{\rho}]$ is the entropy of a given particle density configuration arising due to the fact that the densities do not uniquely specify a particle configuration.
This functional integral can be evaluated in the saddle point approximation. 
The saddle point is found by solving the non-linear integral equation, which is called a thermodynamic Bethe ansatz equation,
\beq
	\label{eq:simple_tba}
	\epsilon(u) - L \tilde E(u) = -\log (1 + e^{-\epsilon}) \star K
\eeq
in terms of the so-called pseudo-energy $\epsilon \equiv \log \tilde{\rho} / \rho$. 
Once it is found the ground state energy is given by
\beq
	E_0(L) = - \int \frac{du}{2\pi} (\partial_u \tilde p) \log \(1 + e^{-\epsilon(u)}\).
\eeq
Excited states can also be reached in this way via analytic continuation in some parameter, for example the volume of the system \cite{Dorey:1996re}.
Basically one has to introduce singularities in the integrand of \eq{eq:simple_tba} in such a way that the large $L$ limit of the new solution coincides with the asymptotic Bethe ansatz solution.

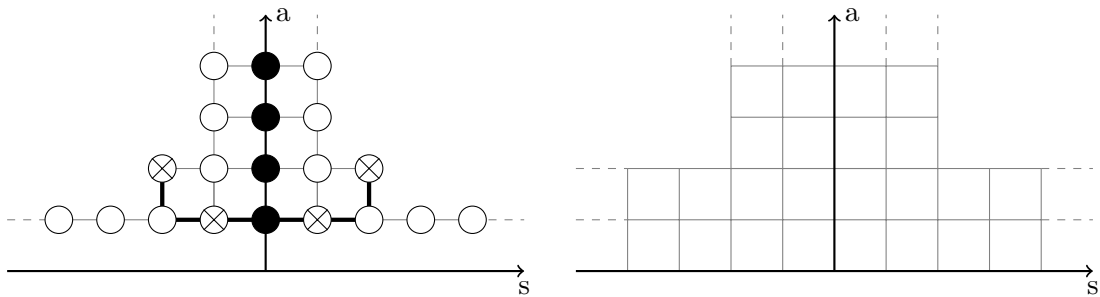
\begin{figure}[t]
\begin{tikzpicture}[scale=0.68]
	
	\tikzset{hooknode/.style={draw,circle,minimum size=1.5ex,fill=#1}}
	\tikzset{hooknodecross/.style={draw,cross out,minimum size=1.5ex}}
	
	
	\begin{scope}[shift={(11,0)}]
	
		\draw[gray] (1,0) grid (9,2); 
		\draw[gray] (3,2) grid (7,4); 		
		
		\draw[gray,dashed] (0,1) -- (1,1);  \draw[gray,dashed] (0,2) -- (1,2);
		\draw[gray,dashed] (9,1) -- (10,1); \draw[gray,dashed] (9,2) -- (10,2);
		\draw[gray,dashed] (3,4) -- (3,5);  \draw[gray,dashed] (4,4) -- (4,5); 
		\draw[gray,dashed] (6,4) -- (6,5);  \draw[gray,dashed] (7,4) -- (7,5); 
		
		\draw[->,thick] (0,0) -- (10,0); \draw[->,thick] (5,0) -- (5,5);
		\node[below] at (10,0) {s}; \node[right] at (5,5) {a};
	
	\end{scope}
	
	
	\begin{scope}[shift={(0,0)}]
	
		\draw[gray] (4,1) grid (6,4); 
		\draw[gray] (3,1) grid (4,2); \draw[gray] (6,1) grid (7,2);
		\draw[gray] (1,1) -- (9,1);
		\draw[black,ultra thick] (3,2) -- (3,1) -- (7,1) -- (7,2);
		
		\draw[gray,dashed] (0,1) -- (1,1); \draw[gray,dashed] (9,1) -- (10,1); 
		\draw[gray,dashed] (4,4) -- (4,5); \draw[gray,dashed] (6,4) -- (6,5); 
		
		\draw[->,thick] (0,0) -- (10,0); \draw[->,thick] (5,0) -- (5,5);
		\node[below] at (10,0) {s}; \node[right] at (5,5) {a};
		
		\node[hooknode=black] at (5,1) {};
		\node[hooknode=black] at (5,2) {};
		\node[hooknode=black] at (5,3) {};
		\node[hooknode=black] at (5,4) {};
		
		\node[hooknode=white] at (4,1) {}; \node[hooknodecross] at (4,1) {};
		\node[hooknode=white] at (4,2) {};
		\node[hooknode=white] at (4,3) {};
		\node[hooknode=white] at (4,4) {};
		
		\node[hooknode=white] at (6,1) {}; \node[hooknodecross] at (6,1) {};
		\node[hooknode=white] at (6,2) {};
		\node[hooknode=white] at (6,3) {};
		\node[hooknode=white] at (6,4) {};
		
		\node[hooknode=white] at (3,1) {}; 
		\node[hooknode=white] at (3,2) {}; \node[hooknodecross] at (3,2) {};
		
		\node[hooknode=white] at (7,1) {}; 
		\node[hooknode=white] at (7,2) {}; \node[hooknodecross] at (7,2) {};
		
		\node[hooknode=white] at (8,1) {}; 
		\node[hooknode=white] at (9,1) {}; 
		
		\node[hooknode=white] at (2,1) {}; 
		\node[hooknode=white] at (1,1) {}; 
	
	\end{scope}
	
\end{tikzpicture}
\caption[The domains of the $Y_{a,s}$ and $T_{a,s}$ functions]{The domains of the $Y_{a,s}$ and $T_{a,s}$ functions (left and right respectively). The $Y_{a,s}$ functions are defined on the nodes, where the type of node signifies the type of excitation. The $T_{a,s}$ functions are defined on the latttice points of the grid.}
\label{fig:yt_hooks}
\end{figure}

The high level ideas outlined above have been successfully applied to the spectral problem in $\N=4$ super Yang-Mills.
The TBA equations schematically read \cite{Gromov:2009bc, Arutyunov:2009ur}
\beq
	\label{eq:tba_final}
	\log Y_{a,s}(u) = \delta^0_s\, i L\, p_a(u) + \int dv \, K_{a,s}^{a',s'}(u,v) \log \( 1+Y_{a',s'}(v) \),
\eeq
where the Y-functions are related to the pseudo-energies as $Y(u) = \exp(-\epsilon(u))$.
A key difference from the example above is that here we can have different types of excitations which are labelled by the indices $a,s$. 
The hook diagram on the left of figure \ref{fig:yt_hooks} indicates the ranges of values they can take on,
here the type of node also distinguishes the type of particle.
Finally the energy of the state is given by
\beq
	E = \sum_j \epsilon_1(u_{4,j}) + \sum_{a=1}^\infty \int_{-\infty}^\infty \frac{du}{2\pi i} \frac{\pd \epsilon_a^*}{\pd u} \log \(1 + Y_{a,0}^* (u) \),
\eeq
where the first term is basically the result of analytic continuation for excited states.
It is highly non-trivial to actually find the correct analytic continuation that describes a specific excited state thus the applicability of this method is difficult and only a handful of operators can be studied, one particular example is the Konishi operator \cite{Gromov:2009zb}.
Another obvious difficulty is the fact that the TBA equations \eq{eq:tba_final} are an infinite set of coupled non-linear integral equations, which in practice limits their applicability to numeric calculations only.

\subsubsection{Y/T/$\mathcal{Q}$-systems}

The Y-functions introduced in \eq{eq:tba_final} are technically defined on an infinite sheeted Riemann surface and have a complicated analytic structure, this is done so that one could rewrite the TBA equations as \cite{Gromov:2009bc, Cavaglia:2010nm}
\beq
	\label{eq:ysystem}
	\frac{Y_{a,s}^{+} Y_{a,s}^{-}}{Y_{a+1,s} Y_{a-1,s}} = \frac{\(1+Y_{a,s+1}\)\(1+Y_{a,s-1}\)}{\(1+Y_{a+1,s}\)\(1+Y_{a-1,s}\)},
\eeq
which are called the Y-system equations, here $Y^\pm_{a,s}(u) \equiv Y_{a,s}(u \pm i/2)$.
The advantage of this form is that instead of an infinite set of integral equations we have a still infinite set of functional equations.
Another notable feature of this system is that its form is universal for any excited state, which is now encoded in the analyticity properties of the Y-functions. 
It can also be cast into yet another form by introducing T-functions as
\beq
	\label{eq:tyrelation}
	Y_{a,s} = \frac{T_{a,s+1} \, T_{a,s-1}}{T_{a+1,s} \, T_{a-1,s}},
\eeq
thus reducing the Y-system to the T-system
\beq
	\label{eq:tsystem}
	T_{a,s}^+ \, T_{a,s}^- = T_{a+1,s} \, T_{a-1,s} + T_{a,s+1} \,  T_{a,s-1},
\eeq
also known as the Hirota discrete bilinear equation, a well studied object in the context of classical integrability.
The T-functions are defined on the lattice points of the T-hook shown on the right of figure \ref{fig:yt_hooks}.
The advantage of this reformulation is that solutions of the T-system can be parametrized in terms of Wronskians built from eight independent $\mathcal{Q}$-functions, which have much simpler analytic properties than the Y-functions \cite{Gromov:2010km}. 
The exact formulae expressing the $T$s in terms of the $\mathcal{Q}$s are very involved, see \cite{Gromov:2010km} for details.
One can build up a total of $2^8$ $\mathcal{Q}$--functions denoted by $\mathcal{Q}_{A|J}(u)$ where $A,J \in \{1,2,3,4\}$ are two ordered subsets of indices.
These functions are defined through the QQ-relations by reshuffling indices, denoted by $a, b, i, j$, as
\begin{subequations}\label{definingQQ}
    \begin{align}
       \label{QQbb}
       \mathcal{Q}_{A|I}\mathcal{Q}_{A ab|I} &=\mathcal{Q}_{A a|I}^{+} \mathcal{Q}_{A b|I}^{-}-
       \mathcal{Q}_{A a|I}^{-}
       \mathcal{Q}_{A b|I}^{+}\,,\\
       \label{QQff}
       \mathcal{Q}_{A|I}\mathcal{Q}_{A|I ij} &=\mathcal{Q}_{A|I i}^{+} \mathcal{Q}_{A|I j}^{-}-
       \mathcal{Q}_{A|I i}^{-} \mathcal{Q}_{A|I j}^{+}\,,\\
       \label{QQbf}
       \mathcal{Q}_{A a|I}\mathcal{Q}_{A|I i} &= \mathcal{Q}_{A a|I i}^{+}\mathcal{Q}_{A|I}^{-}-
       \mathcal{Q}_{A|I}^{+} \mathcal{Q}_{A a|I i}^{-}, \,
     \end{align}
\end{subequations} 
where $\pm$ superscripts denote shifts in the spectral parameter by $\pm i/2$.
In addition we impose the constraints $\mathcal{Q}_{\emptyset |\emptyset}=\mathcal{Q}_{1234|1234}=1$ coming from normalization and the unimodularity of the superconformal group.
A dual QQ system can then be introduced by $\mathcal{Q}^{A|J}\equiv (-1)^{|A|\;|J|}\mathcal{Q}_{\bar A|\bar J}$ which satisfies the same QQ-relations.
Here  the bar over a subset means the subset complementary with respect to the full~set $\{1,2,3,4\}$ and  $|X|$ denotes the number of indexes in $X$. 
It was later shown that with a good choice of the eight basis $\mathcal{Q}$-functions it is possible to reduce the problem to a finite set of non-linear integral equations (FiNLIE) \cite{Gromov:2011cx}, which is better suited for numerical calculations.

\subsection{Quantum spectral curve}
\label{sec:pmu_system}

The so far discussed gradually improving formulations of the solution to the spectral problem seem to be pointing to some ultimate simplification.
In this section we discuss the quantum spectral curve (QSC) approach which many believe finally unveils the long anticipated beauty of the spectral problem.

\subsubsection{Emergence from the $\mathcal{Q}$-system}

The $\mathcal{Q}$-functions defined in the previous section have a disadvantage of having a rather complicated analytic structure, namely they are defined on infinite sheeted Riemann surfaces with square root type cuts parallel to the real axis between the branch points $\pm 2g + i n$ with $n \in \mathbb{Z}$ going either through infinity (long cuts) or through the imaginary axis (short cuts).
Apart from the branch cuts the $\mathcal{Q}$ functions do not have other singularities and are otherwise regular functions. 
However in order to fully define them one has to specify their analytic continuations through all of these cuts.
Let us now introduce the notation 
\beqa
	\bP_a &\equiv& \mathcal{Q}_{a|\emptyset}, \;\;\; \bP^a \;\; \equiv \;\; \mathcal{Q}^{a|\emptyset}, \\
	\bQ_j &\equiv& \mathcal{Q}_{\emptyset|j}, \;\;\; \bQ^j \;\; \equiv \;\; \mathcal{Q}^{\emptyset|j},
\eeqa
where $a,j=1,2,3,4$. 
The nice thing about these is that $\bP_i$ and $\bP^i$ have only a single short cut $u \in [-2g, 2g]$ on their main sheet of the Riemann surface whereas $\bQ_i$ and $\bQ^i$ only have long cuts $u \in (-\infty,-2g]\cup [2g, \infty)$.
Using the QQ-relations one can always use either the eight $\bP$ or the eight $\bQ$ functions to find all of the remaining ones. 
Take the $\bP$ functions, we define their analytic continuation through the short cuts as
\beq
	\label{Pt}
	\tilde \bP_a=\mu_{ab}(u)\,\bP^b\;\;,\;\;\tilde \bP^a=\mu^{ab}(u)\,\bP_b
\eeq
where $\mu_{ab}(u)$ is an antisymmetric matrix with components having infinitely many branch points at $u\in\pm2 g+i\mathbb{Z}$. 
The matrix itself has a unit Pfaffian which translates to the constraints on the components
\beqa
\label{constraint}
\mu_{12}\,\mu_{34}-\mu_{13}\,\mu_{24}+\mu_{14}^2&=&1\;,\\
\label{constraint2}
\mu_{14}=\mu_{23}
\eeqa
and the inverse matrix is defined by $\mu^{ab}=-\frac{1}{2}\epsilon^{abcd}\mu_{cd}$. 
These functions can also be shown to be periodic
\beq
	\check\mu_{ab}(u+i)=\check\mu_{ab}(u),
\eeq
where we distinguish between the short/long cut version of the same function by adding a hat/check over the symbol.
For the short cut versions $\mu_{ab}(u) \equiv \hat{\mu}_{ab}(u)$ of the functions this condition reads
\beq
	\label{muper}
	\tilde{\mu}_{ab}(u) = \mu_{ab}(u+i).
\eeq
\begin{figure}[t]
\begin{tikzpicture}[scale=0.5]
	
	
	\tikzset{pole/.style={draw,circle,inner sep=0pt,minimum size=3pt,fill=black}}
	
	
	\begin{scope}[shift={(6,6)}]
	
		\draw (5,5) -- (5,4) -- (6,4); \node[above right] at (5,4) {$u$};
	
		\draw[thick] (-5,0) -- (5,0); 
		\draw[thick] (0,-5) -- (0,5);
		
		\def\cutx{3}
		
		\foreach \y in {1,2,3} {
			\draw[dashed,black] (-\cutx,-\y) -- (\cutx,-\y);  
			\node[pole] at (-\cutx,-\y) {}; 
			\node[pole] at (\cutx,-\y) {}; 
			
			\draw[dashed,black] (-\cutx,\y) -- (\cutx,\y);  
			\node[pole] at (-\cutx,\y) {}; 
			\node[pole] at (\cutx,\y) {}; 
		}
		
		\draw[black,ultra thick] (-\cutx,0) -- (\cutx,0);  \node[pole] at (-\cutx,0) {}; \node[pole] at (\cutx,0) {};

		\node[above left] at (-\cutx,0) {$-2g$}; \node[above right] at (\cutx,0) {$2g$};	
		\node[left] at (-\cutx,-2) {$-2g-2 i$}; \node[right] at (\cutx,-2) {$2g-2 i$};		
		\node[left] at (-\cutx,2) {$-2g+2 i$}; \node[right] at (\cutx,2) {$2g+2 i$};	
	
	\end{scope}
	
	
	\begin{scope}[shift={(21,6)}]
	
		\draw (5,5) -- (5,4) -- (6,4); \node[above right] at (5,4) {$u$};
		
		\draw[thick] (-5,0) -- (5,0); 
		\draw[thick] (0,-5) -- (0,5);
		
		\def\cutx{3}
		
		\foreach \y in {1,2,3} {
			\draw[black,very thick] (-\cutx,-\y) -- (\cutx,-\y);  
			\node[pole] at (-\cutx,-\y) {};
			\node[pole] at (\cutx,-\y) {};

			\draw[black,very thick] (-\cutx,\y) -- (\cutx,\y);  
			\node[pole] at (-\cutx,\y) {}; 
			\node[pole] at (\cutx,\y) {};
		}
		
		\draw[black,very thick] (-\cutx,0) -- (\cutx,0);  \node[pole] at (-\cutx,0) {}; \node[pole] at (\cutx,0) {};

		\node[above left] at (-\cutx,0) {$-2g$}; \node[above right] at (\cutx,0) {$2g$};	
		\node[left] at (-\cutx,-2) {$-2g-2 i$}; \node[right] at (\cutx,-2) {$2g-2 i$};		
		\node[left] at (-\cutx,2) {$-2g+2 i$}; \node[right] at (\cutx,2) {$2g+2 i$};

	\end{scope}
	
\end{tikzpicture}
\caption[The location of branch cuts in $u$ for the functions $\bP_a(u)$ and $\hat{\mu}_{ab}(u)$]{The location of branch cuts in $u$ for the functions $\bP_a(u)$ (left) and $\hat{\mu}_{ab}(u)$ (right). The infinitely many cuts of $\tilde{\bP}_a$ are shown on the left picture with dashed lines.}
\label{fig:pmu_cuts}
\end{figure}
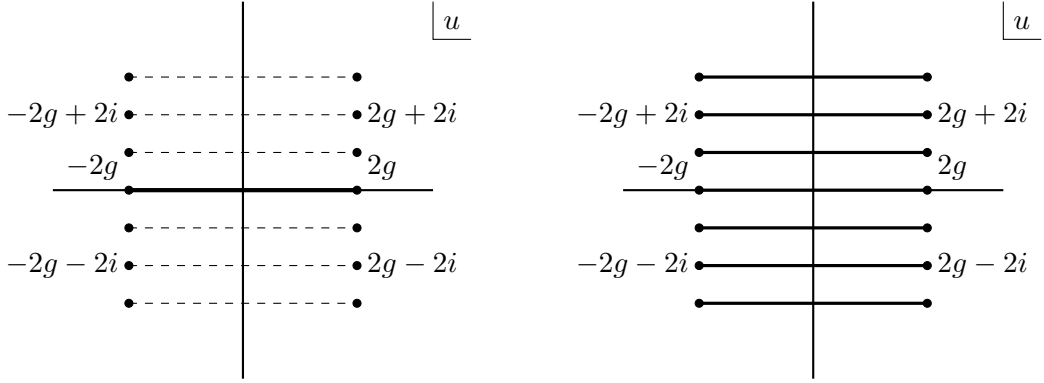
The magic of the whole quantum spectral curve construction is that the analytic continuations of the monodromies $\mu_{ab}$ through the cuts are again given by the same functions \cite{Gromov:2014caa}
\beq
	\label{muPPt-eq}
	\tilde\mu_{ab}-\mu_{ab}=\bP_a\tilde \bP_b-\bP_b\tilde \bP_a\;
\eeq
and the $\pmu$-system of eight $\bP$ and five $\mu$ functions closes!
The cut structure is shown in figure \ref{fig:pmu_cuts}.
An analogous closure can be implemented for the $\bQ$ functions by denoting the analytic continuations
\begin{equation}
\label{omegaQ}
\tilde \bQ_i=\omega_{ij} \bQ^j,
\end{equation}
where the monodromy $\omega$ is related to $\mu$ by
\beq
\label{muomega}
\omega_{ij}= \tilde{Q}_{a|i}^- \tilde{Q}_{b|j}^-\, \mu^{ab} \,.
\eeq
and itself has the following analytic continuation 
\begin{equation}\label{tQ=omegaQ}
\tilde\omega_{ij}-\omega_{ij}=\bQ_i\tilde{\bQ}_j-\bQ_j\tilde{\bQ}_i
\end{equation}
thus closing the $\bQ\omega$-system.
Both the $\pmu$ and the $\bQ\omega$-systems are complete in the sense they describe all solutions in the theory. 
Since they are related it is a matter of convenience to use one or the other and for the remainder of the thesis we will stick with the $\pmu$-system.

\subsubsection{Asymptotics}

Tracing back from the $\mathcal{Q}$-system to the Y-system one can find the following relation
\beq
	\label{eq:y_asymp}
	Y_{11} Y_{22} = 1 + \frac{\bP_1 \tilde{\bP_2} - \bP_2 \tilde{\bP_1}}{\mu_{12}} = \frac{\mu_{12}(u+i)}{\mu_{12}(u)},
\eeq
and from the T-system one can find
\beq
	T_{1,s}(u) = \bP_1 \(u+\frac{is}{2}\) \, \bP_2 \(u-\frac{is}{2}\) - \bP_2 \(u+\frac{is}{2}\) \, \bP_1 \(u-\frac{is}{2}\),
\eeq
thus given a solution to the $\pmu$-system one can fully reconstruct all of the T-functions using the T-system equations \eq{eq:tsystem} and then reconstruct the Y functions using \eq{eq:tyrelation}.
The Y-functions encode the global charges in their asymptotics, for example in the $\alg{sl}{2}$ sector one has \cite{Gromov:2011cx} 
\beq
	\label{eq:y1122_sl2_asymptotics}
	\log Y_{11} Y_{22} \; \simeq \; i \frac{\Delta - J}{u},
\eeq
implying through \eq{eq:y_asymp} that the asymptotics of the $\bP$ and $\mu$ functions also encode the global charges of the state/operator being described.

The picture described so far strongly resembles the classical spectral curve construction outlined in section \ref{sec:spectral_curve}.
Quite expectedly the classical spectral curve turns out to be the strong coupling limit of the quantum spectral curve as implied by the naming \cite{Gromov:2014caa}.
More precisely one can think of the classical spectral curve as an WKB approximation to the quantum case, namely the quasi-momenta are related to the $\bP$ and $\bQ$ functions as
\beq
	\bP_i \simeq e^{i\int^u \tilde{p}_i(u) \, du}, \;\;\; \bQ_i \simeq e^{i\int^u \hat{p}_i(u) \, du}
\eeq
where the four $\bP_i$ functions correspond to the $S^5$ quasi-momenta and the four $\bQ_i$ functions to the $AdS_5$ part.
The emergence of the cuts can be seen by recalling that the quasi-momenta are the eigenvalues of the monodromy matrix \eq{eq:lax_monodromy}, which develops the cuts after changing variables from $x$ to $u$ via the Zhukovsky map
\beq
	x + \frac{1}{x} = \frac{u}{g},
\eeq
which is the classical analogue of the rapidity map \eq{eq:xox_rapidity} from the asymptotic Bethe ansatz.
Similarly how the classical quasi-momenta encoded the global charges in their asymptotics as shown in \eq{eq:quasi_asymptotics}, so do their quantum analogues $\bP$ \cite{Gromov:2014caa}
\vspace{5pt}
\beq
\label{eq:pmu_large_u}
\left(
\bea{c}
\bP_1\\
\bP_2\\
\bP_3\\
\bP_4
\eea
\right)\simeq
\left(
\bea{l}
A_1\; u^\frac{-J_1+J_2-J_3}{2}\\
A_2\; u^\frac{-J_1-J_2+J_3-2}{2}\\
A_3\; u^\frac{+J_1+J_2+J_3}{2}\\
A_4\; u^\frac{+J_1-J_2-J_3-2}{2}
\eea
\right)\;\;,\;\;
\left(
\bea{c}
\bP^1\\
\bP^2\\
\bP^3\\
\bP^4
\eea
\right)\simeq
\left(
\bea{l}
A^1\; u^\frac{+J_1-J_2+J_3-2}{2}\\
A^2\; u^\frac{+J_1+J_2-J_3}{2}\\
A^3\; u^\frac{-J_1-J_2-J_3-2}{2}\\
A^4\; u^\frac{-J_1+J_2+J_3}{2}
\eea
\right)\; \vspace{3pt}
\eeq
and similarly for $\bQ$
\vspace{3pt}
\beq
\left(
\bea{c}
\bQ_1\\
\bQ_2\\
\bQ_3\\
\bQ_4
\eea
\right)\simeq
\left(
\bea{l}
B_1\; u^\frac{+\Delta+S_1+S_2-2}{2}\\
B_2\; u^\frac{+\Delta-S_1-S_2}{2}\\
B_3\; u^\frac{-\Delta+S_1-S_2-2}{2}\\
B_4\; u^\frac{-\Delta-S_1+S_2}{2}
\eea
\right)\;\;,\;\;
\left(
\bea{c}
\bQ^1\\
\bQ^2\\
\bQ^3\\
\bQ^4
\eea
\right)\simeq
\left(
\bea{l}
B^1\; u^\frac{-\Delta-S_1-S_2}{2}\\
B^2\; u^\frac{-\Delta+S_1+S_2-2}{2}\\
B^3\; u^\frac{+\Delta-S_1+S_2}{2}\\
B^4\; u^\frac{+\Delta+S_1-S_2-2}{2}
\eea
\right).\vspace{5pt}
\eeq
Since the $\pmu$ and $\bQ\omega$-systems are coupled via the QQ-relations the above asymptotics have to be compatible thus producing algebraic relations between the coefficients $A$ and the global charges.
We will be using simplified relations for the solutions we will consider later, the most general case can be found in \cite{Gromov:2014caa}.

\subsection{Revisiting the slope function}
\label{sec:slope_pmu}

In section \ref{sec:slope_function_aba} we derived the slope function which is the leading small spin expansion coefficient of the anomalous dimension for the generalized Konishi operator in the $\alg{sl}{2}$ sector.
We used the asymptotic Bethe ansatz as the starting point of the derivation which is justified by the fact that finite size effects are irrelevant for the slope function.
In this section we will derive the slope function \eq{eq:slope_function} using the $\pmu$-system, which is not only more concise but also generalizes to higher orders as we shall demonstrate in the next section when we derive the next coefficient in the small spin expansion, the curvature function.

\subsubsection{$\pmu$-system for the $\alg{sl}{2}$ sector}
\label{sec:pmu_sl2}

The first simplification we can employ is the fact that solutions in the $\alg{sl}{2}$ sector are symmetric under the left-right exchange of the Y-functions $Y_{a,s}=Y_{a,-s}$, which implies the following relations for the $\bP$ and $\bQ$ functions
\begin{eqnarray}
&&\bP^a=\chi^{ac} \bP_c,\qquad  \bQ^i=\chi^{ij} \bQ_j
\end{eqnarray}
and thus the analytic continuation for the $\pmu$-system reads
\beq
\tilde \bP_a=-\mu_{ab}\chi^{bc}\bP_c,\; \ \ \ \text{with}\;\ \ \chi^{ab}=\left(
            \begin{array}{cccc}
              0 & 0 & 0 & -1 \\
              0 & 0 & 1 & 0 \\
              0 & -1 & 0 & 0 \\
              1 & 0 & 0 & 0 \\
            \end{array}
          \right),
\label{eq:Pmu}
\eeq
and
\beq
\tilde \mu_{ab}-\mu_{ab}=\bP_a \tilde\bP_b-  \bP_b \tilde\bP_a\;.
\label{eq:mudisc}
\eeq
Explicitly the $\pmu$-system equations are now given by
\beqa
\label{eq:pmuexpanded1}
&&\tilde \bP_1= -\bP_3 \mu_{12}+\bP_2 \mu_{13}-\bP_1 \mu_{14} \\
&&\tilde \bP_2= -\bP_4 \mu_{12}\hspace{16mm}+\bP_2 \mu_{14}-\bP_1 \mu_{24} \\
&&\tilde \bP_3= \hspace{16mm}-\bP_4 \mu_{13}+\bP_3 \mu_{14}\hspace{16mm}-\bP_1 \mu_{34} \\
&&\tilde \bP_4= \hspace{16mm}\hspace{15.5mm}-\bP_4 \mu_{14}+\bP_3 \mu_{24}-\bP_2 \mu_{34}\;.
\label{eq:pmuexpanded}
\eeqa
The above equations ensure that the branch points of $\bP_a$ and $\mu_{ab}$ are of the square root type, i.e. $\tilde{\tilde{\bP}}_a=\bP_a$ and $\tilde{\tilde{\mu}}_{ab}=\mu_{ab}$, which is also true in the most general case.
Finally, we require that $\bP_a$ and $\mu_{ab}$ do not have any singularities except these branch points.

The large $u$ asymptotics for the $\bP_a$ functions are given by \eq{eq:pmu_large_u} and can be uniquely fixed for the $\mu_{ab}$ functions using the $\pmu$-system \eq{eq:Pmu}-\eq{eq:mudisc}.
For the $\alg{sl}{2}$ sector they are given by \cite{Gromov:2013pga}
\beq
\bP_a\sim(A_1u^{-J/2},A_2u^{-J/2-1},A_3u^{J/2},A_4u^{J/2-1})
\label{eq:asymptotics}
\eeq
\beq
	\(\mu_{12},\ \mu_{13},\ \mu_{14},\ \mu_{24},\ \mu_{34}\)\sim
	\(u^{\Delta-J},\ u^{\Delta+1},\ u^{\Delta},\ u^{\Delta-1},\ u^{\Delta+J}\)
\label{eq:muasymptotics}
\eeq
where $J$ is the twist of the gauge theory operator, and $\Delta$ is its conformal dimension. 
With these asymptotics, the equations \eq{eq:Pmu}-\eq{eq:mudisc} form a closed system which fixes $\bP_a$ and $\mu_{ab}$.
Lastly, the spin $S$ of the operator is related \cite{Gromov:2013pga} to the leading coefficients $A_a$ of the $\bP_a$ functions
\beqa
&&A_1 A_4=\frac{\((J+S-2)^2-\Delta^2\)\((J-S)^2-\Delta^2\)}{16 i J(J-1)} \label{AA1} \\
&&A_2 A_3=\frac{\((J-S+2)^2-\Delta^2\)\((J+S)^2-\Delta^2\)}{16 i J(J+1)} \label{AA2},
\eeqa
which as discussed in the last section comes from the compatibility condition of the asymptotics for the $\pmu$ and $\bQ\omega$-systems.

Additionally the simplified $\bP\mu$-system enjoys a symmetry preserving all of its essential features. 
It has the form of a linear transformation of $\bP_a$ and $\mu_{ab}$ which leaves the system \eq{eq:Pmu}, \eq{eq:mudisc} and the asymptotics \eq{eq:asymptotics}, \eq{eq:muasymptotics} invariant. 
Indeed, consider a general linear transformation $\bP_a'={R_a}^b \bP_b$ with a non-degenerate constant matrix $R$. In order to preserve the $\pmu$-system equations, $\mu$ should at the same time transform as
\beq
\mu'=-R \mu \chi R^{-1}\chi.
\label{gammaP}
\eeq
Such a transformation also preserves the form of \eq{eq:mudisc} if
\beq
R^T\chi R\chi=-1\;,
\label{eq:sxsx}
\eeq
which also automatically ensures antisymmetry of $\mu_{ab}$ and the constraints (\ref{constraint}), (\ref{constraint2}).
However in general this transformation will spoil the asymptotics of $\bP_a$.
These asymptotics are ordered as $|\bP_2|<|\bP_1|<|\bP_4|<|\bP_3|$, which implies that the matrix $R$ must have the following structure
 \beq
R=\left(
\begin{array}{cccc}
 * & * & 0 & 0 \\
 0 & * & 0 & 0 \\
 * & * & * & * \\
 * & * & 0 & * \\
\end{array}
\right).
\eeq
This matrix would of course be lower triangular if we ordered $\bP_a$ by their asymptotics.
The general form of $R$ which satisfies \eq{eq:sxsx} and does not spoil the asymptotics generates a 6-parametric transformation, which we will call a $\gamma$-transformation. 
The simplest \text{$\gamma$-transformation} is the following rescaling:
\beq
\bP_1 \to \alpha \, \bP_1,\;\;
\bP_2 \to \beta \, \bP_2,\;\;
\bP_3 \to 1/\beta \, \bP_3,\;\;
\bP_4 \to 1/\alpha \, \bP_4,\;\;
\label{eq:alphabeta}
\eeq
\beq
\mu_{12} \to \alpha\beta \, \mu_{12},\;\;
\mu_{13} \to \frac{\alpha}{\beta} \, \mu_{13},\;\;
\mu_{14} \to \mu_{14}\;\;,\;\;
\mu_{24} \to \frac{\beta}{\alpha}\,\mu_{24},\;\;
\mu_{34} \to \frac{1}{\alpha\beta}\,\mu_{34},\;\;
\eeq
with $\alpha,\beta$ being constants.
In all the solutions that we consider all functions $\bP_a$ turn out to be functions of definite parity with respect to the spectral parameter $u$, so it makes sense to consider $\gamma$-transformations which preserve parity. 
$\bP_1$ and $\bP_2$  always have opposite parity as one can see from from \eq{eq:asymptotics} and thus should not mix under such transformations, the same is true about $\bP_3$ and $\bP_4$. 
Thus depending on parity of $J$ the parity-preserving $\gamma$-transformations are either
\beqa
\label{gammatransform2}
&\bP_3\rightarrow\bP_3+\gamma_3\bP_2,\ \bP_4\rightarrow\bP_4+\gamma_2\bP_1,\\
\nn&\mu_{13}\rightarrow\mu_{13}+\gamma_3\mu_{12},\ \mu_{24}\rightarrow\mu_{24}-\gamma_2\mu_{12},\ \mu_{34}\rightarrow\mu_{34}+\gamma_3\mu_{24}-\gamma_2\mu_{13}-\gamma_2\gamma_3\mu_{12}
\eeqa
for odd $J$ or
\beqa
\label{gammatransform1}
&\bP_3\rightarrow\bP_3+\gamma_1\bP_1,\ \bP_4\rightarrow\bP_4-\gamma_1\bP_2,\\
\nn&\mu_{14}\rightarrow\mu_{14}-\gamma_1\mu_{12},\ \mu_{34}\rightarrow\mu_{34}+2\gamma_1\mu_{14}-\gamma^2_1\mu_{12}\;,
\eeqa
for even $J$.

\subsubsection{Solving the system}
\label{sec:slope_pmu_solve}

The description of the $\bP\mu$-system in the previous section was done for physical operators with the charges $S$ and $J$ being integers. 
Our goal is to take some peculiar limit when the (integer) number of covariant derivatives $S$ goes to zero.  
As we will see this requires some extension of the asymptotic requirement for $\mu$ functions.
In this section we will be guided by principles of naturalness and simplicity to deduce these modifications which we will summarize in section~\ref{sec:ancont}. 
There we also give a concrete prescription for analytical continuation in $S$, which we then use to derive the curvature function.
We also note that the solution of the $\pmu$-system is a little simpler for even $J$, because for odd $J$ extra branch points at infinity will appear in $\bP_a$ due to the asymptotics \eq{eq:asymptotics}, thus we will start with the $J$ even case.

We start solving the $\pmu$-system by first finding $\mu_{ab}$. 
Recalling that $\Delta=J+{\cal O}(S)$, from \eq{AA1}, \eq{AA2} we see that $A_1A_4$ and $A_2A_3$ are of order $S$ for small $S$, so we can take the functions $\bP_a$ to be of order $\sqrt{S}$. 
This is a key simplification, because now \eq{eq:mudisc} indicates that the discontinuities of $\mu_{ab}$ on the cut are small when $S$ goes to zero. 
Thus at leading order in $S$ all $\mu_{ab}$ are just periodic entire functions without cuts.
For power-like asymptotics of $\mu_{ab}$ like in \eq{eq:muasymptotics} the only possibility is that they are all constants.
However, we found that in this case there is only a trivial solution, i.e. $\bP_a$ can only be zero.
The reason for this is that for physical states $S$ must be integer and thus cannot be arbitrarily small, nevertheless, it is a sensible question how to define an analytical continuation from integer values of $S$.

Thus we have to relax the requirement of power-like behaviour at infinity. The first possibility is
to allow for $e^{2\pi u}$ asymptotics at $u\to +\infty$.
We should, however, remember about the constraints \eq{constraint} and \eq{constraint2} which restrict our choice and the fact that we can also use $\gamma$-symmetry.
Let us show that by allowing $\mu_{24}$ to have exponential behaviour and setting it to $\mu_{24}=C\sinh(2\pi u)$ we arrive to the correct result. 
We analyse the reason for this choice in detail in section~\ref{sec:ancont}.

To simplify the constant part of $\mu_{ab}$ let us now make use of the $\gamma$-transformation, described in the last subsection.
This allows us to set $\mu_{12}=1,\;\mu_{34}=0$ and the constant $C$ to $1$ then the constraint \eq{constraint} imposes $\mu_{13}=0$ and $\mu_{14}=-1$.
Having fixed all $\mu$'s at leading order we get the following system of equations for $\bP_a$:
\beqa
&&\tilde \bP_1= -\bP_3 +\bP_1, \label{eq:P1L2} \\
&&\tilde \bP_2= -\bP_4 -\bP_2 -\bP_1 \sinh(2\pi u), \label{eq:P2L2}\\
&&\tilde \bP_3= \hspace{10mm}-\bP_3,\hspace{16mm} \label{eq:P3L2} \\
&&\tilde \bP_4= \hspace{10mm}+\bP_4+\bP_3 \sinh(2\pi u).\label{eq:P4L2}
\eeqa
Recalling that the functions $\bP_a$ only have a single short cut, we see from these equations that $\tilde{\bP}_a$ also have only this cut! 
This means that we can take all $\bP_a$ to be infinite Laurent series in the Zhukovsky variable $x(u)$, which rationalizes the Riemann surface with two sheets and one cut. 
It is defined by the ubiquitous mapping
\beq
	x+\frac{1}{x}=\frac{u}{g}
\eeq
where we pick the solution with a short cut, i.e.
\beq
x(u)=\frac{1}{2}\left(\frac{u}{g}+\sqrt{\frac{u}{g}-2}\;\sqrt{\frac{u}{g}+2}\;\right)\;\;.\;\;
\eeq
Solving the equations \eq{eq:P1L2} and \eq{eq:P3L2} with the asymptotics \eq{eq:asymptotics} we uniquely fix $\bP_1=\epsilon \, x^{-J/2}$ and $\bP_3=\epsilon\(x^{-J/2}-x^{+J/2}\)$, where $\epsilon$ is a constant yet to be found which we expect to be proportional to $\sqrt{S}$.
Thus the equations \eq{eq:P2L2} and \eq{eq:P4L2} become
\beqa
\label{eq:P2eq}
\tilde \bP_2+\bP_2&=& -\bP_4 -\epsilon x^{-J/2}\sinh(2\pi u)\;, \\
\label{eq:P4eq}
\tilde \bP_4-\bP_4&=& \epsilon(x^{-J/2}-x^{+J/2}) \sinh(2\pi u)\;.
\eeqa
We will first solve the second equation.
It is useful to introduce operations $[f(x)]_+$ and $[f(x)]_-$, which take parts of Laurent series with positive and negative powers of $x$ respectively.  
Taking into account that
\beq
	\sinh(2\pi u)=\sum\limits_{n=-\infty}^{\infty}I_{2n+1}  x^{2n+1},
\eeq
where $I_k\equiv I_{k}(4 \pi g)$ is the modified Bessel function of the first kind, we can write $\sinh(2\pi u)$ as
\beq
 \sinh(2\pi u)= \sinh_++\sinh_-,
\eeq
 where explicitly
\beqa
&& \sinh_+=[\sinh(2\pi u)]_+=\sum\limits_{n=1}^\infty I_{2n-1}x^{2n-1} \\
\label{defshm}
&& \sinh_-=[\sinh(2\pi u)]_-=\sum\limits_{n=1}^\infty I_{2n-1}x^{-2n+1}\;.
\eeqa
We now take the following ansatz for $\bP_4$
\beq
\bP_4=\epsilon(x^{J/2}-x^{-J/2})\sinh_-+Q_{J/2-1}(u),
\eeq
where $Q_{J/2-1}$ is a polynomial of degree $J/2-1$ in $u$. 
It is easy to see that this ansatz solves \eq{eq:P4eq} and has correct asymptotics. 
The polynomial $Q_{J/2-1}$ can be fixed from the equation \eq{eq:P2eq} for $\bP_2$. 
Indeed, from the asymptotics of $\bP_2$ we see that the lhs of \eq{eq:P2eq} does not have powers of $x$ from $-J/2+1$ to $J/2-1$. 
This fixes
\beq
Q_{J/2-1}(x)=-\epsilon\sum\limits_{k=1}^{J/2}I_{2k-1}\(x^{\frac{J}{2}-2k+1}+x^{-\frac{J}{2}+2k-1}\).
\eeq
Once $Q_{J/2-1}$ is found, we set $\bP_2$ to be the part of the right hand side of \eq{eq:P2eq} with powers of $x$ less than $-J/2$, which gives
\beq
\bP_2=-\epsilon x^{+J/2} \sum_{n=\frac{J}{2}+1}^\infty I_{2n-1}x^{1-2n}.
\eeq
This completes the solution for even $J$, we summarize it below:
\beqa
\label{eq:musolLOevenL}
&&\mu_{12}=1,\ \mu_{13}=0,\ \mu_{14}=-1,\ \mu_{24}=\sinh(2\pi u),\ \mu_{34}=0,\\
\label{eq:P1solLOevenL}
&&\bP_1=\epsilon x^{-J/2}\\
\label{eq:P2solLOevenL}
&&\bP_2=-\epsilon x^{+J/2} \sum_{n={J/2}+1}^\infty I_{2n-1}x^{1-2n}\\
\label{eq:P3solLOevenL}
&&\bP_3=\epsilon \(x^{-J/2}-x^{+J/2}\)\\
\label{eq:P4solLOevenL}
&&
	\bP_4=\epsilon \(x^{J/2}-x^{-J/2}\)\sinh_- -\epsilon \sum\limits_{n=1}^{J/2}I_{2n-1}\(x^{\frac{J}{2}-2n+1}+x^{-\frac{J}{2}+2n-1}\)\;.
\label{solutionevenL}
\eeqa
In the next subsection we fix the remaining parameter $\epsilon$ of the solution in terms of $S$ and find the energy, but now
let us briefly discuss the solution for odd $J$. 
As we mentioned above the main difference is that the functions $\bP_a$ now have a branch point at $u=\infty$, which is dictated by the asymptotics \eq{eq:asymptotics}. 
In addition, the parity of $\mu_{ab}$ is different according to the asymptotics of these functions \eq{eq:muasymptotics}. 
The solution is still very similar to the even $J$ case, and we discuss it in detail in Appendix \ref{sec:oddL}. 
Let us present the result here:
\beq
	\mu_{12}=1,\ \mu_{13}=0,\ \mu_{14}=0,\  \mu_{24}=\cosh(2\pi u),\ \mu_{34}=1
\eeq
\beqa
\label{P1oddL}
&&   \bP_1=\epsilon  x^{-J/2}, \\
&&   \bP_2=-\epsilon  x^{J/2}\sum\limits_{k=-\infty}^{-\frac{J+1}{2}}I_{2k}x^{2k},\\
&&   \bP_3=-\epsilon  x^{J/2}, \\
\label{P4oddL}
&&    \bP_4=\epsilon  x^{-J/2}\cosh_--\epsilon  x^{-J/2}\sum\limits_{k=1}^{\frac{J-1}{2}}I_{2k}x^{2k}-\epsilon  I_0 x^{-J/2}.
\eeqa
Note that now $\bP_a$ include half-integer powers of $x$.

\subsubsection{Fixing the global charges of the solution.}
\label{sec:LOresultevenL}

Finally, to fix our solution completely we have to find the value of $\epsilon$ and find the energy in terms of the spin using \eq{AA1} and \eq{AA2}.
For this we first extract the coefficients $A_a$ of the leading terms for all $\bP_a$, c.f. \eq{eq:asymptotics}.
From \eq{eq:P1solLOevenL}-\eq{eq:P4solLOevenL} or \eq{P1oddL}-\eq{P4oddL}
we get
\beqa
\label{Aexp1}
&& A_1= g^{J/2} \epsilon , \\
&& A_2=-g^{J/2+1} \epsilon  I_{J+1}, \\
\label{eq:A3LOL3}
&& A_3=-g^{-J/2} \epsilon , \\
\label{Aexplast}
&& A_4=-g^{-J/2+1}\epsilon  I_{J-1}.
\eeqa
Expanding \eq{AA1}, \eq{AA2} at small $S$ with $\Delta=J+S+\gamma$, where $\gamma={\cal O}(S)$, we find at linear order
\beqa
&& \gamma=i(A_1 A_4-A_2 A_3) \\
&& S=i(A_1A_4+A_2A_3)\;.
\eeqa
Plugging in the coefficients \eq{Aexp1}-\eq{Aexplast} we find that
\beq
\label{epss}
	\epsilon=\sqrt{\frac{2\pi i S}{JI_J(\sqrt\lambda)}}
\eeq
and we obtain the anomalous dimension at leading order,
\beq
\gamma=\frac{\sqrt{\lambda}I_{J+1}(\sqrt{\lambda})}{JI_J(\sqrt{\lambda})}S+{\cal O}(S^2),
\label{eq:resultLO}
\eeq
where the leading coefficient is precisely the slope function \eq{eq:slope_function} for the mode number $n=1$.
In fact the above calculation can be easily generalized for higher mode numbers. 
In the asymptotic Bethe ansatz for such operators we have two symmetric cuts formed by Bethe roots, with corresponding mode numbers being $\pm n$ (for the ground state $n=1$). 
To describe these operators within the $\bP\mu$-system we found that we should take $\mu_{24}=C\sinh(2\pi n u)$ instead of $\mu_{24}=C\sinh(2\pi u)$ (and for odd $J$ we similarly use $\mu_{24}=C\cosh(2\pi n u)$ instead of $\mu_{24}=C\cosh(2\pi u)$). 
Then the solution is very similar to the one above, and we find
\beq
\gamma=\frac{n\sqrt{\lambda}I_{J+1}(n\sqrt{\lambda})}{JI_J(n\sqrt{\lambda})}S\;,
\label{slopen}
\eeq
which exactly reproduces \eq{eq:slope_function} with $\Lambda \equiv n \sqrt{\lambda}$. 
In Appendix \ref{sec:Sanyn} we also show how using the $\bP\mu$-system one can reproduce the slope function for a configuration of Bethe roots with arbitrary mode numbers and filling fractions.

\subsubsection{Prescription for analytical continuation}
\label{sec:ancont}

To deduce the general prescription for the asymptotics of $\mu_{ab}$ for non-integer $S$ from our analysis, we first study the possible asymptotics of $\mu_{ab}$ for given $\bP_a$ in more detail. 
For that we combine the periodicity constraint \eq{muper} with \eq{eq:Pmu} and \eq{eq:mudisc} to write a finite difference equation on $\mu_{ab}$,
\beq\label{5bax}
\mu_{ab}(u+i)=\mu_{ab}(u)-\mu_{bc}(u)\chi^{cd}\bP_d\bP_a+\mu_{ac}(u)\chi^{cd}\bP_d\bP_b.
\eeq
As there are $5$ linear independent components of $\mu_{ab}$ this is a 5th order finite-difference equation.
Given the asymptotics of $\bP_a$ in \eq{eq:asymptotics} and \eq{AA1}, \eq{AA2} there are exactly $5$ different asymptotics a solution of \eq{5bax} could have as discussed in \cite{Gromov:2013pga}.
We denote these $5$ independent solutions of \eq{5bax} as $\mu_{12,A}$ where $A=1,\dots,5$ and summarize their leading asymptotics at large $u>0$ in the table below
\beq
\bea{c||l|l|l|l|l}
A=&1&2&3&4&5\\ \hline\hline
\mu_{12,A}\sim & u^{\Delta-J}& C_{1,2} u^{-S+1-J}& C_{1,3} u^{-J}& C_{1,4} u^{S-1-J}& C_{1,5} u^{-\Delta-J}\\
\mu_{13,A}\sim & C_{2,1} u^{\Delta+1}& C_{2,2} u^{-S+2}& C_{2,3} u^{+1}& { u^{S}}& C_{2,5} u^{-\Delta+1}\\
\mu_{14,A}\sim & C_{3,1} u^{\Delta}& C_{3,2} u^{-S+1}& 1 & C_{3,4} u^{S-1}& C_{3,5} u^{-\Delta}\\
\mu_{24,A}\sim & C_{4,1} u^{\Delta-1}& u^{-S}& C_{4,3} u^{-1}& C_{4,4} u^{S-2}& C_{4,5} u^{-\Delta-1}\\
\mu_{34,A}\sim & C_{5,1} u^{\Delta+J}& C_{5,2} u^{-S+1+J}& C_{5,3} u^{+J}& C_{5,4} u^{S-1+J}& { u^{-\Delta+J}}
\label{tablemu}
\eea
\eeq
where we fix the normalization of our solutions so that some coefficients are set to one.
The coefficients $C_{a,A}$ are some rational functions of $S,\Delta,J$ and $A_1,A_2$ and in the small $S$ limit all $C_{a,A}\to 0$ in our normalization.
As it was pointed out in \cite{Gromov:2013pga} the asymptotics for different $A's$ are obtained by replacing $\Delta$ in \eq{eq:muasymptotics} by $\pm \Delta,\pm (S-1) $ and $0$.
We label these solutions so that in the small $S$ regime these asymptotics are ordered $\Delta> 1-S>0>S-1>-\Delta$.
Of course any solution of \eq{5bax} multiplied by an $i$-periodic function will still remain a solution of \eq{5bax}.
The true $\mu_{ab}$ is thus a linear combination of the partial solutions $\mu_{ab,A}$ with some constant or periodic coefficients.
This particular combination should in addition satisfy the analyticity condition \eq{muper} which is not guaranteed by \eq{5bax}.

The prescription for analytical continuation in $S$ which we propose here is based on the large $u$ asymptotics of these periodic coefficients.
As we discussed in the previous section the assumption that all these coefficients are asymptotically constant is too constraining already at the leading order in $S$, and we must assume that at least some of these coefficients grow exponentially as $e^{2\pi u}$.
To get some extra insight into the asymptotic behaviour of these coefficients it is very instructive to go to the weak coupling regime.
It is known that at one loop the equation \eq{5bax} reduces to a second order equation. When written as a finite difference equation for $\mu_{12}$ it coincides exactly with the Baxter equation for the non-compact $\alg{sl}{2}$ spin chain. 
For $J=2$ it reads
\beq\label{oneloopbaxter}
\(2u^2-S^2-S-\frac{1}{2}\)Q(u)=(u+\tfrac{i}{2})^2 Q(u+i)
+(u-\tfrac{i}{2})^2 Q(u-i)
\eeq
where $Q(u)=\mu_{12}(u+i/2)$.
This equation is already very well studied and all its solutions are known explicitly \cite{Derkachov:2002wz} -- in particular it is easy to see that one of the solutions must have $u^S$ asymptotics at infinity, while the other behaves as $1/u^{S+1}$.
It is also known that at one loop and for any integer $S$ \eq{oneloopbaxter} has a polynomial solution which gives the energy as 
\beq
	\Delta=J+S+\left.2ig^2\d_u\log\frac{Q(u-i/2)}{Q(u-i/2)}\right|_{u=0}=S+J+8g^2 H_S.
\eeq
At the same time, for non-integer $S$ there are of course no polynomial solutions, and according to \cite{Janik:2013nqa} and \cite{Alfimov:2014bwa} the solution which produces the energy $S+J+8g^2 H_S$ cannot even have power-like asymptotics, instead the correct large $u$ behaviour must be
\beq
Q(u)\sim \(u^S+\dots\)+(A+B e^{2\pi u})\(\frac{1}{u^{S+1}}+\dots\)\;\;,\;\;u\to +\infty\;.
\eeq
Furthermore, there is a unique entire $Q$ function with the above asymptotics.
For $S>-1/2$ we can reformulate the prescription by saying that the correct solution has power-like asymptotics, containing all possible solutions, plus a small solution reinforced with an exponent.
In this form we can try to translate this result to our case. 
We notice that for $g\to 0$ we have $\mu_{12,1}\sim u^{S}$ and $\mu_{12,2}\sim u^{-S-1}$, which tells us that at least the second solution must be allowed to have a non-constant periodic coefficient in the asymptotics. 
We also assume that the coefficient in front of $\mu_{ab,3}$ tends to a constant.
This extra condition does not follow from the one loop analysis we deduced from our solution.
We will show how this prescription produces the correct known result for the leading order in $S$.
From our analysis it is hard to make a definite statement about the behaviour of the periodic coefficients in front of $\mu_{12,4}$ and  $\mu_{12,5}$, but due to the expected $\Delta \to-\Delta$ symmetry, which interchanges $\mu_{12,5}$ and $\mu_{12,1}$, one may expect that the coefficient of $\mu_{12,5}$ should also go to a constant. 
To summarize we should have
\beq\label{prescr}
\mu_{ab}(u)=\sum_{A=1}^5 c_{A}\mu_{ab,A}(u)
+\sum_{A=2,4,5} p_{A}(u)\mu_{ab,A}(u)
\eeq
where $c_A$ are constants whereas $p_{A}(u)$ are some linear combinations of $e^{\pm 2\pi u}$. 
It could also be that some of the coefficients of $p_{A}$ should be zero due to the constraint \eq{constraint}.

In the small $S$ limit, $\bP_a\to 0$ and the finite difference equation \eq{5bax} simply tells us that $\mu_{ab}(u+i)=\mu_{ab}(u)$ which implies that our $5$ independent solutions are just constants at the leading order in $S$.
We begin by noticing that in this limit $\mu_{12}$ must be entirely coming from $\mu_{12,1}$ as all the other solutions could only produce negative powers and thus cannot contribute at the leading order. 
So we start by imposing $\mu_{ab}=C_{ab}+D_{ab}\sinh(2\pi u)+E_{ab}\cosh(2\pi u)$ for some constants $C_{ab},D_{ab},E_{ab}$ such that $D_{12}=E_{12}=0$. 
Thus we have $5$ different $C$'s, $4$ different  $D$'s and $4$ different $E$'s.
We notice that this general form of $\mu_{ab}$ can be significantly simplified.
First, using the Pfaffian constraint \eqref{constraint} and the \text{$\gamma$-transformation} \eqref{gammaP} any generic $\mu_{ab}$ of this form can be reduced to one belonging to the following two-parametric family inside the original 13-parametric space
\beqa
&&\mu_{12}=1,\ \mu_{14}=a^2 \sinh{2\pi u}+\frac{a}{2}\cosh{2\pi u}\;,\\\
&&\mu_{24}=b\sinh{2\pi u}+\sinh{2\pi u}\;,\
\mu_{34}=\frac{a^2}{4}\frac{(1-2ab)^2}{b^2-1}+1\;,
\label{mufamilyab}
\eeqa
where $\mu_{13}$ is found from the Pfaffian constraint.
Second, recall that according to our prescription the 1st and 3rd solutions (columns in the table \eqref{tablemu}) cannot contain exponential terms.
Consider $\mu_{14}$ and $\mu_{24}$, we again see that the 4th and 5th solutions could only contain negative powers of $u$ and thus only the 2nd solution can contribute to the parts of $\mu_{14}$ and $\mu_{24}$ that are non-decaying at infinity.
This means that these components can be represented in the following form
\beqa
\mu_{14}=(a_1\sinh{2\pi u}+ a_2\cosh{2\pi u})\mu_{14,2}(u)+\mathcal{O}\(e^{2\pi u}/u\)\;,\\
\mu_{24}=(a_1\sinh{2\pi u}+a_2\cosh{2\pi u})\mu_{24,2}(u)+\mathcal{O}\(e^{2\pi u}/u\)\;,
\eeqa
for $u\to+\infty$.
The $\mathcal{O}\(e^{2\pi u}/u\)$ terms contain contributions from all of the solutions except for the 2nd. 
One can see that \eqref{mufamilyab} can be of this form only in two cases: if $a=0$ or if $a=\frac{1}{2b}$.
Both of these cases can be brought to the form
\beq
\label{muresan}
\mu_{12}=1,\ \mu_{13}=0,\ \mu_{14}=0,\ \mu_{24}=d_1\sinh{2\pi u}+d_2\cosh{2\pi u},\ \mu_{34}=1
\eeq
by a suitable $\gamma$-transformation \eqref{mufamilyab}. 
However, we found that there is an additional constraint which follows from compatibility of $\mu_{ab}$ with the decaying asymptotics of $\bP_2$. 
As we show in appendix \ref{sec:Sanyn} for even $J$ one must set $d_2=0$. 
For odd $J$ we must set $d_1=0$ as a compatibility requirement. 
This justifies the choice of $\mu_{ab}$ used in the previous section.
In the next section we will show how the same prescription can be applied at the next order in $S$ and leads to non-trivial results which we subjected to intensive tests later in the text.

\subsection{The curvature function}
\label{sec:curvature}

In this section we use the $\bP\mu$-system to compute the next coefficient in the small spin expansion of the folded string solution after the slope, which we call the curvature function $\gamma^{(2)}(g)$. 
First we will discuss the case $J=2$ in detail and then describe the modifications of the solution for the cases $J=3$ and $J=4$, more details on which can be found in appendix \ref{sec:NLOapp}.
Initially we assume the mode number to be $n=1$ and mention the issues encountered while generalizing the result in subsection \ref{sec:curvature_mode_numbers}.
The notations used in this section are summarized in appendix \ref{sec:notations}.

For convenience let us repeat the leading order solution \eq{eq:musolLOevenL}-\eq{eq:P4solLOevenL} of the $\bP\mu$-system for the slope function in the case when $J=2$
\beqa
{\bf P}^{(0)}_1=\epsilon\frac{1}{x}\;\;&,&\;\;{\bf P}^{(0)}_2=+\epsilon I_1-\epsilon x[\sinh(2\pi u)]_-\;\;,\\
{\bf P}^{(0)}_3=\epsilon\(\frac{1}{x}-x\)\;\;&,&\;\;
{\bf P}^{(0)}_4=
-2\epsilon I_1-
\epsilon \(\frac{1}{x}-x\)[\sinh(2\pi u)]_-.
\label{P10P40}
\eeqa
Here $\epsilon$ is a small parameter, proportional to $\sqrt{S}$ as seen in \eq{epss} and by $\bP_a^{^{(0)}}$ we denote the $\bP_a$ functions at leading order in $\epsilon$.
The key observation is that the $\bP\mu$-system can be solved iteratively order by order in $\epsilon$. 
Let us write $\bP_a$ and $\mu_{ab}$ as an expansion in this small parameter
\beq
	\bP_a=\epsilon\bP_a^{(0)}+\epsilon^3\bP_a^{(1)}+\epsilon^5\bP_a^{(2)}+\dots
\eeq
\beq
	\mu_{ab}=\mu_{ab}^{(0)}+\epsilon^2\mu_{ab}^{(1)}+\epsilon^4\mu_{ab}^{(2)}+\dots \;.
\eeq
This structure of the expansion is dictated by the equations \eq{eq:Pmu} and \eq{eq:mudisc} of the $\bP\mu$-system as we will soon see explicitly. 
Since the leading order $\bP_a$ are of order $\epsilon$, equation \eq{eq:mudisc} implies that the discontinuity of $\mu_{ab}$ on the cut is of order $\epsilon^2$. 
Thus to find $\mu_{ab}$ in the next to leading order (NLO) we only need the functions $\bP_a$ at leading order. 
After this, we can find the NLO correction to $\bP_a$ from equations \eq{eq:mudisc}. 
This will be done below, and having thus the full solution of the $\bP\mu$-system at NLO we will find the energy at order $S^2$.

\subsubsection{Correcting $\mu_{ab}$}
\label{sec:muNLOL2}
In this subsection we find the NLO corrections $\mu^{(1)}_{ab}$ to $\mu_{ab}$. 
As follows from \eq{eq:mudisc} and \eq{muper} they should satisfy the equation
\beq
 \mu_{ab}^{(1)}(u+i)-\mu_{ab}^{(1)}(u)=\bP_a^{(0)} \tilde\bP_b^{(0)}-  \bP_b^{(0)} \tilde\bP_a^{(0)},
\label{eq:mudiscNLO}
\eeq
in which the right hand is known explicitly. 
For that reason let us define an apparatus for solving equations of this type, i.e.
\beq
f(u+i)-f(u)=h(u).
\label{eqperiod}
\eeq
More precisely, we consider functions $f(u)$ and $h(u)$ with one cut in $u$ between $-2g$ and $2g$, and no poles. 
Such functions can be represented as infinite Laurent series in the Zhukovsky variable $x(u)$, and we additionally restrict ourselves to the case where for $h(u)$ this expansion does not have a constant term. 
Note that the right hand side of \eq{eq:mudiscNLO} has the form $F(u)-\tilde F(u)$ and therefore indeed does not have a constant term in its expansion, as the constant in $F$ would cancel in the difference $F(u)-\tilde F(u)$.
One can see that the general solution of \eq{eqperiod} has a form of a particular solution plus an arbitrary $i$-periodic function, which we also call a zero mode. 
First we will describe the construction of the particular solution and later deal with zero modes. 

The linear operator which gives the particular solution of \eq{eqperiod} described below will be denoted as $\Sigma$.
Notice that given the explicit form \eq{P10P40} of $\bP^{(0)}_a$, the right hand side of \eq{eq:mudiscNLO} can be represented in a form
\beq
\alpha(x)\sinh(2\pi u)+\beta(x),
\label{alphabetasinh}
\eeq
where $\alpha(x),\beta(x)$ are power series in $x$ growing at infinity not faster than polynomially. 
Thus for such $\alpha$ and $\beta$ we define
\beq
\Sigma\cdot\[\alpha(x)\sinh(2\pi u)+\beta(x)\]\equiv \sinh(2\pi u) \Sigma\cdot \alpha(x)+\Sigma\cdot \beta(x).
\eeq
We also define $\Sigma\cdot x^{-n}=\Gamma'\cdot x^{-n}$ for $n>0$, where the integral operator $\Gamma'$ defined as
\beq
\(\Gamma'\cdot h\)(u)\equiv \oint_{-2g}^{2g}\frac{dv}{{4\pi i}}\partial_u \log \frac{\Gamma[i (u-v)+1]}{\Gamma[-i (u-v)]}h(v),
\label{Gammaprime}
\eeq
where we integrate around the cut. 
This requirement is consistent because of the following relation
\beq
\(\Gamma'\cdot h\)(u+i)-\(\Gamma'\cdot h\)(u)
=
-\frac{1}{2\pi i}\oint_{-2g}^{2g}\frac{h(v)}{u-v}dv=h_-(u)-\widetilde{h_+}(u).
\label{eq:Gammaproperty}
\eeq
We remind that $f_+$ and $f_-$ stand for the part of the Laurent expansion with, respectively, positive and negative powers of $x$, while $\tilde f$ is the analytic continuation around the branch point at $u=2g$ (which amounts to replacing $x\to 1/x$).
What is left is to define $\Sigma$ on positive powers of $x$, which we do by requiring
\beq
\Sigma\cdot\left[x^a+1/x^a\right]\equiv 2 p_a'(u) 
\label{paprime}
\eeq
where $p_a'(u)$ is a polynomial in $u$ of degree $a+1$, which is a solution of
\beq
p_a'(u+i)-p_a'(u)=\frac{1}{2}\(x^a+1/x^a\)
\eeq
and satisfies the following additional properties: $p_a'(0)=0$ for odd $a$  and $p_a'(i/2)=0$ for even $a$. 
One can check that this definition is consistent and defines of $p'_a(u)$ uniquely. 
Explicit form of the first few $p_a'(u)$, which we call periodized  Chebyshev polynomials, can be found in appendix \ref{sec:appPeriodized}.
From this definition of $\Sigma$ one can see that the result of its action on expressions of the form \eq{alphabetasinh} can again be represented in this form - what is important for us is that no exponential functions other than $\sinh(2\pi u)$ appear in the result.

A good illustration of how the definitions above work would be the following two simple examples.
Suppose one wants to calculate $\Sigma\cdot\(x-\frac{1}{x}\)$, then it is convenient to split the argument of $\Sigma$ in the following way:
\beq
\Sigma\cdot\(x-\frac{1}{x}\)=\Sigma\cdot\(x+\frac{1}{x}\)-2\,\Sigma\cdot\frac{1}{x}.
\eeq
In the first term we recognize $p_1'(u)=\frac{i u(u-i)}{2g}$, whereas in the second the argument of $\Sigma$ is decaying at infinity, thus $\Sigma$ is equivalent to $\Gamma'$ in this context. 
Notice also that $\Gamma'\cdot \frac{1}{x}=-\Gamma'\cdot x$. 
All together we get
\beq
\Sigma\cdot\(x-\frac{1}{x}\)=\Sigma\cdot\(x+\frac{1}{x}\)-2\,\Sigma\cdot\frac{1}{x}=2\,p_1'(u)+ 2 \,\Gamma'\cdot x.
\eeq
In a similar way, in order to calculate $\Sigma\cdot \(\sinh_--\sinh_+\) / 2$, one can write 
\beq
	\frac{\sinh_--\sinh_+}{2}=\sinh_- \, - \, \frac{1}{2}\sinh(2\pi u).
\eeq
Notice that since $\sinh_-$ decays at infinity, thus
\beq
\Sigma\cdot\sinh_-=\Gamma'\cdot\sinh_-.
\eeq
 Also, since $i$-periodic functions can be factored out of $\Sigma$,
\beq
\Sigma\cdot\sinh(2\pi u)=\sinh(2\pi u)\Sigma\cdot 1=\sinh(2\pi u)p_0'(u)/2.
\eeq
Finally,
\beq
\Sigma\cdot\frac{\sinh_--\sinh_+}{2}=\Gamma'\cdot(\sinh_-)-\frac{1}{2}\sinh(2\pi u)p_0'(u).
\eeq
As an example we present the particular solution for two components of $\mu_{ab}$.
\beqa
\label{muexpl1}
\mu_{13}^{(1)}-\pi_{13}&=&\Sigma\cdot\({\bf P}_1 \tilde{\bf P}_3-{\bf P}_3 \tilde{\bf P}_1\)=\epsilon^2\,\Sigma\cdot\(x^2-\frac{1}{x^2}\) =\epsilon^2\;\(\Gamma'\cdot x^2+p_2'(u)\),\\
\nonumber
\mu_{12}^{(1)}-\pi_{12}&=&\Sigma\cdot\({\bf P}_1
   \tilde{\bf P}_2-{\bf P}_2
   \tilde{\bf P}_1\)= \\ &=&
   -\epsilon^2\[2 I_1\Gamma'\cdot x-\sinh(2\pi u)\;\Gamma'\cdot x^2-\Gamma'\cdot\(\sinh_-\(x^2+\frac{1}{x^2}\)\)
   \].\;\;\;\;\;\;\;\qquad\label{muexpl2}
\eeqa
Below we will argue that $\pi_{12}$ and $\pi_{13}$ can be chosen to be zero, as seen in \eq{eq:periodicpart}.
Now let us apply $\Sigma$ defined above to \eq{eq:mudiscNLO}, writing that its general solution is
\beq
\mu^{(1)}_{ab}=\Sigma\cdot(\bP_a^{(0)} \tilde\bP_b^{(0)}-  \bP_b^{(0)} \tilde\bP_a^{(0)})+\pi_{ab},
\label{eq:sol13}
\eeq
where the zero mode $\pi_{ab}$ is an arbitrary $i$-periodic entire function, which can be written similarly to the leading order as $c_{1,ab}\cosh{2\pi u}+c_{2,ab}\sinh{2\pi u}+c_{3,ab}$. 
Again, many of the coefficients $c_{i,ab}$ can be set to zero. 
First, the prescription from section \ref{sec:ancont} implies that non-vanishing at infinity part of coefficients of $\sinh(2\pi u)$ and $\cosh(2\pi u)$ in $\mu_{12}$ is zero. 
As one can see from the explicit form \eq{muexpl2} of the particular solution which we choose for $\mu_{12}$, it does not contain $\cosh(2\pi u)$ and the coefficient of $\sinh(2\pi u)$ is decaying at infinity. 
So in order to satisfy the prescription, we have to set $c_{2,12}$ and $c_{2,12}$ to zero. 
Second, since the coefficients $c_{n,ab}$ are of order $S$, we can remove some of them by making an infinitesimal  $\gamma$-transformation with $R=1 + {\cal O}(S)$ in \eq{gammaP}.
Furthermore, the Pfaffian constraint \eq{constraint} imposes 5 equations on the remaining coefficients, which leaves the following 2-parametric family of zero modes
\beqa
\pi_{12}= \pi_{13} = 0,&&\ \pi_{14}=\frac{1}{2}c_{1,34}\cosh{2\pi u},\\
\pi_{24}= c_{1,24}\cosh{2\pi u},&&\ \pi_{34}= c_{1,34}\cosh{2\pi u}.
\eeqa
Let us now look closer at the exponential part of $\mu_{14}$ and $\mu_{24}$. 
Combining the leading order \eq{eq:musolLOevenL} and the perturbation \eq{eq:sol13} and taking into account the fact that operator $\Sigma$ does not produce terms proportional to $\cosh{2 \pi u}$, we obtain
\beqa
&&\mu_{14}=\frac{1}{2}c_{1,34}\cosh{2\pi u}+{\cal O}(\epsilon) \sinh{2\pi u}+\mathcal{O}(\epsilon^2)+\dots, \\
&&\mu_{24}=\frac{1}{2}c_{1,24}\cosh{2\pi u}+(1+{\cal O}(\epsilon)) \sinh{2\pi u}+\mathcal{O}(\epsilon^2)+\dots,
\eeqa
where dots stand for powers-like terms or exponential terms suppressed by powers of $u$.
As we remember from section \ref{sec:ancont}, only the 2nd solution of the 5th order Baxter equation \eq{5bax} can contribute to the exponential part of $\mu_{14}$ and $\mu_{24}$, which means that $\mu_{14}$ and $\mu_{24}$ are proportional to the same linear combination of $\sinh{2\pi u}$ and $\cosh{2\pi u}$. 
From the second equation one can see that this linear combination can be normalized to be $\frac{1}{2}c_{1,24}\cosh{2\pi u}+(1+{\cal O}(\epsilon)) \sinh{2\pi u}$ and thus 
\beq
	\mu_{14}=C\(\frac{1}{2}c_{1,24}\cosh{2\pi u}+(1+{\cal O}(\epsilon)) \sinh{2\pi u}\),
\eeq 
where $C$ is some constant, which is of order ${\cal O}(\epsilon)$, because the coefficient of $\sinh{2\pi u}$ in the first equation is ${\cal O}(\epsilon)$.
Taking into account that $c_{1,24}$ is ${\cal O}(\epsilon)$ itself, we find that $c_{1,34}=\mathcal{O}(\epsilon^2)$, i.e. it does not contribute at the order which we are considering. 
So the final form of the zero modes in \eq{eq:sol13} is
\beq
	\pi_{12} = \pi_{13} = \pi_{14} = \pi_{34}=0, \;\;\; \pi_{24}=c_{1,24}\cosh{2\pi u}.
	\label{eq:periodicpart}
\eeq
In this way, using the particular solution given by $\Sigma$ and the form of zero modes \eq{eq:periodicpart} we have computed all the functions $\mu_{ab}^{(1)}$. 
The details and the results of the calculation can be found in appendix \ref{sec:appmu2}.

\subsubsection{Correcting $\bP_{a}$}
\label{sec:CalculationofPa}

Now that we found the NLO part of $\mu_{ab}$ we can use the iterative procedure described at the beginning of the section to write a closed system of equations for $\bP_a^{(1)}$.
Indeed, expanding the system \eq{eq:pmuexpanded} to NLO we get
\beqa
\label{eq:P1eqNLOL2}
&&\tilde \bP^{(1)}_1
- \bP^{(1)}_1
= -\bP^{(1)}_3+r_1,  \\
\label{eq:P2eqNLOL2}
&&\tilde \bP^{(1)}_2+\bP_2^{(1)}= -\bP^{(1)}_4  -\bP^{(1)}_1 \sinh(2\pi u)+r_2, \\
\label{eq:P3eqNLOL2}
&&\tilde \bP^{(1)}_3+\bP_3^{(1)}=r_3,\\
\label{eq:P4eqNLOL2}
&&\tilde \bP^{(1)}_4-\bP_4^{(1)}=\bP_3^{(1)} \sinh(2\pi u)+r_4,
\eeqa
where the free terms are given by
\beq
r_a=-\mu_{ab}^{(1)}\chi^{bc}\bP_c^{(0)}.
\label{eq:ra}
\eeq
Notice that $r_a$ does not change if we add a matrix proportional to $\bP_a^{(0)} \tilde\bP_b^{(0)}-  \bP_b^{(0)} \tilde\bP_a^{(0)}$ to $\mu^{(1)}_{ab}$ due to the relations
\beq
	\bP_a \chi^{ab}\bP_b=0,\;\bP_a\chi^{ab}\tilde\bP_b=0,
\eeq	
which follow from the $\bP\mu$-system equations. 
In particular we can use this property to do the following replacement in \eq{eq:ra}
\beq
	\mu_{ab}^{(1)} \to \mu_{ab}^{(1)}+\frac{1}{2}\(\bP_a^{(0)} \tilde\bP_b^{(0)}-  \bP_b^{(0)} \tilde\bP_a^{(0)}\).
\eeq	
This will be convenient for us, since in expressions for $\mu^{(1)}_{ab}$ in terms of $p_a$ and $\Gamma$  as seen in \eq{muexpl1}, \eq{muexpl2} and appendix \ref{sec:appmu2}, this change amounts to simply replacing $\Gamma'$ by a convolution with a more symmetric kernel $\Gamma' \rightarrow  \Gamma$ defined by
\beq
\(\Gamma\cdot h\)(u)\equiv \oint_{-2g}^{2g}\frac{dv}{{4\pi i}}\partial_u \log \frac{\Gamma[i (u-v)+1]}{\Gamma[-i (u-v)+1]}h(v),
\label{Gamma}
\eeq
while at the same time replacing
\beq
	 p_a'(u)\rightarrow  p_a(u) = p_a'(u)+\frac{1}{2}\(x^a(u)+x^{-a}(u)\).
\label{pa}
\eeq

Having made this comment, we will now develop tools for solving the equations \eq{eq:P1eqNLOL2} - \eq{eq:P4eqNLOL2}.
Notice first that if we solve them in the order \eq{eq:P3eqNLOL2}, \eq{eq:P1eqNLOL2}, \eq{eq:P4eqNLOL2}, \eq{eq:P2eqNLOL2}, substituting into each subsequent equation the solution of all the previous, then at each step the problem we have to solve has a form
\beq
 \tilde f+f=h\;\; \text{or}\;\; \tilde f-f=h\;\;,
 \label{eq:eqs}
\eeq
where $h$ is known, $f$ is unknown and both the right hand side and the left hand side are power series in $x$. 
Since all cuts are of the square root type, namely $\tilde{\tilde{f}} = f$, it is obvious that equations \eq{eq:eqs} have solutions only for $h$ such that $h=\tilde h$ and $h=-\tilde h$ respectively.
On the class of such $h$ a particular solution for $f$ can be written as
\beq
f= [h]_-+[h]_0/2\equiv H\cdot h\;\; \Rightarrow\;\; \tilde f+f=h
\label{eq:solfh1}
\eeq
and
\beq
f= [h]_-\equiv K\cdot h\;\; \Rightarrow\;\; \tilde f-f=h,
\label{eq:solfh2}
\eeq
where $[h]_0$ is the constant part of Laurent expansion of $h$ (it does not appear in the second equation, because $h$ such that $h=-\tilde h$ does not have a constant part).
The operators $K$ and $H$ introduced here can be also defined by their integral kernels
\beqa
H(u,v)&=&-\frac{1}{4\pi i}\frac{\sqrt{u-2g}\sqrt{u+2g}}{\sqrt{v-2g}\sqrt{v+2g}}\frac{1}{u-v}, \\
K(u,v)&=&+\frac{1}{4\pi i}\frac{1}{u-v},
\label{eq:HK}
\eeqa
which are equivalent to \eq{eq:solfh1}, \eq{eq:solfh2} of the classes of $h$ such that $h=\tilde h$ and $h=-\tilde h$ respectively.
The particular solution $f=K\cdot h$ of the equation $\tilde f+ f=h$ is unique in the class of functions $f$ decaying at infinity, and the solution $f=H \cdot h$ of $\tilde f- f=h$ is unique for non-growing $f$. 
In all other cases the general solution will include zero modes, which, in our case are fixed by asymptotics of $\bP_a$.
Now it is easy to write the explicit solution of the equations
\eq{eq:P1eqNLOL2}-\eq{eq:P4eqNLOL2}:
\beqa
\bP_3^{(1)}&=&H\cdot r_3,\\
\bP_1^{(1)}&=&\frac{1}{2}\,\bP^{(1)}_3+K\cdot \(r_1-\frac{1}{2} r_3\),\\
\bP_4^{(1)}&=&K\cdot\(-\frac{1}{2}\(\tilde\bP_3^{(1)}-\bP_3^{(1)}\) \sinh(2\pi u)+
\frac{2r_4+r_3 \sinh(2\pi u)}{2}\)-2\,\delta,\;\;\qquad\\
\bP_2^{(1)}&=&H\cdot\(-\frac{1}{2}
\(
{\bf P}^{(1)}_4+\sinh(2\pi u){\bf P}^{(1)}_1+\tilde{\bf P}^{(1)}_4+\sinh(2\pi u)\tilde{\bf P}^{(1)}_1
\)+\right.\\ \nn
&&\left.
+\frac{r_4+\sinh(2\pi u) r_1+2r_2}{2}\)+\delta,
\label{eq:P4solNLOL2}
\eeqa
where $\delta$ is a constant fixed uniquely by requiring $\mathcal{O}(1/u^2)$ asymptotics for $\bP_2$. 
This asymptotic also sets the last coefficient $c_{1,24}$ left in $\pi_{12}$ to zero. 
Thus in the class of functions with asymptotics \eq{eq:asymptotics} the solution for $\mu_{ab}$ and $\bP_a$ is unique up to a $\gamma$-transformation.

\subsubsection{Result for $J=2$}
\label{sec:resultL2}

In order to obtain the result for the anomalous dimension, we again use the formulas \eq{AA1}, \eq{AA2} which connect the leading coefficients of $\bP_a$ with $\Delta,\ J\ $ and $S$. 
After plugging in $A_i$ which we find from our solution, we obtain the result for the $S^2$ correction to the anomalous dimension:
\beqa
\label{gamma2L2}
\gamma^{(2)}_{J=2}&=&\frac{\pi}{g^2(I_1-I_3)^3}\oint \frac{du_x}{2\pi i}\oint \frac{du_y}{2\pi i}\[\frac{8  I_1^2(I_1+I_3) \left(x^3-\left(x^2+1\right) y\right) }{ \left(x^3-x\right) y^2}\right.\\ \nn
&&   +\frac{8  \sh_-^x \sh_-^y
   \left(x^2 y^2-1\right) \left(I_1 (x^4 y^2+1)-I_3x^2(y^2+1)\right)}{ x^2 \left(x^2-1\right)
   y^2}\\ \nn
&&-\frac{4  (\sh_-^y)^2 x^2 \left(y^4-1\right) \left( I_1(2x^2-1)-I_3 \right)}{ \left(x^2-1\right) y^2}\\ \nn
&&+\frac{8
   I_1^2 \sh_-^y x  \left(2 \(x^3-x\) \left(y^3+y\right)-2 x^2
   \left(y^4+y^2+1\right)+y^4+4 y^2+1\right)}{ \left(x^2-1\right) y^2}\\  \nn
&&-\frac{8 (I_1-I_3)
   I_1 \sh_-^y x   (x-y) (x
   y-1)}{ \left(x^2-1\right) y}\\ \nn
&&\left.-\frac{4 (I_1-I_3) (\sh_-^x)^2 \left(x^2+1\right)
   y^2}{ \left(x^2-1\right)}\right]
	\frac{1}{4\pi i}\partial_u \log\frac{\Gamma (i u_x-i u_y+1)}{\Gamma (1-i u_x+i u_y)}\;.
\eeqa
Here the integration contour goes around the branch cut at $(-2g,2g)$. 
We also denote
$\sh_-^x=\sinh_-(x) ,\ \sh_-^y=\sinh_-(y)$, recall that $\sinh_-$ was defined in \eq{defshm}. 
This is our final result for the curvature function at any coupling.

It is interesting to note that our result contains the combination $\log\frac{\Gamma (i u_x-i u_y+1)}{\Gamma (1-i u_x+i u_y)}$ which plays an essential role in the construction of the BES dressing phase, namely they enter the $\beta_{r,s}$ terms in \eq{eq:bes_phase}. 
We will use this identification in section \ref{sec:strong_curvature} to compute the integral in \eq{gamma2L2} numerically with high precision.

 \subsubsection{Results for higher $J$}
\label{sec:SolvingPmuL3}

Solving the $\bP\mu$-system for $J=3$ is similar to the $J=2$ case described above, except for several technical complications, which we will describe here, leaving the details for the appendix \ref{sec:appnlo3}.
As in the previous section, the starting point is the LO solution of the $\bP\mu$-system, which for $J=3$ reads
\beq
	\bP_1=\epsilon x^{-3/2},\ \bP_3=-\epsilon x^{3/2},
\label{P1P3LOsolL3}
\eeq
\beq
	\bP_2=-\epsilon x^{3/2}\cosh_- +\epsilon x^{-1/2}I_2,
\label{P2LOsolL3}
\eeq
\beq
	\bP_4=-\epsilon x^{1/2}I_2-\epsilon x^{-3/2}I_0-\epsilon x^{-3/2}\cosh_-,
\label{P4LOsolL3}
\eeq
\beq
	\mu_{12}=1,\ \mu_{13}=0,\ \mu_{14}=0,\  \mu_{24}=\cosh(2\pi u),\ \mu_{34}=1\;.
\eeq
The first step is to construct $\mu^{(1)}_{ab}$ from its discontinuity given by the equation \eq{eq:mudiscNLO}. 
The full solution consists of a particular solution and a general solution of the corresponding homogeneous equation, i.e. zero mode $\pi_{ab}$. 
In our case the zero mode can be an $i$-periodic function, i.e. a linear combination of $\sinh(2\pi u)$, $\cosh(2\pi u)$ and constants. 
As in the case of $J=2$, we use a combination of the Pfaffian constraint, prescription from section \ref{sec:ancont} and a $\gamma$-transformation to reduce all the parameters of the zero mode to just one, sitting in $\mu_{24}$:
 \beq
\pi_{12}=0,\;\pi_{13}=0,\;\pi_{14}=0,\;\pi_{24}=c_{24,2} \sinh\(2\pi u\),\;\pi_{34}=0.
\label{eq:periodicpartL3}
\eeq
As in the previous section, the next step is to find $\bP_a^{(1)}$ from the $\bP\mu$-system expanded to the first order, namely from
\beqa
\label{eq:P1L3}
&&\tilde \bP_1^{(1)}+\bP_3^{(1)}=r_1,\\
&&\tilde \bP_2^{(1)}+\bP_4^{(1)}+\bP_1^{(1)} \cosh(2\pi u)=r_2,\\
&&\tilde \bP_3^{(1)}+\bP_1^{(1)}=r_3,\\
&&\tilde \bP_4^{(1)}+\bP_2^{(1)}-\bP_3^{(1)}\cosh(2\pi u)=r_4,
\label{eq:P4L3}
\eeqa
where $r_a$ are defined by \eq{eq:ra} and for $J=3$ are given explicitly in appendix \ref{sec:appnlo3}.
In attempt to solve this system, however, we encounter another technical complication. 
As one can see from \eq{P1P3LOsolL3} - \eq{P4LOsolL3}, the LO solution contains half-integer powers of $J$, meaning that the $\bP_a$ now have an extra branch point at infinity.
However, the operations $H$ and $K$ defined by \eq{eq:HK} work only for functions which have Laurent expansion in integer powers of $x$. 
In order to solve equations of the type \eq{eq:mudiscNLO} on the class of functions which allow Laurent-like expansion in $x$ with only half-integer powers $x$, we introduce operations $H^*,K^*$:
\beqa
&&H^*\cdot f\equiv\frac{x+1}{\sqrt{x}}H\cdot\frac{\sqrt{x}}{x+1} f, \\
&&K^*\cdot f\equiv\frac{x+1}{\sqrt{x}}K\cdot\frac{\sqrt{x}}{x+1} f.
\eeqa
In terms of these operations the solution of the system \eq{eq:P1L3} - \eq{eq:P4L3} is
\beqa
\label{eq:P1J3}
\bP_{1}^{(1)}&=&\frac{1}{2}\(H^*(r_1+r_3)+ K^*(r_1-r_3)\)+\bP_1^{\text{zm}},\\
\bP_{3}^{(1)}&=&\frac{1}{2}\(H^*(r_1+r_3)- K^*(r_1-r_3)\)+\bP_2^{\text{zm}},\\
\bP_{2}^{(1)}&=&\frac{1}{2}\(H^*(r_2+r_4)+ K^*(r_2-r_4)\,-\right.\nonumber\\
&-&\left.H^*\(\cosh(2\pi u)K^*(r_1-r_3)\)- K^*\(\cosh(2\pi u)H^*(r_1+r_3)\)\right)+\bP_3^{\text{zm}},\qquad\;\;\\
\label{eq:P4J3}
\bP_{4}^{(1)}&=&\frac{1}{2}\(H^*(r_2+r_4)- K^*(r_2-r_4)\,-\right.\nonumber\\
&-&\left.H^*\(\cosh(2\pi u)K^*(r_1-r_3)\)+ K^*\(\cosh(2\pi u)H^*(r_1+r_3)\)\right)+\bP_4^{\text{zm}},
\eeqa
where $\bP_a^{\text{zm}}$ is a solution of the system \eq{eq:P1L3} - \eq{eq:P4L3} with the right hand side set to zero, whose explicit form $\bP_a^{\text{zm}}$ is given in Appendix \ref{sec:appnlo3}, expressions \eq{P1J3zm} - \eq{P4J3zm} and which is parametrized by four constants $L_1,L_2,L_3,L_4$, e.g.
\beqa
\bP_1^{\text{zm}}=L_1 x^{-1/2}+L_3x^{1/2}.
\eeqa
These constants are fixed by requiring correct asymptotics of $\bP_a$, which also fixes the parameter $c_{24,2}$ in the zero mode \eq{eq:periodicpartL3} of $\mu_{ab}$, which is actually fixed to be zero. 
Indeed, a priori $\bP_2$ and $\bP_1$ have wrong asymptotics. 
Imposing a constraint that $\bP_2$ decays as $u^{-5/2}$ and $\bP_1$ decays as $u^{-3/2}$ produces five equations, which fix all the parameters uniquely.
Skipping the details of the intermediate calculations, we present the final result for the anomalous dimension,

\footnotesize
\beqa
&&\gamma^{(2)}_{J=3}=\oint \frac{du_x}{2\pi i}\oint \frac{du_y}{2\pi i}
i \frac{1}{g^2(I_2-I_4)^3} \left[\frac{2 \left(x^6-1\right) y (\ch_-^y)^2 (I_2-I_4)}{x^3 \left(y^2-1\right)}-\right.\\ \nn
&&-\frac{4 \ch_-^x
   \ch_-^y \left(x^3 y^3-1\right) \left(I_2 x^5 y^3+I_2-I_4 x^2 \left(x y^3+1\right)\right)}{x^3
   \left(x^2-1\right) y^3}+\\ \nn
&& +\frac{(y^2-1)  (\ch_-^y)^2 I_2 \left( (x^8+1) \left(2 y^4+3 y^2+2\right)-(x^6+x^2)
   \left(y^2+1\right)^2\right)}{x^3 \left(x^2-1\right) y^3}-
   \\ \nn
   && -\frac{(y^2-1)  (\ch_-^y)^2 I_4 \left((x^8+1) y^2+(x^6+x^2) \left(y^4+1\right)\right)}{x^3 \left(x^2-1\right) y^3}-
   \\ \nn
&&-\frac{4 I_2 \ch_-^y (x-y) (x y-1) \left(I_2
   \left(\(x^6+1\) \left(y^3+y\right)+\(x^5+x\) \left(y^4+y^2+1\right)-x^3 \left(y^4+1\right)\right)+I_4 x^3
   y^2\right)}{x^3 \left(x^2-1\right) y^3}\\ \nn
&& \left.-\frac{I_2^2 (y^2-1)  (x-y) (x y-1) \left(I_2 \left(\(x^6 +x^4 +x^2 +1\)y+2 x^3
   \left(y^2+1\right)\right)+I_4 \left(x^5+x\right) \left(y^2+1\right)\right)}{x^3 \left(x^2-1\right) y^3}\right]\\ \nn
	&& \frac{1}{4\pi i}\partial_u \log\frac{\Gamma (i u_x-i u_y+1)}{\Gamma (1-i u_x+i u_y)}.
\label{gamma2L3}
\eeqa
\normalsize
We defined $\ch_-^x=\cosh_-(x)$ and $\ch_-^y=\cosh_-(y)$, where $\cosh_-(x)$ is the part of the Laurent expansion of $\cosh\(g(x+1/x)\)$ vanishing at infinity, i.e.
\beq
\cosh_-(x)=\sum_{k=1}^{\infty} I_{2k}x^{-2k}.
\eeq
The result for $J=4$ is given in appendix \ref{sec:SolvingPmuL4}.

\subsubsection{Weak coupling expansion}
\label{sec:curvature_weak}

Our results for the curvature function $\gamma^{(2)}(g)$ at $J=2,3,4$ given in \eq{gamma2L2}, \eq{gamma2L3} and \eq{gamma2L4} are straightforward to expand at weak coupling. 
We give expansions to 10 loops in \text{appendix \ref{sec:weakS3}}. Let us start with the $J=2$ case, for which we found
\beqa
\label{weak22}
\gamma_{J=2}^{(2)}&=&-8 g^2 \zeta_3+g^4 \left(140 \zeta_5-\frac{32 \pi ^2 \zeta_3}{3}\right)+g^6 \left(200 \pi ^2 \zeta_5-2016
   \zeta_7\right)
	\\ \nn
	&+&g^8 \left(-\frac{16 \pi ^6 \zeta_3}{45}-\frac{88 \pi ^4 \zeta_5}{9}-\frac{9296 \pi ^2 \zeta_7}{3}+27720 \zeta_9\right)
	\\ \nn
	&+&g^{10} \left(\frac{208 \pi ^8 \zeta_3}{405}+\frac{160 \pi ^6 \zeta_5}{27}+144
   \pi ^4 \zeta_7+45440 \pi ^2 \zeta_9-377520 \zeta_{11}\right)
	+\dots
\eeqa
Remarkably, at each loop order all contributions have the same transcendentality and only simple zeta values (i.e. $\zeta_n$) appear. 
This is also true for the $J=3$ and $J=4$ cases.
We can check this expansion against known results, as the anomalous dimensions of twist two operators have been computed up to five loops for arbitrary spin \cite{Kotikov:2001sc,Kotikov:2003fb,Kotikov:2004er,Moch:2004pa,Staudacher:2004tk,Kotikov:2007cy,Bajnok:2008qj,Lukowski:2009ce} (see also \cite{Velizhanin:2013vla} and the review \cite{Freyhult:2010kc}).
To three loops they can be found solely from the ABA equations, while at four and five loops wrapping corrections need to be taken into account which was done in \cite{Bajnok:2008qj,Lukowski:2009ce} by utilizing generalized Luscher formulas. 
All these results are given by linear combinations of harmonic sums
\beq
	S_a(N) = \sum_{n=1}^N\frac{(\mathrm{sign}(a))^n}{n^{|a|}}, \ \
	S_{a_1,a_2,a_3,\dots}(N)=\sum_{n=1}^N\frac{(\mathrm{sign}(a_1))^n}{n^{|a_1|}}S_{a_2,a_3,\dots}(n)
\eeq
with argument equal to the spin $S$. 
To make a comparison with our results we expanded these predictions in the $S\to 0$ limit. 
For this lengthy computation, as well as to simplify the final expressions, we used the \verb"Mathematica" packages HPL \cite{HPL}, the package \cite{VolinPackage} provided with the paper \cite{Leurent:2013mr}, and the HarmonicSums package \cite{Ablinger}.
In this way we have confirmed the coefficients in \eq{weak22} to four loops. 
Let us note that expansion of harmonic sums leads to multiple zeta values (MZVs), which however cancel in the final result leaving only $\zeta_n$.
Importantly, the part of the four-loop coefficient which comes from the wrapping correction is essential for matching with our result. 
This is a strong confirmation that our calculation based on the $\bP\mu$-system is valid beyond the ABA level. 
Additional evidence that our result incorporates all finite-size effects is found at strong coupling, as we shall see in section \ref{sec:strong_curvature}.

For operators with $J=3$, our prediction at weak coupling is
\beqa
	\gamma_{J=3}^{(2)}&=&-2g^2\zeta_3+g^4\(12 \zeta_5-\frac{4 \pi ^2 \zeta_3}{3}\)
	+g^6\(\frac{2 \pi ^4 \zeta_3}{45}+8 \pi ^2 \zeta_5-28
   \zeta_7\)\\ \nn
   &+&
   g^8\(-\frac{4 \pi ^6 \zeta_3}{45}-\frac{4 \pi ^4 \zeta_5}{15}-528 \zeta_9\)
   +\dots
\eeqa
The known results for any spin in this case are available at up to six loops, including the wrapping correction which first appears at five loops \cite{Beccaria:2007cn,Beccaria:2009eq,Velizhanin:2010cm}.
Expanding them at $S\to 0$ we have checked our calculation to four loops.
For future reference, in appendix \ref{sec:weakS3} we present an expansion of known results for $J=2,3$ up to order $S^3$ at first several loop orders. 
In particular, we found that multiple zeta values appear in this expansion, which did not happen at lower orders in $S$.

\begin{figure}[h]
\centering
\includegraphics[scale=0.9]{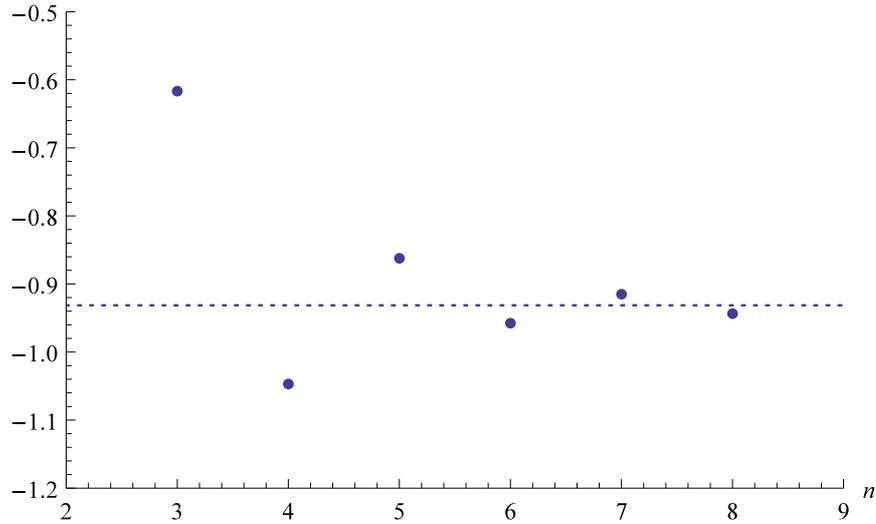}
\caption[$\bP\mu$ and ABA predictions for the curvature at 1-loop for $J=4$]{The dashed line shows the result from the $\bP\mu$-system for the coefficient of $S^2$ in the 1-loop energy at $J=4$, i.e. $-\frac{14 \zeta_3}{5}+\frac{48 \zeta_5}{\pi ^2}-\frac{252 \zeta_7}{\pi
   ^4}\approx-0.931$ (see \eq{gamma42weak}). The dots show the Bethe ansatz prediction \eq{gammaJ} expanded to orders $1/J^3,1/J^4,\dots,1/J^8$ (the order of expansion $n$ corresponds to the horizontal axis), and it appears to converge to the $\bP\mu$-system result.}
\label{fig:j4plot}
\end{figure}

For $J=4$ the expansion reads
\beqa
\label{gamma42weak}
	\gamma_{J=4}^{(2)}&=&g^2 \left(-\frac{14 \zeta_3}{5}+\frac{48 \zeta_5}{\pi ^2}-\frac{252 \zeta_7}{\pi
   ^4}\right)	\\ \nn
	&+&g^4 \left(-\frac{22 \pi ^2 \zeta_3}{25}+\frac{474 \zeta_5}{5}-\frac{8568 \zeta_7}{5 \pi
   ^2}+\frac{8316 \zeta_9}{\pi ^4}\right)\\ \nn
	&+&g^6 \left(\frac{32 \pi ^4 \zeta_3}{875}+\frac{3656 \pi ^2 \zeta_5}{175}-\frac{56568 \zeta_7}{25}+\frac{196128 \zeta_9}{5 \pi ^2}-\frac{185328 \zeta_{11}}{\pi ^4}\right)	\\ \nn
	&+&g^8 \left(-\frac{4 \pi ^6 \zeta_3}{175}-\frac{68 \pi ^4 \zeta_5}{75}-\frac{55312 \pi ^2 \zeta_7}{125}+\frac{1113396 \zeta_9}{25}-\frac{3763188 \zeta_{11}}{5 \pi ^2}\right.
	\\ \nn
	 && \ \ \ \ \ \  + \left.\frac{3513510 \zeta_{13}}{\pi ^4} \right)+\dots
\eeqa
Unlike for the $J=2$ and $J=3$ cases, we could not find a closed expression for the energy at any spin $S$ in literature even at one loop, however there is another way to check our result. 
One can expand the asymptotic Bethe ansatz equations at large $J$ for fixed values of $S=2,4,6,\dots$ and then extract the coefficients in the expansion which are polynomial in $S$. 
This was done in \cite{Beccaria:2012kp} (see appendix C there) where at one loop the expansion was found up to order $1/J^6$:
\beq
\label{gammaJ}
	\gamma(S,J) \simeq g^2\(\frac{S}{2\,J^{2}}-\Big(\frac{S^{2}}{4}+\frac{S}{2}\Big)\,\frac{1}{ J^{3}}
+\Big[
\frac{3 S^3}{16}+\Big(\frac{1}{8}-\frac{\pi ^2}{12}\Big) S^2+\frac{S}{2}
\Big]\,\frac{1}{ J^{4}} +\dots\)
\eeq
Now taking the part proportional to $S^2$ and substituting $J=4$ one may expect to get a numerical approximation to the 1-loop coefficient in our result \eq{gamma42weak}, i.e. $-\frac{14 \zeta_3}{5}+\frac{48 \zeta_5}{\pi ^2}-\frac{252 \zeta_7}{\pi^4}$. 
To increase the precision we extended the expansion in \eq{gammaJ} to order $1/J^8$. 
Remarkably, in this way we confirmed the 1-loop part of the $\bP\mu$ prediction \eq{gamma42weak} with about $1\%$ accuracy! 
In figure \ref{fig:j4plot} one can also see that the ABA result converges to our prediction when the order of expansion in $1/J$ is being increased.
Also, in contrast to $J=2$ and $J=3$ cases we see that negative powers of $\pi$ appear in \eq{gamma42weak}, although still all the contributions at a given loop order have the same transcendentality. 
It would be interesting to understand why this happens  from the gauge theory perspective, especially since expansion of the leading $S$ term \eq{eq:resultLO} has the same structure for all $J$,
\beq
	\gamma_{J}^{(1)}=\frac{8 \pi ^2 g^2}{J (J+1)}-\frac{32 \pi ^4 g^4}{J (J+1)^2
   (J+2)}+\frac{256 \pi ^6 g^6}{J (J+1)^3 (J+2) (J+3)}
	+\dots
\eeq
The change of structure at $J=4$ might be related to the fact that for $J\geq 4$ the ground state anomalous dimension even at one loop is expected to be an irrational number for integer $S>0$ \cite{Beccaria:2008pp,Belitsky:2008mg}, and thus cannot be written as a linear combination of harmonic sums with integer coefficients.

\subsubsection{Strong coupling expansion}
\label{sec:strong_curvature}

To obtain the strong coupling expansion of our exact results for the curvature function, we evaluated it numerically with high precision for a range of values of $g$ and then made a fit to find the expansion coefficients. 
We should note that the expansion has been carried out analytically in \cite{Beccaria:2014rca} using some clever tricks that are not obvious to justify.

For numerical study it is convenient to write our exact expressions \eq{gamma2L2}, \eq{gamma2L3}, \eq{gamma2L4} for $\gamma^{(2)}(g)$, which have the form
\beq
\label{ints}
	\gamma^{(2)}(g)=\oint {du_x}\oint {du_y} f(x,y) \partial_{u_x} \log\frac{\Gamma (i u_x-i u_y+1)}{\Gamma (1-i u_x+i u_y)}
\eeq
where the integration goes around the branch cut between $-2g$ and $2g$, in a slightly different way (we remind that we use notation $x+\frac{1}{x}=\frac{u_x}{g}$ and $y+\frac{1}{y}=\frac{u_y}{g}$). 
Namely, by changing the variables of integration to $x,y$ and integrating by parts one can write the result as
\beq
\label{ints2}
	\gamma^{(2)}(g)=\oint {dx}\oint {dy} F(x,y) \log\frac{\Gamma (i u_x-i u_y+1)}{\Gamma (i u_y-i u_x+1)}
\eeq
where $F(x,y)$ is some polynomial in the following variables: $x,\;1/x,\;y,\;1/y,\;\sh_-^x$ and $\sh_-^y$ (for $J=3$ it includes $\ch_-^x,\;\ch_-^y$ instead of the $\sh_-$ functions). 
The integral in \eq{ints2} is over the unit circle.
The advantage of this representation is that plugging in $\sh_-^x$, $\sh_-^y$ as series expansions (truncated to some large order), we see that it only remains to compute integrals of the kind
\beqa
C_{r,s}&=&\frac{1}{i}\oint\frac{dx}{2\pi}\oint\frac{dy}{2\pi}x^r y^s\log\frac{\Gamma(i u_x-iu_y+1)}{\Gamma(i u_y-iu_x+1)}
\eeqa
These are nothing but the coefficients of the BES dressing phase \cite{Beisert:2006ib, Beisert:2006ez, Vieira:2010kb, Dorey:2007xn}. 
They can be conveniently computed using the strong coupling expansion \cite{Beisert:2006ez}
\small
\beq
C_{r,s}=\sum_{n=0}^\infty\[-\frac{2^{-n-1} (-\pi )^{-n} g^{1-n} \zeta_n
   \left(1-(-1)^{r+s+4}\right) \Gamma \left(\frac{1}{2}
   (n-r+s-1)\right) \Gamma \left(\frac{1}{2}
   (n+r+s+1)\right)}{\Gamma (n-1) \Gamma \left(\frac{1}{2}
   (-n-r+s+3)\right) \Gamma \left(\frac{1}{2} (-n+r+s+5)\right)}\]
\eeq
\normalsize
However this expansion is only asymptotic and does not converge. 
For fixed $g$ the terms will start growing with $n$ when $n$ is greater than some value $N$, and we only summed the terms up to $n=N$ which gives the value of $C_{r,s}$ with very good precision for large \text{enough $g$}.
Using this approach we computed the curvature function for a range of values of $g$ (typically we took $7\leq g \leq 30$) and then fitted the result as an expansion in $1/g$. This gave us only numerical values of the expansion coefficients, but in fact we found that with very high precision the coefficients are as follows. 
For $J=2$
\beqa
\label{eq:ssj2}
\gamma^{(2)}_{J=2}&=&-\pi ^2 g^2+\frac{\pi  g}{4}+\frac{1}{8}-\frac{1}{\pi g}\(\frac{3 \zeta_3}{16}+\frac{3}{512}\)-\frac{1}{\pi^2g^2}\(\frac{9 \zeta_3}{128}+\frac{21}{512}\)\quad\quad
\\ \nn
&+&
\frac{1}{\pi^3g^3}\(\frac{3 \zeta_3}{2048}+\frac{15 \zeta_5}{512}-\frac{3957}{131072}\) + \dots\;,
\eeqa
then for $J=3$
\beqa
\label{eq:ssj3}
\gamma^{(2)}_{J=3}&=&-\frac{8 \pi ^2 g^2}{27}+\frac{2 \pi  g}{27}+\frac{1}{12}-
\frac{1}{\pi g}\(
\frac{1}{216}
+\frac{\zeta_3}{8}
\)-
\frac{1}{\pi^2g^2}\(\frac{3 \zeta_3}{64}+\frac{743}{13824}\)\quad\quad
\\ \nn
&+&
\frac{1}{\pi^3g^3}\(\frac{41 \zeta_3}{1024}+\frac{35 \zeta_5}{512}-\frac{5519}{147456}\) + \dots\;,
\eeqa
and finally for $J=4$
\beqa
\gamma^{(2)}_{J=4}&=&-\frac{\pi ^2 g^2}{8}+\frac{\pi  g}{32}+\frac{1}{16}-\frac{1}{\pi g}\(\frac{3 \zeta_3}{32}+\frac{15}{4096}\)-\frac{0.01114622551913}{g^2}\quad\quad
\\ \nn
&+&\frac{0.004697583899}{g^3}+ \dots\;.
\eeqa
To fix coefficients for the first four terms in the expansion we were guided by known analytic predictions and found that our numerical result matches these predictions with high precision. 
Then for $J=2$ and $J=3$ we extracted the numerical values obtained from the fit for the coefficients of $1/g^2$ and $1/g^3$, and plugging them into the online calculator EZFace \cite{ezface} we obtained a prediction for their exact values as combinations of $\zeta_3$ and $\zeta_5$. 
Fitting again our numerical results with these exact values fixed, we found that the precision of the fit at the previous orders in $1/g$ increased. 
This is a highly non-trivial test for the proposed exact values of $1/g^2$ and $1/g^3$ terms. 
For $J=2$ we confirmed the coefficients of these terms with absolute precision $10^{-17}$ and $10^{-15}$ at $1/g^2$ and $1/g^3$ respectively (at previous orders of the expansion the precision is even higher). 
For $J=3$ the precision was correspondingly $10^{-15}$ and $10^{-13}$.
For $J=4$ we were not able to get a stable fit for the $1/g^2$ and $1/g^3$ coefficients from EZFace, so above we gave their numerical values (with uncertainty in the last digit). 
However in the next section on short strings we will see that based on $J=2$ and $J=3$ results one can make a prediction for these coefficients, which we again confirmed by checking that precision of the fit at the previous orders in $1/g$ increases. 
The precision of the final fit at orders $1/g^2$ and $1/g^3$ is $10^{-16}$ and $10^{-14}$ respectively.

\subsubsection{Higher mode numbers}
\label{sec:curvature_mode_numbers}

The starting point for finding the curvature function was considering the leading order solution of the $\pmu$-system \eqref{eq:P1solLOevenL} - \eqref{eq:P4solLOevenL} or \eqref{P1oddL} - \eqref{P4oddL}, which resulted in the slope function. 
One of the assumptions for constructing the leading order solution was to allow $\mu_{ab}$ to have only $e^{\pm 2\pi u}$ in asymptotics at infinity, which in turn led to all $\mu$'s being constant except $\mu_{24}$ which is equal to $\sinh\({2\pi u}\)$ or $\cosh\({2\pi u}\)$.
In principle requiring $\mu_{ab}$ to be periodic one could also allow to have $e^{2n\pi u}$ with any integer $n$.
Thus a natural generalization of the leading order solution is to consider $\mu_{24}=\sinh\({2\pi n u}\)$ or $\mu_{24}= \cosh\({2\pi n u}\)$, where $n$ is an arbitrary integer. 
As discussed in section \ref{sec:LOresultevenL}, we believe that at the leading order in $S$ such solutions correspond to states with mode numbers equal to $n$, and they reproduce the slope function for this case.

Proceeding to order $S^2$, the calculation of the curvature function $\gamma^{(2)}(g)$ with $\mu_{24}=\sinh\({2\pi n u}\)$ or $\mu_{24}=\cosh\({2\pi n u}\)$ can be done following the same steps as for $n=1$.
The final results for $J=2,3$ and $4$ are given by exactly the same formulas as for $n=1$ (\eqref{gamma2L2}, \eqref{gamma2L3} and \eqref{gamma2L4} respectively) -- the only difference is that now one should set in those expressions
\beqa
\label{nrep1}
&&I_k=I_k(4\pi n g ),\\
&&\sh_-^x=\[\sinh\({2 \pi n u_x}\)\]_-, \\
&&\sh_-^y=\[\sinh\({2 \pi n u_y}\)\]_-, \\
&&\ch_-^x=\[\cosh\({2 \pi n u_x}\)\]_-, \\
\label{nreplast}
&&\ch_-^y=\[\cosh\({2 \pi n u_y}\)\]_-.
\eeqa
It would be natural to assume that this solution of the $\bP\mu$-system describes anomalous dimensions for states with mode number $n$ at order $S^2$. 
However we found some peculiarities in the strong coupling expansion of the result. 
The strong coupling data available for comparison in the literature for states with $n>1$ also relies on some conjectures (see \cite{Basso:2011rs,Gromov:2011bz}), so the interpretation of this solution is not fully clear to us.

The weak coupling expansion for this case turns out to be related in a simple way to the $n=1$ case. 
One should just replace $\pi\to n\pi$ in the expansions for $n=1$ which are given in \eq{weak22long}, \eq{weak23long}, \eq{weak24long}. 
For example,
\beqa
\nn
	\gamma_{J=2}^{(2)}&=&-8 g^2 \zeta_3+g^4 \left(140 \zeta_5-\frac{32 n^2\pi ^2 \zeta_3}{3}\right)+g^6
   \left(200 n^2\pi ^2 \zeta_5-2016 \zeta_7\right)
	+\dots
\eeqa
It would be interesting to compare these weak coupling predictions to results obtained from the asymptotic Bethe ansatz (or by other means) as it was done for the $n=1$ case in section \ref{sec:curvature_weak}.

\subsection{Update on short strings}
\label{sec:konishi_three_loops}

Having just found the curvature function we recall section \ref{sec:short_strings} where we were able to boost the folded string semi-classical energy one loop further by utilizing the knowledge of the slope function.
We also argued that knowing the curvature function would enable us to repeat the procedure and boost the result by one more loop.

The key idea was to utilize the structure of the small spin expansion \eq{eq:delta_squared_basso} of the folded string energy.
Re-expanding it at fixed global charges $S$ and $J$ we found the energy in terms of the coefficients $A_i, B_i, C_i$, etc. as seen in \eq{eq:delta_abc}.
We also found that in principle the curvature function should fix all of the $B_i$ coefficients, which we can indeed do by first expanding the curvature function at strong coupling in terms of $A$'s and $B$'s
\small
\beqa
	\label{eq:ss_abc}
	\gamma^{(2)}_{J}(g) \!\!\!\!&=&\!\!\!\! -\frac{2 \pi ^2 g^2 A_1^2 }{J^3} - \frac{\pi g A_1 A_2 }{J^3}-\frac{A_2^2+2 A_1 A_3-4 B_1 J^2}{8 J^3} - \frac{A_2 A_3+A_1 A_4-2 B_2 J^2}{16 \pi g J^3} \\
	&-&  \frac{A_3^2+2 A_2 A_4+2 A_1 A_5-4 B_3 J^2}{128 \pi^2 g^2 J^3} - \frac{A_3 A_4 + A_2 A_5 + A_1 A_6 - 2 B_4 J^2}{256 \pi^3 g^3 J^3} + \mc{O}\left(\frac{1}{g^4}\right) \nonumber
\eeqa
\normalsize
and plugging in the $A_i$ coefficients \eq{eq:bassos_as} found from the slope function.
A slight complication here is that we only have strong coupling expansions of the curvature function for $J=2,3$ and for $J=4$ we only have a numeric result. 
By comparing the $J=2,3$ expansions \eq{eq:ssj2} and \eq{eq:ssj3} to the above general form we can extract the values of the $B$'s for those two cases only.
However as discussed in section \ref{sec:short_strings} we expect all series of coefficients to have a certain power-like dependencies on $J$, namely we expect $B_1$, $B_2$ to be constant and $B_3$, $B_4$ to have the form $a J^2 + b$ with $a$ and $b$ constant.
Having just found two data points for $J=2,3$ we immediately find $a$ and $b$ and thus deduce that
\beqa
\label{eq:curvature_bs}
\arraycolsep=1.6pt
 \begin{array}{rcrlrlrcl}
B_1 &=& &3/2, &B_2& &=& &-3\,\zeta_3+\frac{3}{8},  \\
B_3 &=& &-\frac{J^2}{2}-\frac{9 \, \zeta_3}{2}+\frac{5}{16},\;\; &B_{4}& &=& &\frac{3}{16} J^2 (16 \, \zeta_3+20 \, \zeta_5-9)-\frac{15 \, \zeta_5}{2}-\frac{93 \, \zeta_3}{8}-\frac{3}{16}. 
\end{array}
\eeqa
Having fixed all the unknowns we can now write the strong coupling expansion of the curvature function for arbitrary values of $J$ as
\small
\beqa
 \gamma^{(2)}_{J}(g) &=& -\frac{8 \pi ^2 g^2}{J^3}+\frac{2 \pi  g}{J^3}+\frac{1}{4 J}+\frac{1-J^2 (24 \, \zeta_3 +1)}{64 \pi  g J^3} - \frac{8 J^4+J^2 (72 \, \zeta_3 +11)-4}{512 g^2 \left(\pi ^2 J^3\right)}  \nonumber \\
  &+& \frac{3 \left(8 J^4 (16 \, \zeta_3 +20 \, \zeta_5-7)-16 J^2 (31 \, \zeta_3 +20 \, \zeta_5+7)+25\right)}{16384 \pi ^3 g^3 J^3} + \mc{O}\left(\frac{1}{g^4}\right).\qquad\qquad
\eeqa
\normalsize
Expanding $\gamma^{(2)}_{J=4}$ defined in \eq{gamma2L4} at strong coupling numerically we were able to confirm the above result with high precision.

\begin{table}[t]
\begin{tabular}{|l||rl|l|l||l|l|l|}
  \hline
  $(S,J)$ & \multicolumn{2}{|l|}{$\lambda^{-5/4}$ prediction} & $\lambda^{-5/4}$ fit & error & fit order\\
  \hline
  $(2,2)$ & $\frac{15 \, \zeta_5}{2} + 6 \, \zeta_3-\frac{1}{2}$&$= 14.48929958$ & $14.12099034$ & $2.61\%$ & 6\\
  $(2,3)$ & $\frac{15 \, \zeta_5}{2} + \frac{63 \, \zeta_3}{8} - \frac{1131}{512}$&$= 15.03417190$ & 14.88260078 & $1.02\%$ & 5 \\
  $(2,4)$ & $\frac{21 \, \zeta_3}{2} + \frac{15 \, \zeta_5}{2} - \frac{25}{8}$&$= 17.27355565$ & $16.46106336$ & $4.94\%$ & 7\\
  \hline
\end{tabular}
\caption[Comparisons of TBA numerics with fits from strong coupling]{Comparisons of strong coupling expansion coefficients for $\lambda^{-5/4}$ obtained from fits to TBA data versus our predictions for various operators. The fit order is the order of polynomials used for the rational fit function (see \cite{Gromov:2011bz} for details).}
\label{tab:coefficients}
\end{table}

Now that we know the strong coupling expansion of the curvature function and thus all the coefficients $B_i$, we can do the same trick and find the three loop strong coupling scaling dimension coefficient $\Delta^{(3)}$, which now depends on $A_{1;2;3;4}$, $B_{1,2,3}$, $C_{1,2}$, $D_1$. We find it to be
\beqa
	\Delta^{(3)} &=& \frac{187\,S^6 + 2\,(624\,\zeta_3 + 480\,\zeta_5-193)\,S^5 +\left(-146\,J^2 - 4\,(336\,\zeta_3-41)\right)S^4 }{512 \sqrt{2}\,S^{5/2}} + \nonumber \\
	&+& \frac{\left(32\,(6\,\zeta_3+7)\,J^2-88\right)S^3 + \left(-28\,J^4 + 40\,J^2\right) S^2 - 24\,J^4 S + 8\,J^6}{512 \sqrt{2}\,S^{5/2}},
\eeqa
for $S=2$ it simplifies to
\beq
	\Delta^{(3)}_{S=2} = \frac{1}{512} \left(J^6-20 J^4+48 J^2 (4 \zeta_3 - 1)+192 (12 \, \zeta_3+20 \, \zeta_5+1)\right)
\eeq
and finally for the Konishi operator, which has $S=2$ and $J=2$ we get
 \beq
  \Delta^{(3)}_{S=2,J=2} = \frac{15 \, \zeta_5}{2} + 6 \, \zeta_3-\frac{1}{2}.
 \eeq
In order to compare our predictions with numeric data available from explicit Y-system calculations \cite{Frolov:2010wt}, we employed Pad\'{e} type fits as explained in \cite{Gromov:2011bz}. 
The fit results are shown in table \ref{tab:coefficients}, we see that our predictions are within $5\%$ error bounds, which is a rather good agreement. 
However we must be honest that for the $J=3$ and especially $J=4$ states we did not have as many data points as for the $J=2$ state and the fit is somewhat shaky.

\subsubsection{Issues with higher mode numbers}
\label{sec:curvature_inconsistencies}

In section \ref{sec:small_spin_structure} we conjectured a particular structure \eq{eq:delta_squared_basso} for the small spin expansion of generalized Konishi anomalous dimension, which we successfully utilized up to now.
In section \ref{sec:inconsistencies} however we saw first hints that the naive replacement $\lambda \to \mu \equiv n^2 \lambda$ we use for the conjecture might not actually be justified.
Namely we were not able to match the structure of the semi-classical one-loop string energy \eq{delta_oneloop_sj} with the conjectured structure \eq{sec:small_spin_structure} for mode numbers $n>1$. 

Let us now generalize the calculations in the previous subsection for the case when $n=2$.
First we use find the strong coupling expansion of the curvature function in terms of the coefficients $A_i$ and $B_i$, generalizing \eq{eq:ss_abc} to
\beq
\label{gamma2basso}
	\gamma^{(2)}_{J}(g)=-\frac{8 \pi ^2g^2 n^2}{J^3}+\frac{2 \pi  g n}{J^3}+\frac{B_1-1}{2 J}+\frac{8 B_2 J^2-4 J^2+1}{64 \pi  g J^3 n}+\dots
\eeq
This expansion assumes the conjectured structure \eq{eq:delta_squared_basso} with the naive replacement $\lambda \to \mu$.
Now we compare the expansion \eq{gamma2basso} to our results from the $\bP\mu$-system, which we generalized to arbitrary mode numbers in section \ref{sec:curvature_mode_numbers} by also doing some naive replacements.
To compute the strong coupling expansions of our results, similarly to the $n=1$ case, we evaluated $\gamma_{J}^{(2)}(g)$ numerically for many values of $g$, and then fitted the result by powers of $g$. 
As for $n=1$ we found with high precision (about $\pm10^{-16}$) that the first several coefficients involve only rational numbers and powers of $\pi$. 
Our results for $n=2,3$ and $J=2,3,4$ are summarized below:
\beqa
\label{gamma2L22}
\gamma^{(2)}_{J=2,n=2} &=& -4\pi^2g^2+\frac{\pi g}{2}+\frac{17}{8}-\frac{0.29584877037648771(2)}{g}+\dots \\
\gamma^{(2)}_{J=3,n=2} &=& -\frac{32}{27} \pi ^2 g^2+\frac{4 \pi  g}{27}+\frac{17}{12}-\frac{0.2928304112866493(9)}{g}+\dots \\
\gamma^{(2)}_{J=4,n=2} &=& -\frac{1}{2} \pi ^2 g^2+\frac{\pi  g}{16}+\frac{17}{16}-\frac{0.319909936615448(9)}{g}+\dots \\
\gamma^{(2)}_{J=2,n=3} &=& -9 \pi ^2 g^2+\frac{3 \pi  g}{4}+\frac{23}{4}-\frac{0.8137483(9)}{g}+\dots \\
\gamma^{(2)}_{J=3,n=3} &=& -\frac{8}{3} \pi ^2 g^2+\frac{2 \pi  g}{9}+\frac{23}{6}-\frac{0.892016609(2)}{g}+\dots \\
\gamma^{(2)}_{J=4,n=3} &=& -\frac{9}{8} \pi ^2 g^2+\frac{3 \pi  g}{32}+\frac{23}{8}-\frac{1.035945580(6)}{g}+\dots
\eeqa
Here in the coefficient of $1/g$ the digit in brackets is the last known one within our precision.
Comparing to \eq{gamma2basso} we find full agreement in the first two terms (of order $g^2$ and of order $g$). The next term in \eq{gamma2basso} (of order $g^0$) is determined by $B_1$, which we found to be $3/2$  for all $n,J$, based on consistency with the classical energy. 
However, comparing our results with \eq{gamma2basso} we find a different value:
\beqa
	B_1&=&\frac{19}{2} \ \  \text{for }\ n=2\;,\\ \nn
	B_1&=&23 \ \ \, \text{for }\ n=3\;.
\eeqa
For both $n=2$ and $n=3$ this prediction for $B_1$ is independent of $J$.
The next term is of order $1/g$ and is determined by $B_2$, which we also fixed in \eq{BB}.
Unfortunately it does not agree with our numerical predictions for $n=2$ and $3$. 
Furthermore, for $n=2$ we extracted the coefficient of $1/g$ with high precision, but were unable to fit it as a combination of simple zeta values using the EZ-Face calculator \cite{ezface}.

Presently we do not have a definite explanation for these inconsistencies.
Although our solution of the $\bP\mu$-system for $n>1$ looks fine at order $S$, it may be that to capture anomalous dimensions at order $S^2$ some other solution should be used.
Another option is that the ansatz for the structure of anomalous dimensions at strong coupling may need to be modified when $n>1$, as we already suspected in section \ref{sec:inconsistencies}.

\subsection{Cusped Wilson line}
\label{sec:wilson_line}

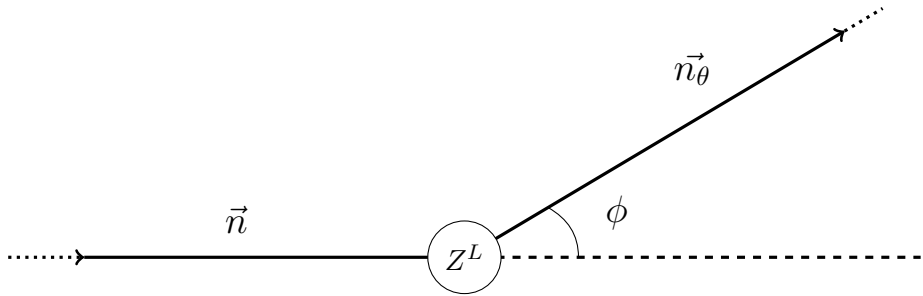
\begin{figure}[t]
\centering
\begin{tikzpicture}
	
	
	\draw[->,dotted,very thick] (1,1) -- (2,1);
	\draw[->,very thick] (2,1) -- (7,1) -- (12,4);
	\draw[dotted,very thick] (12,4) -- (12.5,4.3);
	\draw[dashed,very thick] (7,1) -- (13,1);
	
	\node[draw,circle,fill=white] at (7,1) {$Z^L$};
	
	\draw[] (8.5,1) to [out=90,in=-30] (8.1,1.66);

	\node at (4,1.5) {\Large $\vec{n}$};
	\node at (10,3.5) {\Large $\vec{n_\theta}$};
	\node at (9,1.6) {\Large $\phi$};
	
\end{tikzpicture}
\caption[The cusped Wilson line with an operator insertion]{The cusped Wilson line with an operator insertion.}
\label{fig:wilson_line}
\end{figure}

In this section we will consider an observable illustrated in figure \ref{fig:wilson_line}, it consists of two rays of a supersymmetric Wilson line forming a cusp with the angle $\phi$ and an operator $Z^L$ inserted at the cusp, where $Z$ is a complex scalar of ${\cal N}=4$ super Yang-Mills. 
To completely define a supersymmetric Wilson line we should also specify the coupling to scalars, which is parametrized by a six-dimensional unit vector $\vec n(t)$ at each point of the line, where $t$ is a parameter on the line. 
In our case $\vec n(t)$ is constant and equal $\vec n$ on one ray and $\vec n_\theta$ on another ray, so that $\vec n\cdot \vec n_\theta=\cos\theta$. 
Because of R-symmetry the observable depends on $\vec n,\vec n_\theta$ only through $\theta$.
Explicitly the observable is defined as
\beq
\label{WilsL}
	W_L={\rm P}\exp\!\int\limits_{-\infty}^0\! dt\(i  A\cdot\dot{x}_q+\vec\Phi\cdot\vec n\,|\dot x_q|\)\times Z^L\times {\rm P}\exp\!\int\limits_0^\infty\!dt\(i A\cdot\dot x_{\bar q}+\vec\Phi\cdot\vec n_\theta\,|\dot x_{\bar q}|\).
\eeq
Due to the cusp the expectation value of such an observable diverges as
\beq
\left\langle W_L\right\rangle \sim \(\frac{\Lambda_{IR}}{\Lambda_{UV}}\)^{\Gamma_L(\lambda)},
\eeq
where $\Lambda_{IR}$ and $\Lambda_{UV}$ are the infra-red and ultraviolet cut-offs respectively \cite{Polyakov:1980ca,Correa:2012at}. 
The quantity $\Gamma_{L}$, which we will call the cusp anomalous dimension, will be the main object of our study in this section.
The AdS/CFT duality allows one to relate the observable in ${\cal N}=4$ SYM described above to an open string in $\adsfive$ which ends on a cusped line on the boundary of AdS. In particular, in the classical scaling limit when $L$ and $\lambda$ are both taken to infinity with $L/\sqrt{\lambda}$ fixed, we can match $\Gamma_{L}$ with the energy of the classical string \cite{Correa:2012hh, Gromov:2012eu}.

Wilson lines are non-local observables in the sense that they are parametrized by a line in spacetime instead of just a point.
Technically this renders the whole machinery of spin chains and spectral curves unsuitable for studying these observables as we assumed to be working with local operators. 
Remarkably it turns out that cusped Wilson lines are very similar to local single trace operators.
They key idea is to consider the insertion of the local operator $Z^L$ at the cusp as an open spin chain with the Wilson lines imposing certain boundary conditions for the magnons inside the chain \cite{Drukker:2012de}.
One can then write down Bethe ansatz equations with a modified $S$-matrix incorporating the boundary conditions and find the cusp anomalous dimension at weak coupling.
Remarkably a set of thermodynamic Bethe ansatz equations have also been found for this observable, which also go by the name of twisted boundary thermodynamic Bethe ansatz equations (abbreviated by TBTBA, BTBA or TBA) \cite{Correa:2012hh}.
In the small $\phi$ and $\theta=0$ limit the equations can be dramatically simplified and cast in a FiNLIE form, which even admits an analytic solution \cite{Gromov:2012eu}, later generalized \cite{Gromov:2013qga} to the case of arbitrary but close to each other angles $\theta^2-\phi^2\rightarrow 0$. 
In this general case the cusp anomalous dimension was found to be
\beq
\label{eq:mainresultIntro}
\Gamma_L(\lambda)=\frac{\phi-\theta}{4}\partial_\theta\log\frac{\det{\cal M}_{2L+1}}{\det{\cal M}_{2L-1}}+{\cal O}((\phi-\theta)^2),
\eeq
where ${\cal M}_N$ is an $(N+1)\times (N+1)$ sized matrix defined as
\begin{align}
\label{eq:M}
&\({\cal M}_N\)_{ij} =I^\theta_{i-j+1},\\
&I_n^\theta =i^{n+1}I_n\(\frac{\sqrt{\lambda}}{\sin\beta}\)\sin{n\beta}, \;\;\;\; \mathrm{with} \;\; \sin{\beta}=\frac{1}{\sqrt{1-\theta^2/\pi^2}},
\nonumber
\end{align}
and $I_n(x)$ are modified Bessel functions of the first kind. 
When $\theta^2-\phi^2=0$ the observable $W_L$ becomes BPS and the cusp anomalous dimension vanishes \cite{Drukker:2006xg}. 

\subsubsection{$\pmu$-system solution}

Since a Y-system of equations can be written down for the cusped Wilson line it is then hardly a surprise that it can also be studied using the quantum spectral curve construction, as was shown in \cite{Gromov:2013pga}. 
We will briefly review the solution for the small $\phi$ case.
The $\pmu$-system has the same general form as given in \eq{Pt}, \eq{muPPt-eq} and what distinguishes this case from the $\alg{sl}{2}$ sector are the asymptotics.
They can be read off from the Y-system solution and shown to be \cite{Gromov:2013pga}
\beq
	\label{eq:cusp_pmu_asymptotics}
	\bP_a \simeq \( A_1 u^{-L+1/2}, A_2 u^{-L-1/2}, A_3 u^{+L+3/2}, A_4 u^{+L+1/2} \).
\eeq
The algebraic constraints on $A_i$'s suggest that at $\phi=0$ one has $A_2 A_3 \simeq A_4 A_1 \to 0$, which in turn implies that $\bP_a = 0$ to leading order in $\phi$.
This can be seen by analogy with the $\alg{sl}{2}$ sector constraints \eq{AA1} - \eq{AA2}, which although not technically valid for this case still capture the correct behaviour at small $\phi$ by setting the spin $S=0$ and $\Delta = L +\ord{\phi^2}$.
Since to leading order $\bP$'s are small we find from the $\pmu$-system that $\tilde{\mu}_{ab} = \mu_{ab}$ and thus it has no cuts and is just a periodic function as follows from \eq{muper}.
Another feature of the Y-system in this case is that due to the boundary dressing phase the product $Y_{11} Y_{22}$ contains poles at $u= i n/2$ for integer $n \neq 0$ \cite{Correa:2012hh}, thus from \eq{eq:y_asymp} we deduce that 
\beq
	\mu_{12} = C \sinh (2 \pi u),
\eeq
where we also used the fact that $\mu_{12}$ has to be odd for vacuum states, the origin of which lies in the Y-system equations.
Such parity requirements also apply to other $\mu_{ab}$'s which we assume to be constant, together with the constraints \eq{constraint} - \eq{constraint2} thus fixing them to be
\beq
	\mu_{13} = \mu_{24} = \mu_{34} = 0, \;\; \mu_{14} = -1.
\eeq
The $\pmu$-system finally reduces to the following set of equations
\beqa
	\tilde{\bP}_1 - \bP_1 &=& -C \sinh \( 2\pi u \) \bP_3, \label{eq:cusp_p1} \\
	\tilde{\bP}_2 + \bP_2 &=& -C \sinh \( 2\pi u \) \bP_4, \label{eq:cusp_p2} \\
	\tilde{\bP}_3 + \bP_3 &=& 0, \label{eq:cusp_p3} \\ 
	\tilde{\bP}_4 - \bP_4 &=& 0. \label{eq:cusp_p4} 
\eeqa
Let us consider the $L=0$ case for simplicity.
First thing we notice is that \eq{eq:y_asymp} should not have a pole at $u=0$, thus we have to set $\bP_1 \bP_2 = 0$ at $u=0$, which we do by introducing factors of $\sqrt{u}$ in order to comply with the asymptotics \eq{eq:cusp_pmu_asymptotics}.
From \eq{eq:cusp_p4} we see that
\beq
	\bP_4 = A_4 \sqrt{u},
\eeq	
as it does not change sign when crossing the cut, meanwhile from \eq{eq:cusp_p3} we get
\beq
	\bP_3 = A_3 \sqrt{u} \sqrt{u^2 - 4g^2},
\eeq
as it does change sign when crossing the cut $u \in [-2g, 2g]$.
From \eq{eq:cusp_p2} we can find $\bP_2$ which is given by a Hilbert transform of $\sinh(2\pi u)$,
\beq
	\frac{-\bP_2}{C A_4 \sqrt{u}} = \oint_{-2g}^{2g} dv \frac{\sqrt{u^2-4g^2}}{\sqrt{v^2-4g^2}} \frac{\sinh(2\pi v)}{4 \pi i (v-u)} = \sum_{n=1}^\infty \frac{I_{2n-1}(4\pi g)}{x^{2n-1}},
\eeq
furthermore we find that $A_2 = -g\, C A_4\, I_1 \( 4 \pi g \)$.
Finally the solution for $\bP_1$ can be easily verified to be
\beq
	\bP_1 = -\frac{A_3}{A_4} \sqrt{u^2 - 4g^2} \, \bP_2 + \( A_1 + \frac{A_3 A_2}{A_4} \) \sqrt{u}.
\eeq
Now we introduce the angle $\phi$ and energy $\Delta$ by comparing to the Y-system asymptotics when $u \to \infty$,
\beqa
	Y_{11} + 1 &\simeq& -\frac{\phi^2}{2} \\
	Y_{11} Y_{22} - 1 &\simeq& \frac{2 i \Delta}{u},
\eeqa
where the last equation only differs by the factor of 2 from the $\alg{sl}{2}$ case \eq{eq:y1122_sl2_asymptotics}.
First we notice that $A_1 A_4 = A_2 A_3$, since otherwise $Y_{11} Y_{22} - 1$ would grow linearly, then we find $A_1 A_4 = i \phi^2$, thus fixing the energy to be
\beq
	\Delta = -\phi^2 g^2 \(1 - \frac{I_3(4\pi g)}{I_1(4\pi g)} \),
\eeq
which exactly reproduces the TBA result \eq{eq:mainresultIntro} for $L=0$ and $\theta=0$.
The general case can also be reproduced with just a bit more effort. 
We should also note that this result was also found in a completely different manner by utilizing localization techniques \cite{Correa:2012at,Fiol:2012sg}, thus confirming the validity of the $\pmu$-system approach.

\subsubsection{Matrix model formulation}

An obvious complication in the formulation of the result \eq{eq:mainresultIntro} is that it contains determinants of $(2L+1)\times (2L+1)$ sized matrices, which is very problematic if one wants to take the large $L$ limit. 
It becomes considerably easier once we realize that the cusp anomalous dimension can be expressed in terms of an expectation value of some operator in a matrix model. 
One can check that the quantities $I^\theta_n$ defined in \eq{eq:M} can be rewritten in the following integral representation
\begin{equation}
	I_n^\theta = \frac{1}{2\pi i} \oint \frac{dx}{x^{n+1}} \sinh(2\pi g \, (x + 1/x)) \, e^{2 g \theta (x - 1/x)},
\end{equation}
where the integration contour is the unit circle and $g=\frac{\sqrt\lambda}{4\pi}$. 
This makes it possible to write the determinant of ${\cal M}_{N}$ as
\beq
\det \M_{N}=
\oint\prod_{i=1}^{N+1}\frac{dx_i}{2\pi i} e^{2 g \, \theta \left(x_i-\frac{1}{x_i}\right)} \sinh \left(2 \pi g \left( x_i + \frac{1}{x_i} \right) \right) \times \det X,
\eeq
where
\beq
\det X = \left|
\bea{lllll}
x_1^{-2}&x_1^{-1}&\dots&x_1^{N-1}&x_1^{N-2}\\
x_2^{-3}&x_2^{-2}&\dots&x_2^{N}&x_2^{N-1}\\
\vdots&\vdots& \ddots & \vdots& \vdots\\
x_{N}^{-N-1}&x_{N}^{-N}&\dots&x_{N}^{-2}&x_{N}^{-1}\\
x_{N+1}^{-N-2}&x_{N+1}^{-N-1}&\dots&x_{N+1}^{-3}&x_{N+1}^{-2}
\eea
\right| = \frac{\prod_{i<j}^{N+1} (x_i - x_j)}{\prod_{i=1}^{N+1} x_i^{i+1}},
\eeq
and we recognize the numerator as the Vandermonde determinant $\Delta(x_i)$. 
We can further simplify the final result by anti-symmetrizing the denominator, which we can do because everything else in the integrand is anti-symmetric and the integration measure is symmetric with respect to $x_i$, thus under the integral we can replace $\det X$ by
\beq
	\det X' = \frac{\Delta^2(x_i)}{(N+1)!} \, \prod_{i=1}^{N+1} \frac{1}{x_i^{N+2}}.
\eeq
Thus finally we get the following expression
\begin{equation}
  \det \M_{N} = \frac{1}{(2 \pi i)^{N+1}} \oint \prod_{i=1}^{N+1} \frac{d x_i}{x_i^{N+2}} \, \frac{\Delta^2(x_i)}{(N+1)!}  \, \sinh(2\pi g \, (x_i + 1/x_i)) \, e^{2 g \theta (x_i - 1/x_i)},
  \label{eq:Mintegral}
\end{equation}
which indeed has the structure of a partition function of some matrix model.
It now becomes a matter of simple algebra to convince oneself that the cusp anomalous dimension \eq{eq:mainresultIntro} can be written in terms of expectation values in this matrix model, namely
\beq
	\Gamma_L(g) = g \, \frac{\phi-\theta}{2} \left[ \,\, \left< \sum_{i=1}^{2L+1} \( x_i - \frac{1}{x_i}\)  \right>_{2L+1} - \,\,\, \left<  \sum_{i=1}^{2L-1} \(x_i - \frac{1}{x_i}\) \right>_{2L-1} \right],
\eeq
where $\left< \dots \right>_{N}$ denotes the normalized expectation value in the matrix model of size $N$ with the partition function defined in (\ref{eq:Mintegral}). 
Note that this formula is exact and we have not yet taken any limits.

\subsubsection{Classical limit}
\label{sec:wilson_classical}

As already mentioned before, when $L$ and $\lambda$ are both taken to infinity with $L/\sqrt{\lambda}$ fixed the cusp anomalous dimension $\Gamma_L(\lambda)$ can be matched to the energy of a classical open string solution.
In this subsection we will take this limit of the exact result \eq{eq:mainresultIntro} and in the next subsection we will do the matching to the string result.
Alternatively since we already saw that the problem can be solved with the quantum algebraic curve construction, one would expect to be able to start from the string solution and apply the classical spectral curve techniques as discussed in section \ref{sec:spectral_curve}. 
Technically the algebraic curve is defined for closed string solutions, however there is evidence that the construction can be easily generalized for open strings by trivially modifying the definition of the monodromy \eq{eq:monodromy} \cite{Janik:2012ws}.
This evidence is strongly supported by our derivation which looks remarkably similar to the $\alg{sl}{2}$ classical spectral curve construction.
Because of this we will loosely refer to our result as the algebraic curve.

\begin{figure}[t]
	\centering
		\includegraphics[width=0.95\textwidth]{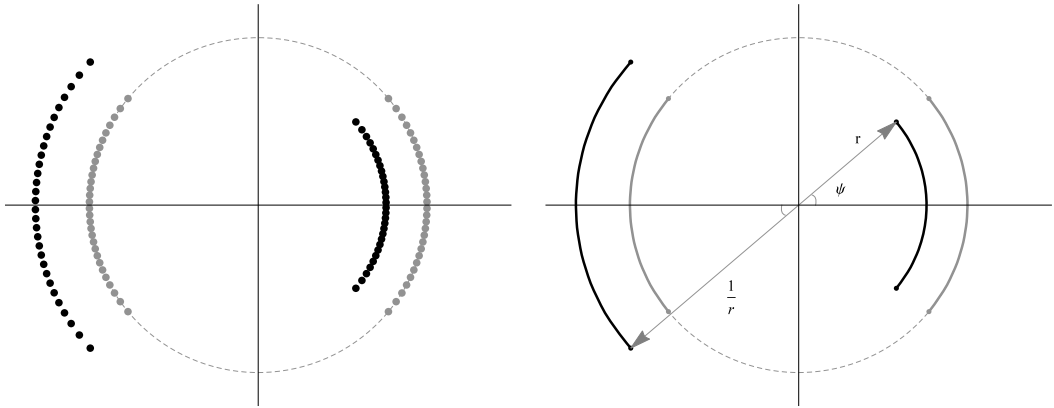}
	\caption[Distributions of roots for the cusped Wilson line]{Distribution of the roots of $P_L(x)$ on the complex plane for $\theta=0$ (gray) and $\theta=1$ (black) on the left and the condensation of the roots to corresponding smooth cuts on the right with the algebraic curve parameters $r$ and $\psi$ identified. The dashed circle is the unit circle.}
	\label{fig:roots}
\end{figure}

We will be utilizing the matrix model reformulation of the cusp anomalous dimension as discussed in the previous subsection. 
As usual in matrix models, when the size of matrices becomes large, the partition function is dominated by the solution of the saddle point equations. 
In the leading order it is just equal to the value of the integrand at the saddle point. 
We start by recasting the partition function \eq{eq:Mintegral} into the form
\beq
	\det \M_{2L} = \frac{1}{(2 \pi i)^{2L+1}} \frac{1}{(2L+1)!} \, \oint \prod_{i=1}^{2L+1} d x_i \, e^{-S(x_1, x_2, \dots, x_{2L+1})},
\eeq
where we take $N=2L$. 
The action is then given by
\begin{eqnarray}
	S &=& \sum_{i=1}^{2L+1} \[ 2 g \theta  \( x_i - \frac{1}{x_i} \)  - \(2L + 2\) \log x_i \] + 2 \sum_{i<j}^{2L+1} \log(x_i - x_j) + \\ \nonumber
	       &+& \sum_{i=1}^{2L+1} \log \sinh \( 2 \pi g \( x_i + \frac{1}{x_i} \) \).
\end{eqnarray}
The saddle point equations $\partial S / \partial x_j = 0$ now read
\begin{equation}
	g \theta \left( 1 + \frac{1}{x_j^2} \right) - \frac{L}{x_j} + \sum_{i \neq j}^{2L+1} \frac{1}{x_j - x_i} + \pi g \left( 1- \frac{1}{x_j^2} \right) \coth \left( 2 \pi g \left( x_j + \frac{1}{x_j} \right) \right) = 0.
\end{equation}
Technically the $x_j^{-1}$ term has a coefficient of $L+1$, but since we are taking $L \rightarrow \infty$ we chose to neglect it for simplicity.
We can further simplify them by noting that a large coupling constant $g$ appears inside the cotangent and since the roots $x_i$ are expected to be of order 1, with exponential precision it is possible to replace
\begin{equation}
	\coth \left( 2 \pi g \left( x_j + \frac{1}{x_j} \right) \right) \approx \mathrm{sgn}(\mathrm{Re}(x_j)).
\end{equation}
Finally we bring the equations to a more canonical and convenient form and get the following result,
\begin{equation}
	-\theta \, \frac{x_j^2 + 1}{x^2_j - 1} + \frac{L}{g} \frac{x_j}{x_j^2 - 1} - \frac{1}{g} \frac{x_j^2}{x_j^2 - 1} \sum_{i \neq j}^{2L+1} \frac{1}{x_j - x_i} = \pi \, \mathrm{sgn}(\mathrm{Re}(x_j)).
	\label{eq:saddlepoint}
\end{equation}
An alternative way of finding these values $x_i$ is to consider the following quantity $P_L(x)$, which played an important role in \cite{Gromov:2012eu},
\beq
P_L(x)=\frac{1}{\det {\cal M}_{2L}}\left|\begin{matrix}
I_1^{\theta}& I_0^{\theta}& \cdots & I_{2-2L}^{\theta}  &I_{1-2L}^{\theta}\\
I_2^{\theta}& I_1^{\theta}& \cdots & I_{3-2L}^{\theta} &I_{2-2L}^{\theta}\\
\vdots      &  \vdots     &\ddots & \vdots            &\vdots           \\
I_{2L}^{\theta}& I_{2L-1}^{\theta}& \cdots & I_{1}^{\theta}  &I_{0}^{\theta}\\
x^{-L}& x^{1-L}& \cdots & x^{L-1} &x^{L}\\
\end{matrix}\right|.
\label{eq:PLrepr}
\eeq
The numerator is the same as $\det \M_{2L}$ except in the last line $I^\theta$ is replaced by $x$ which is not integrated over. 
In the classical limit all integrals are saturated by their saddle point values, i.e. one can remove the integrals by simply replacing $x_i \rightarrow x_i^{cl}$. 
If we replace $x$ with any saddle point value $x_i^{cl}$ the determinant will contain two identical rows and will automatically become zero, thus the zeros of $P_L(x)$ are the saddle point values. 
On the complex plane they are distributed on two arcs as shown in figure \ref{fig:roots}. 
As expected, for the case $\theta=0$ we recover two symmetric arcs on the unit circle \cite{Gromov:2012eu}.
Now, following \cite{Gromov:2012eu,Beisert:2005bm,Gromov:2007aq}, we introduce the quasi-momentum $p(x)$ as
\begin{equation}
	p(x) = -\theta \, \frac{x^2 + 1}{x^2 - 1} + \frac{L}{g} \frac{x}{x^2 - 1} - \frac{2L}{g} \frac{x^2}{x^2 - 1} \, G_L(x),
	\label{eq:quantum_quasimomentum}
\end{equation}
where the resolvent $G_L(x)$ is
\begin{equation}
	G_L(x) = \frac{1}{2L} \sum_{k=1}^{2L+1} \frac{1}{x-x_k}.
	\label{eq:resolvent}
\end{equation}
The motivation for introducing $p(x)$ is that  the saddle point equations (\ref{eq:saddlepoint}) expressed through $p(x)$ take a very simple form
\begin{equation}
	\frac{1}{2} \left( p(x_i + i \epsilon) + p(x_i - i \epsilon) \right) = \pi \, \mathrm{sgn}(\mathrm{Re}(x_i)).
	\label{eq:pepsilon}
\end{equation}
In the classical limit the poles in the quasi-momentum condense and form two cuts. The shifts $\pm i\epsilon$ in the equation above refer to taking the argument of the quasi-momentum to one or the other side of the cut.

In the classical limit the poles of $p(x)$, which we denote as $x_i$, are governed by the saddle-point equation and condense on two cuts in the complex plane, as shown in figure \ref{fig:roots}.
The saddle-point equation \eq{eq:saddlepoint} has a symmetry $x\rightarrow -1/x$, so does the set of poles $x_i$. 
For the quasi-momentum (\ref{eq:quantum_quasimomentum}) this symmetry manifests as the identity $p(x) = -p(-1/x)$. 
Thus we conclude that the two cuts are related by an $x\rightarrow -1/x$ transformation. 
This and the invariance of the saddle-point equation under complex conjugation implies that the four branch points can be parametrized as $\{r\,e^{i\psi},r \, e^{-i\psi},-1/r \, e^{i\psi},-1/r \, e^{-i\psi}\}$. 
Note that in the case $\theta=0$ the symmetry is enhanced to $p(x)=-p(-x)$ and $p(1/x) = p(x)$, which is not true for arbitrary $\theta$.

The crucial point to notice is that while $p(x)$ satisfies the equation \eq{eq:pepsilon} which has different constants on the right hand side for the two different cuts, the corresponding equation for $p'(x)$ has a zero on the right hand side for both cuts, thus we expect $p'(x)$ to have a simpler form than $p(x)$. 
Our strategy is to write down an ansatz for the derivative $p'(x)$ using the symmetries and analytical properties of $p(x)$ and then integrate it. 
The form of the expression we get is analogous to the curve constructed in \cite{Kazakov:2004qf}, which also helps us to construct the ansatz.

First, $p(x)$ has four branch points and according to \eq{eq:pepsilon} its derivative changes sign on each cut, hence all the cuts are of square-root type. One can write $p'(x)\propto 1/y(x)$, where
 \beq
 	y(x) = \sqrt{x-r e^{i \psi}}\sqrt{x-r e^{-i \psi}}\sqrt{x + \frac{1}{r} e^{i \psi}}\sqrt{x + \frac{1}{r} e^{-i \psi}}.
 \label{eq:y}
 \eeq
Second, since the algebraic curve is obtained from \eq{eq:quantum_quasimomentum} in the classical limit, $p(x)$ should have simple poles at $x=\pm 1$. 
Finally, from \eq{eq:quantum_quasimomentum}  we can get the behaviour at infinity:
\beq
	p'(x) \approx \frac{L}{g} \frac{1}{x^2} + \mathcal{O}\(\frac{1}{x^3}\).
\eeq
By using the knowledge about these singularities and asymptotics we can fix $p(x)$ completely. 
Based on what we know up to now we write down our ansatz for the derivative
\beq
	p'(x) = \frac{A_1 x^4 + A_2 x^3 + A_3 x^2 + A_4 x + A_5}{(x^2 - 1)^2 \, \sqrt{x-r e^{i \psi}}\sqrt{x-r e^{-i \psi}}\sqrt{x + \frac{1}{r} e^{i \psi}}\sqrt{x + \frac{1}{r} e^{-i \psi}}}.
\eeq
The polynomial in the numerator is of order four in order to maintain the correct asymptotics, and below we fix its coefficients using the properties of the quasi-momentum. 
Of course, comparing with the asymptotic one can immediately see that $A_1 = L/g$, however our objective is to express $p(x)$ solely in terms of $r$ and $\psi$, which parametrize the algebraic curve.
The $x\rightarrow-1/x$ symmetry for the derivative implies that $A_1 = A_5$ and $A_2 = -A_4$. 
Next, simple poles at $x=\pm 1$ in $p(x)$ require zero residues of $p'(x)$ at $x=\pm1$, which fixes $A_2$ to be
\beq
	A_2 =- \frac{(2A_1 + A_3) \, r \, (r^2 - 1) \cos \psi}{r^4 - 2 \, r^2 \, \cos 2\psi + 1}.
\eeq
We fix the two remaining unknowns $A_1$ and $A_3$ after integrating the $p'(x)$. 
We don't write the intermediate results of the integration as the expressions are enormous without any apparent structure.
Looking back at \eq{eq:pepsilon} we see that at the branchpoints
\beq
	p(x_{bp}) =\pm \pi.
\eeq
We use this condition to fix $A_1$ and we get
\beq
	A_1 = \frac{A_3}{2}\frac{K_1-E_1}{E_1+K_1-2\,a^2 \, K_1\cos^2(\psi)},
\eeq
where
\beq
E_1=\mathbb{E}\(a^2\sin^2(\psi)\),\;K_1=\mathbb{K}\(a^2\sin^2(\psi)\),\;a = \frac{2r}{r^2+1}
\label{eq:E1K1}
\eeq
and $\mathbb{K}, \mathbb{E}$ are complete elliptic integrals of the first and second kind.
Finally we can use the $x \rightarrow -1/x$ symmetry on the quasi-momentum itself, as before we only used it on the derivative. Imposing the symmetry yields
\beq
	A_3 = \frac{8}{a}\(E_1+K_1-2\,a^2\,\cos^2(\psi)K_1\).
\eeq
As expected, after plugging these coefficients into $p(x)$ the whole expression simplifies enormously and we are left with our main result
\begin{align}
	\label{eq:pmainresult}
	p(x) &= \pi - 4\,i\,  E_1\, \mathbb{F}_1(x) + 4\,i\,  K_1\, \mathbb{F}_2(x) - a \left( \frac{x+ r e^{-i\psi}}{x+\frac{1}{r} e^{i\psi}} \right) \left(\frac{2/r\,e^{i \psi}}{x^2 - 1} \right) y(x)\,K_1,
\end{align}
where
\begin{equation}
	\mathbb{F}_1(x) = \mathbb{F}\left( \left. \sin^{-1} \sqrt{ \left( \frac{x -r e^{-i\psi}}{x+ \frac{1}{r} e^{i\psi}} \right) \left( \frac{ e^{ i \psi}}{ia\,r\sin\psi} \right)} \; \right| a^2 \sin^2(\psi) \right),
\end{equation}
\begin{equation}
	\mathbb{F}_2(x) = \mathbb{E}\left( \left. \sin^{-1} \sqrt{ \left( \frac{x -r e^{-i\psi}}{x+ \frac{1}{r} e^{i\psi}} \right) \left( \frac{ e^{ i \psi}}{ia\,r\sin\psi} \right) } \; \right| a^2 \sin^2(\psi) \right)
\end{equation}
and $\mathbb{F}, \mathbb{E}$ are incomplete elliptic integrals of the first and second kind.
We verified this result numerically by comparing it to the extrapolation of the discrete quasi-momentum \eq{eq:quantum_quasimomentum} at large $L$ and got an agreement up to thirty digits. 
We also compared this expression at $\theta = 0$ with the quasi-momentum obtained in \cite{Gromov:2012eu} and the expressions agree perfectly.
The resulting quasi-momentum is parametrized in terms of the branchpoints, i.e. the parameters are the radius $r$ and angle $\psi$. 
They are determined in terms of $L/g$ and $\theta$, which are parameters of the matrix model. 
We already mentioned that $L/g$ is simply the constant $A_1$, which we found to be
\beq
	\frac{L}{g} = 4\,\frac{K_1-E_1}{a} ,
	\label{eq:Lgfixed}
\eeq
and looking back at \eq{eq:quantum_quasimomentum} we see that $\theta = p(0) = -p(\infty)$, hence
\beqa
	\theta &=& -\pi + \frac{2a}{r}\,e^{i\psi}K_1 \nonumber \\
	       &-& \left. 4 \, i \, K_1\,\mathbb{E}\left(\sin ^{-1}\sqrt{\frac{  e^{i\psi}}{ia\,r\sin\psi}} \,\right|\, a^2 \sin ^2(\psi )\right)\nonumber \\
           &+& \left. 4 \, i \, E_1\, \mathbb{F}\left(\sin ^{-1}\sqrt{\frac{  e^{i\psi}}{ia\,r\sin\psi}}\,\right|\, a^2 \sin ^2(\psi )\right).
           \label{eq:thetafixed}
\eeqa

Finally we can find the classical limit of the cusp anomalous dimension from the constructed algebraic curve. 
At large $L$ the formula \eq{eq:mainresultIntro} can be rewritten as
\beq
\Gamma_L(g)=\frac{\phi-\theta}{4}\partial_\theta\partial_L\det{\cal M}_{2L}.
\eeq
Use the integral representation \eq{eq:Mintegral} for $\det{\cal M}_L$ we can notice that
\beq
\partial_\theta \log \det {\cal M}_L=\left\langle 2g\sum\limits_{i=1}^{2L}(x_i-1/x_i)\right\rangle,
\eeq
where by the angular brackets we denoted an expectation value in the matrix model with the partition function \eq{eq:Mintegral}.
In the quasiclassical approximation the expectation value is determined by the saddle-point, i.e. the previous expression is equal to $2g\sum\limits_{i=1}^{2L}(x_i-1/x_i)$, where the roots $x_i$ are the solutions of the saddle-point equation \eq{eq:saddlepoint}.
Since the set of the roots has a $x\rightarrow-1/x$ symmetry, the two terms in the sum give the same contribution. Thus
\beq
\partial_\theta \log \det {\cal M}_L=-4g \sum\limits_{i=1}^{2L}\frac{1}{x_i}=8 \, g \, L \, G(0),
\eeq
where we used the resolvent \eq{eq:resolvent}.
Using the relation \eq{eq:quantum_quasimomentum} between the resolvent and the quasi-momentum  we find 
\beq
	G(0)=\frac{g}{L}\left(p''(0)/4-\theta\right),
\eeq	
so the final expression for the cusp anomalous dimension in terms of the quasi-momentum is given by
\begin{align}
\Gamma_L(g)=-\frac{g^2}{2}\partial_Lp_L''(0),
\label{eq:E2}
\end{align}
where $p(x)$ is given in terms of the parameters of the branch points $r$ and $\psi$ in \eq{eq:pmainresult}. 
They are implicitly defined through $L/g$ and $\theta$ by the equations \eq{eq:Lgfixed} and \eq{eq:thetafixed}. 
In order to get $\Gamma_L$ we express $\partial_L$ though $\partial_r$ and $\partial_\psi$ and then apply \eq{eq:E2} to \eq{eq:pmainresult}. 
Finally we obtain a very simple result in terms of $r$ and $\psi$
\beq
\Gamma_{L}(g)=g(\phi-\theta)\(r-1/r\)\cos\psi.
\label{eq:GammaL}
\eeq
We can now check our formula \eq{eq:GammaL} in the limit $\phi=0$ and $\theta\rightarrow 0$ considered in section E.2 of \cite{Gromov:2012eu}. 
As the angles go to zero, the branch points approach the unit circle $r\rightarrow 1$, thus the formula \eq{eq:GammaL} gives
\beq
	\Gamma_L(g)=2\,g\,\theta (r-1)\cos\psi.
\eeq
In this limit $r-1\propto\theta$, and the coefficient of proportionality can be found by expanding the equation \eq{eq:thetafixed} for $\theta$ around $r=1$,
\beq
	2 (1-r)\frac{\mathbb{E}\left(\sin^2\psi\right)}{\cos\psi}=\theta/2.
\eeq
Plugging it into the formula above we get
\beq
\Gamma_L(g)=g\,\theta^2 \,\frac{\cos^2\psi}{2\mathbb{E}\left(\sin^2\psi\right)}
\eeq
which perfectly agrees with (190) of \cite{Gromov:2012eu}.
We should note that the equation \eq{eq:thetafixed} is written in the approximation $\phi\approx\theta$ and now on the top of it we want to take a limit $\theta\rightarrow 0$. 
Since before we have neglected the terms ${\cal O}(\theta-\phi)^2$, the result, which is now of the order ${\cal O}(\theta)^2$ will not generally be reproduced. 
However, we found that here and in several other formulas correct small angle limit is reproduced if before taking $\theta,\phi$ to zero we replace $\theta$ and $\phi$ by the middle angle $\phi_0=\(\phi+\theta\) / 2$, which is in our case equal to $\theta/2$.

\subsubsection{Matching the string solution}

As we have mentioned before, in the classical $L\sim\sqrt{\lambda}\rightarrow\infty$ limit $\Gamma_L(\lambda)$ can be matched with the energy of an open string. 
The class of string solutions we are interested in was introduced in \cite{Correa:2012hh} and generalized in \cite{Gromov:2012eu}. 
It is a string in $AdS_3\times S^3$ governed by the parameters $\theta,\phi$, $AdS_3$ charge $E$ and $S^3$ charge $L$, furthermore the four parameters are restricted by the Virasoro constraint. 
The ansatz for the embedding coordinates of $AdS^3$ and $S^3$ is
\begin{align}
y_1+iy_2=e^{i\kappa\tau}\sqrt{1+r^2(\sigma)},\;\; y_3+iy_4=r(\sigma) e^{i\phi(\sigma)},\\
x_1+ix_2=e^{i\gamma\tau}\sqrt{1+\rho^2(\sigma)},\;\; x_3+ix_4=r(\sigma) e^{i f(\sigma)}.
\label{eq:embedding}
\end{align}
The range of the worldsheet coordinate is $-s/2<\sigma<s/2$, where $s$ is to be found dynamically. The angles $\theta$ and $ \phi$ parametrizing the cusp enter the string solution through the boundary conditions $\phi(\pm s/2)=\pm (\pi-\phi)/2$ and $f(\pm s/2)=\pm\theta/2$.
The equations of motion and Virasoro constraints lead to the following system of equations (see Appendix E of \cite{Gromov:2012eu} for more details, also \cite{Drukker:2011za}):
\begin{align}
f(\gamma,l_\theta)&=f(\kappa,l_{\phi}),
\label{eq:ff}
\\
h(\gamma,l_{\theta})=\theta, &\;\; h(\kappa,l_{\phi})=\phi,
\label{eq:hh}
\\
g(\gamma,l_\theta)=L, &\;\; g(\kappa,l_\phi)=E,\label{eq:gg}
\end{align}
where
\begin{align}
f(\gamma,l)&=\frac{2\sqrt{2}}{\sqrt{\gamma^2+k^2+1}}\,\mathbb{K}\left(\frac{-k^2+\gamma^2+1}{k^2+\gamma^2+1}\right),
\label{eq:f}
\\
h(\gamma,l)&=\frac{2l}{k(1+k^2-\gamma^2)}\left[(1+\gamma^2+k^2)\,\Pi\left(\frac{k^2-2l^2-\gamma^2+1}{2k^2}\,\vline\,\frac{k^2-\gamma^2-1}{2k^2}\right)-\right.
\nonumber
\\ & \left.-2\gamma^2\,\mathbb{K}\left(\frac{k^2-\gamma^2-1}{2k^2}\right)\right],
\label{eq:h}
\\
g(\gamma,l)&=-2\sqrt{2} \, \frac{\sqrt{\gamma^2+k^2+1}}{\gamma}\left[\mathbb{E}\left(\frac{-k^2+\gamma^2+1}{k^2+\gamma^2+1}
\right)-\mathbb{K}\left(\frac{-k^2+\gamma^2+1}{k^2+\gamma^2+1}
\right)\right],
\label{eq:g}
\\
k^4&=\gamma^4-2 \gamma^2+ 4\, \gamma^2 l^2+1.
\nonumber
\end{align}
One can see that the variables $\theta,l_{\theta},\gamma$ and $L$ are responsible for the $S^3$ part of the solution, while  $\phi,l_{\phi},\kappa$ and $E$ are their analogues for $AdS_3$. 
The two parts of the solution are connected only by the Virasoro condition which leads to \eq{eq:ff}.
We are interested in the limit when $\theta\approx\phi$. 
In this limit the two groups of variables responsible for $S^3$ and $AdS_3$ parts of the solution become close to each other, namely $l_{\theta}\approx l_{\phi}$ and $E\approx L$. 
The cusp anomalous dimension should be compared with the difference $E-L$, because $L$ is the classical part of the dimension of the observable $W_L$. 
To find $E-L$ we linearise the system \eq{eq:f},\eq{eq:h},\eq{eq:g} around $\phi\approx\theta$, which yields
\begin{align}
E-L=(\phi-\theta)\left|\frac{\partial{(g,f)}}{\partial{(l,\kappa)}}\right|/\left|\frac{\partial{(h,f)}}{\partial{(l,\kappa)}}\right|.
\label{eq:ELbig}
\end{align}
Plugging in here the explicit form of $g,f$ and $h$ one gets as a result an extremely complicated expression with a lot of elliptic functions. 
However, there exists a parametrization in which the result looks surprisingly simple, it comes from comparison of the string conserved charges with the corresponding quantities of the algebraic curve. 
One can notice that the equations for $\theta$ and $L/g$ in the end of the last subsection have the same structure as the equations \eq{eq:hh} and \eq{eq:gg}. 
Indeed, it is possible to match them precisely if one chooses the correct identification of parameters of the string solution $l_\theta,\gamma$ with the parameters of the algebraic curve $r,\psi$. 
We used various elliptic function identities to bring the equations to identical form after the following identifications
\beq
\gamma=\frac{2r}{\sqrt{r^4-2r^2\cos2\psi+1}},\; l_{\theta}=\frac{(r^2-1)\cos\psi}{\sqrt{r^4-2r^2\cos2\psi+1}}.
\label{eq:identification}
\eeq
As another confirmation of correctness of this identification, after plugging it into \eq{eq:ELbig} the complicated expression reduces to the following simple formula for the classical energy
\begin{align}
E-L=g(\phi-\theta)(r-1/r)\cos\psi,
\label{eq:E1}
\end{align}
which exactly coincides with the matrix model result \eq{eq:GammaL}.
Notice that this can be rewritten as a sum over the branch points of the algebraic curve
\begin{align}
&E-L=\frac{g}{2}(\phi-\theta)\sum\limits_i a_i,
\end{align}
where $a_i=\{r \, e^{i\psi}, r \, e^{-i\psi},-1/r \, e^{i\psi},-1/r \, e^{-i\psi}\}$.

\newpage 


\section{Developments in ABJM}
\label{sec:abjm}

\begin{chapquote}{Albert Einstein}
Insanity: doing the same thing over and over again and expecting different results.
\end{chapquote}

\noindent
In this last chapter of the thesis we leave $\N=4$ super Yang-Mills behind and switch to the so-called ABJM theory, originally proposed by (and thus named after) Aharony, Bergman, Jafferis and Maldacena in \cite{Aharony:2008ug} following \cite{Schwarz:2004yj,Schwarz:2004yj1,Schwarz:2004yj2,Bagger:2006sk,Gustavsson:2007vu}.
This theory also has a string theory dual and exhibits integrability, as we shall review shortly.
Many of these features are very similar to the ones we presented in the case of $\N=4$ super Yang-Mills.
In order to distinguish the two models we will often refer to ABJM and its string dual as $AdS_4/CFT_3$ and to the $\N=4$ SYM and its dual as $AdS_5/CFT_4$. 
We will keep this chapter short, going into more detail only when discussing the classical spectral curve.

\subsection{Short introduction}
\label{sec:abjm_intro}

The ABJM theory is a three-dimensional superconformal Chern-Simons gauge theory with $\mathcal{N}=6$ supersymmetry. 
The gauge group is $U(N) \times \hat{U}(N)$ at levels $\pm k$, where $A_{\mu}$ and $\hat{A}_{\mu}$ are the associated gauge fields.
The field content consists of four complex scalars $\phi^I$, four Dirac fermions $\psi_I$, their adjoints $\phi_I=\phi^{I\dagger}$, $\psi^I=\psi_I^\dagger$, and two gauge fields, $A_\mu$ and $\hat{A}_\mu$. 
The index $I$ is an $SU(4)$ R-symmetry index.  
The fields $\phi^I$ and $\psi_I$ transform in the $(\bar{N},N)$ representation of the gauge group and their adjoints transform in the $(N,\bar{N})$ representation. 
The global symmetry group is the orhthosymplectic supergroup $OSp(6|4)$ whose bosonic part is the direct product of the R-symmetry group $SO(6) \simeq SU(4)$ and the conformal group $SO(2,3)$.
As before, we are interested in gauge invariant single trace operators.
The Chern-Simons level $k$ acts like the coupling constant with large $k$ corresponding to weak coupling.
The planar limit is defined by $k, N \to \infty$ keeping $\lambda \equiv N/k$ fixed.

This theory was conjectured \cite{Aharony:2008ug} to be dual to M-theory on $AdS_4 \times S^7 / \mathbb{Z}_k$ with four-form flux $F^{(4)} \sim N$ through $AdS_4$. 
Alternatively one can say that it is the effective theory for a stack of N M2 branes moving on a $\mathbb{C}^4 / Z_k$ orbifold point. 
In the limit $k^5 \gg N$ M-theory is well approximated by weakly coupled type IIA string theory on $AdS_4 \times \mathbb{CP}^3$.
Here one also has RR four-form flux $F^{(4)} \sim N $ through $AdS_4$ and RR two-form flux $F^{(2)} \sim k $ through $\mathbb{CP}^1 \subset \mathbb{CP}^3$. 
On the gravity side of the correspondence the string coupling constant and effective tension are given by
\beq
	g_s \sim \( \frac{N}{k^5} \)^{1/4} = \frac{\lambda^{5/4}}{N}, \;\;\;\;\; \frac{R^2}{\alpha'} = 4\pi \sqrt{2\lambda},
\eeq
where $R$ is the radius of $\mathbb{CP}^3$ and twice the radius of $AdS_4$.
From here we see that the planar limit corresponds to small $g_s$, i.e. free strings.
Small t'Hooft coupling $\lambda$ corresponds to a highly curved string background where the string is subject to large quantum fluctuations whereas on the gauge theory side it translates to weak coupling. 
On the opposite end of large t'Hooft coupling we have classical strings and strongly coupled gauge theory.

\subsection{Integrability}
\label{sec:abjm_integrability}

The theories on both ends of the correspondence were shown to be integrable in the planar limit.
At weak coupling in the ABJM theory the spectrum of local operators was shown to be described by long-ranged $OSp(6|4)$ spin chains \cite{Minahan:2008hf}.
Type IIA string theory was formulated as a supercoset sigma model with the target space
\beq
\frac{OSp(2,2|6)}{SO(3,1)\times SU(3)\times U(1)}
\eeq
which has $AdS_4\times CP^3$ as its bosonic part.
The model was and shown to the classically integrable \cite{Stefanski:2008ik,Arutyunov:2008if}. 
We will now review these developments putting off the discussion of exact results and the quantum spectral curve construction for the last section of the chapter. 

\subsubsection{Weak coupling}

Since the matter fields transform in bifundamental representations, local gauge invariant operators are formed by taking traces of fields that transform alternatingly in the representations $(N,\bar{N})$ and  $(\bar{N},N)$, for example
\beq
	\label{eq:abjm_bps}
	\tr \( \phi^1 \phi_4^\dagger \phi^1 \phi_4^\dagger \dots \).
\eeq
These can then be mapped with spin chains of even length having alternating representations in each site.
Hamiltonians for spin-chains have been worked out for different sectors to varying loop levels using a multitude of methods.
The first calculation was performed at two-loops for the $SU(4)$ sector, the Hamiltonian reads \cite{Minahan:2008hf}
\beq
	H = \frac{\lambda^2}{2}\sum_{i=1}^{2L} \( 2-2P_{i,i+2} + P_{i,i+2} K_{i,i+1} + K_{i,i+1} P_{i,i+2} \),
\eeq
where $P_{i,j}$ and $K_{i,j}$ are permutation and contraction operators defined in \eq{eq:spin_perm}, \eq{eq:spin_contract}.
The operator \eq{eq:abjm_bps} is the ground state of this Hamiltonian.
From here on out the story is almost identical as in $\N=4$ super Yang-Mills, namely one can write down the Bethe ansatz equations, find the scattering matrix etc. (see \cite{Klose:2010ki} for a review).
The only complication here is that the coupling constant of the theory $\lambda$ enters all of the integrability constructions (at weak and strong coupling) via the \emph{interpolating function} $h(\lambda)$ appearing in the dispersion relation
\beq
	E(p) = \sqrt{1 + 4h^2(\lambda) \sin^2 \frac{p}{2}}.
\eeq
It is not fixed by symmetry and its asymptotic expansions have the structure
\beq 
\label{eqn:general-h-expansion}
  h(\lambda) = \begin{cases} 
    \lambda \Bigsbrk{ 1 + c_1 \lambda^2 + c_2 \lambda^4 + \ldots }         & \mbox{for $\lambda\ll 1$} \; , \\[2mm]
    \sqrt{\frac{\lambda}{2}} + a_1 + \frac{a_2}{\sqrt{\lambda}} + \ldots   & \mbox{for $\lambda\gg 1$} \;.
  \end{cases}
\eeq
The analogue of this function in $\N=4$ super Yang-Mills is the trivial relation $g = \frac{\sqrt{\lambda}}{4\pi}$. 

\subsubsection{Strong coupling}

The classical spectral curve was constructed for $AdS_4/CFT_3$ much in analogy with the earlier formulated $AdS_5/CFT_4$ curve \cite{Gromov:2008bz}. 
Here we present a compact self-contained summary of the construction using the language of off-shell fluctuation energies \cite{Gromov:2008ec} analogous to the discussion in section \ref{sec:spectral_curve}.
We shall work in the algebraic curve regularization and write all equations in terms of the $\sigma$-model coupling $g$. 
For large $g$, it is  related to the 't Hooft coupling by 
\beq
\lambda = N/k = 8\,g^{2},
\eeq
but, contrary to the $\adsfive$ case, this relation will get corrections at finite $g$, as discussed in the previous subsection.
The classical algebraic curve for $\adscp$ is a 10-sheeted Riemann surface. 
The spectral parameter moves on it and  we shall consider 10 symmetric quasi-momenta $q_{i}(x)$ 
\beq
(q_{1}, q_{2}, q_{3}, q_{4}, q_{5}) = (-q_{10}, -q_{9}, -q_{8}, -q_{7}, -q_{6}).
\eeq
They can have branch cuts connecting the sheets with 
\beq
q_{i}^{+}-q_{j}^{-} = 2\,\pi\,n_{ij}.
\eeq
In the terminology of \cite{Gromov:2008bz}, the physical polarizations $(ij)$ can be split into {\em heavy} and {\em light} ones and are summarized in the following table:
 $$
 \begin{array}{c|ccc}
  & \mbox{AdS${}_{4}$} & \mbox{Fermions} & \mathbb{CP}^{3} \\
  \hline
  \mbox{heavy} & \quad (1,10) (2,9) (1,9)\quad & (1,7) (1,8) (2,7) (2,8) & (3,7) \\
  \mbox{light} & & (1,5) (1,6) (2,5) (2,6) & \quad (3,5) (3,6) (4,5) (4,6)\quad
  \end{array}
 $$
Virasoro constraints require that the poles of the quasi-momenta $q_{i}(x)$ at $x=\pm 1$ are synchronized according to
\beq
(q_{1},  q_{2}, q_{3}, q_{4}, q_{5}) = \frac{\alpha_{+}}{x-1}\,(1,1,1,1,0)+\cdots = 
\frac{\alpha_{-}}{x+1}\,(1,1,1,1,0)+\cdots.
\eeq
Inversion symmetry reads
\beq
q_{1}(x) = -q_{2}(1/x), \qquad
q_{3}(x) = 2\,\pi\,m-q_{4}(1/x), \qquad
q_{5}(x) = q_{5}(1/x),
\eeq
where $m\in\mathbb Z$ is a winding number. 
The asymptotic values of the quasi-momenta for a length $L$ state with energy and spin $E$, $S$ are
\beq
\label{eq:asym}
\left(\begin{array}{c} q_{1}(x) \\ q_{2}(x) \\ q_{3}(x) \\ q_{4}(x) \\ q_{5}(x) \end{array}\right) = 
\frac{1}{2\,g\,x}\,\left(\begin{array}{l} 
L+E+S \\
L+E-S \\
L-M_{r}+M_{s} \\
L+M_{r}-M_{u}-M_{v} \\
M_{v}-M_{u}
\end{array}\right)+\cdots ,
\eeq
where $M_{r,u,v}$ are related to the  $SU(4)$ representation of the state 
\beq
[d_{1}, d_{2}, d_{3}] = [L-2M_{u}+M_{r}, M_{u}+M_{v}-2M_{r}+M_{s}, L-2M_{v}+M_{r}].
\eeq

\normalsize
Semiclassical quantization is achieved by perturbing quasi-momenta introducing extra poles that shift the quasi-momenta $q_{i}\to q_{i}+\delta q_{i}$. 
Virasoro constraints and inversion properties of the variations $\delta q_{i}$ follow from those of the $q_{i}$'s. 
In order to find the asymptotic expression of $\delta q_{i}$ in terms of the number $N_{ij}$ of extra fluctuations we can look at the details of polarized states and obtain 

\small
\beq
\label{eq:asym2}
\left(\begin{array}{c} \delta q_{1}(x) \\ \delta q_{2}(x) \\ \delta q_{3}(x) \\ \delta q_{4}(x) \\ \delta q_{5}(x) \end{array}\right) = 
\frac{1}{2\,g\,x}\,\left(\begin{array}{c} 
\delta E+N_{19}+2\,N_{1, 10} +N_{15}+N_{16}+N_{17}+N_{18}  \\
\delta E+2\,N_{29}+N_{19}  +N_{25}+N_{26}+N_{27}+N_{28}  \\
 -N_{18}-N_{28}  -N_{35}-N_{36}-N_{37} \\
 -N_{17}-N_{27}  -N_{45}-N_{46}-N_{37} 	\\
 +N_{15}-N_{16}+N_{25}-N_{26}  +N_{35}-N_{36}+N_{45}-N_{46}
\end{array}\right).
\eeq
\normalsize
The off-shell frequencies $\Omega^{ij}(x)$ are defined in order to have 
\beq
\delta E = \sum_{n, ij} N^{ij}_{n}\,\Omega^{ij}(x^{ij}_{n}),
\eeq
where the sum is over all pairs $(ij)\equiv (ji)$ of physical polarizations and integer values of $n$ with 
\beq
\label{eq:pole}
q_{i}(x^{ij}_{n})-q_{j}(x^{ij}_{n}) = 2\,\pi\,n.
\eeq
Also, the residues at the extra poles are
\beq
\delta q_{i}(x) = k_{ij}\,N_{n}^{ij}\,\frac{\alpha(x^{ij}_{n})}{x-x_{n}^{ij}},\quad\mbox{with}\quad
\alpha(x) = \frac{1}{2\,g}\,\frac{x^{2}}{x^{2}-1},
\eeq
and $k_{ij}=0, \pm 1, \pm 2$ are the coefficients of $N_{ij}$ in (\ref{eq:asym2}).
%
%
By linear combination of frequencies and inversion (as in the \maldafive case), we can derive all  the off-shell frequencies in terms of two fundamental ones
\beq
\Omega_{A}(x) = \Omega^{15}(x), \qquad \Omega_{B}(x) = \Omega^{45}(x).
\eeq
Their explicit expressions turn out to be 
\ba
\Omega^{29} &=&  2\,\left[-\Omega_{A}(1/x)+\Omega_{A}(0)\right], \nonumber \\
\Omega^{1, 10} &=&  2\,\Omega_{A}(x),\nonumber \\
\Omega^{19} &=&  \Omega_{A}(x)-\Omega_{A}(1/x)+\Omega_{A}(0), \nonumber \\
\Omega^{37} &=&\Omega_{B}(x)-\Omega_{B}(1/x)+\Omega_{B}(0), \nonumber \\
\Omega^{35}  = \Omega^{36} &=& -\Omega_{B}(1/x)+\Omega_{B}(0),\nonumber \\
\Omega^{45} = \Omega^{46} &=& \Omega_{B}(x), \nonumber \\
\Omega^{17} &=& \Omega_{A}(x)+\Omega_{B}(x), \nonumber \\
\Omega^{18} &=& \Omega_{A}(x)-\Omega_{B}(1/x)+\Omega_{B}(0), \nonumber \\
\Omega^{27} &=& \Omega_{B}(x)-\Omega_{A}(1/x)+\Omega_{A}(0), \nonumber \\
\Omega^{28} &=& -\Omega_{A}(1/x)+\Omega_{A}(0)-\Omega_{B}(1/x)+\Omega_{B}(0), \nonumber \\
\Omega^{15} = \Omega^{16} &=& \Omega_{A}(x), \nonumber \\
\Omega^{25} = \Omega^{26} &=& -\Omega_{A}(1/x)+\Omega_{A}(0).
\ea

\subsection{Folded string in $\adscp$}
\label{sec:abjm_folded}

We present the algebraic curve for the folded string in $\adscp$ closely following  the notation of  \cite{Gromov:2008fy}. 
In terms of the semiclassical variables
\beq
\mathcal S = \frac{S}{4\,\pi\,g}, \qquad
\mathcal J = \frac{J}{4\,\pi\,g},
\eeq
the energy of the folded string can be expanded according to 
\beq
E = 4\,\pi\,g\,\,\mc E_{0}(\mc J, \mc S)+E_{1}(\mc J, \mc S)+\mc O\left(\frac{1}{g}\right),
\eeq
where the small $\mc S$ expansion of the classical contribution $\mc E_{0}$ reads
\beq
\label{eq:classical}
 \mathcal E_{0}= \mathcal J+\frac{\sqrt{\mathcal J^{2}+1}}{\mathcal J}\,\mathcal S-\frac{\mathcal J^{2}+2}{4\,\mathcal J^{3}(\mathcal J^{2}+1)}\,\mathcal S^{2}+ 
 \frac{3\,\mathcal J^{6}+13\,\mathcal J^{4}+20\,\mathcal J^{2}+8}{16\,\mathcal J^{5}\,(\mathcal J^{2}+1)^{5/2}}\,\mathcal S^{3}+\cdots.
\eeq
The quasi-momenta are closely related to those of the \ads folded string since motion is still in $AdS_{3}\times S^{1}$
and the $\mathbb{CP}^{3}$ part of the background plays almost no role. 
The only non trivial quasi-momentum is
\ba
q_{1}(x) &=& \pi\,f(x)\,\left\{-\frac{J}{4\,\pi\,g}\,\left(\frac{1}{f(1)\,(1-x)}-\frac{1}{f(-1)(1+x)}\right)+ \right. \\
&& \left. -\frac{4}{\pi\,(a+b)(a-x)(a+x)}\left[
(x-a)\,\mathbb K\left(\frac{(b-a)^{2}}{(b+a)^{2}}\right)+ \right.\right. \nonumber \\
&& \left.\left. + 2\,a\,\Pi\left(\left.
\frac{(b-a)(a+x)}{(a+b)(x-a)} \right| \frac{(b-a)^{2}}{(b+a)^{2}}
\right)
\right]
\right\}-\pi.\nonumber
\ea
where the branch points obey $1<a<b$ and 
\beq
f(x) = \sqrt{x-a}\,\sqrt{x+a}\,\sqrt{x-b}\,\sqrt{x+b},
\eeq
\ba
S &=& 2\,g\,\frac{ab+1}{ab}\,\left[b\,\mathbb E\left(1-\frac{a^{2}}{b^{2}}\right)
-a\,\mathbb K\left(1-\frac{a^{2}}{b^{2}}\right)\right], \nonumber \\
J &=& \frac{4\,g}{b}\,\sqrt{(a^{2}-1)(b^{2}-1)}\,\mathbb K\left(1-\frac{a^{2}}{b^{2}}\right). \\
E &=& 2\,g\,\frac{ab-1}{ab}\,\left[b\,\mathbb E\left(1-\frac{a^{2}}{b^{2}}\right)
+a\,\mathbb K\left(1-\frac{a^{2}}{b^{2}}\right)\right].\nonumber
\ea
The other quasi-momenta are
\ba
&& q_{2}(x) = -q_{1}(1/x), \\
&& q_{3}(x) = q_{4}(x)  = \frac{J}{2\,g}\,\frac{x}{x^{2}-1}.  \\
&& q_{5}(x) = 0.
\ea
The above expressions are valid for a folded string with minimal winding. 
Adding winding is trivial at the classical level, but requires non trivial changes at the one-loop level (see for instance \cite{Gromov:2011bz} for a detailed analysis of the $\adsfive$ case).
The independent off-shell frequencies can be determined by the methods of \cite{Gromov:2008ec}. 
The result is rather simple and reads
\ba
&& \Omega_{A}(x) =  \frac{1}{ab-1}\left(1-\frac{f(x)}{x^{2}-1}\right),   \\
&& \Omega_{B}(x) = \frac{\sqrt{a^{2}-1}\,\sqrt{b^{2}-1}}{ab-1}\,\frac{1}{x^{2}-1}.
\ea
The one-loop shift of the energy is given in full generality by the following sum of zero point energies
\beq
\label{eq:one-loop-correction}
E_{1} = \frac{1}{2}\,\sum_{n=-\infty}^{\infty}\,\sum_{ij}(-1)^{F_{ij}}\,\omega^{ij}_{n},\qquad
\omega_{n}^{ij} = \Omega^{ij}(x^{ij}_{n}),
\eeq
where the sum over $ij$ is over the $8_{B}+8_{F}$ physical polarizations and $x^{ij}_{n}$ is the unique solution 
to the equation (\ref{eq:pole}) under the condition $|x_{n}^{ij}|>1$. 
If it happens that for some $ij$ and $n$ the above equation has no solution, then we shall say that the polarization $(ij)$ has the { missing mode} $n$. 
Missing modes can be treated according to the procedure discussed in \cite{Gromov:2008ec}.
In the same spirit as \cite{Gromov:2011de,Gromov:2011bz}, the infinite sum over on-shell frequencies can be  evaluated by contour integration in the complex plane. 
The result is quite similar to the $\adsfive$ one and reads
\beq
E_{1} = E_{1}^{\rm anomaly, 1}+E_{1}^{\rm anomaly, 2}+E_{1}^{\rm dressing}+E_{1}^{\rm wrapping},
\eeq
with
\ba
E_{1}^{\rm anomaly, 1} &=& 2\,\int_{a}^{b}\,\frac{dx}{2\,\pi\,i}\left[\Omega^{1,10}(x)-\Omega^{1,10}(a)\right]\,
\partial_{x}\,\log\sin q_{1}(x), \\
E_{1}^{\rm anomaly, 2} &=& -2\times 2\,\int_{a}^{b}\,\frac{dx}{2\,\pi\,i}\left[\Omega^{1,5}(x)-\Omega^{1,5}(a)\right]\,\partial_{x}\,\log\sin \frac{q_{1}(x)}{2}, \\
E_{1}^{\rm dressing} &=& \sum_{ij}(-1)^{F_{ij}}\,\int_{-1}^{1}\frac{dz}{2\,\pi\,i}
\,\Omega^{ij}(z)\,\partial_{z}\frac{i\,\left[q_{i}(z)-q_{j}(z)\right]}{2},\\
E_{1}^{\rm wrapping} &=& \sum_{ij}(-1)^{F_{ij}}\,\int_{-1}^{1}\frac{dz}{2\,\pi\,i}
\,\Omega^{ij}(z)\,\partial_{z}\log(1-e^{-i\,(q_{i}(z)-q_{j}(z))}),
\ea 
where $x(z) = z+\sqrt{z^{2}-1}$ and the anomaly contributions are computed integrating on the upper half complex plane.
As in \ads, the labelling of the various contributions denote their physical origin. In particular, dressing and wrapping contributions have been separated in order to split the asymptotic contribution from finite size effects. As in \ads, the anomaly terms are special 
contributions arising from the deformation of contours and ultimately due to the presence of the algebraic curve cuts.
The representation (\ref{eq:one-loop-correction}) is a compact formula for $E_{1}$ and can be evaluated numerically with minor effort. In order to understand it better, we shall now analyse the short and long string limit. In the former case, 
we shall evaluate the explicit sum over frequencies clarifying the relation with the contour integrals. In the latter, we shall extract the analytical expansion at large spin directly from (\ref{eq:one-loop-correction}).

\subsubsection{Short string limit}

The short string limit is generically $\mc S\to 0$. Regarding $\mc J$, we shall consider two cases. The first amounts to 
keeping $\mc J$ fixed, expanding in the end each coefficient of powers of $\mc S$ at small $\mc J$. This is 
precisely the procedure worked out in \cite{Gromov:2011bz} in $\adsfive$. In the second case, we shall keep the ratio
$\rho = \mc J/\sqrt\mc S$ fixed as in \cite{Beccaria:2012tu}. The two expansions are related, but not equivalent and provide useful different information.
After a straightforward computation the result is
\small 
\ba
\label{eq:GV}
&&E_{1} = 
\bigg(
-\frac{1}{2 \mathcal{J}^2}+\frac{\log (2)-\frac{1}{2}}{\mathcal{J}}+\frac{1}{4}+\mathcal{J} \left(-\frac{3 \,\zeta_3}{8}+\frac{1}{2}-\frac{\log (2)}{2}\right)-\frac{3
   \mathcal{J}^2}{16}+\\
   &&+\mathcal{J}^3 \left(\frac{3 \,\zeta_3}{16}+\frac{45 \,\zeta_5}{128}-\frac{1}{2}+\frac{3 \log (2)}{8}\right)+
   \cdots
\bigg)\,\mc S+ \nonumber \\
&& +\bigg(
\frac{3}{4 \mathcal{J}^4}+\frac{\frac{1}{2}-\log (2)}{\mathcal{J}^3}-\frac{1}{8 \mathcal{J}^2}+\frac{\frac{1}{16}-\frac{3 \,\zeta_3}{4}}{\mathcal{J}}-\frac{1}{8}+\mathcal{J} \left(\frac{69 \,\zeta_3}{64}+\frac{165 \,\zeta_5}{128}-\frac{27}{32}+\frac{\log (2)}{2}\right)+\nonumber \\
   && +\frac{3 \mathcal{J}^2}{8}+\mathcal{J}^3 \left(-\frac{163 \,\zeta_3}{128}-\frac{405 \,\zeta_5}{256}-\frac{875 \,\zeta_7}{512}+\frac{235}{128}-\log (2)\right)+\cdots
\bigg)\,\mc S^{2}+ \nonumber \\
&& + 
\bigg(
-\frac{5}{4 \mathcal{J}^6}+\frac{\frac{3 \log (2)}{2}-\frac{3}{4}}{\mathcal{J}^5}+\frac{\frac{9 \,\zeta_3}{16}+\frac{1}{16}}{\mathcal{J}^3}+\frac{1}{16 \mathcal{J}^2}+\frac{\frac{45 \,\zeta_3}{64}+\frac{75 \,\zeta_5}{256}-\frac{7}{32}+\frac{\log (2)}{8}}{\mathcal{J}}+\frac{11}{64}+\nonumber \\
   &&  +\mathcal{J} \left(-\frac{89 \,\zeta_3}{32}-\frac{745 \,\zeta_5}{256}-\frac{3815 \,\zeta_7}{2048}+2-\frac{33 \log (2)}{32}\right)-\frac{465 \mathcal{J}^2}{512}+\nonumber \\
   && + \mathcal{J}^3 \left(\frac{5833 \,\zeta_3}{1024}+\frac{1585 \,\zeta_5}{256}+\frac{98035 \,\zeta_7}{16384}+\frac{259455
   \,\zeta_9}{65536}-\frac{405}{64}+\frac{775 \log (2)}{256}\right)+\cdots
\bigg)\,\mc S^{3}.\nonumber \qquad
\ea
\normalsize
This expansion is rather similar to the one derived in \cite{Gromov:2011bz} for \ads, but there are two remarkable differences.
First, the leading terms at small $\mc J$ are $\mc O(\mc S^{n}/\mc J^{2n})$. 
Instead, they were $\mc O(\mc S^{n}/\mc J^{2n-1})$ in \ads. 
Also, there are terms with all parities in $\mc J$ while in \ads, there appear only terms odd under $\mc J\to -\mc J$. 
The additional terms are important and we shall discuss them in more detail later. Remarkably, they  imply that if one scales $\mc J\sim \sqrt\mc S$ they give a constant contribution in the short string limit. 
This is different from $\adsfive$ where the energy correction vanishes like $\sqrt\mc S$ in this regime.
Second, there are terms proportional to $\log(2)$.
These terms can be removed by expressing the energy correction in terms of the coupling in the 
world-sheet scheme. The scheme dependence is universal and agrees with that found in 
\cite{McLoughlin:2008he} for a circular string solution and in \cite{Abbott:2010yb}
for the giant magnon.

The result in the fixed $\rho = \mc J / \sqrt\mc S$ limit is 
\ba
\label{eq:our-expansion}
&& E_{1} = -\frac{1}{2}\,\mc C(\rho, \mc S)+a_{1,0}(\rho)\,\sqrt\mathcal S+ a_{1,1}(\rho)\,\mathcal S^{3/2}+\mc O(\mc S^{5/2}),
\ea
where
\ba
a_{1,0}(\rho) &=& \frac{2\,\log (2)-1}{2\,\sqrt{\rho^{2}+2}}, \\
a_{1,1}(\rho) &=& -\frac{\log (2)\left(2 \rho ^4+6 \rho ^2+3\right)}{4 \left(\rho ^2+2\right)^{3/2}}+\frac{8 \rho ^4+25 \rho ^2+16}{16 \left(\rho ^2+2\right)^{3/2}}-\frac{3 \left(\rho
   ^2+3\right) \zeta_3}{8 \sqrt{\rho ^2+2}},
\ea
and $\mc C$ is related to the branch cut endpoints by the formula
\beq
\label{eq:theC}
\mc C = \frac{\sqrt{(a^{2}-1)\,(b^{2}-1)}}{1-a\,b}+1.
\eeq
Its expansion at small $\mc S$ with fixed $\rho=\mc J/\sqrt\mc S$ is 
\beq
\mc C = 1-\frac{\rho}{\sqrt{\rho^{2}+2}}-\frac{2\,\rho^{3}+5\,\rho}{4\,(\rho^{2}+2)^{3/2}}\,\mc S + 
\frac{\rho\,(12\,\rho^{6}+68\,\rho^{4}+126\,\rho^{2}+73)}{32\,(\rho^{2}+2)^{5/2}}\,\mc S^{2}+\cdots
\eeq
Expanding $E_{1}$ at large $\rho$ we  partially resum the calculation at fixed $\mathcal J$. Just to give an 
example, from the expansion 
\beq
-\frac{1}{2}\left(1-\frac{\rho}{\sqrt{\rho^{2}+2}}\right) = -\frac{1}{2 \rho ^2}+\frac{3}{4 \rho ^4}-\frac{5}{4 \rho ^6}+\frac{35}{16 \rho
   ^8}-\frac{63}{16 \rho ^{10}}+\cdots,
\eeq
we read the coefficients of all terms $\sim \mc S^{n}/\mc J^{2n}$. The first ones are of course in agreement with \eq{eq:GV}. 
As another non trivial example, the large $\rho$ expansion of $a_{11}(\rho)$ is 
\ba
a_{11}(\rho) &=& \rho  \left(-\frac{3 \zeta_3}{8}+\frac{1}{2}-\frac{\log
   (2)}{2}\right)+\frac{\frac{1}{16}-\frac{3 \zeta_3}{4}}{\rho }+\frac{\frac{9 \zeta_3}{16}+\frac{1}{16}}{\rho ^3}+\\
   &&+\frac{-\frac{3
   \zeta_3}{4}-\frac{1}{32}-\frac{\log
   (2)}{4}}{\rho ^5}+\frac{\frac{75 \zeta_3}{64}-\frac{5}{32}+\frac{15 \log (2)}{16}}{\rho
   ^7}+\cdots, \nonumber
\ea
and allows to read the coefficients of all terms $\sim \mc S^{n} / \mc J^{2n-3}$.

\subsubsection{Summation issues}

The explicit sum over the infinite number of on-shell frequencies requires some care and a definite prescription 
since the sums are not absolutely convergent due to physically sensible cancellations between bosonic and fermionic contributions.
As discussed in \cite{Gromov:2008fy}, the following summation prescription is natural from the point of view
of the algebraic curve (see \cite{Bandres:2009kw} for a different prescription)
\beq
E_{1} = \sum_{n=1}^{\infty} K_{n}, 
\eeq
where $K_{n}$ is a particular grouping of heavy and light modes
\beq
\label{eq:Kdef}
K_{n} = \left\{\begin{array}{cc}
\omega^{\rm heavy}_{n}+\omega^{\rm light}_{n/2} & \quad n\in 2\,\mathbb Z\\ 
\omega^{\rm heavy}_{n} & \quad n\not\in 2\,\mathbb Z,
\end{array}\right.
\eeq
with
\ba
\omega_{n}^{\rm heavy} &=& \omega^{(AdS, 1)}_{n}+\omega^{(AdS, 2)}_{n}+\omega^{(AdS, 3)}_{n}+
\omega^{(\mathbb{CP}, 1)}_{n}-2\,\omega^{(F, 1)}_{n}-2\,\omega^{(F, 2)}_{n}, \\
\omega_{n}^{\rm light} &=& 4\,\omega^{(\mathbb{CP}, 2)}_{n}-2\,\omega^{(F, 3)}_{n}-2\,\omega^{(F, 4)}_{n}.
\ea
The short string expansion of $K_{n}$ takes the form
\ba
K_{p} &=& (-1)^{p}\,\mc C+\widehat K_{p}
\ea
where $\mc C$, given in (\ref{eq:theC}),  is independent of $p$ and the sum of $\widehat K_{p}$ (which start 
at $\mc O(\mc S)$) is convergent.
The alternating constant $\mc C$ poses some problems because we have to give a meaning to 
\beq
-\mc C+\mc C-\mc C+\mc C+\cdots.
\eeq
An analysis of the integral representation shows that it automatically selects the choice
\beq
\label{eq:alternating}
-\mc C+\mc C-\mc C+\mc C+\cdots \equiv -\frac{1}{2}\,\mc C
\eeq
Later, we shall provide various consistency checks of this prescription. In particular, we shall see that  
it is necessary in order to match the
asymptotic Bethe Ansatz equations when wrapping effects are subtracted.
Notice also that the expansion of $\mc C$ at fixed $\mc J$ is 
\beq
\mc C \simeq \frac{\mathcal{S}}{\mathcal{J}^2 \sqrt{\mathcal{J}^2+1}}-\frac{\left(3 \mathcal{J}^4+11 \mathcal{J}^2+6\right) \mathcal{S}^2}{4 \mathcal{J}^4
   \left(\mathcal{J}^2+1\right)^2}+\frac{12\,\mc J^{8}+75\,\mc J^{6}+173\,\mc J^{4}+140\,\mc J^{2}+40}{16\,\mc J^{6}\,(\mc J^{2}+1)^{7/2}}\,\mc S^{3},
 \eeq
 so, upon expanding at small $\mc J$, it provides precisely 
 the terms with even/odd $\mc J$ exponents in the coefficients of the odd/even powers of $\mc S$ in (\ref{eq:GV}).

Apart from the $\mc C$ term, the integral representation implements the Gromov-Mikhaylov (GM) prescription. The reason is that the singularities at $|x|=1$ are avoided by implicitly encircling them by a small circumference. This cut-off on $|x-1|$ translates in a bound on the highest mode $n$ that correlates heavy/light polarizations according to GM. In other words the highest mode for light polarizations is asymptotically half the highest mode for heavy polarizations. 

As a numerical check of the agreement between the integral representation and the series expansion, 
we fix $\rho=1$ in table \ref{tab:check1} and show
the value of $E_{1}$ from our analytical resummation and result from the integral. The agreement is very good already 
at moderately small $\mc S$.

\begin{table}[htb]
\begin{center}
\begin{tabular}{c|ll}
$\mc S$ & $E_{1}$ from (\ref{eq:our-expansion}) & $E_{1}$ \\
\hline
1/10 & -0.18790 & -0.17987  \\
1/50 & -0.19461 & -0.19443 \\
1/100 & -0.19934 & -0.19930 \\
1/300 & -0.20449 & -0.20448 \\
1/500 & -0.206075 & -0.206075
\end{tabular}
\caption[Numerical comparison of integral representation and resummation]{Comparison between resummation at fixed ratio
$\rho=1$ and integral representation. The asymptotic value for $\mc S\to 0$ is $(\sqrt 3-3)/6\simeq -0.211$, but already at $\mc S = 1/500$ we have 6 digit agreement.
}
\label{tab:check1}
\end{center}
\end{table}



\subsubsection{The slope function}
\label{sec:slope}

The one-loop correction $E_{1}$ tends to zero linearly with $\mc S$ when $\mc S\to 0$ at fixed $\mc J$.
The slope ratio
\beq
\sigma(\mc J) = \lim_{\mc S\to 0}\frac{E_{1}(\mc S, \mc J)}{\mc S},
\eeq
is the analogue of the slope function we encountered in $\adsfive$. 
It is known that it does not receive dressing corrections both in $\adsfive$ and in $\adscp$ since such contributions start at 
order $\mc S^{2}$ \cite{Basso:2011rs}. It also 
does not receive wrapping corrections in $AdS_{5}$.  Instead, in the case of $AdS_{4}$ the slope has a non vanishing wrapping contribution. For instance, a rough evaluation at $\mc J=1$ gives  a definitely non zero value of approximately $-0.042$.
Indeed, an analytical calculation shows that  the wrapping contribution to the slope in $\adscp$ is exactly
\beq
\sigma^{\rm wrap}(\mc J) = \sum_{n=-\infty}^{\infty}\sigma_{n} = -\frac{1}{2\,\mc J}\,\sum_{n=-\infty}^{\infty}\frac{(-1)^{n}}{
\sqrt{\mc J^{4}+(n^{2}+1)\,\mc J^{2}+n^{2}}}.
\eeq
This formula is in perfect agreement with numerics since for instance
\beq
\sigma^{\rm wrap}(\mc J=1) = -0.041777654879558824814\dots.
\eeq
The large $\mc J$ limit of this expression is exponentially suppressed as it should
\beq
\sigma^{\rm wrap}(\mc J)  = -\frac{\sqrt 2}{\mc J^{5/2}}\,e^{-\pi\,\mc J}+\cdots
\eeq
To analyze the small $\mc J$ limit it is convenient to split this contribution into the $n=0$ term plus the rest. The result is very intriguing. For the $n=0$ term, 
we find 
\ba
\sigma^{\rm wrap}_{n=0} = -\frac{1}{2\,\mc J^{2}\,\sqrt{\mc J^{2}+1}} = 
-\frac{1}{2 \mathcal{J}^2}+\frac{1}{4}-\frac{3 \mathcal{J}^2}{16}+\frac{5 \mathcal{J}^4}{32}-\frac{35 \mathcal{J}^6}{256}+\frac{63 \mathcal{J}^8}{512}+\cdots\,.
\ea
This is precisely the set of terms even under $\mc J\to -\mc J$ in the full slope which is the first term of  (\ref{eq:GV}).
Similarly, we can consider the rest of $\sigma^{\rm wrap}$ and expand at small $\mc J$. We find 
\ba
\sum_{n\neq 0}\sigma^{\rm wrap}_n &=& \frac{\log (2)}{\mathcal{J}}+\mathcal{J} \left(-\frac{3 \zeta_3}{8}-\frac{\log (2)}{2}\right)+\mathcal{J}^3 \left(\frac{3 \zeta_3}{16}+\frac{45 \zeta_5}{128}+\frac{3 \log (2)}{8}\right)+\nonumber \\
&&+\mathcal{J}^5
   \left(-\frac{9 \zeta_3}{64}-\frac{45 \zeta_5}{256}-\frac{315 \zeta_7}{1024}-\frac{5 \log (2)}{16}\right)+O\left(\mathcal{J}^6\right).
\ea
Comparing again with (\ref{eq:GV}), we see that we are reproducing all the irrational terms of the slope, involving zeta functions or $\log(2)$. The remaining terms 
are the same as in \ads
\beq
\sigma(\mc J) -\sigma^{\rm wrap}(\mc J) = -\frac{1}{2\,\mc J}+\frac{\mc J}{2}-\frac{\mc J^{3}}{2}+\cdots.
\eeq
This is due to the fact that the BAE are essentially the same as for $\mathfrak{sl}(2)$ sector in \ads. 
This is however a nontrivial test that all is done correctly.
Thus, we are led to the following expression for the one-loop full slope 
\beq
\sigma(\mc J) = -\frac{1}{2\,\mc J}\left[\frac{1}{\mc J^{2}+1}+\sum_{n=-\infty}^{\infty}\frac{(-1)^{n}}{
\sqrt{\mc J^{4}+(n^{2}+1)\,\mc J^{2}+n^{2}}}
\right].
\eeq
The above analysis of the slope is a confirmation  that the various terms in (\ref{eq:GV}) are organized in the expected way. The asymptotic contribution is precisely the same as in \ads, while wrapping is different and is exponentially
suppressed for large operators. This is a property of the integral representation and a confirm that the 
prescription (\ref{eq:alternating}) is correct.

\subsubsection{Slope at weak coupling}

It is interesting to evaluate the slope at weak coupling.
 In principle, this  requires the knowledge of the anomalous dimensions of 
short $\mathfrak{sl}(2)$ operators in closed form as a function of the spin at a certain length (i.e. twist, in the gauge theory
dictionary). This information is 
available for the asymptotic contribution, but not for the wrapping, which is only known as a series expansion 
at large spin and low twist \cite{Beccaria:2009ny,Beccaria:2010kd}.
 Nevertheless, if we are interested in the correction to the slope only (so, just the first term at small spin), then the L\"uscher form of the wrapping correction
presented in \cite{Beccaria:2009ny} is enough. 
At twist-1, and following the notation of \cite{Beccaria:2009ny}, the wrapping correction enters at four loops
and is expressed by the following function of the integer spin $N$ of the gauge theory operator
\beq
\gamma_{4}^{\rm wrapping}(N) = \gamma_{2}(N)\,\mc W(N),
\qquad \gamma_{2}(N) = 4\,[S_{1}(N)-S_{-1}(N)].
\eeq
Here, $S_{a}(N)$ are generalized harmonic sums while $\mc W(N)$ is a complicated expression depending on the 
Baxter polynomial $Q_{N}(u)$ associated with the Bethe roots. The first factor $\gamma_{2}(N)$ 
is nothing but the two-loop
anomalous dimension of the twist-1 operators. In the small $N$ limit, it starts at $\mc O(N)$. Thus, the 
factor $\mc W(N)$ can be evaluated at $N=0$ where the Baxter polynomial trivializes $Q_{0}(u)=1$.
After a straightforward calculation, one finds that (on the even $N$ branch),
\beq
\gamma_{4}^{\rm wrapping}(N) = -\frac{\pi^{4}}{3}\,N+\mc O(N^{2}).
\eeq
So, even at weak coupling, we find a correction to the slope coming from the wrapping 
terms.
Notice that  the reason why such a contribution is absent in $\adsfive$ is simply that the factor analogous to $\gamma_{2}(N)$ is squared in the wrapping contribution. 
This leads immediately to a contribution to the slope of order $\mc O(N^{2})$.

\label{eq:short}

\newpage


\section{Conclusions}

In this thesis we addressed exact results in supersymmetric gauge theories, more explicitly results valid at arbitrary values of the coupling constant in the planar limit.
The magic ingredients that helped us do all of this were the AdS/CFT dualities and their integrability.
Let us now summarize the structures we uncovered while exploring these theories and the results they helped us find.

For the major part of the thesis we considered $\N=4$ super Yang-Mills, which is the simplest example of a supersymmetric gauge theory with the ingredients listed above.
First of all it is dual to type IIB strings on $\adsfive$, which is an immensely powerful statement in its own right.
Furthermore it is integrable, which technically means that one can find as much conserved charges of motion as there are degrees of freedom.
We reviewed integrability on the gauge theory side at weak coupling first when in section \ref{sec:integrability_weak} we mapped local gauge invariant operators to long ranged spin chains and the anomalous dimensions of the operators to the energies of the spin chain states.
Remarkably the abstract problem of finding the spectrum of local operators in a conformal field theory is equivalent to finding the spectrum of a spin chain system.
It was at weak coupling where we also found the first exact result, the leading order small spin expansion coefficient of the generalized Konishi anomalous dimension, also known as the slope function. 
Of course it is a lucky discovery because the slope function is not sensitive to finite size effects, which are the major problem plaguing the spin chain picture.
They manifest as spin chain interactions that wrap around the chain.

At strong coupling $\N=4$ super Yang-Mills is more naturally described by type IIB strings on an $\adsfive$ background via the AdS/CFT correspondence. 
We reviewed integrability in this regime as well in section \ref{sec:integrability_strong}.
Here we found a picture of classical spectral curves, which are Riemann surfaces with sheets connected by branch cuts. 
String solutions were then described by quasi-momenta defined on these surfaces and the physical intuition was that each cut corresponded to an excitation of the string, where the sheets connected denoted the polarization, the size of the cut -- the amplitude and the discontinuity going through the branch cut -- the mode number of the excitation.
Amazingly the highly non-trivial motion of strings on a curved background somehow reduced to a description similar to that of a collection of harmonic oscillators.

The highlight of the whole integrability programme is of course the finite coupling regime where the two seemingly different descriptions meet, obviously they have to somehow emerge as limiting cases of some ultimate underlying structure.
In section \ref{sec:pmu_system} we argued that this hidden construction is the quantum algebraic curve. 
By now there are many ways of seeing its emergence from various limits, yet probably the most intuitive explanation is that the classical spectral curve is a type of WKB approximation of the quantum spectral curve, namely the classical quasi-momentum has to be replaced by its quantum version and the analyticity structure of the construction modified.
In a very rough sense the collection of classical harmonic oscillators gets quantized and the quantum quasi-momenta $\bP_a$ can be though of as the wave functions. 

The quantum spectral curve allowed us to rather easily derive the slope function and extend the calculation by finding the next coefficient in the small spin expansion, which we called the curvature function.
Contrary to the slope, the curvature function is sensitive to finite size effects and we indeed found evidence that our result successfully incorporates all of them.
We then used these exact results to find the first three coefficients in the strong coupling expansion of the Konishi anomalous dimension.
We matched them reasonably well with available exact numeric data.
Another observable we discussed was the cusped Wilson line, whose expectation value contains the ubiquitous quantity often called the cusp anomalous dimension.
We showed how easily it can be found in the near BPS limit using the quantum spectral curve, even though technically the observable is non-local and the applicability of the construction is naively questionable.
Our exploration of the classical limit at strong coupling and the corresponding dual open string solution provided further evidence that this class of operators is not that different from local operators, as the result was basically a classical spectral curve.

Finally we touched upon the ABJM theory which is another famous integrable supersymmetric gauge theory with a string dual.
We outlined some parallel developments such as the semiclassical quantization of the classical string and the slope function. 


In conclusion, AdS/CFT dualities and integrability are immensely powerful tools when dealing with certain classes of gauge theories, so powerful that they enable one to find results exact in the coupling constant.
Obviously the ultimate goal of this research programme is to learn something that could be applied to real world theories such as QCD.
If we knew as much about QCD as we do about $\N=4$ super Yang-Mills by now we could analytically find the mass of the proton.
Unfortunately we are not there yet, but one can dream.

\newpage

\appendix


\section{Summary of notation and definitions}
\label{sec:notations}

In this appendix we summarize some notation used ubiquitously throughout the thesis.

\subsection{Laurent expansions in $x$}

We often represent functions of the spectral parameter $u$ as a series in $x$
\beq
f(u)=\sum\limits_{n=-\infty}^{\infty}f_n x^n
\eeq
with
\beq
	u=g(x+1/x).
\eeq
We denote by $[f]_+$ and $[f]_-$ part of the series with positive and negative powers of $x$:
\beqa
&&[f]_+=\sum\limits_{n=1}^{\infty}f_n x^n, \\
&&[f]_-=\sum\limits_{n=1}^{\infty}f_{-n} x^{-n}.
\eeqa
As a function of $u$, $x(u)$ has a cut from $-2g$ to $2g$. For any function $f(u)$ with such a cut we denote another branch of $f(u)$ obtained by analytic continuation (from $\Im \;u>0$) around the branch point $u=2g$ by $\tilde f(u)$. In particular, $\tilde x=1/x$.

\subsection{Functions $\sinh_\pm$ and $\cosh_\pm$}

We define $I_k=I_k(4 \pi g)$, where $I_k(u)$ is the modified Bessel function of the first kind.
Then
\beqa
&& \sinh_+=[\sinh(2\pi u)]_+=\sum\limits_{k=1}^\infty I_{2k-1}x^{2k-1}, \\
&& \sinh_-=[\sinh(2\pi  u)]_-=\sum\limits_{k=1}^\infty I_{2k-1}x^{-2k+1},\\
&& \cosh_+=[\cosh(2\pi u)]_+=\sum\limits_{k=1}^\infty I_{2k}x^{2k}, \\
&& \cosh_-=[\cosh(2\pi u)]_-=\sum\limits_{k=1}^\infty I_{2k}x^{-2k}.
\eeqa
 In some cases we denote for brevity
\beq
	\sh_-^x=\sinh_-(x),\ \ \ \ch_-^x=\cosh_-(x).
\eeq

\subsection{Integral kernels}

In order to solve for $\bP_a^{(1)}$ in section \ref{sec:CalculationofPa} we introduce integral operators $H$ and $K$ with kernels
\beqa
H(u,v)&=&-\frac{1}{4\pi i}\frac{\sqrt{u-2g}\sqrt{u+2g}}{\sqrt{v-2g}\sqrt{v+2g}}\frac{1}{u-v}dv, \\
K(u,v)&=&+\frac{1}{4\pi i}\frac{1}{u-v}dv,
\eeqa
which satisfy
\beq
\tilde f+f=h\;\;,\;\;f=H\cdot h\;\;\;\;{\rm and}\;\;\;\;
\tilde f-f=h\;\;,\;\;f=K\cdot h.
\label{lab123}
\eeq
Since the purpose of $H$ and $K$ is to solve equations of the type \eq{lab123}, $H$ usually acts on functions $h$ such that $\tilde h=h$, whereas $K$ acts on $h$ such that $\tilde h=-h$. 
On the corresponding classes of functions $H$ and $K$ can be represented by kernels which are equal up to a sign
\beqa
H(u,v)&=&-\left.\frac{1}{2\pi i}\frac{1}{x_u-x_v}dx_v\right|_{\tilde h=h},\;\;\\
K(u,v)&=&\left.\frac{1}{2\pi i}\frac{1}{x_u-x_v}dx_v\right|_{\tilde h=-h}.
\eeqa
In order to be able to deal with series in half-integer powers of $x$ in section \ref{sec:SolvingPmuL3} we introduce modified kernels:
\beqa
&&H^*\cdot f\equiv\frac{x+1}{\sqrt{x}}H\cdot\frac{\sqrt{x}}{x+1} f, \\
&&K^*\cdot f\equiv\frac{x+1}{\sqrt{x}}K\cdot\frac{\sqrt{x}}{x+1} f.
\eeqa
Finally, to write the solution to equations of the type \eqref{eq:mudiscNLO}, we introduce the operator $\Gamma'$ and its more symmetric version $\Gamma$
\beq
\(\Gamma'\cdot h\)(u)\equiv \oint_{-2g}^{2g}\frac{dv}{{4\pi i}}\d_u \log \frac{\Gamma[i (u-v)+1]}{\Gamma[-i (u-v)]}h(v),
\eeq
\beq
\(\Gamma\cdot h\)(u)\equiv \oint_{-2g}^{2g}\frac{dv}{{4\pi i}}\d_u \log \frac{\Gamma[i (u-v)+1]}{\Gamma[-i (u-v)+1]}h(v).
\eeq

\subsection{Periodized Chebyshev polynomials}
\label{sec:appPeriodized}

Periodized Chebyshev polynomials appearing in $\mu_{ab}^{(1)}$ are defined as
\beqa
&&p_a'(u)=\Sigma\cdot\left[x^a+1/x^a\right]=2 \Sigma\cdot\left[T_a\(\frac{u}{2g}\)\right],\\
&&p_a(u)=p_a'(u)+\frac{1}{2}\(x^a(u)+x^{-a}(u)\),
\eeqa
where $T_a(u)$ are Chebyshev polynomials of the first kind. Here is the explicit form for the first five of them:	
\beqa
&&p_0'=-i(u-i/2),\\
&&p_1'=-i\frac{u(u-i)}{4g},\\
&&p_2'=-i\frac{(u-i/2)(-6g^2+u^2-iu)}{6g^2},\\
&&p_3'=-i\frac{ u (u-i) \left(-6 g^2+u (u-i)\right)}{8 g^3},\\
&&p_4'=-i\frac{ \left(u-\frac{i}{2}\right) \left(30 g^4-20 g^2 u^2+20 i g^2 u+3 u^4-6 i u^3-2 u^2-i u\right)}{30 g^4}.
\eeqa

\section{Slope function: details}

Here we fill in some of the details and provide generalizations for the $\pmu$-system solution of the slope function discussed in section \ref{sec:slope_pmu}.

\subsection{Solution for odd $J$}
\label{sec:oddL}

Here we give details on solving the $\bP\mu$-system for odd $J$ at leading order in the spin. 
First, the parity of the $\mu_{ab}$ functions is different from the even $J$ case, which can be seen from the asymptotics \eq{eq:muasymptotics}. 
Following arguments similar to the discussion for even $J$ in section \ref{sec:slope_pmu_solve}, we obtain
\beq
	\mu_{12}=1,\ \mu_{13}=0,\ \mu_{14}=0,\  \mu_{24}=\cosh(2\pi u),\ \mu_{34}=1.
\eeq
Plugging these $\mu_{ab}$ into \eqref{eq:pmuexpanded} we get a system of equations for $\bP_a$
\beqa
&&\tilde \bP_1= -\bP_3,  \\
&&\tilde \bP_2= -\bP_4 -\bP_1 \cosh(2\pi u), \\
&&\tilde \bP_3= -\bP_1,\\
&&\tilde \bP_4= -\bP_2+\bP_3 \cosh(2\pi u).
\eeqa
This system can be solved in a similar way to the even $J$ case. 
The only important difference is that due to asymptotics \eq{eq:asymptotics} the $\bP_a$ acquire an extra branch point at $u=\infty$.
Let us first rewrite the equations for $\bP_1,\bP_3$ as
\beqa
\tilde\bP_1+\tilde\bP_3&=&-\(\bP_1+\bP_3\)\\
\tilde\bP_1-\tilde\bP_3&=&\bP_1-\bP_3.
\eeqa
This, together with the asymptotics \eqref{eq:asymptotics} implies $\bP_1=\epsilon  x^{-J/2},\;\bP_3=-\epsilon  x^{J/2}$ where $\epsilon$ is a constant. 
Let us note that these $\bP_1, \bP_3$ contain half-integer powers of $x$, and the analytic continuation around the branch points at $\pm 2g$ replaces $\sqrt{x}\to1/\sqrt{x}$. 
Now, taking the sum and difference of the equations for $\bP_2,\;\bP_4$ we get
\beqa
&&\tilde\bP_2+\tilde\bP_4+\bP_2+\bP_4=-a_1\(x^{J/2}+x^{-J/2}\)\cosh{2\pi u},\\
&&\tilde\bP_2-\tilde\bP_4-\(\bP_2-\bP_4\)=a_1\(x^{J/2}-x^{-J/2}\)\cosh{2\pi u}.
\eeqa
We can split the expansion
\beq
	\cosh{2\pi u}=\sum\limits_{k=-\infty}^{\infty}I_{2k}x^{2k}
\eeq	
into the positive and negative parts according to
\beq
\cosh{2\pi u}=\cosh_-+\cosh_++I_0,
\eeq
where
\beq
\cosh_+=\sum\limits_{k=1}^{\infty}I_{2k}x^{2k},\;\ \ \ \ \cosh_-=\sum\limits_{k=1}^{\infty}I_{2k}x^{-2k}.
\eeq
Then we can write
\beqa
&&\bP_2+\bP_4=-a_1(x^{J/2}+x^{-J/2})\cosh_--a_1 I_0 x^{-J/2}+Q, \\
&&\bP_2-\bP_4=-a_1(x^{J/2}-x^{-J/2})\cosh_-+a_1 I_0 x^{-J/2}+P,
\eeqa
where $Q$ and $P$ are some polynomials in $\sqrt{x},1/\sqrt{x}$ satisfying
\beq\label{QP}
	\tilde Q=-Q,\; \tilde P=P.
\eeq
We get
\beqa
\label{eq:P2tmp}
&&\bP_2=-a_1 x^{J/2}\cosh_- +\frac{Q+P}{2},\\
\label{eq:P4tmp}
&&\bP_4=a_1 x^{-J/2}\cosh_- -a_1 I_0 x^{-J/2}+\frac{Q-P}{2}.
\eeqa
Now imposing the correct asymptotics of $\bP_2$ we find
\beq
\frac{P+Q}{2}=a_1 x^{J/2}\sum\limits_{k=1}^{\frac{J-1}{2}}I_{2k}x^{-2k}.
\eeq
Due to \eq{QP} this relation fixes $Q$ and $P$ completely, and we obtain the solution given in section \ref{sec:slope_pmu_solve},
\beqa
\label{eq:musolLOoddL}
&&	\mu_{12}=1,\ \mu_{13}=0,\ \mu_{14}=0, \ \mu_{24}=\cosh(2\pi u),\ \mu_{34}=1, \\
&&   \bP_1=a_1 x^{-J/2}, \\
&&   \bP_2=-a_1 x^{J/2}\sum\limits_{k=-\infty}^{-\frac{J+1}{2}}I_{2k}x^{2k},\\
&&   \bP_3=-a_1 x^{J/2}, \\
\label{eq:P4solLOoddL}
&&    \bP_4=a_1 x^{-J/2}\cosh_--a_1 x^{-J/2}\sum\limits_{k=1}^{\frac{J-1}{2}}I_{2k}x^{2k}-a_1 I_0 x^{-J/2}.
\eeqa
Notice that the branch point at infinity is absent from the product of any two $\bP$'s, as it should be \cite{Gromov:2013pga,Gromov:2014caa}. 
One can check that this solution gives again the correct result \eqref{eq:resultLO} for the slope function.

\subsection{Generic filling fractions and mode numbers}
\label{sec:Sanyn}

Let us extend the discussion of section \ref{sec:slope_pmu} by considering the state corresponding to a solution of the asymptotic Bethe equations with arbitrary mode numbers and filling fractions.
For simplicity we consider $J$ to be even.
We expect that in the $\bP\mu$-system this should correspond to
\beq
	\mu_{24} = \sum_{n=-\infty}^\infty C_n e^{2\pi n u}.
\eeq
We no longer expect $\mu_{24}$ to be either even or odd, since in the Bethe ansatz description of the state with generic mode numbers and filling fractions the Bethe roots are not distributed symmetrically.
As an example, for the ground state twist operator we have $\mu_{24}=\sinh(2\pi u)$, which is reproduced by choosing $C_{-1} = -1/2, C_1 = 1/2$ and all other $C$'s set to $0$.

It is straightforward to solve the $\pmu$-system in the same way as in section \ref{sec:slope_pmu_solve}, and we find the energy
\beq
	\gamma = \frac{\sqrt{\lambda}}{J} \frac{\sum_n C_n I_{J+1}(n \sqrt{\lambda}) }{\sum_n C_n I_{J}(n \sqrt{\lambda})/n} \, S,
\eeq
which can also be written in a more familiar form as
\beq
	\gamma = \sum_n \alpha_n \frac{n\sqrt{\lambda}}{J} \frac{I_{J+1}(n \sqrt{\lambda})}{I_{J}(n \sqrt{\lambda})} \, S,
\eeq
where
\beq
	\alpha_n = \frac{C_n I_{J}(n \sqrt{\lambda}) / n}{\sum_m C_m I_{J}(m \sqrt{\lambda}) / m}
\eeq
are the filling fractions.
The coefficients $C_n$ are additionally constrained by
\beq
	\sum_n C_n I_J(n\sqrt{\lambda}) = 0,
\eeq
which ensures that the $\bP_a$ functions have correct asymptotics. This constraint implies a relation between the filling fractions,
\beq
	\sum_n n \, \alpha_n = 0,
\eeq
which is also familiar from the asymptotic Bethe ansatz.

\section{Curvature function: details}
\label{sec:NLOapp}

In this appendix we will provide more details on the solution of the $\bP\mu$-system and calculation of curvature function for $J=2,3,4$ which was presented in the main text in section \ref{sec:curvature}.

\subsection{Corrections to $\mu_{ab}$ for $J=2$}
\label{sec:appmu2}
Here we present some details of the calculation of next to leading order corrections to $\mu_{ab}$ for $J=2$ omitted in the main text. 
As described in section \ref{sec:muNLOL2}, $\mu^{(1)}_{ab}$ are found as solutions of \eqref{eq:mudiscNLO} with appropriate asymptotics. 
The general solution of this equation consists of a general solution of the corresponding homogeneous equation (which can be reduced to one-parametric form \eqref{eq:periodicpart}) and a particular solution of the inhomogeneous one. The latter can be taken to be
\beq
\mu_{ab}^{disc}=\Sigma\cdot\(\bP_a^{(1)} \tilde\bP_b^{(1)}- \bP_b^{(1)} \tilde\bP_a^{(1)}\).
\eeq
One can get rid of the operation $\Sigma$, expressing $\mu_{ab}^{disc}$ in terms of $\Gamma'$ and $p_a'$. 
This procedure is based on two facts: the definition \eqref{paprime} of $p'_a$ and the statement that on functions decaying at infinity $\Sigma$ coincides with $\Gamma'$ defined by \eqref{Gammaprime}. 
After a straightforward but long calculation we find
\beqa
\mu_{31}^{disc} &=& \epsilon^2\Sigma\(\frac{1}{x^2}-x^2\)=
-\epsilon^2\;\;\(\Gamma\cdot x^2+p_2\),\\
\mu_{41}^{disc} &=&
   \epsilon^2\[-2I_1p_1-4I_1\Gamma\cdot x+\sinh(2\pi u)\(\Gamma\cdot x^2+p_0\)+
   \Gamma\cdot\sinh_-\(x-\frac{1}{x}\)^2\],\qquad\;\;\\
\mu_{43}^{disc} &=&
   -2{\epsilon^2}\[-2I_1p_1-4 I_1\Gamma\cdot x+\sinh(2\pi u)(p_2-p_0)+\Gamma\cdot\sinh_-\(x-\frac{1}{x}\)^2
   \],\\
\mu_{21}^{disc} &=&
   \epsilon^2\[2 I_1\Gamma\cdot x-\sinh(2\pi u)\;\Gamma\cdot x^2-\Gamma\cdot\sinh_-\(x^2+\frac{1}{x^2}\)
   \],\\
\mu_{24}^{disc}&=&
   \epsilon^2\[2I_1 \Gamma\cdot \sinh_- \(x+\frac{1}{x}\)+I_1^2p_0+ \right.\\
 && \left. +\sinh(2\pi u)\Gamma\cdot\sinh_-\(x^2-\frac{1}{x^2}\)-\Gamma\cdot \sinh_-^2\(x^2-\frac{1}{x^2}\)
   \] \nn.
\eeqa
Here we write $\Gamma$ and $p_a$ instead of $\Gamma'$ and $p'_a$ taking into account the discussion between equations \eqref{eq:ra} - \eqref{pa}.

\subsection{Solution of the $\bP\mu$-system for $J=3$}
\label{sec:appnlo3}

The particular solution of the inhomogeneous equation \eqref{eq:mudiscNLO} which we construct as $\mu_{31}^{disc}=\Sigma\cdot\(\bP_{a}^{(1)}\tilde\bP_{b}^{(1)}-\bP_{b}^{(1)}\tilde\bP_{a}^{(1)}\)$ can be written using the operation $\Gamma$  and $p_a$ defined by \eqref{pa} and \eqref{Gamma} as
\beqa
\mu_{31}^{disc} &=& \Sigma\cdot({\bf P}_3 \tilde{\bf P}_1-{\bf P}_1 \tilde{\bf P}_3)=-2\epsilon^2\[\Gamma x^3+p_3\],\\
\mu_{41}^{disc} &=&
   -\epsilon^2\[2p_2I_2+2I_2\Gamma x^2+2\Gamma\cdot\cosh_-+(I_0-\cosh(2\pi u))p_0\],\\
\mu_{34}^{disc} &=&
   {\epsilon^2}\[2I_2\Gamma x +I_0\Gamma x^3-\Gamma\cdot(x^3+x^{-3})\cosh_-+\cosh(2\pi u)(2p_3+\Gamma x^3)\], \qquad\\
\mu_{21}^{disc} &=&
   \epsilon^2\[2I_2\Gamma x+(I_0-\cosh(2\pi u))\Gamma x^3-\Gamma\((x^3+x^{-3})\cosh(2\pi u)\)\],\\
\mu_{24}^{disc} &=&
   -2\epsilon^2\[-\frac{1}{2}\Gamma\cdot\cosh_-^2\(x^3-x^{-3}\)+\(\frac{\cosh(2\pi u)}{2}-I_0\)\Gamma\cdot\frac{\cosh_-}{x^3}\right. \\
   &&-I_2\Gamma\cdot\(x+\frac{1}{x}\)\cosh_--\frac{1}{2}\cosh(2\pi u)\Gamma\cdot x^3\cosh_-+ \nn \\
   &&+\left.\frac{I_0}{2}\(I_0-\cosh(2\pi u)\)\Gamma\cdot x^3+\frac{I_1 I_2}{2\pi g}\Gamma x-I_2^2 p_1 \] \nn.
\eeqa
Alternatively one can use $p_a'$ and $\Gamma'$ instead of $p_a$ and $\Gamma$, as explained in the discussion between the equations \eqref{eq:ra} - \eqref{pa}.
The zero mode of the system \eqref{eq:P1L3} - \eqref{eq:P4L3}, which we added to the solution in equations \eq{eq:P1J3} - \eq{eq:P4J3} to ensure correct asymptotics, is given by
\beqa
\label{P1J3zm}
\bP_1^{\text{zm}}&=&L_1 x^{-1/2}+L_3x^{1/2}, \nn \\
\nn \bP_2^{\text{zm}}&=&-L_1 x^{1/2}\ch_-+L_2 x^{-1/2}-L_3x^{-1/2}\(\ch_-+\ofrac{2}I_0\)+L_4\(x^{1/2}-x^{-1/2}\),\\ \nn
\bP_3^{\text{zm}}&=&-L_1 x^{1/2}-L_3x^{1/2},\\ \nn
\bP_4^{\text{zm}}&=&-L_1\(I_0 x^{-1/2}+x^{-1/2}\cosh_-\)-L_2x^{1/2}+L_4(x^{1/2}-x^{-1/2})
\\ 
&-&L_3x^{1/2}\(\ch_-+\ofrac{2}I_0\).
\label{P4J3zm}
\eeqa

\subsection{Solution of the $\bP\mu$-system for $J=4$}
\label{sec:appL4}

Solution of the $\bP\mu$-system at next to leading order for $J=4$ is completely analogous to the case of $J=2$. 
The starting point is the leading order solution \eqref{eq:P1solLOevenL} - \eqref{eq:P4solLOevenL}. 
As described in section \ref{sec:appmu2}, from leading order $\bP_a$ we can find $\mu_{ab}$ at the next order. 
Its discontinuous part is
\label{sec:appnlo4}
\beqa
\mu_{31}^{disc} &=& -\epsilon^2\;\;\(\Gamma\cdot x^4+p_4\),\\
\mu_{41}^{disc} &=& \frac{1}{2} \epsilon ^2 \left(\sinh (2 \pi   u) \left(p_0+\Gamma \cdot x^4\right)+2 \left(I_1 p_1+I_3 p_3\right)+ \right.  \\ 
&& \left.+\Gamma \cdot \sinh_- \left(x^2-\frac{1}{x^2}\right)^2-2 \left(I_1+I_3\right) \left(\Gamma \cdot x^3+\Gamma \cdot x\right)\right), \nn \\
\mu_{43}^{disc} &=& \epsilon ^2 \left(\left(p_4-p_0\right) \sinh (2 \pi  u)+2 \left(I_1 p_1+I_3 p_3\right)- \right. \\ 
&&\left.-\Gamma \cdot\sinh_- \left(x^2-\frac{1}{x^2}\right)^2 +2 \left(I_1+I_3\right) \left(\Gamma \cdot x^3+\Gamma \cdot x\right)\right.\Bigg), \nn \\
\mu_{21}^{disc} &=& \epsilon ^2 \left(-\frac{1}{2} \sinh (2 \pi  u) \Gamma \cdot x^4+I_1 p_3+I_3 p_1-\right.  \\ 
&&\left. -\frac{1}{2} \Gamma \cdot \sinh_- \left(x^4+\frac{1}{x^4}\right)+I_1 \Gamma \cdot x^3+I_3 \Gamma \cdot x\right), \nn \\
\mu_{24}^{disc} &=& \epsilon ^2 \left(\frac{1}{2} \sinh (2 \pi u) \Gamma \cdot\sinh_- \left(x^4-\frac{1}{x^4}\right)+I_3^2 p_2+I_1 I_3 p_0- \right. \\ 
&& \left. -\frac{1}{2} \Gamma \cdot \sinh_-^2 \left(x^4-\frac{1}{x^4}\right) +I_1 \Gamma \cdot \sinh_- \left(x^3+\frac{1}{x^3}\right)+ \right. \nn \\ 
&& \left.+I_3 \Gamma \cdot\sinh_- \left(x+\frac{1}{x}\right)+\left(I_3^2-I_1^2\right)\Gamma \cdot x^2\right), \nn
\eeqa
and as discussed for $J=2$ the zero mode can be brought to the form
\beqa
&& \pi_{12}=0,\ \pi_{13}=0,\ \pi_{14}=0,\\
 && \pi_{24}=c_{1,24}\cosh{2\pi u},\ \pi_{34}=0.
 \label{eq:periodicpartL4}
 \eeqa
After that, we calculate $r_a$ by formula \eqref{eq:ra} and solve the expanded to next to leading order $\bP\mu$-system for $\bP_a^{(1)}$ as
\beqa
&&\bP^{(1)}_3=H \cdot r_3, \\
&&\bP^{(1)}_1=\frac{1}{2}\bP^{(1)}_3+K\cdot \(r_1-\frac{1}{2}r_3\),\\
&&\bP^{(1)}_4=K\cdot\[(H\cdot r_3)\sinh(2\pi u)+r_4-\frac{1}{2}r_3\sinh(2\pi u) \]-C(x+1/x),\\
&&\bP^{(1)}_2=H\cdot\[-\bP^{(1)}_4-\bP^{(1)}_1\sinh(2\pi u)+r_2\]+C/x,
\eeqa
where $C$ is a constant which is fixed by requiring correct asymptotics of $\bP_2$.
Finally we find leading coefficients $A_a$ of $\bP^{(1)}_a$ and use expanded up to ${\cal O}(S^2)$ formulas \eqref{AA1}, \eqref{AA2} in the same way as in section \ref{sec:resultL2} to obtain the result \eqref{gamma2L4}.

 \subsection{Result for $J=4$}
 \label{sec:SolvingPmuL4}
The final result for the curvature function at $J=4$ reads
\beqa
\footnotesize
\label{gamma2L4}
&&\gamma^{(2)}_{J=4}=\oint \frac{du_x}{2\pi i}\oint \frac{du_y}{2\pi i}
\frac{1}{i g^2(I_3-I_5)^3} \left.\Bigg[ \right.\\ && \left.\nn
\frac{2 \(\sh_-^x\)^2 y^4 \left(I_3 \left(x^{10}+1\right)-I_5 x^2 \left(x^6+1\right)\right)}{x^4 \left(x^2-1\right)}-\frac{2 \(\sh_-^y\)^2 x^4 \left(y^8-1\right) \left(I_3 x^2-I_5\right)}{\left(x^2-1\right) y^4}+ \right.\\
&&+\frac{4 \sh_-^x \sh_-^y \left(x^4 y^4-1\right) \left(I_3+I_3 x^6 y^4-I_5 x^2 \left(x^2 y^4+1\right)\right)}{x^4 \left(x^2-1\right) y^4}\nn\\
&&+\left.\sh_-^y\(
\left(y^4+y^{-4}\right) x^{-1}\left(\left(I_1 I_5-I_3^2\right) \left(3 x^4+1\right)-2 I_1 I_3 x^6\right)+
\right.\right.\nn\\&&\left.\left.
+\frac{2 I_3 x^2
   \left(I_5 \left(x^2+1\right) x^2+I_1 \left(1-x^2\right)\right)-I_1 I_5 \left(x^2-1\right)^2+I_3^2 \left(-2 x^6+x^4+1\right)}{x(x^2-1)}+
   \right.\right.\nn\\&&\left.\left.
   +2
   \left(y^3+y^{-3}\right) \frac{I_1 I_3 x^6-I_1 I_5 x^4-I_3^2 \left(x^2-1\right)}{x^2-1}-
   \right.\right.\nn\\&&\left.\left.
   -2 I_3 \left(y+y^{-1}\right)
   \frac{ I_1\left(x^2-1\right)-I_3 \left(x^6-x^2+1\right)+I_5 \left(x^4-x^2+1\right)}{x^2-1}
   \)+
       \right.\nn\\&&\left.
   +\frac{4 x^6 y^2 I_3 \left(I_3^2-I_1^2\right)}{x^2-1}+\frac{4 x y I_1 \left(I_3 y^2+I_1\right) \left(I_3+I_5\right)}{x^2-1}+
   \right.\nn\\&&\left.
  \frac{2 y^4 \left(I_1+I_3\right) \left(I_1 I_5-I_3^2\right)}{x^2-1}
-\frac{2 y \left(y^2+1\right) \left(I_1+I_3\right) \left(I_1 I_5-I_3^2\right)}{x \left(x^2-1\right)}-
    \right.\nn\\&&\left.
-\frac{2 x^3 y \left(I_1+I_3\right) \left(I_1 \left(2 I_3+\left(3 \
y^2+1\right) I_5\right)-I_3 \left(2 I_5 y^2+\left(y^2+3\right) I_3\right)\right)}{x^2-1}
    \right.\nn\\&&\left.+
\frac{2 x^2 y^4 \left(-I_3^3-I_1 \left(3 I_3+I_5\right) I_3+I_1^2 I_5\right)}{x^2-1} +\right. \nn \\
&& \left.+
\frac{2 x^4 y \left(I_1^2 \left(2 y I_5-2 y^3 I_3\right)-2 y \left(y^2+1\right) I_3^2 I_5\right)}{x^2-1}+
    \right.\nn\\&&\left.
 +\frac{4 x^5 y I_3 \left(2 I_1^2 y^2+I_3 \left(I_5-I_3\right) y^2+I_1 \left(I_3+I_5\right)\right)}{x^2-1}
  \right] \frac{1}{4\pi i}\d_u \log\frac{\Gamma (i u_x-i u_y+1)}{\Gamma (1-i u_x+i u_y)}\nn
\eeqa
\normalsize
where, similarly to $J=2,3$, the integrals go around the branch between $-2g$ and $2g$.

\newpage
\subsection{Weak coupling expansion}
\label{sec:weakS3}

First, we give the expansion of our results for the curvature functions $\gamma_J^{(2)}$ to 10 loops. 
We start with $J=2$:
\beqa
\label{weak22long}
	\gamma_{J=2}^{(2)}&=&-8 g^2 \zeta_3+g^4 \left(140 \zeta_5-\frac{32 \pi ^2 \zeta_3}{3}\right)+g^6
   \left(200 \pi ^2 \zeta_5-2016 \zeta_7\right)
	\\ \nn
	&+&g^8 \left(-\frac{16 \pi ^6 \zeta_3}{45}-\frac{88 \pi ^4 \zeta_5}{9}-\frac{9296 \pi ^2 \zeta_7}{3}+27720 \zeta_9\right)
	\\ \nn
	&+&g^{10} \left(\frac{208 \pi ^8 \zeta_3}{405}+\frac{160 \pi ^6 \zeta_5}{27}+144 \pi ^4 \zeta_7+45440 \pi ^2 \zeta_9-377520 \zeta_{11}\right)
	\\ \nn
	&+&g^{12}
   \left(-\frac{7904 \pi ^{10} \zeta_3}{14175}-\frac{17296 \pi ^8 \zeta_5}{4725}-\frac{128 \pi ^6 \zeta_7}{15}-\frac{6312 \pi ^4 \zeta_9}{5}
	\right.
	\\ \nn
	&&\Bigl.\ \ \ \ \ \ \
	-653400 \pi
   ^2 \zeta_{11}+5153148 \zeta_{13}\Bigr)
	\\ \nn
	&+&g^{14} \Bigl(\frac{1504 \pi ^{12} \zeta_3}{2835}+\frac{106576 \pi ^{10} \zeta_5}{42525}-\frac{18992 \pi ^8 \zeta_7}{405}
-\frac{16976 \pi ^6 \zeta_9}{15}
	\Bigr.
	\\ \nn
	&& \Bigl. \ \ \ \ \ \ \
	+\frac{25696 \pi ^4 \zeta_{11}}{9}+\frac{28003976 \pi ^2 \zeta_{13}}{3}-70790720 \zeta_{15}\Bigr)
	\\ \nn
	&+&g^{16}
   \Bigl(-\frac{178112 \pi ^{14} \zeta_3}{382725}-\frac{239488 \pi ^{12} \zeta_5}{127575}+\frac{2604416 \pi ^{10} \zeta_7}{42525}+\frac{8871152 \pi ^8 \zeta_9}{4725}
		\Bigr.
	\\ \nn
	&& \Bigl. \ \ \ \ \ \ \
	+\frac{30157072 \pi ^6 \zeta_{11}}{945}+\frac{8224216 \pi ^4 \zeta_{13}}{45}-133253120 \pi ^2 \zeta_{15}
			\Bigr.
	\\ \nn
	&& \Bigl. \ \ \ \ \ \ \
	+979945824 \zeta_{17}\Bigr)
	\\ \nn
	&+&g^{18}
   \Bigl(\frac{147712 \pi ^{16} \zeta_3}{382725}+\frac{940672 \pi ^{14} \zeta_5}{637875}-\frac{490528 \pi ^{12} \zeta_7}{8505}-\frac{358016 \pi ^{10} \zeta_9}{189}
	\Bigr.
	\\ \nn
	&& \Bigl. \ \ \ \ \ \ \
	-\frac{37441312 \pi ^8 \zeta_{11}}{945}-\frac{9616256 \pi ^6 \zeta_{13}}{15}-\frac{16988608 \pi ^4 \zeta_{15}}{3}
		\Bigr.
	\\ \nn
	&& \Bigl. \ \ \ \ \ \ \
	+1905790848 \pi ^2 \zeta_{17}-13671272160 \zeta_{19}\Bigr)
	\\ \nn
	&+&g^{20} \Bigl(-\frac{135748672 \pi ^{18} \zeta_3}{442047375}-\frac{103683872 \pi ^{16} \zeta_5}{88409475}+\frac{1408423616 \pi
   ^{14} \zeta_7}{29469825}
			\Bigr.
	\\ \nn
	&& \Bigl. \ \ \ \ \ \ \
	+\frac{2288692288 \pi ^{12} \zeta_9}{1403325}+\frac{34713664 \pi ^{10} \zeta_{11}}{945}+\frac{73329568 \pi ^8 \zeta_{13}}{105}
				\Bigr.
	\\ \nn
	&& \Bigl. \ \ \ \ \ \ \
	+\frac{305679296 \pi ^6 \zeta_{15}}{27}+121666688 \pi ^4 \zeta_{17}-27342544320 \pi ^2 \zeta_{19}
					\Bigr.
	\\ \nn
	&& \Bigl. \ \ \ \ \ \ \
	+192157325360 \zeta_{21}\Bigr).
\eeqa
Next for $J=3$ we find,
\beqa
\label{weak23long}
\gamma_{J=3}^{(2)}&=&-2 g^2 \zeta_3+g^4 \left(12 \zeta_5-\frac{4 \pi ^2 \zeta_3}{3}\right)+g^6
   \left(\frac{2 \pi ^4 \zeta_3}{45}+8 \pi ^2 \zeta_5-28 \zeta_7\right)
	\\ \nn
	&+&g^8\;
   \left(-\frac{4 \pi ^6 \zeta_3}{45}-\frac{4 \pi ^4 \zeta_5}{15}-528 \zeta_9\right)
	\\ \nn
	&+&g^{10} \left(\frac{934 \pi ^8 \zeta_3}{14175}+\frac{8 \pi ^6 \zeta_5}{9}-\frac{82 \pi ^4 \zeta_7}{9}-900 \pi ^2 \zeta_9+12870 \zeta_{11}\right)
	\\ \nn
	&+&g^{12} \left(-\frac{572 \pi ^{10} \zeta_3}{14175}-\frac{104 \pi ^8
   \zeta_5}{175}-\frac{256 \pi ^6 \zeta_7}{45}+\frac{2476 \pi ^4 \zeta_9}{9}
	\right. \\ \nn
	&&\ \ \ \ \ \ \ \left.+\frac{57860 \pi ^2 \zeta_{11}}{3}-208208 \zeta_{13}\right)
	\\ \nn
	&+&g^{14}
   \left(\frac{2878 \pi ^{12} \zeta_3}{127575}+\frac{404 \pi ^{10} \zeta_5}{1215}+\frac{326 \pi ^8 \zeta_7}{75}+\frac{3352 \pi ^6 \zeta_9}{135}
	\right. \\ \nn
	&&\ \ \ \ \ \ \ -\left.\frac{80806 \pi ^4 \zeta_{11}}{15}
	-316316 \pi ^2 \zeta_{13}+2994992 \zeta_{15}\right)
	\\ \nn
	&+&g^{16} \left(-\frac{159604 \pi ^{14} \zeta_3}{13395375}-\frac{257204
   \pi ^{12} \zeta_5}{1488375}-\frac{14836 \pi ^{10} \zeta_7}{6075}-\frac{71552 \pi
   ^8 \zeta_9}{2025}
	\right.
	\\ \nn
	&&\ \ \ \ \ \ \  \left.+\frac{4948 \pi ^6 \zeta_{11}}{189}+\frac{4163068 \pi ^4 \zeta_{13}}{45}+\frac{14129024 \pi ^2 \zeta_{15}}{3}-41116608 \zeta_{17}\right)
		\\ \nn
	&+&g^{18}
   \left(\frac{494954 \pi ^{16} \zeta_3}{81860625}+\frac{156368 \pi ^{14} \zeta_5}{1819125}+\frac{6796474 \pi ^{12} \zeta_7}{5457375}+\frac{332 \pi ^{10} \zeta_9}{15}
		\right.
	\\ \nn
	&&\ \ \ \ \ \ \ \left.+\frac{1745318 \pi ^8 \zeta_{11}}{4725} -\frac{868088 \pi ^6 \zeta_{13}}{315}-\frac{22594208 \pi ^4 \zeta_{15}}{15}
	\right. \\ \nn
	&&\ \ \ \ \ \ \ \Bigl.-67084992 \pi ^2 \zeta_{17}+553361016
   \zeta_{19}\Biggr)
	\\ \nn
	&+&g^{20} \left(-\frac{940132 \pi ^{18} \zeta_3}{315748125}-\frac{244456 \pi ^{16} \zeta_5}{5893965}-\frac{29637008 \pi ^{14}
   \zeta_7}{49116375}-\frac{11808196 \pi ^{12} \zeta_9}{1002375}
	\right.
	\\ \nn
		&&\ \ \ \ \ \ \ -\left.\frac{2265364 \pi
   ^{10} \zeta_{11}}{8505}-\frac{68767984 \pi ^8 \zeta_{13}}{14175}+\frac{480208 \pi ^6
   \zeta_{15}}{9}
	\right.
	\\ \nn
		&&\ \ \ \ \ \ \ +\left.
	\frac{71785288 \pi ^4 \zeta_{17}}{3}+934787840 \pi ^2 \zeta_{19}-7390666360 \zeta_{21}\right).
\eeqa
Finally, for $J=4$,
\beqa
\label{weak24long}
	\gamma_{J=4}^{(2)}&=&
	g^2 \left(-\frac{14 \zeta_3}{5}+\frac{48 \zeta_5}{\pi ^2}-\frac{252 \zeta_7}{\pi ^4}\right)
	\\ \nn
	&+&g^4
   \left(-\frac{22 \pi ^2 \zeta_3}{25}+\frac{474 \zeta_5}{5}-\frac{8568 \zeta_7}{5 \pi ^2}+\frac{8316
   \zeta_9}{\pi ^4}\right)
\\ \nn
&+&g^6 \left(\frac{32 \pi ^4 \zeta_3}{875}+\frac{3656 \pi ^2 \zeta_5}{175}-\frac{56568 \zeta_7}{25}+\frac{196128 \zeta_9}{5 \pi ^2}-\frac{185328 \zeta_{11}}{\pi
   ^4}\right)
\\ \nn
&+&g^8 \Bigl(-\frac{4 \pi ^6 \zeta_3}{175}-\frac{68 \pi ^4 \zeta_5}{75}-\frac{55312 \pi ^2
   \zeta_7}{125}+\frac{1113396 \zeta_9}{25}-\frac{3763188 \zeta_{11}}{5 \pi ^2}
	\Bigr.
	\\ \nn
	&& \Bigl. \ \ \ \ \ \ \
+\frac{3513510 \zeta_{13}}{\pi ^4}\Bigr)
\\ \nn
&+&g^{10} \Bigl(\frac{176 \pi ^8 \zeta_3}{16875}+\frac{2488 \pi ^6 \zeta_5}{7875}+\frac{2448 \pi ^4 \zeta_7}{125}+\frac{209532 \pi ^2 \zeta_9}{25}-\frac{3969878 \zeta_{11}}{5}
	\Bigr.
	\\ \nn
	&& \Bigl. \ \ \ \ \ \ \
+\frac{13213200 \zeta_{13}}{\pi ^2}-\frac{61261200 \zeta_{15}}{\pi ^4}\Bigr)
\\ \nn
&+&g^{12}
   \Bigl(-\frac{88072 \pi ^{10} \zeta_3}{20671875}-\frac{449816 \pi ^8 \zeta_5}{4134375}-\frac{327212 \pi
   ^6 \zeta_7}{65625}-\frac{338536 \pi ^4 \zeta_9}{875}
	\Bigr.
	\\ \nn
	&& \Bigl. \ \ \ \ \ \ \
-\frac{129520798 \pi ^2 \zeta_{11}}{875}+\frac{66969474 \zeta_{13}}{5}-\frac{220540320 \zeta_{15}}{\pi ^2}
	\Bigr.
	\\ \nn
	&& \Bigl. \ \ \ \ \ \ \
+\frac{1017636048 \zeta_{17}}{\pi ^4}\Bigr)
\\ \nn
&+&g^{14} \Bigl(\frac{795136 \pi ^{12} \zeta_3}{487265625}+\frac{522784 \pi ^{10} \zeta_5}{13921875}+\frac{4021288 \pi ^8 \zeta_7}{2953125}+\frac{1869152 \pi ^6 \zeta_9}{21875}
	\Bigr.
	\\ \nn
	&& \Bigl. \ \ \ \ \ \ \
+\frac{18573952 \pi ^4 \zeta_{11}}{2625}+\frac{62633272 \pi ^2 \zeta_{13}}{25}-\frac{1092799344
   \zeta_{15}}{5}
	\Bigr.
	\\ \nn
	&& \Bigl. \ \ \ \ \ \ \
+\frac{17844607872 \zeta_{17}}{5 \pi ^2}-\frac{16405526592 \zeta_{19}}{\pi ^4}\Bigr)
\\ \nn
&+&g^{16}
   \Bigl(-\frac{30581888 \pi ^{14} \zeta_3}{51162890625}-\frac{43988768 \pi ^{12} \zeta_5}{3410859375}-\frac{446380184 \pi ^{10} \zeta_7}{1136953125}
	\Bigr.
	\\ \nn
	&& \Bigl. \ \ \ \ \ \ \
-\frac{20108936 \pi ^8 \zeta_9}{984375}
-\frac{31755036 \pi ^6 \zeta_{11}}{21875}-\frac{321449336 \pi ^4 \zeta_{13}}{2625}
	\Bigr.
	\\ \nn
	&& \Bigl. \ \ \ \ \ \ \
-\frac{1031925232 \pi ^2 \zeta_{15}}{25}
	+\frac{87296960712 \zeta_{17}}{25}-\frac{283092985656
   \zeta_{19}}{5 \pi ^2}
	\Bigr.
	\\ \nn
	&& \Bigl. \ \ \ \ \ \ \
+\frac{259412389236 \zeta_{21}}{\pi ^4}\Bigr)
\\ \nn
&+&g^{18} \Bigl(\frac{6706432 \pi ^{16}
   \zeta_3}{31672265625}+\frac{816838192 \pi ^{14} \zeta_5}{186232921875}+\frac{2004636572 \pi ^{12} \zeta_7}{17054296875}
	\Bigr.
	\\ \nn
	&& \Bigl. \ \ \ \ \ \ \
+\frac{1950592976 \pi ^{10} \zeta_9}{378984375}
+\frac{2220222512 \pi ^8 \zeta_{11}}{6890625}+\frac{20963856 \pi ^6 \zeta_{13}}{875}
	\Bigr.
	\\ \nn
	&& \Bigl. \ \ \ \ \ \ \
+\frac{254959316 \pi ^4 \zeta_{15}}{125}
+\frac{584553371616 \pi ^2 \zeta_{17}}{875}
	\Bigr.
	\\ \nn
	&& \Bigl. \ \ \ \ \ \ \
-\frac{1375388084412 \zeta_{19}}{25}+\frac{4432313039616 \zeta_{21}}{5 \pi ^2}
-\frac{4049650420200 \zeta_{23}}{\pi ^4}\Bigr)
\eeqa
\beqa
\nn
&+&g^{20}
   \Bigl(-\frac{15308976272 \pi ^{18} \zeta_3}{209512037109375}-\frac{1764947984 \pi ^{16} \zeta_5}{1197211640625}-\frac{18667123736 \pi ^{14} \zeta_7}{517313671875}
	\Bigr.
	\\ \nn
	&& \Bigl. \ \ \ \ \ \ \
-\frac{538293689008 \pi ^{12} \zeta_9}{399070546875}-\frac{657466372 \pi ^{10} \zeta_{11}}{8859375}-\frac{119709052 \pi ^8 \zeta_{13}}{23625}
	\Bigr.
	\\ \nn
	&& \Bigl. \ \ \ \ \ \ \
-\frac{9095498848 \pi ^6 \zeta_{15}}{23625}-\frac{260407748416 \pi ^4 \zeta_{17}}{7875}-\frac{1869110789976 \pi ^2 \zeta_{19}}{175}
	\Bigr.
	\\ \nn
	&& \Bigl. \ \ \ \ \ \ \
+\frac{4293062840352 \zeta_{21}}{5}-\frac{13755955395600 \zeta_{23}}{\pi ^2}+\frac{62673161265000 \zeta_{25}}{\pi
   ^4}\Bigr).
\eeqa
With the help of several Mathematica packages for dealing with harmonic sums (as noted in the main text), we have also computed the weak coupling expansion of the anomalous dimensions at order $S^3$, using the known predictions from ABA which are available for any spin at $J=2$ and $J=3$. 
For $J=2$ we have computed the expansion to three loops:
\beqa
	\gamma_{J=2}^{(3)}&=&g^2\frac{4}{45}\pi^4+g^4 \left(40 \zeta_3^2-\frac{28 \pi ^6}{405}\right)\\ \nn
	&+&g^6 \left(\frac{192}{5} \zeta_{5,3}-\frac{6992 \zeta_3
   \zeta_5}{5}+\frac{280 \pi ^2 \zeta_3^2}{3}+\frac{6962 \pi ^8}{212625}\right)+\mathcal{O}(g^8).
\eeqa
Compared to the $S^2$ part, a new feature is the appearance of multiple zeta values -- here we have $\zeta_{5,3}$, which is defined by
\begin{equation}
	\zeta_{a_1,a_2,\ldots,a_k}=\sum_{0< n_1<n_2<\ldots<n_k<\infty}
	\frac 1{n_{1}^{a_1}n_{2}^{a_2}\ldots n_k^{a_k}}\,
\end{equation}
and cannot be reduced to simple zeta values $\zeta_n$.
For $J=3$ we have obtained the expansion to four loops:
\beqa
\nn
	\gamma_{J=3}^{(3)}&=&\frac{1}{90}\pi^4g^2+g^4\(4 \zeta_3^2+\frac{\pi ^6}{1890}\)
	+g^6\(4\zeta_{5,3}+4 \pi ^2 \zeta_3^2-72 \zeta_3 \zeta_5-\frac{2
   \pi ^8}{675}\)\\ \nn
   &+&g^8\left(-112 \zeta_{2,8}+\frac{20}{3} \pi ^2 \zeta_{5,3}+728 \zeta_3
   \zeta_7+448 \zeta_5^2-\frac{224}{3} \pi ^2 \zeta_3
   \zeta_5\right.\\
	&& \left.+\frac{4 \pi ^4 \zeta_3^2}{5}-\frac{41 \pi
   ^{10}}{133650}\right)
   +\mathcal{O}(g^{10}).
\eeqa
These results would be useful for verifying the next small spin anomalous dimension expansion coefficient after the curvature.

\newpage
\bibliography{bibliography}

\end{document}